%% file: review_delta.tex
\documentclass[seceqn]{elsart}
\usepackage{epsf,graphicx}
\usepackage{times,latexsym,euscript,amstext,amssymb,amsbsy,amsmath,amsfonts}
\usepackage{epsfig}
\usepackage{slashbox}

\hoffset   =- .5 cm
\def\beq{\begin{equation}}
\def\eeq{\end{equation}}
\def\bea{\begin{eqnarray}}
\def\eea{\end{eqnarray}}
\def\beqa{\begin{equation}\begin{array}{l}}
\def\eeqa{\end{array}\end{equation}}
\def\eqlab#1{\label{eq:#1}}
\def\figlab#1{\label{fig:#1}}
\def\tablab#1{\label{tab:#1}}

\def\barr{\left(\begin{array}{c}}
\def\earr{\end{array}\right)}
\def\bmat{\left(\begin{array}{cc}}
\def\emat{\end{array}\right)}
\def\eref#1{(\ref{eq:#1})}
\def\Eqref#1{Eq.~(\ref{eq:#1})}

\def\Figref#1{Fig.~\ref{fig:#1}}

\def\Tabref#1{Table \ref{tab:#1}}

\def\sla#1{#1  \!\!\!\!\slash}
\def\slaa{a \hspace{-2mm} \slash}
\def\slap{p \hspace{-1.9mm} \slash}
\def\slad{\partial \hspace{-2.2mm} \slash}

\def\half{\mbox{\small{$\frac{1}{2}$}}}

\def\thalf{\mbox{\small{$\frac{3}{2}$}}}
\def\quarter{\mbox{\small{$\frac{1}{4}$}}}
\def\third{\mbox{\small{$\frac{1}{3}$}}}

\def\al{\alpha}
\def\be{\beta}
\def\ga{\gamma} \def\Ga{{\it\Gamma}}
\def\de{\delta} \def\De{\Delta}
\def\vDe{\varDelta}
\def\veps{\varepsilon}  \def\eps{\epsilon}

\def\la{\lambda} \def\La{{\Lambda}}

\def\si{\sigma} \def\Si{{\it\Sigma}}
\def\th{\theta}  \def\Th{\Theta}
\def\w{\omega}

\def\pa{\partial}
\def\vrho{\varrho}

\def\pa{\partial}
\def\ra{\rightarrow}
\def\nn{\nonumber}

\def\Tau{{\mathcal T}}
\def\lag{{\mathcal L}}

\def\mathscr{\mathcal}
\def\N{N}

\def\3d{3-D}

\def\ol#1{\overline{#1}}

\def\ceft{$\chi$EFT}
\def\EM{e.m.}

\def\bq{\mathbf{q}}
\def\bvare{\boldsymbol{\varepsilon}}
\def\bsig{\boldsymbol{\sigma}}
\def\OPE{1$\gamma $}
\def\TPE{2$\gamma $}

\def\bk{\mathbf{k}}
\def\pslash{\hspace{-0.11in}}

\begin{document}
\begin{frontmatter}

\title{Electromagnetic excitation of
the $\vDe(1232)$-resonance}

\author[WM,JLAB,ECT]{Vladimir~Pascalutsa},
\author[WM,JLAB]{Marc~Vanderhaeghen},
\author[NTU]{Shin~Nan~Yang}

\address[WM]{Physics Department, The College of William \& Mary,
Williamsburg, VA 23187, USA}

\address[JLAB]{Theory Center, Jefferson Lab, 12000 Jefferson Ave, Newport News,
VA 23606, USA}

\address[ECT]{ECT*, Villa Tambosi, Str.\ delle Tabarelle 286,
I-38050 Villazzano (Trento), Italy}

\address[NTU]{Department of Physics and Center 
for Theoretical Sciences, 
National Taiwan University, Taipei 10617, Taiwan}

\begin{abstract}
We review
the  description of the lowest-energy 
nucleon excitation --- the $\Delta$(1232)-resonance. 
Much of the recent experimental effort has been focused on the
precision measurements of the nucleon-to-$\Delta$
transition by means of electromagnetic probes.
We confront the results of these measurements
with the state-of-the-art calculations based on
chiral effective-field theories (EFT), lattice QCD, large-$N_c$
relations, perturbative QCD, and QCD-inspired models. 
We also discuss the link of the  nucleon-to-$\Delta$ form
factors to generalized parton distributions (GPDs).
Some of the theoretical approaches are reviewed in detail,
in particular,  recent dynamical and unitary-isobar 
models of pion electroproduction, which are
extensively used in the interpretation of experiments. 
Also, the novel extension of chiral EFTs 
to the energy domain of the $\Delta$-resonance is described in detail. 
The two-photon exchange effects in the electroexcitation
of the $\Delta$-resonance are addressed here as well.
\end{abstract}

\begin{keyword}
$\Delta$(1232) \sep 
electromagnetic form factors \sep 
meson production \sep chiral Lagrangians
\PACS 14.20.Gk \sep 13.40.Gp \sep 13.60.Le \sep 12.39.Fe
\end{keyword}

\end{frontmatter}

\vspace{1.5cm}
{\it prepared for Physics Reports}

\newpage

\tableofcontents

\include{chap1_delta}

\include{chap2_delta}

\include{chap3_delta}

\include{chap4_delta}

\include{chap5_delta}

\include{chap6_delta}

\include{chap7_delta}

\begin{appendix}
\include{appendix}
\end{appendix}

\include{ackno}

\include{biblio}
\end{document}

%% file: chap1_delta.tex
\section{Introduction}
\label{sec1}

The $\Delta(1232)$ resonance
--- the first excited state of the nucleon ---  
dominates the pion-production phenomena and plays a prominent role in 
the physics of the strong interaction. 
This resonance, first witnessed more than 50 years ago 
by Fermi and collaborators~\cite{Anderson:1952nw}
in pion scattering off protons at the Chicago cyclotron (now Fermilab),
is a distinguished member of the large family of excited baryons.
It is the lightest
baryon resonance, with a mass of 1232 MeV, less than 300 MeV heavier than the
nucleon. Despite its relatively broad width of 120 MeV
(corresponding with the lifetime of 10$^{-23} s$), the $\Delta$
is very well isolated from other resonances. 
It is almost an ideally elastic $\pi N$ resonance, 99\% of the time it 
decays into the nucleon and pion, $\De \to N\pi$. The only 
other decay channel -- electromagnetic, $\De \to N\ga$, contributes
less than 1\% to the total decay width (even though this channel
 will preoccupy the larger part of this review). The 
$\Delta(1232)$ has isospin 3/2 and as such 
comes in four different charge states: $\Delta^{++}$,
$\Delta^+$, $\Delta^0$, and $\Delta^-$, all with (approximately) the same
mass and width. 
The spin of the $\Delta(1232)$ is also 3/2, and it is 
the lightest known particle with such a spin. 
\newline
\indent
The $\De$-resonance dominates many nuclear phenomena at energies
above the pion-production threshold~\cite{Brown:1975di} (for recent reviews see~\cite{Cattapan:2002rx,Hanhart:2003pg}).
In cosmology it is largely responsible for the 
``GZK cut-off''\cite{GZK}, which occurs due to the 
suppression of the high-energy cosmic ray flux by the cosmic microwave
background (CMB). Once the energy of the cosmic rays is sufficient
to produce the $\De$-resonance in the scattering off the CMB photons,
the rate of observed rays drops dramatically (for a recent calculation see 
\cite{Mucke:1998mk}). 
This effect puts a cutoff on the primary cosmic ray
energy at around $10^{19}$~eV for the rays 
coming from a distance larger than a few tens of Mpc. 
\newline
\indent   
In a laboratory, the $\Delta$'s are produced in scattering 
the pion, photon, or electron beams
off a nucleon target.
High-precision measurements of the $N \to \De$
transition by means of electromagnetic probes became possible with the
advent of the new generation of electron beam facilities,
such as LEGS, BATES, ELSA, MAMI, and Jefferson Lab.
Many such experimental programs devoted to the study of electromagnetic
properties of the $\De$ have been completed in the past few years.
\newline
\indent
The current experimental effort has been accompanied by exciting developments
on the theoretical side, most recently in the fields of lattice QCD 
and chiral effective-field theories. These recent
experimental and theoretical advances in understanding the
$\De$-resonance properties are the main subjects of this review.
\newline
\indent
The {\it electromagnetic} $N \to \Delta$ (or, in short $\ga N \De$)  
transition is predominantly of the magnetic dipole ($M1$) type. 
A first understanding of the $\gamma N \Delta$ transition can be obtained   
based on symmetries of Quantum Chromodynamics (QCD) 
and its large number-of-color ($N_c$) limit. 
The spin-flavor global symmetry of QCD, 
is utilized by many quark models and is exactly realized 
in the large-$N_c$ limit. 
In the quark-model picture, 
the $N\to \De$ transition 
is described by a spin flip of a quark in the $s$-wave state,
which in the $\gamma N \Delta$ case 
leads to the magnetic dipole ($M1$) type of transition.
Any $d$-wave admixture
in the nucleon {\it or} the $\Delta$ wave functions allows also 
for the electric ($E2$) and Coulomb ($C2$)
quadrupole transitions. Therefore, 
by measuring the latter two transitions, one is able to 
assess the presence of the $d$-wave components and hence quantify to 
which extent the nucleon or the $\De$ wave function
deviates from the spherical shape. In this way one can hope to
understand to which extent these particles are 
``deformed''.
\newline
\indent
The  $d$-wave component of $\De$'s wave function can be separately
assessed by measuring the electric quadrupole moment
of the $\De$. However, 
this would be extremely difficult because of the tiny lifetime of the $\De$. 
The small $d$-state probability of the $\De$'s wave function also enters into 
the $\Delta$ magnetic dipole moment, which is being measured, 
but the extraction of a small number from such a quantity is also 
very complicated. 
The $\ga N\Delta$ transition, on the other hand,  was
accurately measured in the pion photo- and electro-production reactions 
in the $\De$-resonance energy region. The $E2$ and $C2$ transitions 
 were found to be relatively small but non-zero 
at moderate  momentum-transfers ($Q^2$), 
the ratios $R_{EM}=E2/M1$ and 
$R_{SM}=C2/M1 $ are at the level of a few percent.  
\newline
\indent
Because the $\Delta$ excitation energy is only around 300 MeV and because 
the $\Delta$-resonance  
almost entirely decays into $\pi N$, pions are expected to play 
a prominent role in the $\Delta$ properties. 
Early calculations of the $\gamma N \Delta$ transition 
within chiral bag models revealed the importance 
of the pion cloud effects.
Recall that pions are the Goldstone bosons of the 
spontaneously broken chiral symmetry of QCD. 
As the strength of the Goldstone boson 
interactions is proportional to their energy, at sufficiently low 
energy a perturbative expansion is possible. 
With the advent of the chiral effective field theory (\ceft) 
of QCD~\cite{Weinberg:1978kz,Gasser:1983yg,BKM}
and its extensions to the $\De$-resonance region, 
it has become possible to study the nucleon and $\Delta$-resonance properties 
in a profoundly different way. 
The advantages of such an approach are apparent: \ceft\ is
a low-energy effective field theory of QCD and as such 
it provides a firm theoretical foundation, with all
the relevant symmetries and scales of QCD built in consistently. 
The $\gamma N \Delta$ transition provides new challenges for 
\ceft \ as it involves the interplay of 
two light mass scales~: the pion mass and the $N - \Delta$ mass difference. 
This topic will preoccupy a large part of this review, and we discuss up to 
which energy/momentum scales such an approach can be expected to hold. 
\newline
\indent
Considerable progress has recently been achieved as well in the 
lattice QCD simulations of hadronic properties. They hold the promise 
to compute non-perturbative properties of QCD 
from first principles. 
The present state-of-the-art results for hadronic 
structure quantities, such as the $\gamma N \Delta$ transition form factors,  
are obtained for pion masses above 300 MeV. Therefore, 
they can only  be confronted with experiment after
an extrapolation  down to the physical pion mass of $140$ MeV.
Such extrapolation can be obtained with the aid of \ceft, 
where the pion mass dependence is systematically calculable. 
The \ceft \ framework thus provides a connection  
between lattice QCD calculations and experiment. 
\newline
\indent
The quark structure of the $N \to \Delta$ transition is accessible 
through the phenomenon of asymptotic freedom of QCD at short distances. 
In a hard scattering process, such as $\gamma^* N \to \gamma \Delta$ where the 
virtual photon $\gamma^*$ transfers a large momentum, the  
QCD factorization theorems allow one to separate the perturbative and 
non-perturbative stages of the interaction. In this way, one accesses in  
experiment the non-perturbative matrix elements parametrized in 
terms of new parton distributions, which  
are generalizations of the quark distributions 
from deep inelastic scattering experiments. 
One obtains in this way quark distribution information for the 
$N \to \Delta$ transition, and the $\gamma N \Delta$ form factors 
are obtained as first moments of such $N \to \Delta$ generalized parton 
distributions. 
\newline
\indent
Traditionally, the resonance parameters are extracted from pion 
electroproduction experiments by using unitary isobar models, which
in essence are unitarized tree-level calculations based on
phenomenological Lagrangians. However, as discussed above, 
at low  $Q^2$ the $\gamma N \Delta$-transition 
shows great sensitivity to the ``pion 
cloud'', which until recently could only be
comprehensively studied within dynamical models, which will 
also be reviewed here. 
\newline
\indent
The outline of this review is as follows. In Sect.~\ref{sec2}, we 
introduce the definitions for the $\gamma N \Delta$  
and $\gamma \Delta \Delta$ form factors 
and review the experimental status of the $\gamma N \Delta$ transition at the 
real photon point. 
We then discuss our present theoretical understanding of the 
$\gamma N \Delta$ transition. Some physical insight is obtained from 
QCD inspired models based on quark degrees of freedom, 
pion degrees of freedom, as well as from the large $N_c$ limit. 
The chiral symmetry of QCD is reviewed 
and the role of pions in the $\gamma N \Delta$ transition highlighted. 
We review the status of the lattice QCD calculations and 
their limitations. 
Subsequently, we discuss the quark structure of the $N \to \Delta$ transition. 
We introduce generalized parton distributions for the $N \to \Delta$ 
transition and discuss our phenomenological information of such  
distributions based on their relation to the $\gamma N \Delta$ form factors. 
Finally, we confront predictions made by perturbative QCD at very large 
momentum transfers with the available data. 
\newline
\indent
The theoretical formalism of dynamical models 
will be discussed in detail in Sect.~\ref{sec3}, 
and compared with other approaches such as unitary isobar models and 
dispersion relations.
\newline
\indent
In Sect.~\ref{sec4}, we will 
review how \ceft\ can be applied to study the $\gamma N \Delta$-transition 
at low momentum transfers.   
It will be discussed that the \ceft\ may provide a theoretically
consistent and phenomenologically viable  framework 
for the extraction of the resonance parameters. 
In particular we will review how the $\Delta$ can be introduced in a 
\ceft\, and discuss the field-theoretic intricacies 
due to the spin-3/2 nature of the $\Delta$. 
We will apply the \ceft \ formalism to pion electroproduction in the 
$\Delta$ region and discuss applications to other complementary 
processes such as Compton scattering and radiative pion photoproduction in the 
$\Delta$ region. The convergence of the perturbative \ceft \ 
expansion will also be addressed. 
\newline
\indent
In Sect.~\ref{sec5}, we compare predictions of both dynamical 
models and \ceft \ to the pion photo- and electroproduction observables. 
\newline
\indent
The information on the $\gamma N \Delta$ transition as discussed above 
is obtained from pion electroproduction assuming 
that the interaction is mediated by a single photon exchange. 
In Sect.~\ref{sec6}, we give a brief description of our understanding 
of corrections to this process beyond the one-photon exchange, which may 
become relevant with increasing $Q^2$. 
\newline
\indent
Finally in Sect.~\ref{sec7}, we give our conclusions and spell out some 
open issues in this field. 
\newline
\indent
Several reviews exist in the literature on various aspects of the 
$\gamma N \Delta$ transition. Among the more recent reviews, 
we refer the reader to Ref.~\cite{DT92} where the formalism to 
extract $\gamma N \Delta$ form factors from pion photo- and electroproduction 
observables is outlined in detail. 
The experimental status of the $\gamma N \Delta$, and $\gamma N N^*$ 
transitions as obtained from meson photoproduction experiments is well 
described in Ref.~\cite{Krusche:2003ik}. 
The status of pion electroproduction experiments in the $\Delta$ region 
as well as a previous review of dynamical models can be 
found in Ref.~\cite{Burkert:2004sk}. 

%% file: chap2_delta.tex
\section{The electromagnetic $N\to \Delta$ transition in QCD}
\label{sec2}

Both the nucleon and the $\Delta(1232)$ can be considered as 
quantum states in the rich and complex spectrum of the
quark-gluon system. It is well established that 
the underlying theory which should describe this spectrum
is Quantum Chromodynamics (QCD). However, direct calculations
in QCD of quantities such as the baryon spectrum, form factors, and parton 
distributions are extremely difficult, 
because they require non-perturbative methods. 
Only the lattice simulations have achieved some limited success to date 
in computing the hadron properties from first principles. Even then,
such calculations are severely limited by the presently
available computing power. 
\newline
\indent
Fortunately, some insight into the properties of the spectrum 
can be obtained based on general principles of QCD,
such as the global symmetries and large number-of-colors ($N_c$) 
limit of QCD. 
After introducing the definitions of $\gamma^* N \Delta$ and 
$\gamma^* \Delta \Delta$ form factors in Sect.~\ref{sec2_def}, 
and reviewing the experimental status on the $\gamma N \Delta$ transition 
at the real photon point in Sect.~\ref{sec2_real}, 
we shall discuss in Sect.~\ref{sec2_models} the 
predictions for the $\gamma N \Delta$ transition 
due to the {\it spin-flavor symmetry},
utilized by many quark models and exactly realized in the large-$N_c$
limit.
Furthermore, we shall compare predictions for the $\gamma N \Delta$ 
transition from varios ``QCD-inspired models'' based on quark 
and/or pion degrees of freedom. 
\newline
\indent
in Sect.~\ref{sec:largenc} we shall review some large-$N_c$
relations relevant to the $\ga N \De$ transition. 
\newline
\indent
In Sect.~\ref{sec2_eft},  
we shall discuss the {\it chiral symmetry} of QCD 
and the role of pions in the electromagnetic 
$N\to \Delta$ transition. The chiral effective field 
theory is able to provide predictions for the low momentum transfer 
behavior of the $\gamma^* N \Delta$ form factors. These predictions
will be confronted with the dynamical models and experiment.
\newline
\indent
In Sect.~\ref{sec2_lattice}, we shall discuss the
lattice QCD calculations of the $\ga N\De$ form factors. Present lattice 
QCD calculations are performed for quark masses  
sizably larger than their values in nature, corresponding with pion mass 
values of around 0.3~GeV or larger. It will be discussed how
the chiral effective field theory can be useful in extrapolating
the present lattice QCD calculations to the physical pion mass.  
\newline
\indent
In Sect.~\ref{sec2_gpd}, we shall review the quark structure of the 
$N \to \Delta$ transition and discuss generalized parton distributions 
(GPDs) for the electromagnetic $N \to \Delta$ transition. 
The GPDs can be accessed in hard exclusive processes such as deeply virtual 
Compton scattering where the hard probe ensures that the process occurs 
on a quark which is taken out of the initial nucleon and inserted into the 
final $\Delta$. We shall see that the $\gamma^* N \Delta$ form factors are 
obtained as the first moment in the struck quark momentum fraction 
of such GPDs, which provide much richer information 
on the $N \to \Delta$ transition at the quark level.   
\newline
\indent
Finally, in Sect.~\ref{sec2_pqcd}, 
we shall consider the $N\to \Delta$ transition 
in the formalism of perturbative QCD (pQCD). These considerations are
valid only at very small distances, where quarks nearly do not interact
--- asymptotic freedom. In this limit, the $\gamma^* N \Delta$ form factors 
correspond with a hard photon which hits a quark in the nucleon. The 
struck quark shares the large momentum with the other two (near collinear) 
valence quarks in such a way that the final three quark state has 
$\Delta$ quantum numbers. We shall discuss the predictions made in this 
limit and confront them with the experimental status of the 
$\gamma^* N \Delta$ form factors at large momentum transfers.

\subsection{Definitions and conventions}
\label{sec2_def}
Throughout this review we shall use 
the following conventions for the metric and $\ga$-matrices:
$g^{\mu\nu} = \mbox{diag} (1, -1,-1,-1)$, 
$\{\ga^\mu, \ga^\nu\} = 2g^{\mu\nu}$, $\ga_5= i\ga^0\ga^1\ga^2\ga^3$,
$\veps_{0123} =+1$. The $\ga N\De$ and $\ga \De\De$ form factors
are introduced as follows.

\subsubsection{The $\gamma^* N \Delta$ vertex and form factors}

\begin{figure}[b,t,p]
\centerline{
\epsfxsize=11cm
\epsffile{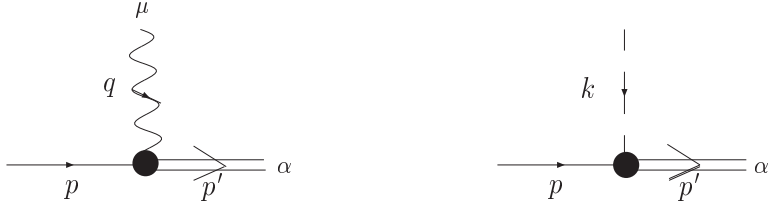}
}
\caption{$\gamma^* N \Delta$ vertex (left panel) and 
$\pi N \Delta$ vertex (right panel). The four-momenta of 
the nucleon ($\Delta$) and of the photon (pion) are given by 
$p$ ($p^\prime$) and $q$ ($k$) respectively. 
The four-vector index of 
the spin 3/2 field is given by $\alpha$, and 
$\mu$ is the four-vector index of the photon field.}
\figlab{treevertex}
\end{figure}

The general  spin structure of the pion- and photon-induced
$N\to \Delta$ transitions, see 
\Figref{treevertex}, can be written as:
\begin{eqnarray}
\eqlab{newform}
\bar u_{\al}(p') \, \Gamma^{\alpha }_{\pi N\De} \, u(p)   & \equiv &
 \frac{h_A}{2f_\pi} \bar u_{\al}(p') \, k^\al \, T_a^\dagger \, u(p)\, 
F_{\pi N\De}(k^2)\,,
\label{eq:pindel} \\
\langle \Delta(p') \,|\, e\,J^\mu(0) \,|\, N(p) \rangle \, &\equiv& \, i \, 
\bar u_{\al}(p') \, \Gamma^{\alpha \mu}_{\ga N\De} \, u(p)   \nonumber \\
& \equiv & i \, \sqrt{\frac{2}{3}} \, 
\frac{3e (M_\Delta + M_N)}{2 M_N [(M_\Delta + M_N)^2+Q^2]} 
\nn\\
&& \times \,  \bar u_{\alpha}(p^\prime) 
\, \left\{\, 
g_M(Q^2) \, \varepsilon^{\alpha \mu \vrho \si} \, 
p^{\prime}_\vrho \,  q_\si  \right. \nn \\ 
&&\hspace{1.5cm} + \,  g_E(Q^2) 
\left( q^\alpha \, {p'}^{\mu} -q \cdot p^\prime  \, g^{\alpha \mu} \right) 
i \gamma_5  \nn \\
&& \left. \hspace{1.5cm} + \,  g_C(Q^2) \,
\left( q^\alpha \, q^\mu - q^2 \, g^{\alpha \mu}  \right) i \gamma_5 
\right\} \; u (p) , 
\label{eq:diagndel1}
\end{eqnarray}
where $k$ is the pion and $q$ is the photon 4-momentum, $p$ ($p'$) is the
nucleon (the $\De$) 4-momentum, $M_N$ ($M_\De$) is the nucleon (the $\De$)
mass, $u$ is the nucleon spinor, and $u_\al$ represents the 
spin-3/2 $\De$ vector-spinor In Eqs.~(\ref{eq:pindel}) 
and (\ref{eq:diagndel1}), 
the spin dependence in the $N$ and $\Delta$ spinors is understood.
The operator $T_a^\dagger$ in Eq.~(\ref{eq:pindel}), with  
$a = 1,2,3$ corresponding with the Cartesian pion fields $\pi^a$, 
is the isospin 1/2 $\to$ 3/2 transition operator.   
 The operator $J^\mu$ is the electromagnetic current operator, 
and the factor $\sqrt{2/3}$ in front of Eq.~(\ref{eq:diagndel1}) 
corresponds with the isospin factor for the photon induced 
$p \to \Delta^+$ transition. 
Furthermore, $f_\pi\simeq 92.4$ MeV is the pion decay constant,
$h_A$ is a dimensionless constant representing the strength of the
$\pi N\to \De$ transition and thus related to the decay width 
of the $\De$ to $\pi N$, see Sect.~\ref{sec4}.5.1. 
\newline
\indent
The strong transition form factor $F_{\pi N \De}$  is normalized
as $F_{\pi N \De}(m_\pi^2)=1$, and  can in principle depend
as well on the invariant mass of the nucleon and $\Delta$, {\it i.e.}, 
$p^2$ and ${p'}^2$. However, unless explicitly specified, we assume that
nucleon and $\Delta$ are on the mass shell and therefore 
$p^2=M_\N^2$, $\slap\, u(p) = M_N \,u(p)$,  ${p'}^2=M_\De^2$, 
$\slap' \,u_\al (p') = M_\De \, u_\al (p')$, and 
$p'_\alpha \, u^\alpha(p')=0=\ga_\alpha \, u^\alpha(p')$. We use 
the covariant normalization convention for the spinors, {\it i.e.} 
$\bar u(p) u(p) = 2 M_N$, 
and $\bar u_\alpha(p') u^\alpha(p') = -2 M_\Delta$.   
\newline
\indent
In Eq.~(\ref{eq:diagndel1}), 
the electromagnetic form factors $g_M$, $g_E$, and $g_C$ 
represent the strength of the 
magnetic dipole, electric quadrupole, and 
Coulomb quadrupole $N\to \Delta$ transitions, respectively, as
a function of the momentum transfer: $Q^2=-q^2$. 
Note that for the real photon case, $Q^2=0$, only the magnetic and 
electric transitions are possible, the Coulomb term drops out.
\newline
\indent
These electromagnetic 
form factors relate to the more conventional magnetic dipole ($G_M^\ast$), 
electric quadrupole ($G_E^\ast$) and Coulomb quadrupole ($G_C^\ast$) 
form factors of Jones and Scadron~\cite{Jones:1972ky} as 
follows,\footnote{The form factors $G_M^\ast, G_E^\ast$, and $G_C^\ast$ 
used throughout this work correspond with the ones also 
used in Ref.~\cite{SL}. Our sign convention leads to 
positive values of all three form factors at $Q^2 = 0$.}
\begin{eqnarray}
\label{eq:JS}
G_M^\ast(Q^2) &=& g_M \,+\frac{1}{Q_+^2} \left[\half(-M_\De^2+M_N^2+Q^2) \,g_E
+  Q^2 g_C\right], \nn\\ 
G_E^\ast(Q^2) &=& \frac{1}{Q_+^2} \left[\half(-M_\De^2+M_N^2+Q^2) \,g_E+  Q^2  g_C\right], \\
G_C^\ast(Q^2) &=&\frac{1}{Q_+^2} \left[ (-M_\De^2+M_N^2+Q^2)\,  g_C -2 M_\De^2 \,
g_E\right],\nn
\end{eqnarray}
where $Q_\pm$ is defined as:
\begin{eqnarray}
Q_\pm \,\equiv\, \sqrt{(M_\De\pm M_N)^2 +Q^2}.
\end{eqnarray}
\begin{figure}
\centerline{  \epsfxsize=9cm%
  \epsffile{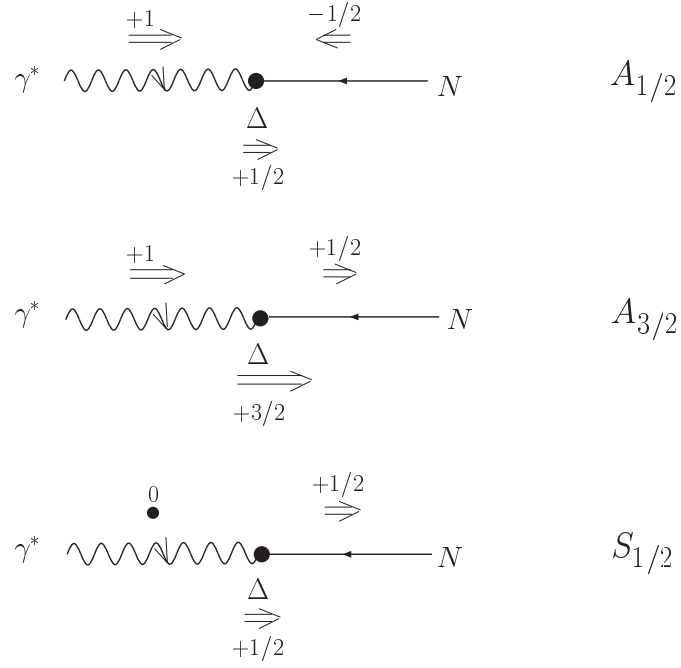} 
}
\caption{$\gamma^\ast N \to \Delta$ helicity amplitudes in the $\Delta$ rest 
frame. The $\gamma$, $N$, and $\Delta$ spin projections onto the $z$-axis, 
which is chosen along the virtual photon direction, are indicated on the 
figure.  
The corresponding helicity amplitudes are indicated on the right. }
\figlab{fig:helndel}
\end{figure}
\newline
\indent
Equivalently, one can also parametrize the $\gamma^\ast N \Delta$ 
transition through three helicity amplitudes $A_{1/2}$, $A_{3/2}$ and 
$S_{1/2}$, which are defined in the $\Delta$ rest frame as illustrated 
in \Figref{fig:helndel}. 
These $\Delta$ rest frame helicity amplitudes are defined 
through the following matrix elements of the electromagnetic current 
operator:
\begin{eqnarray}
A_{3/2} \,&\equiv&\, -\frac{e}{\sqrt{2 q_\Delta}} \; 
\frac{1}{(4 M_N M_\Delta)^{1/2}} \; \langle \; \Delta(\vec 0, \, +3/2) 
\,|\, {\bf J \cdot \epsilon}_{\lambda = +1} \,|\, N(-\vec q, \, +1/2 ) 
\;\rangle, \nn \\
A_{1/2} \,&\equiv&\, -\frac{e}{\sqrt{2 q_\Delta}} \; 
\frac{1}{(4 M_N M_\Delta)^{1/2}} \; \langle \; \Delta(\vec 0, \, +1/2) 
\,|\, {\bf J \cdot \epsilon}_{\lambda = +1} \,|\, N(-\vec q, \, -1/2 ) 
\;\rangle,  \\
S_{1/2} \,&\equiv&\, \frac{e}{\sqrt{2 q_\Delta}} \; 
\frac{1}{(4 M_N M_\Delta)^{1/2}} \; \langle \; \Delta(\vec 0, \, +1/2) 
\,|\, J^0 \,|\, N(-\vec q, \, +1/2 ) 
\;\rangle, \nn
\label{eq:resthel}
\end{eqnarray}
where the spin projections are along the $z$-axis (chosen along the virtual 
photon direction) and where the transverse photon polarization vector 
entering in $A_{1/2}$ and $A_{3/2}$ is 
given by ${\bf \epsilon}_{\lambda = +1} = -1/\sqrt{2} (1, i, 0)$. 
Furthermore in Eq.~(\ref{eq:resthel}), $e$ is the proton 
electric charge, related to the fine-structure constant as 
$\alpha_{em} \equiv e^2 /(4 \pi) \simeq 1/137$, and $q_\Delta$ is
the magnitude of the virtual photon three-momentum in the $\Delta$ rest frame:
\begin{eqnarray}
q_\Delta \,\equiv \, |{\mathbf q}| \,=\, \frac{Q_+ Q_-}{2 M_\Delta} .
\label{eq:gamomdel}
\end{eqnarray}
The helicity amplitudes are functions of the photon virtuality $Q^2$, 
and can be expressed in terms of the Jones-Scadron 
$\gamma^\ast N \Delta$ form factors, by using Eq.~(\ref{eq:diagndel1}), as~:
\begin{eqnarray}
A_{3/2} \,&=&\, - N \, \frac{\sqrt{3}}{2} 
\left\{ G_M^\ast + G_E^\ast \right\} \, , \nn \\ 
A_{1/2} \,&=&\, - N \, \frac{1}{2} 
\left\{ G_M^\ast - 3 \, G_E^\ast \right\} \, , \nn \\
S_{1/2} \,&=&\, N \, \frac{q_\Delta}{\sqrt{2} M_\Delta} \, G_C^\ast \, , 
\label{eq:heljs}
\end{eqnarray}
where $N$ is defined as~:
\begin{eqnarray}
N \,\equiv\, \frac{e}{2} \, \left( \frac{Q_+ Q_-}{2 M_N^3} \right)^{1/2} \, 
\frac{(M_N + M_\Delta)}{Q_+}.
\end{eqnarray}
The above helicity amplitudes are expressed in units GeV$^{-1/2}$, 
and reduce at $Q^2 = 0$ to the photo-couplings quoted by the Particle 
Data Group~\cite{PDG2006}.  
\newline
\indent
Experimentally, the $\gamma^\ast N \Delta$ helicity amplitudes are extracted 
from the $M1$, $E2$, and $C2$ multipoles for 
the $\gamma^* N \to \pi N$ process at the resonance position,  
{\it i.e.} for $\pi N$ {\it c.m.} energy $W = M_\Delta$. 
These pion electroproduction multipoles are denoted by 
$M_{1+}^{(3/2)}$, $E_{1+}^{(3/2)}$, and $S_{1+}^{(3/2)}$ 
(following standard notation, see {\it e.g.} 
Ref.~\cite{Berends:1974zp} and Sect.~\ref{sec3} for more details),  
where the subscript refers to the partial wave 
$l = 1$ in the $\pi N$ system and ``$+$'' indicates that the 
total angular momentum is $J = l + 1/2$, being 3/2. 
The superscript $(3/2)$ in the multipole notation refers to the total  
isospin 3/2. 
The $\gamma^\ast N \Delta$ helicity amplitudes $A_{3/2}$, 
$A_{1/2}$, and $S_{1/2}$ are obtained from the 
$\gamma^\ast N \to \pi N$ resonant multipoles 
as~\cite{Arndt:1990ej}:
\begin{eqnarray}
A_{3/2} \,&=&\, -\frac{\sqrt{3}}{2} \, 
\left\{ \bar M_{1+}^{(3/2)} - \bar E_{1+}^{(3/2)}
\right\} , \nn \\
A_{1/2} \,&=&\, -\frac{1}{2} \, 
\left\{ \bar M_{1+}^{(3/2)} + 3 \, \bar E_{1+}^{(3/2)} 
\right\} , \nn \\ 
S_{1/2} \,&=&\, -\sqrt{2} \, \bar S_{1+}^{(3/2)} ,  
\label{eq:helmult}
\end{eqnarray}
where $\bar M_{1+}^{(3/2)}$, $\bar E_{1+}^{(3/2)}$, and $\bar S_{1+}^{(3/2)}$ 
are obtained from the imaginary parts 
of the resonant multipoles 
at the resonance position $W = M_\Delta$ as:
\begin{eqnarray}
\bar M_{1+}^{(3/2)}(Q^2) \,\equiv \, \sqrt{\frac{2}{3}} \, a_\Delta \, 
\mathrm{Im} M_{1+}^{(3/2)} (Q^2, W = M_\Delta) , 
\end{eqnarray}
where $\sqrt{2/3}$ is an isospin factor, and similar relations define  
$\bar E_{1+}^{(3/2)}$ and $\bar S_{1+}^{(3/2)}$.  
In these relations, $a_\Delta$ is given by:
\begin{eqnarray}
a_\Delta \,&=&\, \left( \frac{4 \pi \, k_\Delta \, M_\Delta \, 
\Gamma_\Delta}{q_\Delta \, M_N}\right)^{1/2},
\label{eq:adel}
\end{eqnarray}
where $\Gamma_\Delta$ denotes the $\Delta$ width, which we take as 
$\Gamma_\Delta = 0.115$~GeV, and 
$k_\Delta$ denotes the magnitude of the pion three-momentum 
in the $\pi N$ {\it c.m.} frame at the resonance position ($W = M_\Delta$):
\begin{eqnarray} 
k_\Delta \,\equiv\, |{\mathbf k}| \,=\, \frac{1}{2 M_\Delta} \,
\left[ (M_\De + M_N)^2 - m_\pi^2 \right]^{1/2} \, \cdot \,
\left[ (M_\De - M_N)^2 - m_\pi^2 \right]^{1/2} ,
\label{eq:pimomdel}
\end{eqnarray}
which yields  $k_\Delta = 0.227$~GeV. Furthermore, for $Q^2 = 0$,  
$q_\Delta = 0.259$~GeV and $a_\Delta \simeq 1.29$~GeV$^{1/2}$. 
\newline
\indent
We can extract the Jones-Scadron $\gamma^\ast N \Delta$ form factors from the 
multipoles at the resonance position. Combining Eqs.~(\ref{eq:heljs}) 
and (\ref{eq:helmult}), we obtain:
\begin{eqnarray}
\bar M_{1+}^{(3/2)} \,&=&\, N \, G_M^\ast \, , \nn \\ 
\bar E_{1+}^{(3/2)} \,&=&\, - N \, G_E^\ast \, , \nn \\
\bar S_{1+}^{(3/2)} \,&=&\, - N \, \frac{q_\Delta}{2 M_\Delta} 
\, G_C^\ast \, . 
\label{eq:multjs}
\end{eqnarray}
Of special interest are the multipole ratios at the position of the 
$\Delta$ resonance. The ratio of electric quadrupole ($E2$) 
over magnetic dipole ($M1$) is denoted by : $R_{EM}=E2/M1$ 
(sometimes also denoted by $EMR$), whereas the ratio of Coulomb quadrupole 
($C2$) over magnetic dipole is denoted by 
$R_{SM}=C2/M1$ (sometimes also denoted by $CMR$):
\begin{eqnarray}
R_{EM} &\equiv& EMR \equiv \frac{\bar E_{1+}^{(3/2)}}{\bar M_{1+}^{(3/2)}} 
= \frac{A_{1/2} - \frac{1}{\sqrt{3}} \, A_{3/2}}{A_{1/2} + \sqrt{3}\, A_{3/2}},
\nn \\
R_{SM} &\equiv& CMR \equiv \frac{\bar S_{1+}^{(3/2)}}{\bar M_{1+}^{(3/2)}} 
= \frac{\sqrt{2} \, S_{1/2}}{A_{1/2} + \sqrt{3} \, A_{3/2}}.
\label{eq:remhel}
\end{eqnarray}
Using the relations of Eq.~(\ref{eq:multjs}), 
these ratios can be expressed in terms of the Jones-Scadron form factors as:
\beq
\eqlab{ratios}
R_{EM} = - \frac{G_E^\ast}{G_M^\ast} \,,\quad \quad \quad \quad
R_{SM} = - \frac{Q_+ Q_-}{4M_\De^2} \, \frac{G_C^\ast}{G_M^\ast}.
\eeq
\newline
\indent
From the values of the $\gamma^\ast N \Delta$ form factors at 
$Q^2 = 0$, one can extract some interesting static quantities. 
For the dominant $M1$ transition, 
one can extract the static $N \to \Delta$ transition magnetic moment 
$\mu_{N \to \Delta}$ from  
the value of $G_M^\ast$ at $Q^2 = 0$ as~\cite{Tiator:2003xr}:
\begin{eqnarray}
\mu_{N \to \Delta} = \sqrt{\frac{M_\Delta}{M_N}} \, G_M^\ast(0),
\label{eq:mundel}
\end{eqnarray}
which is expressed in nuclear magnetons $\mu_N \equiv e / (2 M_N)$.
\newline
\indent 
Furthermore, from the value of $G_E^\ast$ at $Q^2 = 0$, 
one can extract a static $N \to \Delta$ 
quadrupole transition moment $Q_{N \to \Delta}$ as~\cite{Tiator:2003xr}:
\begin{eqnarray}
Q_{N \to \Delta} = - 6 \sqrt{\frac{M_\Delta}{M_N}} \, 
\frac{1}{M_N \, q_\Delta(0)} \, G_E^\ast(0),
\label{eq:qndel}
\end{eqnarray}
where $q_\Delta(0)$ is obtained from Eq.~(\ref{eq:gamomdel}) for $Q^2 = 0$, 
as $q_\Delta(0) = (M_\Delta^2 - M_N^2)/ 2 M_\Delta$. 
\newline
\indent
Finally, we also like to note that in the literature, one sometimes uses 
so-called Ash form factors
for the $\gamma^\ast N \Delta$ transition~\cite{Ash67}. 
They are simply related to the Jones-Scadron form factors as:
\begin{eqnarray}
G^\ast_{M, \, Ash} (Q^2) \,=\, 
\frac{(M_N + M_\Delta)}{Q_+} \, G^\ast_{M}(Q^2),
\label{eq:ash}
\end{eqnarray}
and analogously for the electric and Coulomb quadrupole 
form factors. 

\subsubsection{The $\gamma^* \Delta \Delta$ vertex and form factors} 

\begin{figure}[h,b,t,p]
\centerline{
\epsfxsize=5cm
\epsffile{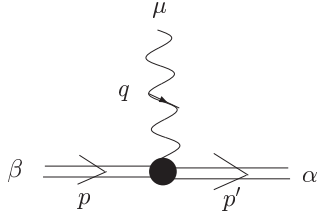}
}
\caption{The $\gamma^* \Delta \Delta$ vertex. The four-momenta of 
the initial (final) $\Delta$ and of the photon are given by 
$p$ ($p^\prime$) and $q$ respectively. 
The four-vector indices of the initial (final) 
spin 3/2 fields are given by $\beta$ ($\alpha$), and 
$\mu$ is the four-vector index of the photon field.}
\figlab{gadeldeltreevertex}
\end{figure}

Consider the coupling of a photon to a $\Delta$, \Figref{gadeldeltreevertex}.  
The matrix element of the electromagnetic current operator $J^\mu$ 
between spin 3/2 states can be decomposed into four multipole transitions:
a Coulomb monopole (C0), a magnetic dipole (M1), 
an electric quadrupole (E2) and 
a magnetic octupole (M3).
We firstly write a Lorentz-covariant decomposition which exhibits
manifest electromagnetic gauge-invariance: 
\begin{eqnarray}
\langle \Delta(p') \,|\,e J^\mu(0) \,|\, \Delta (p) \rangle  &\,\equiv\,&\,  
\bar u_{\alpha}(p') \, \Gamma^{\alpha \beta \mu}_{\gamma \Delta \Delta} (p',p)
\, u_\beta(p) \nn\\
&\,=\,& -e \, \bar u_\alpha (p')\, \left\{ e_\De\,
F_1^\ast(Q^2) \, g^{\alpha \beta} \, \gamma^\mu \right.\nn \\
& + & \,\frac{i}{2M_\De}\left[ F_2^\ast(Q^2) \, g^{\alpha \beta}
+ F_4^\ast(Q^2) \,\frac{q^\al q^\be}{(2M_\De)^2}\right] \si^{\mu\nu} q_\nu \\
&+& \,\left. \frac{F_3^\ast(Q^2)}{(2M_\De)^2}\left[ q^\al q^\be \ga^\mu - 
\half q\cdot\ga\,(g^{\al\mu}q^\be + g^{\be\mu}q^\al) \right] \, \right\}
 u_\beta(p) \, , \nn
\label{eq:gadeldeltree}
\end{eqnarray}
where $F^\ast_i$ are the $\gamma^* \Delta \Delta$ form factors,
$e_\Delta$ is the electric charge in units of $e$ (e.g., $e_{\Delta^+} = +1$), such that
$F^\ast_1(0)=1$. The relation to the multipole 
decomposition~\cite{Weber:1978dh,Noz90} can be written
in terms of the magnetic dipole ($\mu_\De$),
electric quadrupole ($Q_\De$) and magnetic octupole ($O_\De$) moments, 
given by:
\begin{subequations}
\bea
\eqlab{EMmoments}
\mu_\De &=& \frac{e}{2 M_\Delta} \left[e_\De+F_2^\ast (0)\right]  \,,\\
Q_\De & = &  \frac{ e}{ M_\Delta^2}\left[ e_\De - \half F_3^\ast(0)\right]\,,\\
O_\De & = &  \frac{e}{2M_\Delta^3}
\left[ e_\De + F_2^\ast (0) - \half \left(F_3^\ast(0)+F_4^\ast(0)\right)\right]\,.
\eea
\end{subequations}

\subsection{Experimental information at the real photon point}
\label{sec2_real}

In this section we briefly summarize the experimental information on the 
$\gamma N \Delta$ transition at $Q^2 = 0$. 
\newline
\indent
The $\gamma N \Delta$ transition can be studied both in nucleon Compton 
scattering $\gamma N \to \gamma N$ and in the pion photoproduction 
reaction $\gamma N \to \pi N$. The highest precision data on the 
$\gamma N \Delta$ $M1$ and $E2$ amplitudes have been obtained 
in pion photoproduction experiments on a proton target, 
using linearly polarized photons, 
both by the MAMI/A2 Collaboration~\cite{Beck:1997ew,Beck:1999ge} 
and the LEGS Collaboration~\cite{LEGS97,Blanpied:2001ae}. 
A detailed discussion of these $\gamma N \to \pi N$ data will be presented 
in Sect.~\ref{sec51}. 
\newline
\indent
We quote here the real photon 
$\gamma N \Delta$ amplitudes at the resonance position ($W  = M_\Delta$) 
for a proton target resulting from these high precision measurements. 
The MAMI/A2 Collaboration obtained the values~\cite{Beck:1999ge}:
\begin{eqnarray}
A_{1/2} &=& -\left( 131 \pm 1 \right) \quad [10^{-3} \mathrm{GeV}^{-1/2}], 
\nn \\
A_{3/2} &=& - \left( 251 \pm 1 \right) \quad [10^{-3} \mathrm{GeV}^{-1/2}], 
\nn \\
R_{EM} &=& -\left( 2.5 \pm 0.1_{stat.} \pm 0.2_{syst.} \right) \, \%. 
\label{eq:mamireal}
\end{eqnarray}
The LEGS Collaboration obtained the values~\cite{Blanpied:2001ae}:
\begin{eqnarray}
A_{1/2} &=& - \left( 135.7 \pm 1.3_{stat.+syst.} \pm 3.7_{model} \right)
\quad [10^{-3} \mathrm{GeV}^{-1/2}], \nn \\
A_{3/2} &=& - \left( 266.9 \pm 1.6_{stat.+syst.} \pm 7.8_{model} \right) 
\quad [10^{-3} \mathrm{GeV}^{-1/2}], \nn \\
R_{EM} &=& -\left( 3.07 \pm 0.26_{stat.+syst.} \pm 0.24_{model} \right) \, \%. 
\label{eq:legsreal}
\end{eqnarray}
\indent
As summary of the above experiments, the Particle Data Group 
quotes as values~\cite{PDG2006}
\footnote{Note that the central values of  
$A_{1/2}$ and $A_{3/2}$ quoted by PDG\cite{PDG2006} yield 
$R_{EM} = -1.64$~\% when using Eq.~(\ref{eq:remhel}), 
and therefore do not exactly correspond with 
the central value $R_{EM} = -2.5$~\% they quote. 
The photo-couplings are based on an analysis of a larger number of 
experiments, and within their error bars are compatible with 
the quoted value for $R_{EM}$ though.}:
\begin{eqnarray}
A_{1/2} &=& - \left( 135 \pm 6 \right)
\quad [10^{-3} \mathrm{GeV}^{-1/2}], \nn \\
A_{3/2} &=& - \left( 250 \pm 8 \right) 
\quad [10^{-3} \mathrm{GeV}^{-1/2}], \nn \\
R_{EM} &=& - \left( 2.5 \pm 0.5 \right) \, \%.
\label{eq:pdgreal}
\end{eqnarray}
\indent
The experimental information on the $M1$ $\gamma N \Delta$ transition 
can equivalently be expressed in terms of $G_M^\ast(0)$ 
using Eq.~(\ref{eq:multjs}) 
or equivalently in terms of the transition magnetic moment 
$\mu_{p \to \Delta^+}$ using Eq.~(\ref{eq:mundel}). 
Using the experimental values of Eq.~(\ref{eq:mamireal}), 
Ref.~\cite{Tiator:2000iy} extracted the values:
\begin{eqnarray}
G_M^\ast(0) \,&=&\, 3.02 \pm 0.03 , \nn \\
\mu_{p \to \Delta^+} \,&=&\, \left[ 3.46 \pm 0.03 \right ] \mu_N.
\label{eq:mundelexp}
\end{eqnarray} 
\indent
The value for the $E2$ $\gamma N \Delta$ transition 
can equivalently be expressed in terms of a quadrupole transition moment using 
Eq.~(\ref{eq:qndel}). Using the experimental values of 
Eq.~(\ref{eq:mamireal}), Ref.~\cite{Tiator:2003xr} extracted:
\begin{eqnarray}
Q_{p \to \Delta^+} \,&=&\, - \left( 0.0846 \pm 0.0033 \right ) 
\; \mathrm{fm}^2.
\label{eq:qndelexp}
\end{eqnarray}

\subsection{Physical interpretation: quark and pion-cloud models}
\label{sec2_models}

\subsubsection{Constituent quark models}

In a quark model, the nucleon appears as the ground state 
of a quantum-mechanical three-quark system in a confining potential.
In such a picture, the ground state baryons (composed of the 
light up ($u$), down ($d$) and strange ($s$) quark flavors) are 
described by $SU(6)$ spin-flavor symmetric wave functions, supplemented 
by an antisymmetric color wave function.  
\newline
\indent
In the Isgur-Karl model~\cite{Isgur}, 
the constituent quarks  
move in a harmonic oscillator type confining potential. 
For the ground state baryons, 
the three constituent quarks are in the $1s$ 
oscillator ground state, corresponding with the [56]-plet of $SU(6)$. 
The harmonic oscillator states can be represented by 
$| B \; ^{2S+1}L_J \rangle_t$, where $B$ stands for either $N$ or 
$\Delta$ states, $S$ specifies the spin, $L$ the orbital 
angular momentum ($L = S, P, D,...$ in the common spectroscopic notation), 
and $J$ the total angular momentum of the three-quark state. 
Furthermore, $t (= S, M, A)$ refers to the symmetry type : 
symmetric ($S$), mixed symmetric ($M$) or anti-symmetric ($A$) 
under exchange of the quarks 
in both the spin-flavor and space parts of the baryon wave function. 
In the Isgur-Karl model, the long-range confining potential is supplemented by 
an interquark force corresponding with one-gluon exchange. 
The one-gluon exchange leads to a color hyperfine interaction 
between quarks $i$ and $j$ of the form:
\begin{eqnarray}
H_{hyperfine}^{ij} = \frac{2}{3} \frac{\alpha_s}{m_i m_j} \left\{ 
\frac{8 \pi}{3} {\bf S}_i {\bf \cdot} {\bf S}_j \, \delta^3 ({\bf r}_{ij}) 
+ \frac{1}{r_{ij}^3} 
\left[ \frac{3 ({\bf S}_i {\bf \cdot} {\bf r}_{ij}) \, 
({\bf S}_j {\bf \cdot} {\bf r}_{ij})}{r_{ij}^2} 
- {\bf S}_i {\bf \cdot} {\bf S}_j \right] \right\},
\label{eq:hyperfine}
\end{eqnarray}
with $\alpha_S$ the strong coupling constant, 
${\bf S_i} (m_i)$ the spin (mass) of quark $i$, and where 
${\bf r_{ij}} (r_{ij})$ specify the vector (distance) between quarks 
$i$ and $j$. The first term in Eq.~(\ref{eq:hyperfine}) corresponds 
with a zero-range spin-spin interaction, whereas the second term 
corresponds with a tensor force. 
The color hyperfine interaction of Eq.~(\ref{eq:hyperfine})
breaks the $SU(6)$ symmetry and leads to a mass splitting 
between $N(939)$ and $\Delta(1232)$, often referred to as the hyperfine 
splitting. 
It was found that it also predicts well the mass splittings between octet 
and decuplet baryons~\cite{DeRujula:1975ge}. 
Furthermore, the tensor force in Eq.~(\ref{eq:hyperfine}) 
will produce a $D$-state ($L = 2$) admixture in the 
$N$ and $\Delta$ ground states~\cite{Koniuk:1979vy,Isgur:1981yz}. 
Because of this hyperfine interaction, the $N(939)$ and $\Delta(1232)$ 
states are described as superpositions of $SU(6)$ configurations. 
Including configurations up to the $2 \hbar \omega$ oscillator shell, 
they are given by (using the abovementioned spectroscopic notation):
\begin{eqnarray} 
| N (939) \rangle &\,=\,& a_S \, |N \; ^2 S_{1/2} \rangle_S \,+\,
a^\prime_S \, |N \; ^2 S^\prime_{1/2} \rangle_S \,+\,
a_M \, |N \; ^2 S_{1/2} \rangle_M \,+\,
a_D \, |N \; ^4 D_{1/2} \rangle_M , 
\label{eq:cqmnuc} \\
| \Delta (1232) \rangle &=& b_S \, |\Delta \; ^4 S_{3/2} \rangle_S \,+\,
b^\prime_S \, |\Delta \; ^4 S^\prime_{3/2} \rangle_S \,+\,
b_D \, |\Delta \; ^4 D_{3/2} \rangle_S \,+\,
b^\prime_D \, |\Delta \; ^2 D_{3/2} \rangle_M . 
\label{eq:cqmdel}
\end{eqnarray}
By diagonalizing the hyperfine interaction and fitting the results
to the baryon spectrum, Isgur {\it et al.}~\cite{Isgur:1981yz} 
obtained the following values for the wave-function coefficients: 
\begin{eqnarray}
a_S \simeq 0.93, \quad &&a^\prime_S \simeq -0.29, \quad
a_M \simeq -0.23, \quad a_D \simeq -0.04, 
\label{eq:cqmnuc2} \\
b_S \simeq 0.97, \quad &&b^\prime_S \, \simeq  +0.20, \quad
b_D \,\simeq -0.10, \quad b^\prime_D \, \simeq 0.07. 
\label{eq:cqmdel2}
\end{eqnarray} 
From these values it is evident that the $S$-wave component
dominates the $N$ and $\Delta$ wave functions in a constituent quark model. 
The $\Delta(1232)$ resonance is obtained from the nucleon by a 
spin flip of one of the quarks in the $1s$ state 
so as to give a state of total spin 3/2. 
Therefore the electromagnetic $N \to \Delta$ transition is dominantly 
a $M1$ transition~\cite{Becchi:1965}, 
see \Figref{fig:m1e2ndel} (left panel). 
Using $SU(6)$ flavor symmetry, {\it i.e.}, setting $a_S = b_S = 1$ 
and $a^\prime_S = a_M = a_D = b^\prime_S = b_D = b^\prime_D = 0$ in 
Eqs.~(\ref{eq:cqmnuc}) and (\ref{eq:cqmdel}), yields the relation between 
the magnetic moments of proton and $p \to \Delta^+$ as:
\begin{eqnarray}
\mu_{p \to \Delta^+} = \frac{2 \sqrt{2}}{3} \, \mu_p
\,=\, 2.63 \, \mu_N \quad \quad 
\mbox{( SU(6)-symmetric quark model).} 
\end{eqnarray}
This is about 25 \% lower than the experimental number 
Eq.~(\ref{eq:mundelexp}). 
$SU(6)$ breaking effects, due to the hyperfine interaction, decrease   
the quark model prediction even further to a value around 2.3~$\mu_N$.
A very similar small number is also obtained in the  
MIT bag model calculation of Ref.~\cite{Donoghue:1975yg}.
The underestimate of $\mu_{N \to \Delta}$ in quark models is hinting at 
the importance of degrees of freedom beyond constituents 
quarks as will be discussed further on. 
\begin{figure}
\leftline{  \epsfysize=3cm
  \epsffile{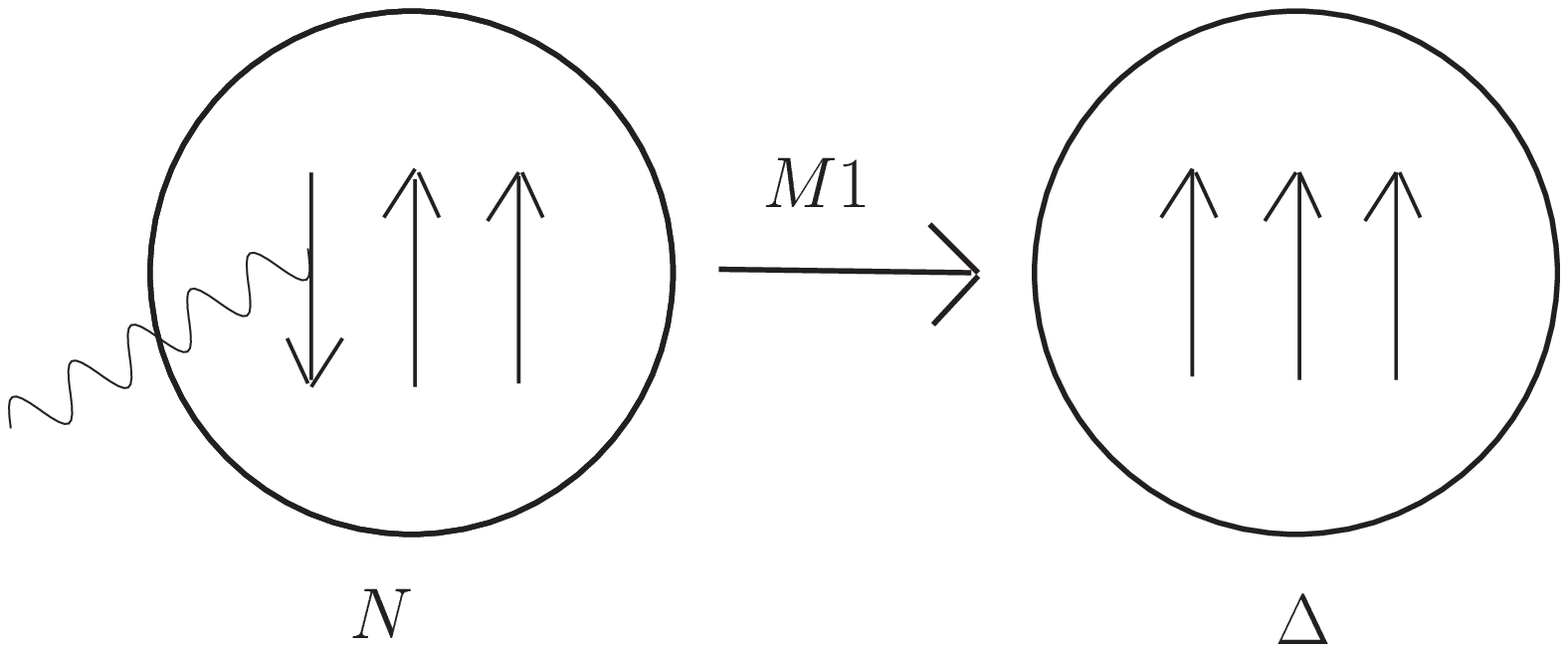} 
}
\vspace{-3cm}
\rightline{  \epsfysize=3cm
  \epsffile{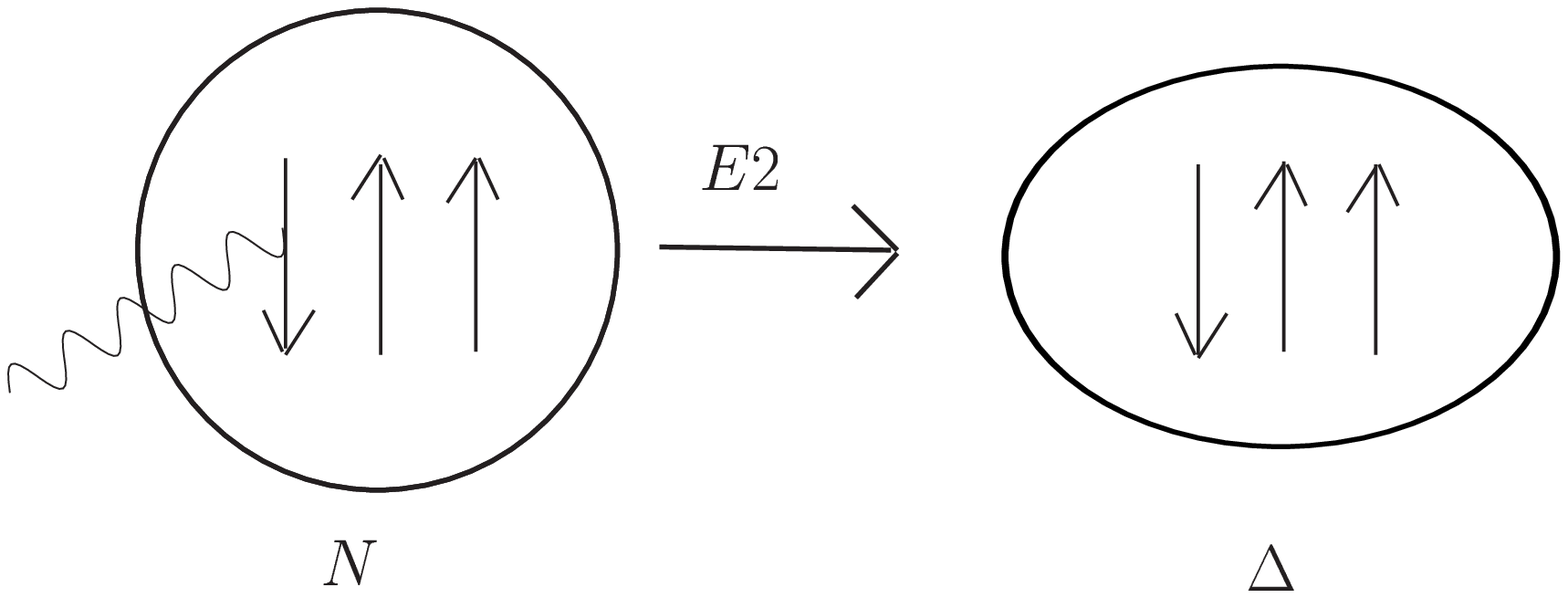} 
}
\caption{Schematic picture within a quark model 
of a $M1$ (left panel) and $E2$ (right panel) 
$N \to \Delta$ transition 
induced by the interaction of a photon with a single quark in the nucleon. 
The $M1$ transition involves a $S$-wave spatial wave function, 
whereas the $E2$ transition in this picture requires $N$ and/or $\Delta$ 
wave functions to have a $D$-wave component (indicated by a non-spherical 
shape).  
}
\figlab{fig:m1e2ndel}
\end{figure}
\newline
\indent
The small non-zero values of $b_D$ and $b^\prime_D$ in 
Eq.~(\ref{eq:cqmdel2}) imply that the $D$-wave 
probability in the $\Delta(1232)$ ground state is around 1 \%. 
As a result of such $D$-wave components, the $N$ and $\Delta$ 
charge densities become non-spherical. 
For a static charge distribution, a measure of  
the non-sphericity (or deformation) is given by its quadrupole moment.  
Since the nucleon has spin 1/2, any intrinsic quadrupole moment of the nucleon 
cannot be directly measured though because angular momentum conservation 
forbids a non-zero matrix element of a ($L = 2$) quadrupole operator 
between spin 1/2 states. 
However this quadrupole deformation may reveal itself in an 
electromagnetically induced transition 
from the spin 1/2 $N$ to the spin 3/2 $\Delta$ state,  
see \Figref{fig:m1e2ndel} (right panel). 
In this way, the tensor force between quarks gives rise to non-zero values 
for the $E2/M1$ ($R_{EM}$) and $C2/M1$ ($R_{SM}$) ratios.\footnote{  
The relation between the tensor force, $D$-wave admixture,
and the electromagnetic $N \to \Delta$ transition 
was already pointed out in the early paper of 
Glashow~\cite{Glashow79}.} 
\newline
\indent
Isgur {\it et al.}~\cite{Isgur:1981yz}, having obtained
the $D$-wave admixture reflected in
Eqs.~(\ref{eq:cqmnuc})--(\ref{eq:cqmdel2}), 
find $R_{EM} \simeq -0.41 \%$. 
Similar small negative values, in the range
 $-0.8 \, \% < R_{EM} < -0.3 \, \%$,   
were found in other non-relativistic 
quark model calculations~\cite{Gershtein:1981zf,Bourdeau:1987ih,Gogilidze87}. 
Such a small value for $R_{EM}$ 
already indicates that any effect of deformation in 
the nucleon and/or $\Delta$ ground state is rather small and very sensitive 
to the details of the wave function. 
\newline
\indent
Drechsel and Giannini~\cite{Drechsel:1984ie} 
demonstrated that 
a calculation of the $N \to \Delta$ quadrupole transition within the 
quark model can be very sensitive to a truncation in the quark model basis. 
Due to such truncation (e.g. to a $2 \hbar \omega$ oscillator basis as in 
\cite{Isgur:1981yz})
the charge-current operator for the quadrupole $N \to \Delta$ 
transition violates gauge invariance which can lead to $R_{EM}$ ratios 
which differ by an order of magnitude when calculated either from the charge 
or current operator.  Drechsel and Giannini showed that 
when performing a $2 \hbar \omega$ truncation in the quark model basis, 
a calculation based on the charge operator is more stable than a calculation 
based on the current operator. By using $N$ and $\Delta$ 3-quark 
wave functions which have a $D$-state probability of around 1\%,  
they obtained a value of $R_{EM} \simeq -2 \%$ and a $\Delta$ quadrupole 
moment $Q_\Delta \simeq -0.078$~fm$^2$~\cite{Drechsel:1984ie,Giannini:1990pc}.
\newline
\indent
The error induced due to the truncation in the quark model basis has 
been further investigated in the relativized quark model calculation of 
Capstick and Karl~\cite{Capstick:1989ck,Capstick:1992uc}. Using 
wave functions which have been expanded in a large harmonic 
oscillator basis up to $6 \hbar \omega$ states, and are solutions of a 
relativized Hamiltonian, they obtained an even smaller  
negative value:~$R_{EM} \simeq -0.21 \%$.

\subsubsection{Pion cloud models}

Even though the constituent quark model, despite its simplicity, 
is relatively successful in predicting the structure and spectrum of 
low-lying baryons, it under-predicts $\mu_{N \to \Delta}$ by more than 
25 \% and leads to values for the $R_{EM}$ ratio which are typically 
smaller than experiment. 
More generally, constituent quark models do not satisfy the symmetry 
properties of the QCD Lagrangian. In nature, the up and down (current) quarks 
are nearly massless. In the exact massless limit, the QCD Lagrangian is 
invariant under $SU(2)_L \times SU(2)_R$ rotations of left ($L$) and 
right ($R$) handed quarks in flavor space. This {\it chiral symmetry} 
is spontaneously broken in nature leading to the appearance of massless 
Goldstone modes. For two flavors, there are three Goldstone bosons ---
pions, 
which acquire a mass due to the explicit breaking of chiral 
symmetry by the current quark masses. 
\newline
\indent
Since pions are the lightest hadrons, they 
dominate the long-distance 
behavior of hadron wave functions and yield characteristic 
signatures in the low-momentum transfer behavior of hadronic 
form factors. Furthermore, as the 
$\Delta$(1232) resonance nearly entirely decays into $\pi N$ 
pions are of particular relevance 
to the electromagnetic $N \to \Delta$ transition. 
Therefore, a natural way to qualitatively improve 
on the above-mentioned constituent 
quark models is to include the pionic degrees of freedom. 
\newline
\indent
An early investigation of the $\gamma N \Delta$ transition including 
pionic effects was performed by
Kaelbermann and Eisenberg~\cite{Kaelbermann:1983zb}
within the {\it chiral bag model}.
The chiral (or, cloudy) bag model 
improves the MIT bag model by introducing 
an elementary, perturbative pion which couples to quarks in the bag 
in such a way that chiral symmetry is restored~\cite{Thomas:1982kv}. 
\newline
\indent
Bermuth~{\it et al.}~\cite{Bermuth:1988ms} 
partially corrected and repeated the calculation of 
Ref.~\cite{Kaelbermann:1983zb}
for two formulations of the chiral bag model: 
one with pseudoscalar (PS) surface 
coupling and one with pseudovector (PV) volume coupling of the pions.
It was found that using the charge operator 
to calculate the quadrupole amplitude 
decreases the sensitivity to truncation effects which manifest themselves 
in a violation of gauge invariance in a model calculation. 
For a bag radius around $R \simeq 1.0$~fm, the photo-couplings in the 
chiral bag model are~\cite{Bermuth:1988ms}:
\begin{eqnarray} 
A_{3/2} \,&=&\, -198 \; \; [10^{-3} \;\mathrm{GeV}^{-1/2}], 
\quad \quad R_{EM} = -1.8 \%, \quad \quad (\; \mathrm{PS} \;),  \nn \\
A_{3/2} \,&=&\, -171 \; \; [10^{-3} \;\mathrm{GeV}^{-1/2}], 
\quad \quad R_{EM} = -2.0 \%, \quad \quad (\; \mathrm{PV} \;).  \nn 
\end{eqnarray}
A comparison with experimental data 
and the predictions of quark models, see Table~\ref{table_emr}, shows that 
the chiral bag model calculation for PS coupling yields 
an $A_{3/2}$ amplitude that is larger than 
quark models, but still smaller than experiment. 
For PV coupling, there is a seagull term with opposite sign which yields 
a partial cancellation and deteriorates the agreement with experiment. 
\newline
\indent
Lu {\it et al.}~\cite{Lu:1996rj} in a later calculation within the 
chiral bag model improved upon the previous calculations by 
applying a correction for the center-of-mass motion of the bag. 
It was found that with a smaller bag radius, 
$ R \approx 0.8$ fm, one is able to obtain a reasonable size for the magnitude 
of the helicity amplitudes, see Table~\ref{table_emr}. 
For such a small bag radius, the pionic effects 
are crucial as they account for around 75 \% of the total strength of the 
amplitude $A_{3/2}$. 
\newline
\indent
For the $R_{EM}$ ratio, Lu {\it et al.}~\cite{Lu:1996rj} 
obtained the value $R_{EM} \simeq -0.03 \%$, in
disagreement with  experiment. 
Both Bermuth {\it et al.} and Lu {\it et al.}\ found 
severe cancellations between
the $A_{1/2}$ and $A_{3/2}$ amplitudes in Eq.~(\ref{eq:remhel}), 
leading to a very small value for $R_{EM}$, despite the
`reasonable' values for the helicity amplitudes.
Bermuth {\it et~al.}~\cite{Bermuth:1988ms} therefore calculated $R_{EM}$ 
in the Siegert limit ($|{\bf q}| \to 0$) where it can be obtained 
from $R_{SM}$ using the charge operator, and obtained 
the value $R_{EM} \simeq -2 \%$, close to experiment. 
\newline
\indent
In the chiral bag model  the quadrupole amplitude $E2$ 
has no contribution from the bare bag, which is 
a pure $S$-state. Thus, the transition is  mediated solely
by the pion cloud.
To get a more realistic value for $R_{EM}$ within the chiral bag model, one 
probably must either deform the bag or carry out a 
diagonalization like in the quark model 
(configuration mixing)~\cite{Luprivate}.
\newline
\indent
The pion-cloud contributions also play an important role in 
the {\it linear $\sigma$-model} and 
{\it chiral chromodielectric model} of 
Ref.~\cite{Fiolhais:1996bp}, even though these models still
 under-predict 
the magnitude of the helicity amplitudes, see Table~\ref{table_emr}.
\newline
\indent
The $R_{EM}$ ratio has also been calculated in 
{\it Skyrme models}~\cite{Wirzba:1986sc,Abada:1995db,Walliser:1996ps}, 
which have only pionic degrees of freedom. 
In the Skyrme model, the nucleon appears 
as a soliton solution of an effective nonlinear meson field theory.  
Wirzba and Weise~\cite{Wirzba:1986sc} performed
 a modified Skyrme model calculation,  
at leading order in the number of colors $N_c$, 
based on the chiral effective Lagrangian, which
corresponds with 
the standard Skyrme model supplemented by stabilizing fourth and sixth 
order terms in the pion fields.
This calculation obtained
$R_{EM}$ values 
between $-2.5$\% and $-6$\%, depending on the 
coupling parameters of the stabilizing terms~\cite{Wirzba:1986sc}. 
Certainly, the sign 
and order of magnitude of these results 
are consistent with the empirical result.  
Walliser and Holzwarth~\cite{Walliser:1996ps} included rotational corrections, 
which are  of order $1/N_c$, and lead to a quadrupole distortion 
of the classical soliton solution. 
Including such corrections, one finds a very good description 
of the photo-couplings 
(see Table~\ref{table_emr}) and obtains a ratio $R_{EM} = -2.3 \%$, 
consistent with experiment. 
\newline
\indent
A model which has both quark and pion degrees of freedom 
and interpolates between a constituent quark model and the Skyrme model is 
the {\it chiral quark soliton model} ($\chi$QSM). 
This model is based on the interaction 
of quarks with Goldstone bosons resulting from the spontaneous breaking of 
chiral symmetry. 
As for the Skyrme model, 
the $\chi$QSM is essentially based on the $1/N_c$ expansion.  
Its effective chiral action has been
derived from the instanton model of the QCD vacuum \cite{Dia86}, which
provides a natural mechanism of chiral symmetry breaking 
and enables one to generate dynamically the constituent 
quark mass.  
\newline
\indent
Although in reality the number of colors $N_c=3$,
the extreme limit of large $N_c$ is 
known to yield useful insights. At
large $N_c$ the nucleon is heavy and can 
be viewed as $N_c$ ``valence" quarks bound by a self-consistent pion
field (the ``soliton") whose energy coincides with the aggregate
energy of the quarks of the negative-energy Dirac continuum~\cite{Dia88}.
A successful description of static properties of baryons, 
such as mass splittings, axial constants, magnetic moments, 
form factors, has been achieved (typically at the 30 \% level or better, 
see Ref.~\cite{Chr96} for a review of early results). 
After reproducing masses and decay constants in the mesonic sector, 
the only free parameter left to be fixed in the baryonic sector
is the constituent quark mass
The good agreement of the $\chi$QSM with the empirical
situation is achieved for quark mass $M_q \simeq 420$~MeV.
\newline
\indent
The $\chi$QSM was also applied to the calculation of the 
$\gamma N \Delta$ transition~\cite{Watabe:1995xy,Silva:1999nz}. 
In this model, the $\Delta$ is a bound 
state which corresponds to a soliton rotating in flavor space. 
The rotational energy is responsible for the $N - \Delta$ mass splitting.
With a constituent quark mass $M_q = 420$~MeV, the empirical 
$N - \Delta$ mass splitting is well reproduced. 
The $\gamma N \Delta$ amplitudes 
are obtained taking rotational ($1/N_c$) corrections into account. In this 
way, Silva {\it et al.}~\cite{Silva:1999nz} obtained as result: 
$R_{EM} = -2.1 \%$ (in the two-flavor case), 
fairly close to the experimental ratio, considering that 
in the $\chi$QSM calculation no parametrization adjustment has been made 
to the $N \to \Delta$ transition.    
However, the value of the $M1$ $\gamma N \Delta$ amplitude is   
largely underpredicted in the $\chi$QSM, 
which is also reflected in an underprediction of the magnitude of the 
photo-couplings, see Table~\ref{table_emr}.
\newline
\indent
The above calculations incorporating the chiral symmetry of QCD 
(to lowest order in the pion fields) highlight the 
role of the pionic degrees of freedom in the $\gamma N \Delta$ transition.  
A number of subsequent works have therefore revisited 
quark models, which break chiral symmetry, by including 
two-body exchange currents.
As an example of such approach we consider the work
of Buchmann, Hernandez and Faessler~\cite{Buchmann:1996bd}. 
\newline
\indent
When the one-gluon exchange potential between quarks is 
complemented by a one-pion and one-sigma exchange potential between the 
quarks, the pion exchange gives rise to an additional 
tensor interaction between the quarks. 
The one-pion exchange  then requires  
the presence of two-body exchange currents between the quarks
due to current conservation. 
Within the non-relativistic framework, Buchmann 
{\it et al.}~\cite{Buchmann:1996bd} found that the 
overall effect of the exchange 
currents on the dominant $M1$ $\gamma N \Delta$ 
transition are relatively small due to cancellation. As a result, 
the transition magnetic moment comes out 
to be underpredicted: $\mu_{p \to \Delta^+} \simeq 2.5 \, [\mu_N]$, 
similar to the non-relativistic quark model without exchange currents. 
This is also reflected in the 
helicity amplitudes obtained in~\cite{Buchmann:1996bd} and
quoted in Table~\ref{table_emr}. 
They are smaller in the magnitude than the experimental values.
\newline
\indent 
However, the two-body exchange 
currents lead to non-vanishing $\gamma N \Delta$ quadrupole 
($E2$ or $C2$) amplitudes~\cite{Buchmann:1996bd}, 
even if the quark wave functions have no $D$-state 
admixture. In this picture, the $\Delta$ 
is excited by flipping the spins of two quarks, 
see \Figref{fig:e2ndel2}. According to Buchmann 
{\it et al.}, this mechanism yields $R_{EM} \simeq -3.5 \%$. 
\begin{figure}
\centerline{  \epsfysize=3cm
  \epsffile{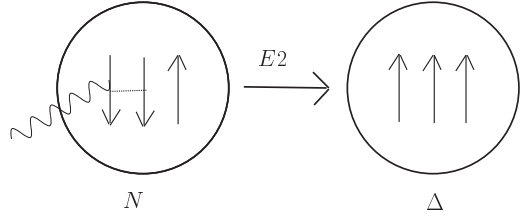} 
}

\caption{Schematic picture within a quark model of a $E2$  
$N \to \Delta$ transition 
induced by the interaction of a photon with two quarks in the nucleon 
(correlated through, {\it e.g.}, one-pion exchange). Both $N$ and $\Delta$ wave 
functions are $S$-states.
}
\figlab{fig:e2ndel2}
\end{figure}
\newline
\indent
Also within their quark model, Buchmann 
{\it et al.} related 
the $N \to \Delta$ and $\Delta^+$ quadrupole moments to the neutron 
charge radius as\footnote{See also 
Refs.~\cite{Grabmayr:2001up,Buchmann:2004ia} for 
an analogous relation 
between the $N \to \Delta$ quadrupole form factor $G_C^\ast$ 
and the neutron electric form factor. 
Possible higher order corrections to Eq.~(\ref{eq:buchrel}) 
allowed by QCD were recently investigated 
in Ref.~\cite{Dillon:1998zm} within 
a general parametrization method, see also Ref.~\cite{Blanpied:2001ae}.}: 
\begin{eqnarray}
Q_{p \to \Delta^+} = \frac{1}{\sqrt{2}} \, r_n^2 , \quad \quad \quad \quad 
Q_{\Delta^+} = r_n^2 \, . 
\label{eq:buchrel}
\end{eqnarray}
The origin of these relations in the model calculation of 
Ref.~\cite{Buchmann:1996bd} is the dominance of two-body exchange currents 
in both $Q_{p \to \Delta^+}$, $Q_{\Delta^+}$ and $r_n^2$. 
Using the experimental neutron charge radius, $r_n^2 = -0.113 (3)$~fm$^2$, 
the relations of Eq.~(\ref{eq:buchrel}) yield:
\begin{eqnarray}
Q_{p \to \Delta^+} = - 0.08 \; \mathrm{fm}^2, \quad \quad \quad \quad 
Q_{\Delta^+} = - 0.113 \; \mathrm{fm}^2. 
\end{eqnarray}
The value of $Q_{p \to \Delta^+}$ is close to the empirical determination of 
Eq.~(\ref{eq:qndelexp}). 
\newline
\indent
Recently, the $\gamma^* N \Delta$ transition form factors were calculated 
in a Lorentz covariant chiral quark approach~\cite{Faessler:2006ky}, 
where baryons are modeled as 
bound states of constituent quarks dressed by a cloud of pseudoscalar mesons. 
To calculate baryon matrix elements, a phenomenological hadron-quark vertex 
form factor was used, 
which parametrizes the distribution of quarks inside a given baryon. 
In Ref.~\cite{Faessler:2006ky}, a Gaussian form was adopted for this 
hadron-quark form factor which involves a size parameter $\Lambda_B$, which 
was used as a free parameter. Using a value $\Lambda_B \simeq 0.8$~GeV, 
it was found that the contribution of the 
meson cloud to the static properties of light baryons is up to 20 \% and, 
together with the relativistic corrections, helps to explain how the 
shortfall in the SU(6) prediction for $\mu_{N \to \Delta}$ is ameliorated. 
\newline
\indent
As a summary, we list in Table~\ref{table_emr} 
the $\gamma N \Delta$ photo-couplings $A_{1/2}$ and $A_{3/2}$ as well as the 
ratio $R_{EM}$ in the various models discussed above.
\begin{table}[h]
{\centering \begin{tabular}{|c|c|c|c|}
\hline
&&& \\ &
$A_{1/2}$ $[10^{-3}~\mathrm{GeV}^{-1/2}]$ & 
$A_{3/2}$ $[10^{-3}~\mathrm{GeV}^{-1/2}]$ & $R_{EM}$ [\%]  \\
&&& \\
\hline 
\hline
experiment &&& \\
Ref.~\cite{PDG2006} & $-135 \pm 6$ & $-250 \pm 8$ & $-2.5 \pm 0.5$  \\
\hline
\hline 
SU(6) symmetry & -107 & -185 & 0 \\
\hline 
non-relativistic quark models &&& \\ 
Refs.~\cite{Koniuk:1979vy,Isgur:1981yz,Gershtein:1981zf,Bourdeau:1987ih,Gogilidze87,Drechsel:1984ie}  
& -103 & -179 & $-2 $  to  0 \\
\hline 
relativized quark model &&& \\
Refs.~\cite{Capstick:1989ck,Capstick:1992uc} 
& -108 & -186 & $-0.2$ \\
\hline 
MIT bag model, Ref.~\cite{Donoghue:1975yg}
& -102 & -176 & 0 \\
\hline 
chiral (cloudy) bag models &&& \\
Ref.~\cite{Bermuth:1988ms} : (PS, R = 1 fm)
& -106 & -198 & $-1.8$ \\
Ref.~\cite{Bermuth:1988ms} : (PV, R = 1 fm)
& -91 & -171 & $-2.0$ \\
Ref.~\cite{Lu:1996rj} : (+ recoil, R = 0.8 fm)
& -134 & -233 & $-0.03$ \\
\hline 
linear $\sigma$-model, Ref.~\cite{Fiolhais:1996bp} 
& -107 & -199 & -1.8 \\
\hline 
chromodielectric model, Ref.~\cite{Fiolhais:1996bp} 
& -70 & -131 & -1.9 \\
\hline 
Skyrme models
& & &  \\
Ref.~\cite{Wirzba:1986sc} & & & $-6$ to $-2.5$ \\
Ref.~\cite{Walliser:1996ps} : (+ $1/N_c$ corrections) & -136 & -259 & $-2.3$ \\
\hline 
chiral quark soliton model 
& & &  \\
Ref.~\cite{Silva:1999nz} : SU(2) flavor & -70.5 & -133 & -2.1 \\
\hline 
quark model + $\pi, \sigma$ exchange 
& & &  \\
Ref.~\cite{Buchmann:1996bd} & -91 & -182 & -3.5 \\
\hline 
chiral quark model ($\Lambda_B = 0.8$~GeV)  
& & &  \\
Ref.~\cite{Faessler:2006ky} & -124.3 & -244.7 & -3.1 \\
\hline 
\hline
\end{tabular}\par}
\caption{Summary of the values of the $\gamma N \Delta$ helicity amplitudes 
$A_{1/2}$ and $A_{3/2}$ at $Q^2 = 0$, and the ratio $R_{EM} (Q^2 = 0)$ 
in different models compared with experiment.}
\label{table_emr}
\end{table}

\subsubsection{Intrinsic quadrupole moment and more about shape}

The relations of Eq.~(\ref{eq:buchrel}) were further interpreted by
Buchmann and Henley~\cite{Buchmann:2001gj} in terms of an intrinsic 
quadrupole moment of the nucleon and $\Delta$ states. 
The intrinsic quadrupole moment $Q^0$ of a static charge distribution 
$\rho(\vec r)$ is given by:
\begin{eqnarray}
Q^0 \,=\, \int d^3 \vec r \, \rho(\vec r) \, ( 3 \, z^2 - r^2 ),
\label{eq:qo}
\end{eqnarray}
which is defined w.r.t. the body fixed frame. A charge distribution 
concentrated along the $z$-axis (symmetry axis of the system) corresponds 
with $Q^0 > 0$ (prolate deformation), whereas a charge distribution 
concentrated in the equatorial $xy$-plane corresponds with $Q^0 < 0$ 
(oblate deformation). This intrinsic quadrupole moment has to be 
distinguished from a measured (or spectroscopic) 
quadrupole moment $Q$. As an example, 
for a rigid rotor (which was considered  
within the context of the collective nuclear shell model~\cite{BM75}) 
these quantities are related as:
\begin{eqnarray}
Q \,=\, \frac{3 J_z^2 - J (J + 1)}{(J + 1)(2 J + 3)} \, Q^0,
\label{eq:qoqlab}
\end{eqnarray}
where $J_z$ is the projection of the nucleon total spin $J$ onto its symmetry 
axis ($z$-axis in a body fixed frame). The difference between $Q^0$ and 
$Q$ represents the averaging of the nonspherical charge distribution 
due to its rotational motion as seen in the laboratory frame. 
One verifies from Eq.~(\ref{eq:qoqlab}) that the multiplication factor is 
zero for a spin 1/2 particle, yielding $Q_p = 0$ for the proton. 
Eq.~(\ref{eq:qoqlab}) does not preclude however that the proton 
has an intrinsic quadrupole moment $Q^0_p$.
\newline
\indent
Within a model where 
the nucleon consists of a spherically symmetric 
quark core surrounded by a pion with orbital angular momentum $l = 1$, 
Buchmann and Henley obtain~\cite{Buchmann:2001gj}:
\begin{eqnarray}
Q^0_{\Delta^+} \,=\, r_n^2 \,=\, - Q^0_{p}\,.
\label{eq:qopdel}
\end{eqnarray}
Thus, the proton and $\Delta^+$ 
have respectively a prolate and an oblate 
intrinsic deformation. In this hybrid (quark/pion-cloud) 
model, the pion cloud is fully
responsible for the non-zero values of the quadrupole moments 
$Q^0_p$ and $Q^0_{\Delta^+}$, and hence for the non-spherical shape
of these particles, see \Figref{shapepion}. 
\begin{figure}
\centerline{  \epsfxsize=10cm
  \epsffile{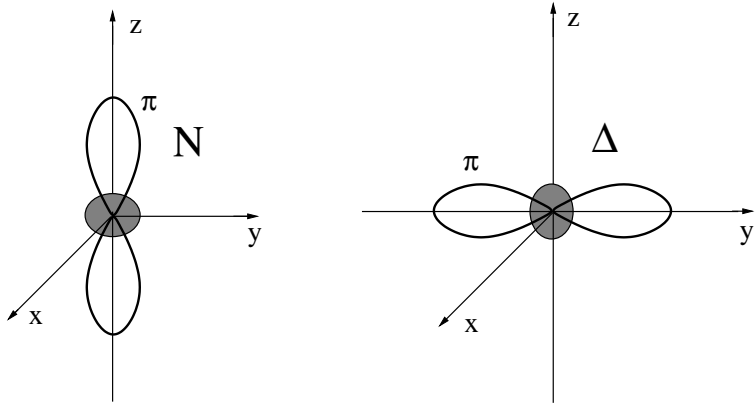} 
}
\caption{Model picture of the intrinsic quadrupole deformation of the nucleon 
(left) and $\Delta$ (right) within the pion cloud model of 
Ref.~\cite{Buchmann:2001gj}. In the $N$, the $p$-wave pion cloud 
is concentrated along the polar (symmetry) axis leading to a prolate 
deformation, whereas in the $\Delta$, the pion cloud is concentrated 
in the equatorial plane, corresponding to an oblate deformation.  
Figure from Ref.~\cite{Buchmann:2001gj}.
}
\figlab{shapepion}
\end{figure}
\newline
\indent
Estimates such as Eq.~(\ref{eq:qopdel}) of intrinsic quadrupole moments are 
surely useful to reveal details of a given model calculation and gain
physical insight. One should keep in mind, however,
that, even though a specific model such as the rigid rotor
relates the intrinsic and total quadrupole moments,
only the latter is directly related to observables.  

\subsection{Large $N_c$ limit}
\label{sec:largenc}

Though the results from the different QCD inspired models discussed above may 
provide us with physical insight on the $\gamma N \Delta$ transition, 
they are not a rigorous consequence of QCD. In the following subsections, we 
will discuss what is known on the $\gamma N \Delta$ transition 
from approaches which are a direct consequence 
of QCD in some limit, such as chiral effective field theory (chiral limit 
of small pion masses or momentum transfers) or 
lattice QCD simulations (continuum limit). 
\newline
\indent
Another first-principle technique is the 
$1/N_c$ expansion of QCD proposed by 't Hooft~\cite{'tHooft:1973jz}
and Witten~\cite{Witten:1979kh}. 
The large-$N_c$ limit is unique in that
it provides a perturbative parameter at all energy scales.
This expansion has proved quite useful in describing the properties 
of baryons, such as, ground-state and excited baryon masses, magnetic moments, 
electromagnetic decays, 
see {\it e.g.}, Refs.~\cite{Jenkins:1998wy,Lebed:1998st} 
for reviews.
\newline
\indent
In the large $N_c$ limit, the baryons are infinitely heavy and 
can be treated as static. In this limit 
the baryon sector of QCD has an exact contracted 
$SU(2 N_f)$ spin-flavor symmetry, where $N_f$ is the number of light quark 
flavors. 
The large $N_c$ limit thus validates many of the quark model 
$SU(6)$ spin-flavor symmetry results without making any 
model assumption (such as assuming non-relativistic quark dynamics 
for the baryon wave functions).
For example,  the ratio of proton to neutron magnetic 
moments in the large-$N_c$ limit is predicted to be given by 
$\mu_p / \mu_n = -3/2$ in agreement with the naive quark model. 
Likewise at leading order in $1/N_c$, the $N \to \Delta$ transition 
magnetic moment $\mu_{N \Delta}$ is related to the isovector combination 
of proton and neutron magnetic moments as~\cite{Jenkins:1994md}:
\begin{eqnarray}
\mu_{p \to \Delta^+} = \frac{1}{\sqrt{2}} \, \left( \mu_p - \mu_n \right),
\label{eq:mulargenc}
\end{eqnarray}
up to a correction of order $1/N_c^2$.
Using the empirical values for $\mu_p$ and $\mu_n$, one obtains in the large 
$N_c$ limit $\mu_{N \to \Delta} = 3.23 \; \mu_N$, within 10 \% of the 
experimental value of Eq.~(\ref{eq:mundelexp}). 
\newline
\indent
The $R_{EM}$ ratio for the $\gamma N \Delta$ transition is shown to 
be of order $1/N_c^2$~\cite{Jenkins:2002rj}. Thus, 
the smallness of the $\gamma N \Delta$ 
$R_{EM}$ ratio is naturally explained in the large $N_c$ limit.
Using Eq.~(\ref{eq:remhel}), 
this can equivalently be expressed as a 
large $N_c$ prediction for the $\gamma N \Delta$ helicity amplitudes:
\begin{eqnarray}
\frac{A_{3/2}}{A_{1/2}} \,=\, \sqrt{3} \,+\, 
{\mathcal O} \left(\frac{1}{N_c^2} \right).
\end{eqnarray}
\newline
\indent 
The large $N_c$ limit also allows to obtain relations between the 
$\Delta$ and $N \to \Delta$ quadrupole moments~\cite{Buchmann:2002mm}:
\begin{eqnarray}
\frac{Q_{\Delta^+}}{Q_{p \to \Delta^+}} \,=\, \frac{2 \sqrt{2}}{5} \,+\, 
{\mathcal O} \left(\frac{1}{N_c^2} \right).
\label{eq:qlargenc}
\end{eqnarray}
This result for $Q_{\Delta^+}$ is different from the quark model 
ratio of Eq.~(\ref{eq:buchrel}) as explained in \cite{Buchmann:2002mm}.
Using the phenomenological value of Eq.~(\ref{eq:qndelexp}) 
for $Q_{p \to \Delta^+}$, the large $N_c$ relation of Eq.~(\ref{eq:qlargenc}) 
yields for the $\Delta^+$ quadrupole moment:
\begin{eqnarray}
Q_{\Delta^+} \,&=&\, - \left( 0.048 \pm 0.002 \right ) \; \mathrm{fm}^2,
\label{eq:qdellargenc}
\end{eqnarray} 
accurate up to corrections of order $1/N_c^2$.
\newline
\indent
Buchmann, Hester and Lebed~\cite{Buchmann:2002mm} 
derived recently another relation by 
making the additional assumption that the baryon charge radii
and quadrupole moments arise from the same contributions.
This assumption holds true when, {\it e.g.}, the 
interaction between quarks arises from one-gluon or one-pion exchange, for 
which the ratio of spin-spin and tensor terms in the Hamiltonian is fixed. 
Under this assumption, one finds a large-$N_c$ relation between 
the $N \to \Delta$ quadrupole moment and the neutron charge radius 
$r_n^2$~\cite{Buchmann:2002mm}:
\begin{eqnarray}
Q_{p \to \Delta^+} \,=\, \frac{1}{\sqrt{2}} \, r_n^2 \, 
\frac{N_c}{N_c + 3} \, \sqrt{\frac{N_c + 5}{N_c - 1}},
\label{eq:qpdelrnlargenc}
\end{eqnarray}
where both $Q_{p \to \Delta^+}$ and $r_n^2$ are of order $1/N_c^2$. 
\newline
\indent
The factor on the {\it rhs} of Eq.~(\ref{eq:qpdelrnlargenc}) 
after $r_n^2$ is unity both for the cases 
$N_c = 3$ and $N_c \to \infty$, and deviates from unity by only about 1 \% 
for all values in between. Therefore the large $N_c$ limit predicts to good 
approximation the same relation of Eq.~(\ref{eq:buchrel}), 
$Q_{p \to \Delta^+} \,=\, 1/\sqrt{2} \, r_n^2 $, which  
was also derived in the quark model and pion cloud model, 
and which was found to 
be consistent with the empirical value of Eq.~(\ref{eq:qndelexp}) 
for $Q_{p \to \Delta^+}$.
\newline
\indent
More recently, Cohen and Lebed~\cite{Cohen:2002sd,Cohen:2006up} 
have argued that 
the relations, such as in Eqs.~(\ref{eq:qlargenc}) and
(\ref{eq:qpdelrnlargenc}), may be significantly altered in the real world 
where the pions are 
light. 
When taking $N_c \to \infty$ at a fixed value of $m_\pi$, the $N$-$\Delta$ 
mass difference, which goes as $1/N_c$, vanishes. Therefore, the ``usual'' 
large-$N_c$ limit implies that one is in the region 
$M_\Delta - M_N \ll m_\pi$, where, for instance, the
$\De$ is stable. Cohen points out~\cite{Cohen:2002sd}
that the large-$N_c$ limit 
and the chiral limit ($m_\pi \to 0$) do not commute and 
for quantities which diverge in 
the chiral limit, such as the charge radii, one expects 
chiral corrections to dominate over the large-$N_c$ 
predictions. 
\newline
\indent
Although the region  
$M_\Delta - M_N \ll m_\pi$ is not accessible in nature, 
lattice QCD results, which are currently obtained in this regime 
(see Sect.~\ref{sec2_lattice}), may provide an interesting testing ground 
for the above large $N_c$ predictions. 
\newline
\indent
Let us conclude the discussion of the large-$N_c$ relations by observing
a new one, {\it i.e.}:
\beq
\eqlab{newlnc}
R_{SM}=R_{EM} \,, \,\,\,\, \mbox{for $Q^2 = 0$ and $N_c\to\infty$}\,.
\eeq
To derive this relation we follow the arguments given in the Appendix
of  Ref.~\cite{Pascalutsa:2003zk}. 
Namely, we first make use of the expression
of the ratios in terms of the little $g$'s of \Eqref{diagndel1}, 
at $Q^2=0$:
\begin{subequations}
\eqlab{REMlargeNc}
\bea
R_{EM}&=&\frac{1}{g_M 
-  \frac{\vDe}{2(M_N+M_\De)}g_E}\,\frac{\vDe}{2(M_N+M_\De)}g_E\,,\\
R_{SM}&=&\frac{1}{g_M 
-  \frac{\vDe}{2(M_N+M_\De)}g_E}\,
\left[\frac{\vDe}{2(M_N+M_\De)}g_E + \frac{\vDe^2}{4M_\De^2} g_C\right]\,,
\eea
\end{subequations}
where $\vDe=M_\De-M_N$ is the $N$-$\Delta$ mass difference.
These are general expressions, which can, {\it e.g.}, be obtained 
by substituting \Eqref{JS} into \Eqref{ratios}. 
Now, the little $g$'s are the coupling constants from
 an effective Lagrangian (see Sect.~\ref{sec:sec4ss3}) and, 
as can be seen from the
similar operator structure of the corresponding terms, these constants
have the same large-$N_c$ scaling. Finally, we only need to recall
that the baryon masses go as $N_c$, while the $N$-$\De$ mass difference
as $1/N_c$:
\beq
M_N,\, M_\De \sim N_c,\,\,\, \vDe \sim 1/N_c \,\,\, \mbox{ for $N_c\to\infty$},
\eeq
 to see that the $g_C$ term in $R_{SM}$ is 
of higher $1/N_c$ order, and hence the relation \Eqref{newlnc} holds.
From \Eqref{REMlargeNc}, one immediately verifies that both of these 
quantities are of order $1/N_c^2$.

\subsection{Chiral effective field theory}
\label{sec2_eft}

The $\ga N\to \De$ transition has also been studied within
the chiral effective field theory ($\chi$EFT) 
expansions based on chiral Lagrangians
with nucleon and $\De$-isobar fields. In this
framework the expansion for each of the $\ga^\ast N\De$-transition form factors
begins with a low-energy constant (LEC), which then receives the
chiral loop corrections of the type depicted in \Figref{gandelabs}.
\newline
\indent
A first such study was performed by 
Butler, Savage and Springer~\cite{Butler:1993ht} 
in the framework of heavy-baryon $\chi$PT~\cite{JeM91a}. At leading order
they obtained a ``chiral log'' ({\it i.e.}, $\ln m_\pi$) enhancement
of the $E2$ transition, which lead to relatively large and positive
values of $R_{EM}$. Their result thus showed that the chiral correction
to the $E2$ transition diverges in the chiral limit. 
\begin{figure}[t,p]
\centerline{\epsfxsize=10cm
\epsffile{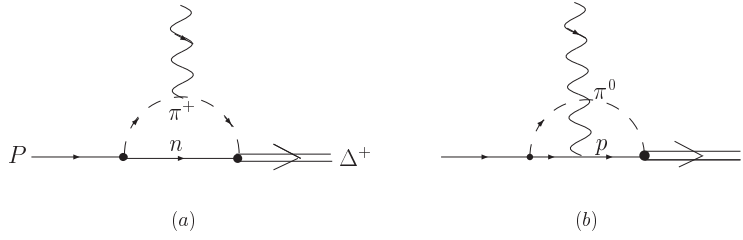}
}
\caption{The $\pi N$ loop corrections to the $\gamma p \Delta^+$ vertex 
as calculated in $\chi$EFT. 
Diagram (a) : $\pi^+ n$ loop where the photon couples to the $\pi^+$; 
diagram (b) : $\pi^0 p$ loop where the photon couples to the charge of the 
proton. } 
\figlab{gandelabs}
\end{figure}
\newline
\indent
A more comprehensive study was subsequently carried out by
Gellas {\it et al.}~\cite{Gellas:1998wx} using
the so-called ``small scale expansion'' (SSE)~\cite{HHK97}, also called 
``$\eps$-expansion''.
In the SSE scheme the two light scales in the problem:
the pion mass ($m_\pi$) and the $\De$-resonance excitation
energy ($\vDe \equiv M_\Delta - M_N$) are counted
as having the same size. In addition, the heavy-baryon
(semi-relativistic) expansion is performed. The calculation of
Ref.~\cite{Gellas:1998wx} was performed to order $p^3$, which is the
next-to-leading order for $M1$ and leading order for the $E2$ and $C2$ 
transitions. 
More recently, Gail and Hemmert~\cite{Gail:2005gz}  have 
updated this work and analyzed some of the higher-order contributions.
In overall, they show a good agreement with the low $Q^2$ data for the
$\ga N\to \De$ transition, as will be illustrated below. However, 
as a note of caution, one should point out that
the $R_{EM}$ and $R_{SM}$ ratios are reasonably well described only upon
adding a nominally higher-order ($p^5$) counter term with an
 unnaturally large LEC ($C_s\simeq -17$~GeV$^{-2}$ in the 
notation of Ref.~\cite{Gail:2005gz}). 
\newline
\indent 
Two of us~\cite{Pascalutsa:2005ts,Pascalutsa:2005vq} have recently
computed the
$\ga^\ast N\De$ form factors in a manifestly covariant
(no heavy-baryon expansions) $\chi$EFT expansion.
In the chiral limit, it was found that
the $C2$ transition diverges for $Q^2=0$, which is also 
in agreement with the result of
Ref.~\cite{Gail:2005gz}\footnote{This result of 
both Refs.~\cite{Gail:2005gz} and 
\cite{Pascalutsa:2005ts,Pascalutsa:2005vq} is in disagreement with  
Butler {\it et al.}~\cite{Butler:1993ht} 
where a chiral log enhancement was reported for $E2$ at $Q^2 = 0$ 
instead of $C2$.}.
The coefficient of the 
chiral log turns out to be small. Only for
tiny pion masses one starts to see the dominance of the chiral log
in the $C2$ transition, see \Figref{chirallog}. 
\begin{figure}[t,p]
\centerline{\epsfxsize=10cm
\epsffile{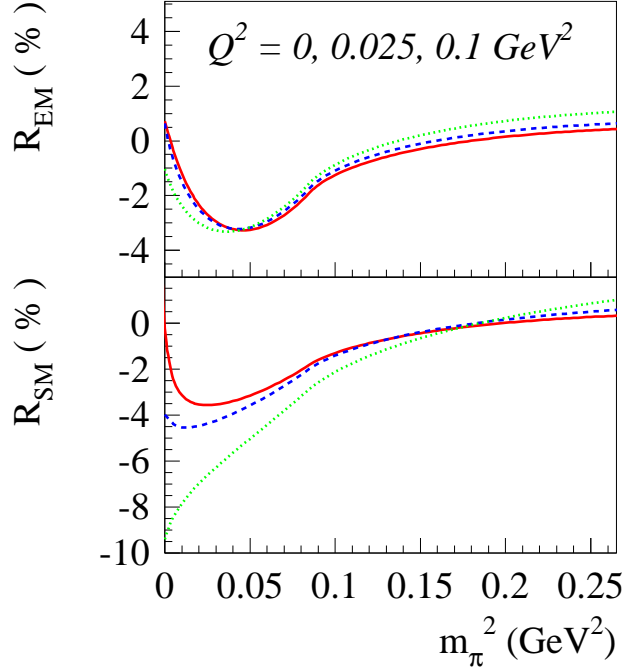}
}
\caption{The pion mass dependence of $R_{EM}$ and $R_{SM}$ 
for three fixed values of $Q^2$: 0 (solid curves), 
0.025~GeV$^2$ (dashed curves), and 0.1~GeV$^2$ (dotted curves), resulting
from the $\chi$EFT calculation 
of Refs.~\cite{Pascalutsa:2005ts,Pascalutsa:2005vq}. The figure shows
the absence (presence) of the $\ln m_\pi$
divergence for $R_{EM}$ ($R_{SM}$) at $Q^2=0$.}
\figlab{chirallog}
\end{figure}
\newline
\indent
Our $\chi$EFT study of the $\ga N\De$ transition~\cite{Pascalutsa:2005ts,Pascalutsa:2005vq} 
utilizes the ``$\delta$-expansion'' scheme of Ref.~\cite{Pascalutsa:2002pi}. 
In this scheme the two light scales 
$\delta \equiv \vDe / \Lambda_{\chi \mathrm{SB}}$ and 
$\eps \equiv m_\pi / \Lambda_{\chi \mathrm{SB}}$, 
with $\Lambda_{\chi \mathrm{SB}} \sim 1$~GeV 
the chiral symmetry breaking scale, are treated differently. 
In contrast, the SSE assumes $\eps \sim \delta$, which leads to an
unsatisfactory feature that the $\Delta$-resonance contributions are always 
estimated to be of the same size as the nucleon contributions (hence,  
overestimating the $\Delta$ contribution at lower energies and 
underestimating some of them at the resonance energies). 
In the $\delta$-expansion, one counts $\eps = \delta^2$, 
which is the closest integer-power relation
between these parameters in the real world.
\newline
\indent 
At the physical pion mass, the $\de$-expansion provides an 
energy-dependent power-counting scheme designed to take into
account the large variation of the $\De$-resonance contributions
with energy. As such it allows for an efficient calculation of
observables in the resonance region. The relevant LECs can in this 
fashion be directly extracted from observables. 
In Refs.~\cite{Pascalutsa:2005ts,Pascalutsa:2005vq} the three
LECs corresponding to the three $\ga N\De$ transitions
are all extracted from a next-to-leading (NLO) calculation of 
pion photo- and electro-production observables (see Sect.~4.5.2). 
The resulting prediction of the $Q^2$ dependence is
then shown to be in a qualitative agreement with phenomenological 
extractions.
\begin{figure}
\centerline{  \epsfxsize=15cm
  \epsffile{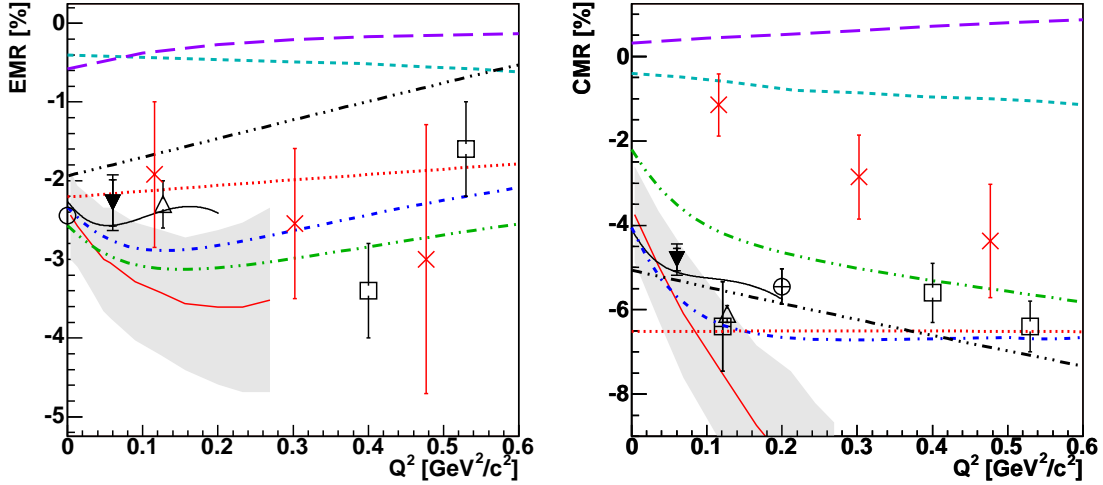} 
}
\caption{
The low $Q^2$ dependence of the quadrupole ratios 
$EMR$ (left panel) and $CMR$ (right panel) for the $\gamma^* p \to \Delta$ 
transition. 
The data are from MAMI : Refs.~$\bigcirc$~\cite{Beck:1999ge}, 
$\bigoplus$~\cite{Elsner:2005cz}, $\boxplus$~\cite{Pospischil:2000ad}, 
$\blacktriangledown$~\cite{Stave:2006ea}; 
from Bates $\triangle$ \cite{Sparveris:2004jn}; 
and from CLAS $\square$ \cite{Joo:2001tw}.  
The lattice QCD calculations with linear pion mass extrapolations 
are shown as $\times$ \cite{Alexandrou:2004xn}.  
Two chiral EFT calculations are shown: the $\delta$-expansion result from 
Ref.~\cite{Pascalutsa:2005ts,Pascalutsa:2005vq} (red solid curves with error 
estimates),   
and the $\varepsilon$-expansion result of Ref.~\cite{Gail:2005gz} 
(black solid curves). 
The dynamical model predictions from SL~\cite{SL} 
(green dashed-double-dotted curves) 
and DMT~\cite{KY99} (blue dashed-dotted curves) are shown 
alongside the phenomenological MAID2003~\cite{MAID98} (red dotted curves) 
and SAID~\cite{Arndt:2002xv} (black dashed-triple-dotted curves) models. 
The hypercentral (long dashed curves) \cite{DeSanctis:2005vq} and
relativistic (short dashed curves) \cite{Capstick:1989ck}  
constituent quark models have been included. 
Figure from Ref.~\cite{Stave:2006ea}.
}
\figlab{ratiosqsqr}
\end{figure}
\newline
\indent
Namely, in \Figref{ratiosqsqr}, we show the present experimental situation 
for the $R_{EM}$ and $R_{SM}$ ratios at low $Q^2$ and compare the data 
with $\chi$EFT calculations, dynamical and phenomenological 
model predictions,  as well as quark model predictions. 
One sees from \Figref{ratiosqsqr}, that the $R_{EM}$ ratio stays small 
and negative, around $-2$ to $-3$ \% up to momentum-transfer of
around 0.5 GeV$^2$. 
\newline
\indent
The $R_{SM}$ ratio displays a steep slope at low $Q^2$ (Siegert limit), and 
seems to level off to a negative value around $-6$ to $-7$ \% 
above $Q^2 \simeq 0.1$~GeV$^2$. 
Both the $\chi$EFT calculations are consistent with the low $Q^2$ 
dependence of the $R_{EM}$ and $R_{SM}$ ratios. 
The $\delta$-expansion results are quoted with a theoretical uncertainty 
band as will be explained in Sect.~\ref{sec4}. 
Furthermore, one sees from \Figref{ratiosqsqr} that the constituent quark 
models largely under-predict the $R_{EM}$ and $R_{SM}$ ratios, as discussed 
above. The dynamical models of Refs.~\cite{SL,KY99}, 
which include pionic degrees of freedom, 
are in qualitative agreement with the empirical $Q^2$ dependence. 
\begin{figure}
\centerline{  \epsfxsize=9cm
  \epsffile{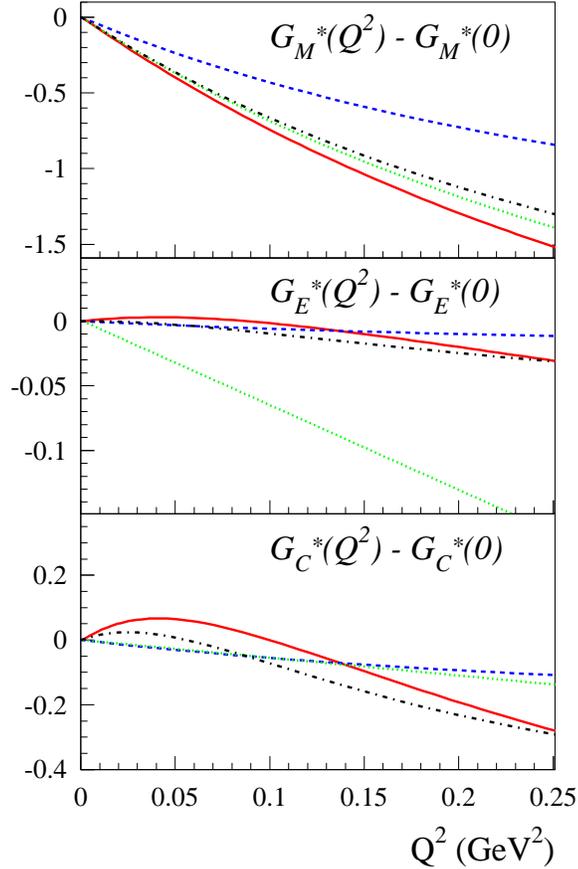} 
}
\caption{
Comparison of the pion cloud in the 
dynamical model and chiral EFT  
calculations for the low $Q^2$ behavior 
of the Jones Scadron form factors 
$G_M^*$, $G_E^*$, and $G_C^*$ (with value at $Q^2 = 0$ subtracted). 
The dynamical model predictions from SL~\cite{SL} are shown for both the 
'bare' version of the model (blue dashed curves) where the pion loop diagrams  
are not accounted for, 
and 'dressed' version of the model (black dashed dotted curves) 
which include the pion loop diagrams.  
The chiral EFT calculation of 
Ref.~\cite{Pascalutsa:2005ts,Pascalutsa:2005vq} are shown 
without (green dotted curves) and with (red solid curves) the 
pion loop diagrams. 
}
\figlab{pioncloud_eftsl}
\end{figure}
\newline
\indent
In \Figref{pioncloud_eftsl}, we compare $Q^2$-dependence
of the the pion cloud contributions 
to the $\gamma^\ast N \Delta$ form factors in 
dynamical models versus $\chi$EFT. 
In $\chi$EFT the values of the form factors 
$G_M^\ast$, $G_E^\ast$, and $G_C^\ast$ at the real photon point 
are related to three LECs $g_M$, $g_E$, and $g_C$ 
appearing in the chiral Lagrangian (see Sect.~\ref{sec4}), and 
determined from a fit to the data.  
Similarly, in the dynamical models the values at 
$Q^2 = 0$ are determined from a fit to the same data. However,
the renormalization of the pion loop contributions is 
done in a dramatically different fashion. In $\chi$EFT one
uses the dimensional regularization and absorbs infinities and
other power-counting violating pieces into the LECs. In dynamical
models, the loops are regularized by a cutoff, which is then
fitted to the data. As the result the pion cloud contributions
do not satisfy the chiral power-counting, and are usually larger
than the ones obtained from $\chi$EFT.
\newline
\indent
Therefore, to compare the dynamical model and $\chi$EFT results directly
we consider  the difference 
$G_M^\ast(Q^2) - G_M^\ast(0)$, and analogously for $G_E^\ast$ 
and $G_C^\ast$, such that the above renormalization-scheme
dependence drops out. In \Figref{pioncloud_eftsl}, we compare
the results of the dynamical model of Ref.~\cite{SL} 
with the $\chi$EFT results 
of Ref.~\cite{Pascalutsa:2005ts,Pascalutsa:2005vq}. 
For both the dynamical model and $\chi$EFT we show
the results with and without the 
pion-loop effects, which for these difference-quantities
appear to be similarly renormalized.
\newline
\indent
From this comparison, we conclude that  
the full result, including the pion-loop contributions, 
is very similar in both the dynamical model and the $\chi$EFT
approach,
for all the three form factors. 
The pion cloud contribution to
$G_M^\ast$ appears to be larger in the dynamical model than in the 
$\chi$EFT calculation. This difference in size is mainly due to 
the photon coupling to the anomalous magnetic moments 
of the nucleons in the loops, which are included 
in the dynamical model, but not in $\chi$EFT since they
are of higher than NLO 
considered in Refs.~\cite{Pascalutsa:2005ts,Pascalutsa:2005vq}.  
For the Coulomb quadrupole form factor $G_C^\ast$, 
one sees a very similar role of the pion cloud in both approaches, 
giving rise to a structure around $Q^2 \simeq 0.03 - 0.05$~GeV$^2$. 
For the electric quadrupole form factor $G_E^\ast$, on the 
other hand, one notices a 
qualitatively different behavior. Whereas the effect of the pion cloud 
is very small in the dynamical model calculation, 
a strong, nearly linear $Q^2$ dependence of the pion loops arises 
in $\chi$EFT. 
\newline
\indent
The role of the pion cloud will further be 
elucidated within the dynamical models in Sect.~\ref{sec3}, and  
$\chi$EFT in Sect.~\ref{sec4}.

\subsection{Lattice QCD and chiral extrapolation}
\label{sec2_lattice}

\subsubsection{Lattice simulations}

Lattice QCD calculations of nucleon structure quantities have matured 
considerably in the recent past. They provide an {\it ab initio} calculation 
of quantities such as the $\gamma^* N \Delta$ transition form factors 
from the underlying theory of the strong interaction. 
The first such calculation was performed by Leinweber, Draper and 
Woloshyn~\cite{Leinweber:1992pv}.
\newline
\indent
The calculation of the 
$\gamma^* N \Delta$ form factors requires the evaluation of three-point 
functions, which involve the computation of a sequential propagator, see 
\Figref{ndelta_lattice}. Leinweber {\it et al.}~\cite{Leinweber:1992pv} 
evaluated these in the so-called 
fixed current approach, which requires the current to have a fixed direction 
and to carry a fixed momentum. The initial and final states, on the other 
hand, can vary without requiring further inversions, which are the 
time-consuming part of the evaluation of three-point functions. 
For rather large quark masses, corresponding with pion masses 
in the range $m_\pi \simeq 0.65 - 0.95$~GeV, they obtained~\cite{Leinweber:1992pv}: $R_{EM} = (-3 \pm 8) \%$. 
This initial result clearly 
indicates the need for high statistics to establish a non-zero value for a 
quantity such as $R_{EM}$.
\begin{figure}
\centerline{  \epsfxsize=7cm 
  \epsffile{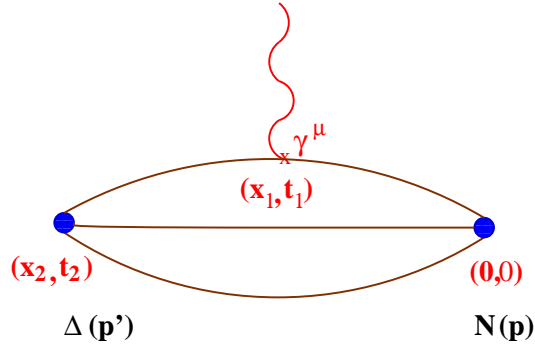} 
}
\caption{
The $\gamma^* N \Delta$ matrix element as evaluated in lattice QCD. 
The photon couples to one of the quarks in the nucleon at a fixed time 
$t_1$ to produce a $\Delta$. 
}
\figlab{ndelta_lattice}
\end{figure}
\newline
\indent
Alexandrou, de Forcrand and Tsapalis~\cite{Alexandrou:2002nn} studied
the deformation of baryons with spin higher 
than 1/2 in lattice QCD via three-density correlators. 
Such three-density correlators, when considered at a fixed time $t$, 
are function of the two relative distances 
${\bf r_1}$ and ${\bf r_2}$ between the three quarks in a baryon $B$ and 
are defined as: 
\begin{eqnarray}
C({\bf r_1}, {\bf r_2}, t) = \int d^3 {\bf r^\prime} 
\, \langle B \,|\, \rho^d({\bf r^\prime}, t) \; 
\rho^u({\bf r^\prime} + {\bf r_1}, t) \;
\rho^u({\bf r^\prime} + {\bf r_2}, t) \, | \, B \rangle ,
\label{eq:3denscorr}
\end{eqnarray}
where the density insertion $\rho^q$ ($q = u, d$) into an up ($u$) 
or down ($d$) quark line is given by the normal order product: 
\begin{eqnarray}
\rho^q ({\bf r}, t) \,=\, : \, \bar q({\bf r}, t) \, \gamma^0 \, 
q({\bf r}, t) \, : \; 
\end{eqnarray}
so that disconnected graphs are excluded. 
These correlation functions of quark densities, which, being expectation 
values of local operators, are gauge invariant, reduce in the 
non-relativistic limit to the wave function squared.  
The three-density correlator for a baryon is shown schematically 
in\footnote{In addition to one-density insertions into each 
$u$-quark line, one can also have two-density insertions into 
the same $u$-quark line. 
The latter contribution was however checked to be 
small~\cite{Alexandrou:2002nn}.}~\Figref{correlator}. 
For simplicity, the baryon source and sink are taken 
at the same spatial location. Denoting the lattice extent in Euclidean 
time by $T$ and taking periodic boundary conditions, yields a maximum time 
separation between source and sink of $T/2$. 
To isolate the ground state baryon, both the time distances 
between the source and the insertions ($t$) and between 
the sink and the insertions ($T/2 - t$) are taken to be 
as large as possible in such calculations. 
\begin{figure}
\centerline{  \epsfxsize=8cm \epsfysize=5cm
  \epsffile{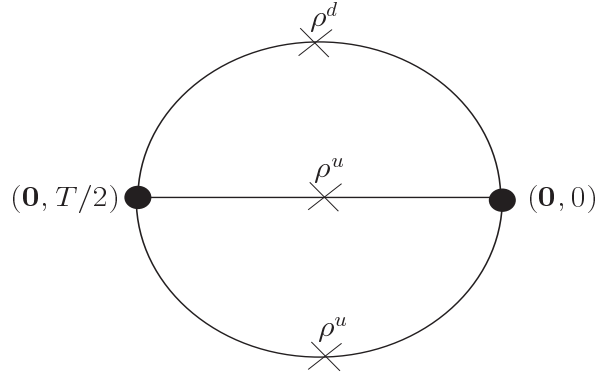} 
}
\caption{
Three-density correlator of a baryon as defined in Eq.~(\ref{eq:3denscorr}) 
and used in the lattice gauge calculations of Ref.~\cite{Alexandrou:2002nn}.
The source (right) and sink (left) are taken at the same spatial location,  
and are separated by a Euclidean time distance $T/2$. The crosses 
indicate the density insertions $\rho^u$ ($\rho^d$) into the up (down) quark 
lines respectively.
}
\figlab{correlator}
\end{figure}
\newline
\indent
In Ref.~\cite{Alexandrou:2002nn}, the correlator has been calculated in both 
quenched and unquenched lattice QCD for  
a $\Delta^+$ in a spin +3/2 state. The quenched calculation was performed 
for pion masses in the range $m_\pi \simeq 0.35 - 0.65$~GeV, and 
no signal of deformation has been observed. This may not come as a surprise, 
because in a quenched calculation, 
where graphs with disconnected quark loops are neglected, pion cloud 
effects are only partially accounted for. 
It was discussed above in the context of phenomenological models that 
pion cloud effects are partly responsible for 
hadron deformation. This was checked by 
performing an unquenched lattice QCD calculation with two 
heavy dynamical quarks, 
using the Wilson Dirac operator in the same quark mass range as the 
quenched calculations~\cite{Alexandrou:2002nn}. The resulting correlator for the $\Delta^+$ 
is displayed in a three-dimensional contour plot in \Figref{deltashape}. 
The unquenched lattice QCD calculation indicates a 
slightly oblate deformation for the $\Delta^+$ in a spin +3/2 state, in 
agreement with the negative value of the $\Delta^+$ quadrupole moment 
obtained from the large $N_c$ estimate of Eq.~(\ref{eq:qdellargenc}).  
It will be very interesting to check if this signal 
can be consolidated by higher statistics 
results using lighter quark masses and a larger lattice. 
\begin{figure}
\centerline{  \epsfxsize=8cm
  \epsffile{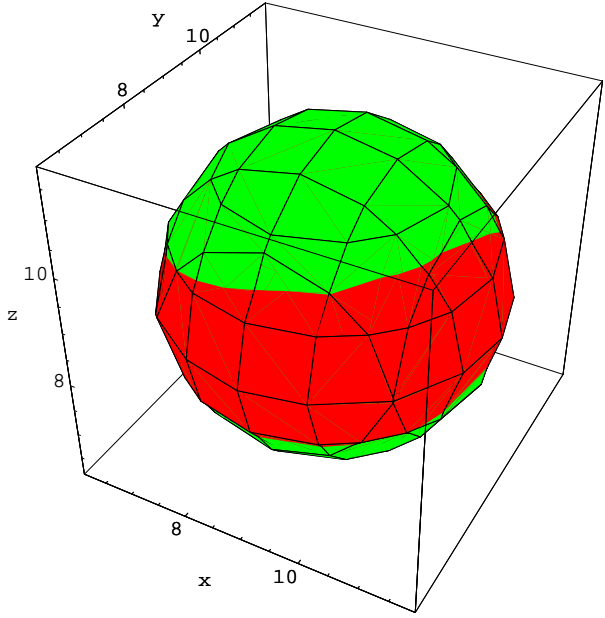} 
}
\caption{
Three-dimensional contour plot of the correlator (red) 
for the $\Delta^+$ state with spin projection +3/2 (the $z$-axis denotes the 
spin direction) as it follows 
from lattice QCD calculations~\cite{Alexandrou:2002nn} 
for two heavy dynamical quarks, corresponding 
with a pion mass $m_\pi > 0.6$~GeV. 
The contour (red), corresponding with a value (0.8) for the correlator which  
has been chosen to show large distances, indicates a slightly oblate 
deformation of the $\Delta^+$ in the +3/2 spin state.  
For comparison, the contour of a sphere is also shown (green). 
Figure from Ref.~\cite{Alexandrou:2002nn}.
}
\figlab{deltashape}
\end{figure}
\newline
\indent
A direct calculation of the $\gamma^* N \Delta$ transition form factors 
on the lattice has also been performed recently. 
Based on the method of Ref.~\cite{Leinweber:1992pv}, 
the Nicosia-MIT group~\cite{Alexandrou:2003ea}
has performed a high-statistics calculations 
of the $\gamma^\ast N \Delta$ form factors. 
The Nicosia-MIT group has 
also implemented several improvements of the algorithm, 
such as smearing techniques to effectively 
filter the ground state, check on the volume dependence of the results, 
use of a larger lattice such as to simulate smaller quark masses.
Both quenched and unquenched 
results have been obtained.  
The quenched calculation was performed for pion masses in the 
range $m_\pi \simeq 0.4 - 0.9$~GeV. 
The unquenched calculation, using dynamical Wilson fermions, 
was performed in the range of $m_\pi \simeq 0.5 - 0.8$~GeV.   
At such pion mass values, both the quenched and unquenched 
calculations give $R_{EM}$ values in the range from $-1$ to $-5$~\%. 
For pion masses in the range $0.5 - 0.8$~GeV unquenching effects were 
found to be within statistical errors, which means that pion cloud 
contributions, expected to drive the $R_{EM}$ value more negative 
are suppressed for these large quark (pion) masses. 
\newline
\indent
In order to compare the calculated $R_{EM}$ values with experiment, 
the Nicosia-MIT group uses 
a linear fit in $m_\pi^2$ (corresponding with a linear fit  
in the quark mass) was used  
to extrapolate down to the physical pion mass. 
Such a naive extrapolation can not be justified,
as will be discussed below. 
Nonetheless, using such a linear extrapolation, the $R_{EM}$ value at 
$Q^2 = 0.4$~GeV$^2$ was found to be~\cite{Alexandrou:2003ea} : 
\begin{eqnarray} 
R_{EM}(Q^2 = 0.4 \; \mathrm{GeV}^2) \,&=&\, (-4.4 \pm 1.7) \; \% 
\quad \quad \quad \quad 
\mathrm{(quenched)} , \nn \\
R_{EM}(Q^2 = 0.4 \; \mathrm{GeV}^2) \,&=&\, (-3.7 \pm 1.2) \; \% 
\quad \quad \quad \quad 
\mathrm{(unquenched) } , \nn 
\end{eqnarray}
which is to be compared with the experimental number 
of Ref.~\cite{Joo:2001tw}: $R_{EM} = (-3.4 \pm 0.4 \pm 0.4) \, \%$. 
One sees that the lattice QCD calculation clearly supports a negative value 
of $R_{EM}$, in basic agreement with experiment. 
\newline
\indent
In the subsequent work of the Nicosia-MIT group~\cite{Alexandrou:2004xn}, the 
$\gamma^* N \Delta$ form factors were studied within a fixed sink method 
in which the initial state, created at time zero, has the nucleon quantum 
numbers, and the final state, annihilated at a later time $t_2$ has 
the $\Delta$ quantum numbers. The current can couple to a quark line 
at any time slice $t_1$, see \Figref{ndelta_lattice}, 
carrying any possible value of the lattice momentum. 
Implementing further improvements, the method of Ref.~\cite{Alexandrou:2004xn} 
is superior to the fixed current approach discussed above, 
yielding a more accurate evaluation of all three $\gamma^* N \Delta$ form 
factors. The calculations in Ref.~\cite{Alexandrou:2004xn} 
were performed in the quenched approximation for pion masses: 
$m_\pi = 0.51$, $0.45$, and $0.37$~GeV, 
which are also smaller than in the previous calculations. 
\begin{figure}
\centerline{  \epsfysize=6.5cm
  \epsffile{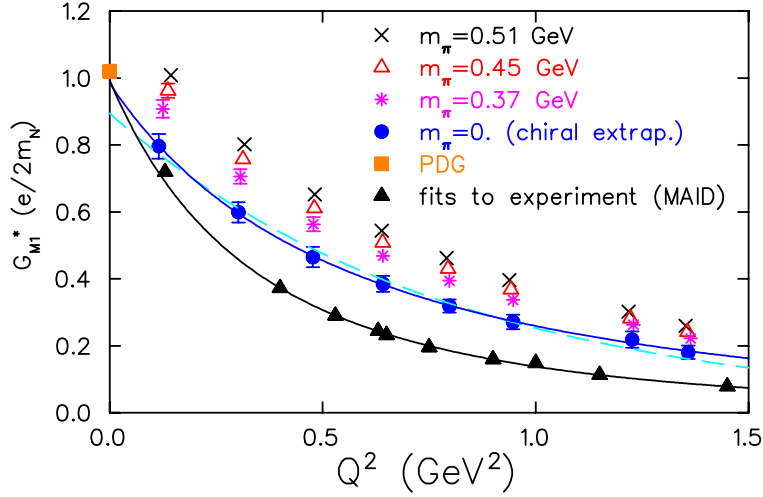} 
}
\caption{
Quenched lattice QCD results for the $Q^2$ dependence of the 
(Ash) form factor $G^\ast_{M1} = G^\ast_{M, Ash} /3$, 
as defined in Eq.~(\ref{eq:ash}) 
for three values of $m_\pi$ as indicated on the figure. 
The (blue) solid circles are the result of a linear extrapolation 
in $m_\pi^2$. The experimental value at $Q^2 = 0$ (red square, PDG) 
corresponds with the value given in Eq.~(\ref{eq:mundelexp}).  
The lower solid curve corresponds with a fit to the data, according to the 
MAID analysis~\cite{Tiator:2000iy}. 
Figure from Ref.~\cite{Alexandrou:2004xn}. 
}
\figlab{gmlattice}
\end{figure}
\newline
\indent
The quenched lattice QCD results of Ref.~\cite{Alexandrou:2004xn} 
for the magnetic $\gamma^* N \Delta$ form factor 
are shown in \Figref{emrcmrlattice} together with a linear extrapolation 
in $m_\pi^2$ of the lattice results. The lattice results show a decrease 
of $G^\ast_M$ with decreasing pion mass. However the linear extrapolated 
values in $m_\pi^2$ are still lying significantly above the experimental 
results. The most likely explanation of this discrepancy is 
the inadequacy in the linear extrapolation. 
At low momentum transfer, where chiral effective field theory can 
be applied, one expects chiral logs to appear in the form factors, as will 
be discussed further on. 
\begin{figure}
\leftline{  \epsfysize=7.25cm
  \epsffile{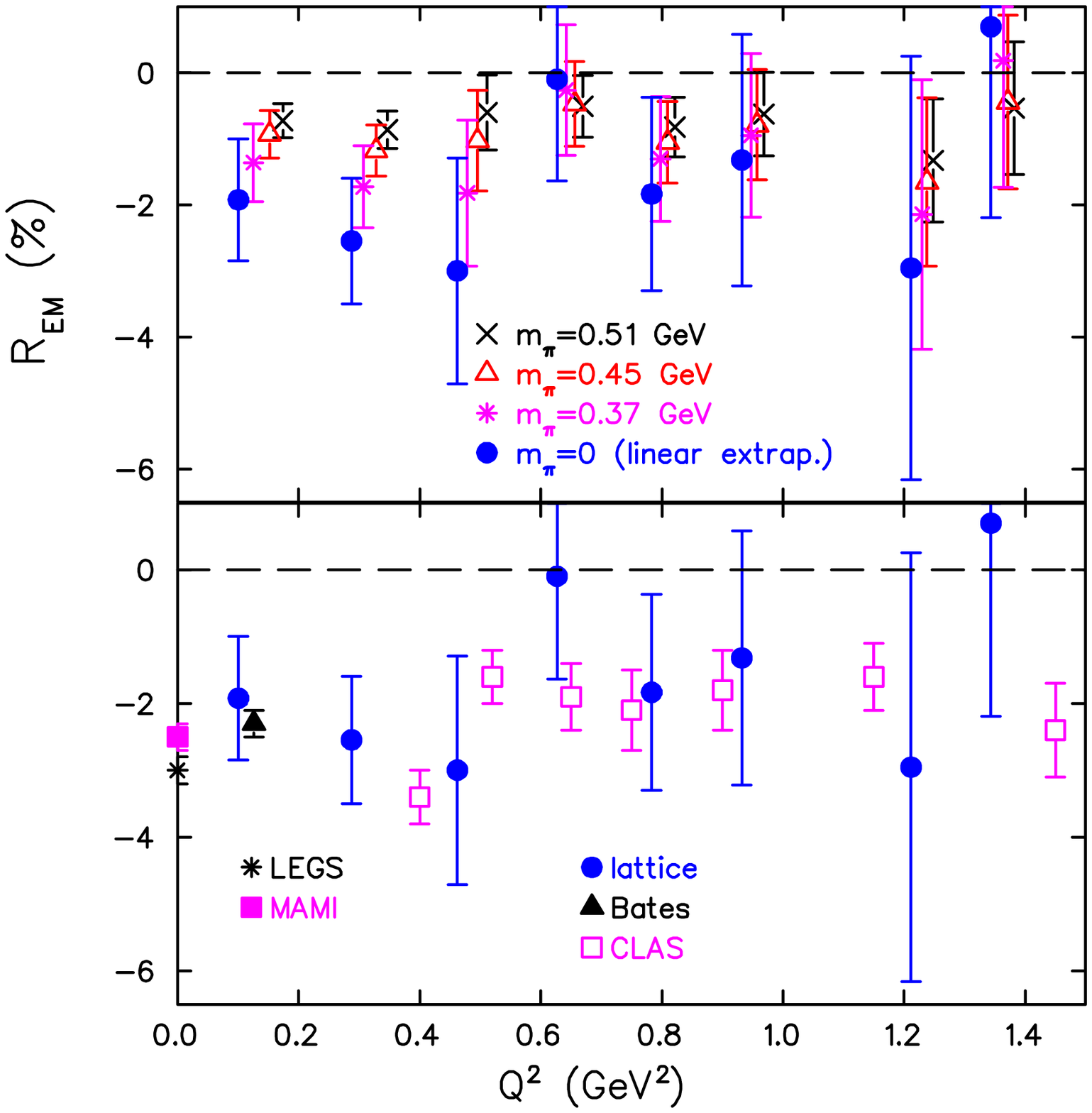} 
}
\vspace{-7.25cm}
\rightline{  \epsfysize=7.25cm
  \epsffile{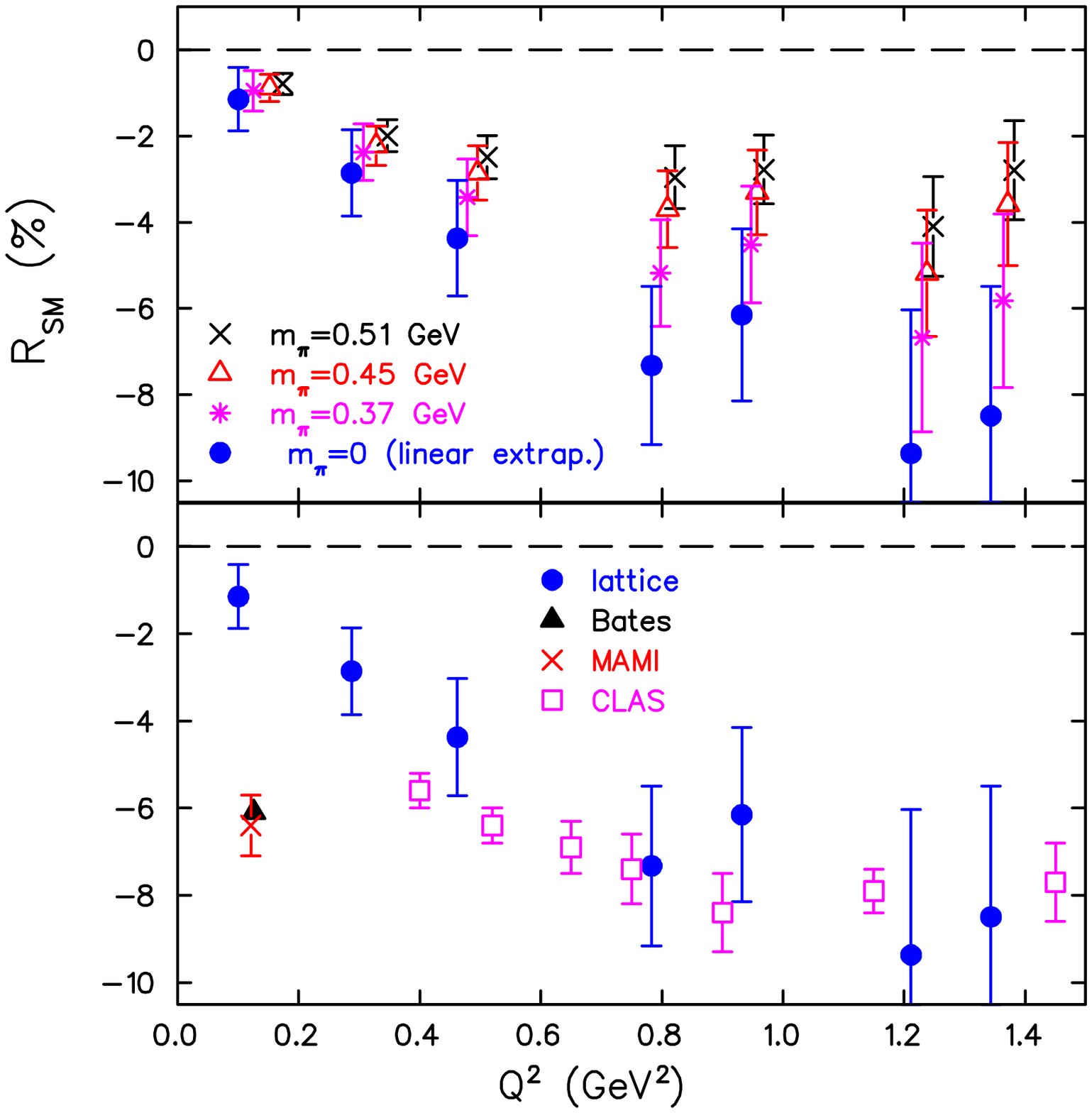} 
}
\caption{
Quenched lattice QCD results 
for the $Q^2$ dependence of the $\gamma^* N \Delta$ 
ratios $R_{EM}$ (left panels) and $R_{SM}$ (right panels),
for three values of $m_\pi$ as indicated on the figure. 
The (blue) solid circles are the result of a linear extrapolation 
in $m_\pi^2$. 
In the lower panels a comparison with experimental data is shown: 
real photon points from MAMI~\cite{Beck:1999ge} and 
LEGS~\cite{Blanpied:2001ae};  
finite $Q^2$ values are from BATES~\cite{Mertz:1999hp,Sparveris:2004jn} 
(solid triangles), 
MAMI~\cite{Pospischil:2000ad} (cross), 
and JLab/CLAS~\cite{Joo:2001tw} (open boxes). 
Figure from Ref.~\cite{Alexandrou:2004xn}. 
}
\figlab{emrcmrlattice}
\end{figure}
\newline
\indent
In \Figref{emrcmrlattice}, 
the quenched lattice QCD results of Ref.~\cite{Alexandrou:2004xn} are shown 
for the $\gamma^* N \Delta$ ratios $R_{EM}$ and $R_{SM}$, together 
with a linear extrapolation in $m_\pi^2$. 
For the $R_{EM}$ ratio, the lattice results are accurate enough to show 
a negative value, which becomes more negative 
as one approaches the chiral limit. 
The linearly extrapolated lattice results 
seem to be in good agreement with the experiment, although the lattice 
results at larger $Q^2$ values show some scatter.  
The $R_{SM}$ ratio is clearly negative over the whole $Q^2$ range. 
For $Q^2 \gtrsim 0.5$~GeV$^2$, the linearly extrapolated lattice 
results are in agreement with experiment. At lower $Q^2$ on the other hand, 
the linearly extrapolated lattice results fall increasingly short of the data, 
and cannot explain the large negative value of $R_{SM}$ established 
in experiment. The present empirical results for $R_{SM}$ at low $Q^2$ 
have reached a high level of accuracy and have been cross-checked 
by several experiments at both 
BATES~\cite{Mertz:1999hp,Kunz:2003we,Sparveris:2004jn} 
and MAMI~\cite{Pospischil:2000ad,Elsner:2005cz}, 
all obtain a rather large negative $R_{SM}$ ratio of around -6\%  
at low $Q^2$. This puzzle was studied in 
Refs.~\cite{Gail:2005gz,Pascalutsa:2005ts,Pascalutsa:2005vq} 
within the framework of $\chi$EFT, yielding some interesting
results as is discussed in the following.

\subsubsection{Chiral extrapolations}

The extrapolation in the quark mass $m_q$ is not straightforward,
because the non-analytic dependencies, such as $\sqrt{m_q}$
and $\ln m_q$, become important as one approaches the
small physical value of $m_q$. Therefore naive extrapolations 
often fail, while spectacular non-analytic effects
are found in a number of different quantities, 
see {\it e.g.}, Refs.~\cite{Leinweber:2001ui,Hemmert:2003cb}. 
The $\chi$EFT, discussed in the previous section, provides 
a framework to compute these non-analytic terms. 
We will first address the quark mass dependence of the 
nucleon and $\Delta$ masses within $\chi$EFT, 
which have been discussed extensively in the literature, see 
{\it e.g.} Refs.~\cite{Banerjee:1994bk,Leinweber:1999ig,Ross,Bernard:2003xf,Frink:2005ru,Bernard:2005fy,HWGS05,Pascalutsa:2005nd}. 
Subsequently, we will review the predictions 
for the pion mass dependence of the $\gamma^* N \Delta$ form factors. 
\begin{figure}[tb]
\centerline{  \epsfxsize=7cm
  \epsffile{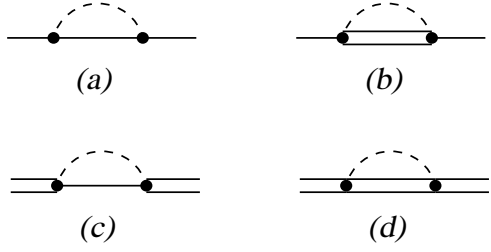} 
}
\caption{The nucleon and $\De$ self-energy contributions
in $\chi$EFT. Dashed lines represent pion propagators, 
solid lines represent nucleon propagators, 
whereas double lines represent $\Delta$ propagators.}
\figlab{selfenergy_diagrams}
\end{figure}
\newline
\indent
In the $\chi$EFT including $\Delta$ degrees of freedom, 
the pion mass dependence of $N$ and $\Delta$ masses 
is calculated from the self-energy diagrams 
of \Figref{selfenergy_diagrams}. Near the chiral limit, i.e. for 
$m_\pi < \vDe \equiv M_\Delta - M_N$, 
the pion mass dependence for the $N$ and $\Delta$ masses goes 
as~\cite{Banerjee:1994bk}:
\begin{eqnarray}
M_N \,&=&\, M_{N}^{(0)} - 4 \, c_{1N}\, m_\pi^2 
- \frac{3}{32 \pi f_\pi^2} g_A^2 \, m_\pi^3 
+ \frac{2}{(8 \pi f_\pi)^2} h_A^2\, \frac{m_\pi^4}{\vDe} 
\ln m_\pi + O(m_\pi^4),
\label{eq:nucmass} \\
M_\Delta \,&=&\, M_{\Delta}^{(0)} - 4 \, c_{1\Delta}\, m_\pi^2 
- \frac{3}{32 \pi f_\pi^2} \frac{25}{81} H_A^2 \, m_\pi^3 
- \frac{1}{2 (8 \pi f_\pi)^2} h_A^2\, 
\frac{m_\pi^4}{\vDe} \ln m_\pi + O(m_\pi^4).
\label{eq:delmass}
\end{eqnarray}
The non-analytic terms (proportional to $m_\pi^3$, 
$m_\pi^4/\vDe \ln m_\pi$,...) on the {\it rhs} 
of Eqs.~(\ref{eq:nucmass},\ref{eq:delmass}) are predictions of 
$\chi$EFT obtained from the loop diagrams of \Figref{selfenergy_diagrams}.  
These loop expressions depend on the coupling constants 
appearing in the lowest order chiral Lagrangian (see Sect.~\ref{sec4} 
for details):
$g_A = 1.267$ is the axial coupling of the nucleon, 
$h_A \simeq 2.85$ is the $\pi N \Delta$ coupling constant  
(in the notation of Ref.~\cite{Pascalutsa:2005nd}), 
and $H_A$ is the axial $\pi \Delta \Delta$ coupling constant, 
which is related with $g_A$ through the $SU(6)$ relation, which 
coincides with the large $N_c$ relation: $H_A = (9/5) g_A \simeq 2.28$. 
The analytic terms on the {\it rhs} of 
Eqs.~(\ref{eq:nucmass},\ref{eq:delmass}) are low-energy constants 
which have to be determined from experiment or 
from a fit to lattice QCD results. 
In particular $M_N^0$ ($M_\Delta^0$) is the $N$ ($\Delta$) mass 
in the chiral limit, and the term proportional to 
$c_{1N}$ ($c_{1 \Delta}$) is the quark mass contribution  
to the $N$ ($\Delta$) mass. For the nucleon, it is 
obtained from the experimental information on the 
pion-nucleon $\sigma$-term.  
\newline
\indent
The above formulas of Eqs.~(\ref{eq:nucmass},\ref{eq:delmass}) can be 
fitted to full lattice QCD results for the $N$ and $\Delta$ masses. 
Care has to be taken however when fitting to quenched lattice 
QCD results, where sea quark loop effects are neglected. 
The quenched approximation also modifies the leading chiral expansion 
of baryon masses in the corresponding effective field theory, see 
Ref.~\cite{Labrenz:1996jy}. For instance, in the quenched approximation, the 
pion mass dependence of $N$ and $\Delta$ masses contain a term linear in 
$m_\pi$, whereas such a term is absent in the full QCD 
expansions of Eqs.~(\ref{eq:nucmass},\ref{eq:delmass}).
\begin{figure}
\centerline{  \epsfxsize=11cm
  \epsffile{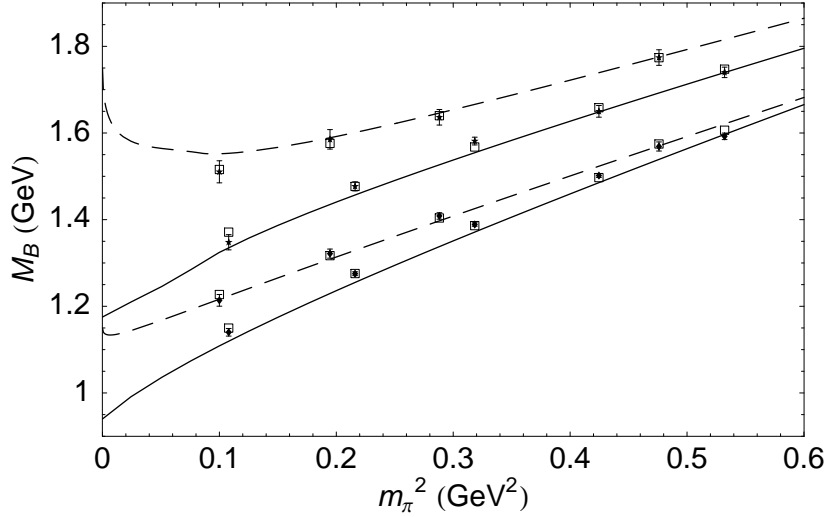} 
}
\caption{
Pion-mass dependence of the 
$N$ and $\Delta$ masses according to Ref.~\cite{Ross} compared 
with lattice QCD results.
The solid symbols are lattice results from the MILC 
Collaboration~\protect\cite{Bernard:2001av}. 
Solid symbols from bottom to top correspond with the nucleon unquenched, 
nucleon quenched, $\Delta$ unquenched, and $\Delta$ quenched lattice results. 
The open boxes are the fit of Ref.~\cite{Ross}  
to $N$ and $\Delta$ masses accounting for finite volume and lattice spacing 
artifacts, as described in the text.  
The curves describe the infinite-volume limit for
dynamical (solid curves) and quenched (dashed curves) $N$ and $\Delta$ masses.
}
\figlab{rossfig}
\end{figure}
\indent
In Ref.~\cite{Ross}, the $m_\pi$ dependence of 
$N$ and $\Delta$ masses have been fitted to lattice results using  
both the quenched theory and full QCD. 
The non-analytic terms have been calculated from the one-loop diagrams 
of \Figref{selfenergy_diagrams}, including a 
phenomenological form factor at the 
$\pi NN$, $\pi N \Delta$ and $\pi \Delta \Delta$ vertices to account for the 
finite size of the pion source. The parameters in the analytic terms were 
treated as free parameters and fit to the lattice results. Such a fit, 
is shown in \Figref{rossfig}. 
One clearly sees that this procedure is able to successfully account for the 
different behavior of the $m_\pi$ dependence of 
$N$ and $\Delta$ masses in quenched as compared to full QCD. 
One notices that the effect of unquenching is larger for the $\Delta$ as 
compared to the $N$. It will be interesting to test the more singular behavior 
of the quenched result for the $\Delta$ as lattice results become available 
for smaller pion masses. 
\newline
\indent
A comparison of both quenched and full QCD  
also allows to draw interesting conclusions on the physical nature 
of the $N-\Delta$ mass splitting, because the effect of the pion cloud is 
only partly accounted for in the quenched theory. Ref.~\cite{Ross} 
found that only about 50 MeV of the observed 300 MeV 
$N-\Delta$ mass splitting arises from the pion cloud, 
while the rest arises from short distance processes, such as 
gluon exchange. This short distance contribution is 
responsible for the finite $N-\Delta$ mass splitting 
in the chiral limit (i.e. when $m_\pi \to 0$).   
\begin{figure}
\centerline{  \epsfxsize=12cm%
  \epsffile{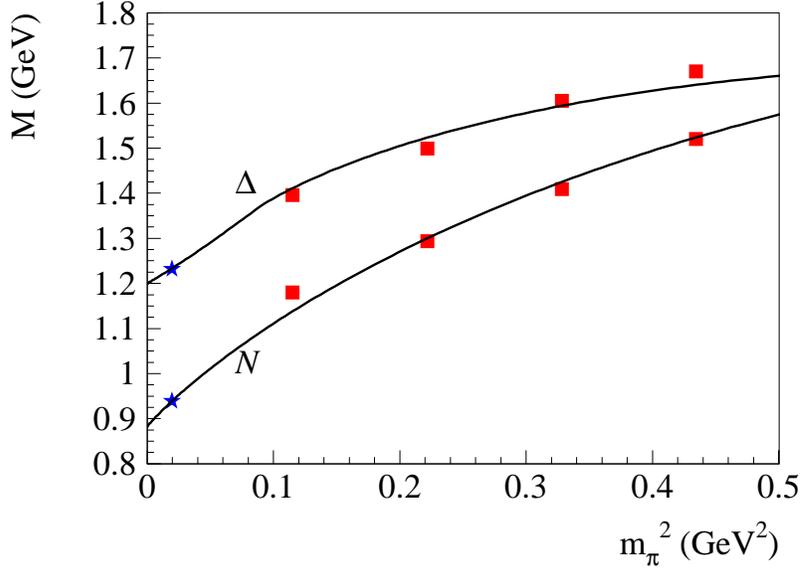} 
}
\caption{
Pion-mass dependence of the $N$ and $\Delta$ masses.  
The curves are two-parameter expressions for $M_N$ and $M_\Delta$ 
according to the manifestly covariant $\chi$EFT framework 
of Ref.~\cite{Pascalutsa:2005nd}.  
For both $N$ and $\Delta$, the chiral limit mass values $M_N^{(0)}$ 
($M_\Delta^{(0)}$) and the parameters $c_{1 N}$ ($c_{1 \Delta}$) 
are fitted. 
The red squares are lattice results from the MILC 
Collaboration~\protect\cite{Bernard:2001av} in full QCD. 
The stars represent the physical mass values.}
\figlab{nucdelmass}
\end{figure}
\newline
\indent
To account for higher order terms in the chiral expansion, 
the $m_\pi$ dependence of $M_N$ and $M_\Delta$ has been studied 
by two of us~\cite{Pascalutsa:2005nd} in a manifestly covariant $\chi$EFT 
framework consistent with analyticity. 
The resulting relativistic loop corrections 
obey the chiral power-counting, after renormalizations of the available 
counter-terms are done. The relativistic expressions also contain 
the nominally higher-order terms, which are necessary to satisfy the 
analyticity constraint. 
In such approach, the analytic terms in the quark mass arising from the 
one-loop pion diagrams of \Figref{selfenergy_diagrams}
are partially resummed. We found~\cite{Pascalutsa:2005nd} 
that the convergence of the chiral expansion for $M_N$ and $M_\Delta$ 
can be improved without introducing additional parameters. 
In \Figref{nucdelmass}, the results for the 
$m_\pi$ dependence of $N$ and $\De$-resonance
masses in this approach are compared with full lattice QCD results. 
For both $N$ and $\Delta$, the chiral limit mass values $M_N^{(0)}$ 
($M_\Delta^{(0)}$) and the parameters $c_{1 N}$ ($c_{1 \Delta}$) 
are fitted. 
As is seen from the figure, with this two-parameter form for $M_N$ and 
$M_\Delta$, a good description of lattice results is obtained 
up to $m_\pi^2 \simeq 0.5$~GeV$^2$. 
\newline
\indent
Having discussed the $m_\pi$ dependence of $M_N$ and $M_\Delta$, we 
turn to the $m_\pi$ dependence of the 
$\gamma^* N \Delta$ transition form factors. 
The study of the $m_\pi$-dependence 
is crucial to connect to lattice QCD results, which at present 
can only be obtained for larger pion masses 
(typically $m_\pi \gtrsim 0.3$ GeV) as discussed above  
in \Figref{gmlattice} for $G_M^\ast$, and in \Figref{emrcmrlattice} 
for $R_{EM}$ and $R_{SM}$. 
\begin{figure}
\centerline{ \epsfxsize=10cm%
\epsffile{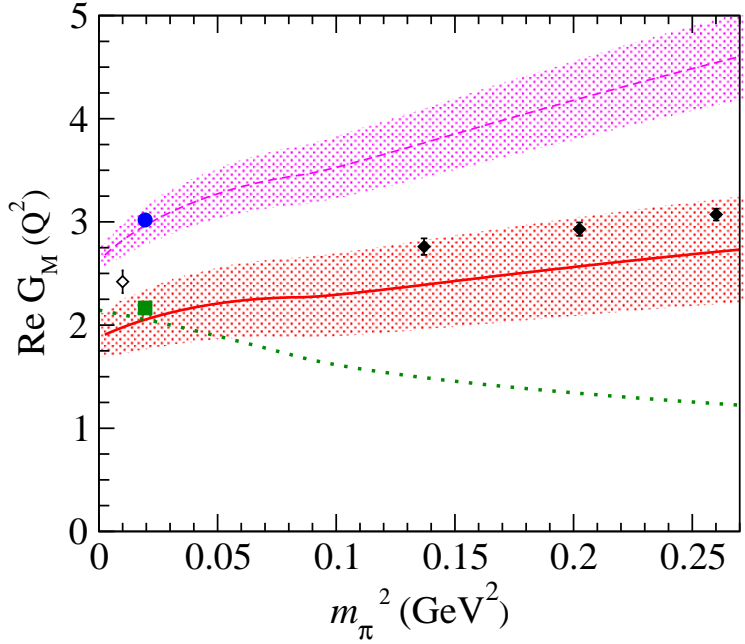}
}
\caption{
The pion mass dependence of the real part of 
the Jones-Scadron $\gamma^* N \Delta$ form factor $G_M^\ast$ for $Q^2 = 0$ and 
$Q^2$ = 0.127 GeV$^2$ in the relativistic $\chi$EFT framework of 
Refs.~\cite{Pascalutsa:2005ts,Pascalutsa:2005vq}.
The solid (dashed) curves are the NLO results for $Q^2 = 0.127$~GeV$^2$
($Q^2 = 0$) respectively, including
the $m_\pi$ dependence of $M_N$ and $M_\Delta$. 
The green dotted curve is 
the corresponding result for $Q^2 = 0.127$~GeV$^2$ where   
the $m_\pi$ dependence of $M_N$ and $M_\Delta$ is not included. 
The blue circle for $Q^2 = 0$ 
is a data point from MAMI~\protect\cite{Beck:1999ge}, and 
the green square for $Q^2 = 0.127$~GeV$^2$ is 
a data point from BATES~\protect\cite{Mertz:1999hp,Sparveris:2004jn}. 
The three filled black diamonds at larger $m_\pi$   
are lattice calculations~\protect\cite{Alexandrou:2004xn} for $Q^2$ values of 
0.125, 0.137, and 0.144 GeV$^2$ respectively, 
whereas the open diamond near $m_\pi \simeq 0$ represents their  
extrapolation assuming linear dependence in $m_\pi^2$. 
Figure from Ref.~\cite{Pascalutsa:2005vq}. 
}
\figlab{fig:regmmpi}
\end{figure}
\newline
\indent
In \Figref{fig:regmmpi}, the $m_\pi$-dependence of the 
{\it magnetic} $\gamma^* N \Delta$-transition (Jones-Scadron) 
form factor $G_M^\ast$ is shown in the relativistic $\chi$EFT framework 
of Refs.~\cite{Pascalutsa:2005ts,Pascalutsa:2005vq}. 
It is calculated from the one-loop diagrams of \Figref{gandelabs}. 
Recall that the value of $G_M^\ast$ at $Q^2 = 0$ is 
determined by the low-energy constant $g_M$.
The $Q^2$-dependence then follows as a prediction of the NLO
result, and \Figref{fig:regmmpi} shows that
this prediction is consistent with the experimental 
value at $Q^2 = 0.127$~GeV$^2$ and physical pion mass. 
The $m_\pi$-dependence  
of $G_M^\ast$ is also completely fixed at NLO, no new parameters appear. 
\newline
\indent
In \Figref{fig:regmmpi}, the result for $G_M^\ast$ at
$Q^2 = 0.127$~GeV$^2$ is shown both when the $m_\pi$-dependence of 
the nucleon and $\Delta$ masses is included and when it is not.
Accounting for the $m_\pi$-dependence in $M_N$ and $M_\Delta$, 
as shown in \Figref{nucdelmass}, 
changes the result for $G_M^\ast$ quite significantly. 
The \ceft\  calculation,  
with the $m_\pi$ dependence of $M_\N$ and $M_\Delta$ included, is in
a qualitatively good agreement with the lattice data shown in the figure. 
The \ceft\ result also follows 
an approximately linear behavior in $m_\pi^2$, 
although it falls about 10 - 15 \% below the lattice data.  
This is just within the uncertainty of the NLO results.
One should also keep in mind that the lattice simulations 
of Ref.~\cite{Alexandrou:2004xn} are not done in full QCD, 
but are ``quenched'', so discrepancies are not unexpected.
\begin{figure}
\centerline{  \epsfxsize=9cm%
  \epsffile{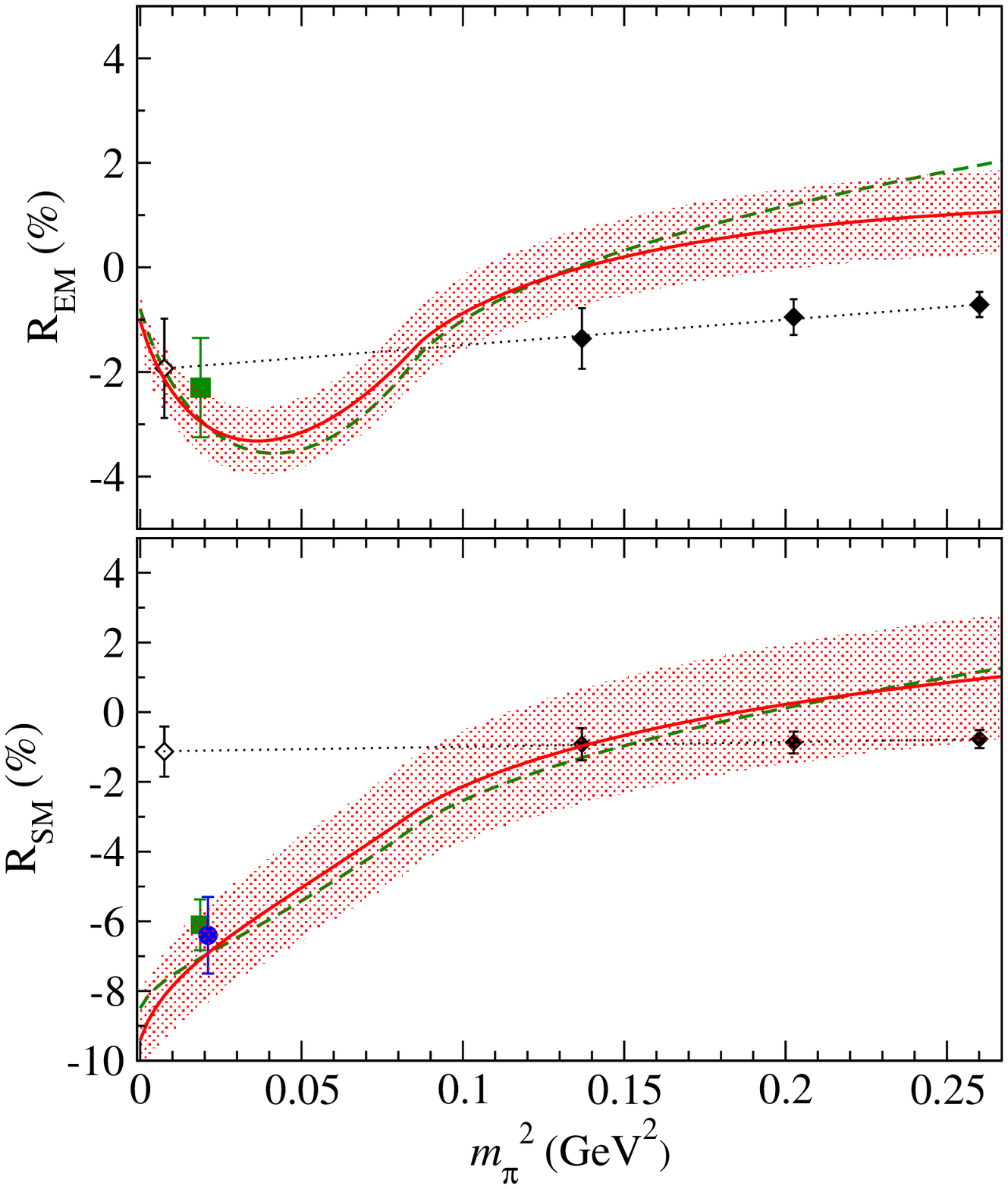} 
}
\caption{
The pion mass dependence of
 $R_{EM}$ (upper panel) and 
$R_{SM}$ (lower panel), at $Q^2=0.1$ GeV$^2$.
The blue circle is a data point from MAMI~\protect\cite{Pospischil:2000ad}, 
the green squares are data points from 
BATES~\protect\cite{Mertz:1999hp,Sparveris:2004jn}. 
The three filled black diamonds at larger $m_\pi$   
are lattice calculations~\protect\cite{Alexandrou:2004xn}, 
whereas the open diamond near $m_\pi \simeq 0$  
represents their extrapolation assuming linear dependence in $m_\pi^2$. 
Red solid curves: NLO result when accounting for the $m_\pi$ dependence in 
$M_N$ and $M_\Delta$; 
green dashed curves: NLO 
result of Refs.~\cite{Pascalutsa:2005ts,Pascalutsa:2005vq}, where   
the $m_\pi$-dependence of $M_N$ and $M_\Delta$ was not accounted for.
The error bands represent the estimate of theoretical uncertainty for
the NLO calculation. 
Figure from Ref.~\cite{Pascalutsa:2005vq}. 
}
\figlab{ratios}
\end{figure}
\newline
\indent
In \Figref{ratios}, the $m_\pi$-dependence of the ratios 
$R_{EM}$ and $R_{SM}$ is shown within the same relativistic $\chi$EFT 
framework and compared to the lattice QCD calculations.  
As discussed above, the recent state-of-the-art lattice calculations of 
$R_{EM}$ and $R_{SM}$~\cite{Alexandrou:2004xn} use a {\it linear}, 
in the quark mass ($m_q\propto m_\pi^2$), {\it extrapolation}
to the physical point,  
thus assuming that the non-analytic $m_q$-dependencies are  negligible. 
The thus obtained value for $R_{SM}$ at the physical 
$m_\pi$ value displays a large 
discrepancy with the  experimental result, as seen in \Figref{emrcmrlattice}. 
The relativistic $\chi$EFT calculation~\cite{Pascalutsa:2005ts,Pascalutsa:2005vq}, 
on the other hand, shows  that the non-analytic dependencies 
are {\it not} negligible. While at larger values of $m_\pi$, 
where the $\Delta$ is stable, the ratios display a smooth 
$m_\pi$ dependence, at $m_\pi =\vDe $ there is an inflection point, and 
for  $m_\pi \leq \vDe$ the non-analytic effects are crucial, 
as was also observed for the $\De$-resonance
magnetic moment~\cite{Cloet03,PV05}. 
\newline
\indent
One also sees from \Figref{ratios} that, unlike the result for 
$G_M^\ast$, there is only little difference in the ratios between the \ceft\ 
calculations with the $m_\pi$-dependence of 
$M_N$ and $M_\Delta$ accounted for, and where this $m_\pi$ 
dependence of the masses is neglected. 
The \ceft\ framework of Refs.~\cite{Pascalutsa:2005ts,Pascalutsa:2005vq} 
thus shows that the assumption of a linear 
extrapolation in $m_\pi^2$ is not valid for $R_{EM}$ and $R_{SM}$. 
Once the non-analytic dependencies on the quark mass as 
they follow from $\chi$EFT are accounted for,  
there is no apparent discrepancy between the lattice results of 
Ref.~\cite{Alexandrou:2004xn} and the experimental results for $R_{SM}$. 
\newline
\indent
To test the difference between quenched and full lattice QCD results 
for the $\gamma^* N \Delta$ form factors, 
new lattice calculations, within full QCD, 
are underway~\cite{Alexandrou:2005em}. 
The full QCD results obtained so far \cite{Alexandrou:2005em} 
using dynamical Wilson fermions 
are more noisy but in agreement with those obtained in the quenched theory. 
It will be interesting to test the predicted strong non-analytic effects 
in the $R_{EM}$ and $R_{SM}$ 
ratios as shown in \Figref{ratios} once high statistics 
full lattice QCD results for pion masses smaller than 0.3 GeV become 
available.

\subsection{Generalized parton distributions (GPDs)}
\label{sec2_gpd}

\subsubsection{Definition of $N \to N$ and $N \to \Delta$ GPDs and sum rules}

So far we have discussed the $N \to \Delta$ transition 
as revealed with the help of the electromagnetic probe. By measuring 
the response of the hadron to a virtual photon, one measures the matrix 
element of a  well-defined quark-gluon operator (in this case the vector 
operator $\bar q \gamma^\mu q$) over the hadronic state. This matrix element  
can be parametrized in terms of the $\gamma^* N \Delta$ 
transition form factors, revealing the quark-gluon structure of the 
hadron. We are however not limited in nature to probes such as photons 
(or $W$, $Z$ bosons for the axial transition). The phenomenon of asymptotic 
freedom of QCD, meaning that at short distances the interactions between 
quarks and gluons become weak, provides us with more sophisticated 
QCD operators to explore the structure of hadrons. Such operators can 
be accessed by selecting a small size configuration of quarks and gluons, 
provided by a hard reaction, such as deep inelastic scattering (DIS), or 
hard exclusive reactions such as deeply virtual Compton scattering (DVCS).  
We will be mostly interested here in DVCS reactions which are of the type 
$\gamma^*(q_h) + N(p) \to \gamma(q^\prime) + B(p^\prime)$, where the 
virtual photon momentum $q_h$ is the hard scale, 
and where the final state $B$ stands for either the nucleon $N$ 
or the $\Delta$ state. 
The common important feature of such hard reactions is the possibility
to separate clearly the perturbative and nonperturbative stages of
the interactions, this is the so-called factorization property. 
\newline
\indent
The all-order factorization theorem for the DVCS process on the 
nucleon has been proven in Refs.~\cite{Ji98a,Col99,Rad98}.
Qualitatively one can say that the hard reactions allow
one to perform a ``microsurgery'' of a nucleon by removing in a
controlled way a quark of one flavor and spin and implanting
instead another quark (in general with a different flavor and
spin) in the final baryon. It is illustrated in \Figref{fig:ndelta_dvcs} 
for the case of the DVCS process with a $\Delta$ in the final state. 
The non-perturbative stage of such hard exclusive 
electroproduction processes is described by 
universal objects, so-called generalized parton distributions 
(GPDs)~\cite{Muller:1998fv,Ji:1996ek,Radyushkin:1996nd}, 
see Refs.~\cite{Ji:1998pc,Goeke:2001tz,Diehl:2003ny,Belitsky:2005qn} 
for reviews and references. 
\begin{figure}
\centerline{  \epsfxsize=8cm
  \epsffile{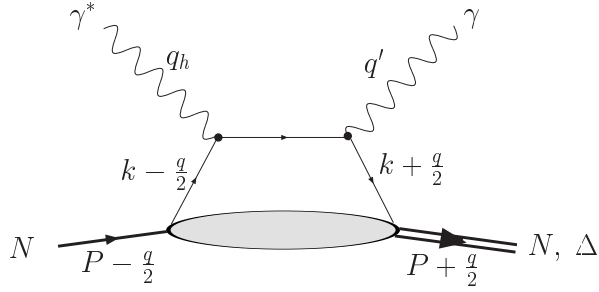} 
}
\caption{
The ``handbag'' diagram for the $N \to N$ and $N \to \Delta$ DVCS processes. 
Provided the virtuality of the initial photon (with momentum $q_h$) 
is sufficiently large, the 
QCD factorization theorem allows to express the 
total amplitude as the convolution  
of a Compton process at the quark level and a non-perturbative 
amplitude parametrized in terms of generalized parton distributions 
(lower blob). The diagram with the photon lines crossed is also understood.    
}
\figlab{fig:ndelta_dvcs}
\end{figure}
\newline
\indent
Before discussing the $N \to \Delta$ GPDs, we start by 
briefly reviewing the $N \to N$ GPDs. 
The nucleon structure information entering the nucleon DVCS process, 
can be parametrized at leading twist-2 level, in
terms of four (quark chirality conserving) GPDs\footnote{
We do not consider chirally odd GPDs,
which are also discussed in the literature, 
see e.g. the review of Ref.~\cite{Diehl:2003ny}}. 
The GPDs depend on three variables: the quark longitudinal 
momentum fractions $x$ and $\xi$, and the momentum transfer 
$Q^2 = - q^2$ to the hadron.
The light-cone momentum\footnote{We use the definition
$a^{\pm}= (a^0 \pm a^3)/\sqrt{2}$ for the light-cone
components.} fraction $x$ is defined by $k^+ = x P^+$,
where $k$ is the quark loop momentum and
$P$ is the average nucleon momentum 
$P = (p + p^{\ \prime})/2$, where $p (p^{\ \prime})$
are the initial (final) nucleon four-momenta respectively,  
see \Figref{fig:ndelta_dvcs}.
The skewedness variable $\xi$ is
defined by $q^+ = - 2 \xi \,P^+$, where 
$q = p^{\ \prime} - p$ is the
overall momentum transfer in the process, and where
$2 \xi \rightarrow x_B/(1 - x_B/2)$ in the Bjorken limit: 
$x_B = Q_h^2/(2 p \cdot q_h)$ is the usual Bjorken scaling variable, 
with $Q_h^2 = -q_h^2 > 0$ the virtuality of the hard photon.
Furthermore, the third variable entering the GPDs
is given by the invariant $Q^2 = - q^2$, being the total
squared momentum transfer to the nucleon. 
\newline
\indent
The DVCS process 
corresponds with the kinematics $Q_h^2 \gg Q^2, M_N^2$, 
so that at twist-2 level, terms proportional to $Q^2 / Q_h^2$ 
or $M_N^2 / Q_h^2$ are neglected in the amplitude.  
In a frame where the virtual photon momentum \( q_h^{\mu } \) and the average
nucleon momentum \(  P^{\mu } \) are collinear
along the \( z \)-axis and in opposite directions, one can parametrize
the non-perturbative object entering the nucleon DVCS process as 
(following Ji~\cite{Ji:1996ek})\footnote{In all non-local
expressions we always assume the gauge link:
P$\exp(ig\int dx^\mu A_\mu)$, ensuring the color gauge
invariance.}:
\begin{eqnarray}
&& \frac{1}{2\pi} \, \int dy^{-}e^{ix  P^{+}y^{-}}
\left. \langle N(p^\prime)|\bar{\psi } (-y/2) \; \gamma \cdot n \; \psi (y/2)
| N(p) \rangle \right|_{y^{+}=\vec{y}_{\perp }=0} \nonumber \\
&&=\; H^{q}(x,\xi ,Q^2)\; \bar{u}(p^{'}) \; \gamma \cdot n \; u(p)
\,+\, E^{q}(x,\xi ,Q^2)\; \bar{u}(p^{'}) \; i\sigma^{\mu \nu} 
\frac{n_\mu \, q_\nu}{2 M_N} \; u(p) ,
\label{eq:qsplitting}
\end{eqnarray}
where \( \psi  \) is the quark field
of flavor $q$, \( u \) the nucleon spinor, and $n^\mu$ is a 
light-cone vector along the negative $z$-direction which can be 
expressed at twist-2 level\footnote{For kinematical expressions including 
correction terms proportional to $Q^2/Q_h^2$ and $M_N^2/Q_h^2$, 
which are formally of higher twist, see e.g. Ref.~\cite{Goeke:2001tz}.}
in terms of the external momenta as:
\begin{eqnarray}
q_h \,=\, - 2 \xi \, P \,+\, \frac{Q_h^2}{4 \xi} \, n  , \hspace{1.5cm}
q   \,=\, - 2 \xi \, P \,+\, q_\perp , \hspace{1.5cm}
q^\prime \,=\, \frac{Q_h^2}{4 \xi} \, n  \,-\, q_\perp , 
\end{eqnarray}
where $q_\perp = (0, {\bf q_\perp},0)$ 
is the transverse component of the momentum transfer $q$, 
satisfying $q_\perp \cdot n = q_\perp \cdot P = 0$. Furthermore the 
light-cone vector $n$, satisfying $n^2 = 0$ is normalized in 
such a way that $n \cdot P = 1$. 
The {\it lhs} of Eq.~(\ref{eq:qsplitting}) can be interpreted as a Fourier
integral along the light-cone distance $y^-$ of a quark-quark
correlation function, representing the process where
a quark is taken out of the
initial nucleon (having momentum $p$) at the space-time point $y/2$, and
is put back in the final nucleon (having momentum $p^{\ \prime}$) 
at the space-time
point $-y/2$. This process takes place at equal light-cone time ($y^+
= 0$) and at zero transverse separation ($\vec y_\perp = 0$) between
the quarks. The resulting one-dimensional Fourier integral along the
light-cone distance $y^-$ is with respect to the quark light-cone
momentum $x  P^+$.
The {\it rhs} of Eq.~(\ref{eq:qsplitting}) parametrizes this
non-perturbative object in terms of the GPDs $H^q$ and $E^q$ 
for a quark of flavor $q$.  
The quark vector operator ($\gamma \cdot n$) 
corresponds at the nucleon side to a vector transition 
(parametrized by the function $H^q$) and
a tensor transition (parametrized by the function $E^q$). 
Analogously, there are two GPDs corresponding with 
a quark axial vector operator ($\gamma \cdot n \gamma_5$), which are 
commonly denoted by the polarized GPDs $\tilde H^q$ and $\tilde E^q$. 
\newline
\indent
The variable $x$ in the GPDs runs from $-1$ to 1.
Therefore, the momentum fractions of the
active quarks ($x + \xi$) for the initial quark and ($x - \xi$) for the final 
quark can either be positive or negative. Since positive
(negative) momentum fractions correspond to quarks (antiquarks), it
has been noted in \cite{Radyushkin:1996nd} that in this way, one can
identify two regions for the GPDs:
when $x > \xi$ both partons represent quarks, whereas for
$x < - \xi$ both partons represent antiquarks. In these regions,
the GPDs are the generalizations of the usual parton distributions from
DIS. Actually, in the forward direction, the GPD $H$ 
reduces to the quark (anti-quark) density distribution $q(x)$ 
($\bar q(x)$) obtained from DIS:
 \begin{eqnarray}
\label{eq:dislimit}
H^{q}(x,0,0)\,
&=& \left\{
\begin{array}{cr}
q(x),& \hspace{.5cm} x \; > \; 0\,, \\
- \bar q(-x),& \hspace{.5cm} x \; < \; 0 \,.
\end{array}
\right.
\label{eq:dislimitp}
\end{eqnarray}
The function $E$ (and likewise the function $\tilde E$ for the axial operator) 
are not measurable through DIS because the associated tensor 
in Eq.~(\ref{eq:qsplitting}) vanishes in the forward limit ($q \to 0$).
Therefore, $E$ is a new leading twist function, which
is accessible by measuring hard exclusive electroproduction reactions, such 
as DVCS.
\newline
\indent
As the momentum fractions of initial and final quarks are
different, one accesses quark momentum correlations in the
nucleon. Furthermore, 
in the region $ -\xi < x < \xi$, one parton connected to the lower
blob in \Figref{fig:ndelta_dvcs} represents a
quark and the other one an antiquark. In this region, the GPDs
behave like a meson distribution amplitude and contain completely new
information about nucleon structure, because the region
$ -\xi < x < \xi$ is absent in DIS, which corresponds to the limit
$\xi \to 0$.
\newline
\indent
Besides coinciding with the quark distributions at vanishing momentum
transfer, the generalized parton distributions have interesting 
links with other
nucleon structure quantities. The first moments of the GPDs are related to
the elastic form factors
of the nucleon through model independent sum rules.
By integrating Eq.~(\ref{eq:qsplitting}) over \( x \), one
obtains for any $\xi$
the following relations for a particular quark flavor \cite{Ji:1996ek} :
\begin{eqnarray}
\int_{-1}^{+1}dx\, H^{q}(x,\xi ,Q^2)\,=\, F_{1}^{q}(Q^2)\, ,
\hspace{2.cm}
\int _{-1}^{+1}dx\, E^{q}(x,\xi ,Q^2)\,=\, F_{2}^{q}(Q^2)\, ,
\label{eq:ffsumrulehe}
\end{eqnarray}
where $F_1^q(Q^2)$ represents the elastic Dirac form factor for the
 quark flavor $q$ in the nucleon. These quark form factors are expressed, 
using $SU(2)$ isospin, 
as flavor combinations of the proton and neutron elastic form factors as:
\begin{eqnarray}
F_{1}^u \,=\, 2\,F_{1}^{p}\,+\,F_{1}^{n}\,+\,F_{1}^{s}\; , 
\hspace{2.5cm}
F_{1}^d \,=\, 2\,F_{1}^{n}\,+\,F_{1}^{p}\,+\,F_{1}^{s}\; ,
\label{eq:vecff}
\end{eqnarray}
where \( F_{1}^{p} \) and \( F_{1}^{n} \) are the proton and
neutron electromagnetic form factors respectively, with $F_1^p(0) = 1$
and $F_1^n(0) = 0$.
In Eq.~(\ref{eq:vecff}) \( F_{1}^{s} \)
is the strangeness form factor of the nucleon (which is neglected in 
the practical calculations below).
Relations similar to Eq.~(\ref{eq:vecff}) hold for the
Pauli form factors \( F_{2}^{q} \).
At $Q^2 = 0$, the normalizations of the Dirac form factors are given by:
$F_{1}^{u}(0) = 2$ ($F_{1}^{d}(0) = 1$) 
so as to yield the normalization of 2 (1) for the
$u$ ($d$)-quark distributions in the proton. 
The normalizations of the Pauli form factor at $Q^2 = 0$ are 
given by $F_{2}^{q}(0) = \kappa_q$ (for $q = u, d$), where 
$\kappa_u, \kappa_d$ can be expressed 
in terms of the proton ($\kappa_p$) 
and neutron ($\kappa_n$) anomalous magnetic moments as:
\begin{eqnarray}
\kappa_u \,\equiv \, 2 \kappa_p + \kappa_n = +1.673, 
\quad \quad \quad \quad 
\kappa_d \,\equiv \, \kappa_p + 2 \kappa_n = -2.033. 
\label{eq:kappaud}
\end{eqnarray}
The sum rules of Eq.~(\ref{eq:ffsumrulehe}) also satisfy the condition that 
they are independent of $\xi$, which is a consequence of Lorentz 
invariance\footnote{This is the simplest example
of a so-called polynomiality condition when calculating moments of GPDs.}.  
\newline
\indent
We next discuss the \( N \rightarrow \Delta  \) matrix element 
for the vector twist-2 operator, which was worked out in  
Refs.~\cite{Frankfurt:1999xe,Goeke:2001tz}.  
There are four $N \to \Delta$ helicity
amplitudes for the vector operator (as well as four for the 
axial-vector operator). 
However, the electromagnetic gauge invariance leads to
only three electromagnetic form factors. Hence, there are
four GPDs for the vector $N \to \Delta$ transition, 
of which one has a vanishing first moment. 
In the following we shall
neglect the GPD which has a vanishing first moment.
This approximation is justified in the large $N_c$ 
limit (discussed below) where  this GPD is subdominant. 
The non-perturbative object entering the 
$N \to \Delta$ DVCS process (lower blob in \Figref{fig:ndelta_dvcs}) 
can then be expressed as~\cite{Goeke:2001tz,Frankfurt:1999xe}:
\begin{eqnarray}
&& \frac{1}{2\pi} \, \int dy^{-}e^{ix  P^{+}y^{-}}
\langle \Delta(p_\Delta)|\bar{\psi } (-y/2) \; \gamma \cdot n 
\, \tau_3 \; \psi (y/2)
 | N(p) \rangle {\Bigg |}_{y^{+}=\vec{y}_{\perp }=0} \nonumber \\
&& =  \sqrt{\frac{2}{3}} \; u^{\alpha }(p_{\Delta })\; 
\left\{ \, H_{M}(x,\xi ,Q^2)\; 
\left( -{\mathcal{K}}_{\alpha \mu }^{M} \right) \, n^{\mu }\right. 
\;+\; H_{E}(x,\xi ,Q^2)\; 
\left( -{\mathcal{K}}_{\alpha \mu }^{E} \right) \, n^{\mu } \nonumber \\
&&\hspace{3cm} \left. +\; H_{C}(x,\xi ,Q^2)\; 
\left(- {\mathcal{K}}_{\alpha \mu }^{C} \right) \, 
n^{\mu }\right\} \, u(p), 
\label{eq:ndelvec} 
\end{eqnarray}
where \( u ^{\alpha }(p_{\Delta }) \) is the Rarita-Schwinger spinor
for the \( \Delta  \)-field, 
$\tau_3 /2$ is the third isospin generator for quarks, 
and $\sqrt{2/3}$ is the isospin
factor for the $p \to \Delta^+$ transition. 
\newline
\indent
Furthermore, in Eq.~(\ref{eq:ndelvec}), 
the covariants \( {\mathcal{K}}^{M,E,C}_{\alpha \mu } \)
are the magnetic dipole, electric quadrupole, and Coulomb quadrupole
Jones-Scadron covariants \cite{Jones:1972ky}:
\begin{eqnarray}
{\mathcal{K}}_{\alpha \mu }^{M} & = & 
-i\frac{3(M_{\Delta }+M_N)}{2M_N Q_+^2} 
\varepsilon_{\alpha \mu \lambda \sigma} 
P^{\lambda } q^{\sigma}\; ,\nonumber \\
{\mathcal{K}}_{\alpha \mu }^{E} & = & -{\mathcal{K}}_{\alpha \mu }^{M}-\frac{6(M_{\Delta }+M_N)}{M_N Q_+^2 Q_-^2 }
\varepsilon _{\alpha \sigma \lambda \rho }P^{\lambda } q^{\rho } \, 
\varepsilon^{\sigma}_{\, \, \mu \kappa \delta }P^{\kappa} q^{\delta}\gamma_5\;,
\label{K-def} \\
{\mathcal{K}}_{\alpha \mu }^{C} & = & 
-i\frac{3(M_{\Delta}+M_N)}{M_N Q_+^2 Q_-^2} 
q_{\alpha }(q^2 P_{\mu }- q \cdot P q_{\mu })\gamma_5 \; .
\nonumber 
\end{eqnarray}
Here \( P=(p_{\Delta }+p)/2 \), \( q = p_{\Delta }-p \), 
and \( p_{\Delta }^{2}=M_{\Delta }^{2} \). 
In Eq.~(\ref{eq:ndelvec}), the GPDs \( H_{M} \), \( H_{E} \),
and \( H_{C} \) for the \( N\rightarrow \Delta  \) vector transition
are linked with the three \( N\rightarrow \Delta  \) vector current
(Jones-Scadron) transition form factors 
\( G_{M}^{*} \), \( G_{E}^{*} \), and \( G_{C}^{*} \) introduced in 
Eq.~(\ref{eq:JS}) through the sum rules: 
\begin{eqnarray}
\int _{-1}^{1}dx\, \, H_{M,E,C}(x,\xi ,Q^2)
=2\, \, G_{M,E,C}^{*}(Q^2)\; ,
\label{eq:gmgegcsumrule} 
\end{eqnarray}
where the factor 2 arises because the electromagnetic form factors are 
conventionally defined with isospin generator $\tau_3 / 2$ in the 
current operator in contrast to the operator $\tau_3$ in 
Eq.~(\ref{eq:ndelvec}) adopted in 
Refs.~\cite{Frankfurt:1999xe,Goeke:2001tz} to define GPDs. 
\newline
\indent
The above sum rules allow us to make a prediction for the $N \to \Delta$ form 
factors provided we have a model for the $N \to \Delta$ GPDs. Such a model 
will be discussed in the following sections. Conversely, the existing 
precise experimental information on the $N \to \Delta$ vector form factors 
provides a strong constraint on the $N \to \Delta$ GPDs. 
As discussed above, the GPDs are however much richer observables and 
provide us with quark distribution information in the $\Delta$ resonance. 
They can be accessed by the $N \to \Delta$ DVCS process as discussed in 
Ref.~\cite{GMV03}.  
First experiments which are sensitive to the $N \to \Delta$ GPDs 
have recently been reported~\cite{Guidal:2003ji}.

\subsubsection{Model for the magnetic dipole $N \to \Delta$ GPD}
\label{sec:gpdmodel}

Here we will be guided by the large
\( N_{c} \) limit, which allows to 
connect the \( N\rightarrow \Delta  \) GPD \( H_{M} \), 
to the \( N\rightarrow N \) isovector GPDs. 
For the magnetic $N \to \Delta$ transition, 
it was shown by Frankfurt {\it et al.}~\cite{Frankfurt:1999xe} that, 
in the large-$N_c$ limit, the 
relevant  $N \to \Delta$ GPD $H_M$ can be expressed in terms of the
nucleon isovector GPD $E^u - E^d$: 
\begin{eqnarray}
H_{M}(x,\xi ,Q^2) & = & 2  \frac{G_M^*(0)}{\kappa_V}  
 \left\{ E^{u}(x,\xi,Q^2) - E^{d}(x, \xi ,Q^2) \right\} ,
\label{eq:hmparam} 
\end{eqnarray}
where $\kappa_V = \kappa_p - \kappa_n = 3.70$. 
Within the large $N_c$ approach used in  Ref.~\cite{Frankfurt:1999xe},
the value $G_M^{*}(0)$ is  given by\footnote{Note the typo 
in the formula for $H_M$ in 
Ref.~\cite{Frankfurt:1999xe}. Due to a different choice of isospin 
factors for the vector and axial vector transitions chosen there, one should 
correct Eq.~(7) in Ref.~\cite{Frankfurt:1999xe} to be 
$H_M = \sqrt{2} (E^u - E^d)$ instead of 
$H_M = \sqrt{\frac{2}{3}} \sqrt{2} (E^u - E^d)$.}
$G_M^{*}(0) = \kappa_V / \sqrt{2}$, 
which  is about  20\% smaller than the
experimental number\footnote{In the large-$N_c$ limit, 
the isovector combination $H^u - H^d$ is suppressed, 
therefore one could as well give as estimate 
$\mu_{p \to \Delta^+} \simeq 
G_{M}^{*}(0)\simeq {1 \over {\sqrt{2}}} \, (\mu_p - \mu_n) \simeq 3.32 $ 
(where the magnetic moments are expressed in nuclear magnetons whereas  
$G_M^\ast$ is dimensionless),  
corresponding with Eq.~(\ref{eq:mulargenc}) 
where $M_\Delta \simeq M_N$, in the large-$N_c$ limit. 
This estimate is accurate at the 10 \% level.}. 
In order to give more realistic estimates, we will therefore use 
in the following calculations 
the phenomenological value $G_M^{*}(0) \approx 3.02$ of 
Eq.~(\ref{eq:mundelexp}). 
All other (sub-dominant) GPDs for the vector $N \to \Delta$ transition 
vanish at leading order in the 1/\( N_{c} \) expansion, consistent with the 
large $N_c$ limit for the $\gamma N \Delta$ form factors discussed in 
Subsect.~\ref{sec:largenc}.
\newline
\indent
Using the large $N_c$ estimate of Eq.~(\ref{eq:hmparam}), the sum rule 
Eq.~(\ref{eq:gmgegcsumrule}) for $G_M^\ast$ can be written as:    
\begin{eqnarray}
G_M^{*}(Q^2) \,&=&\, {{G_M^{*}(0)} \over {\kappa_V}} \; \int _{-1}^{+1}dx\; 
\biggl \{ E^{u}(x,\xi ,Q^2) \,-\, E^{d}(x,\xi ,Q^2) \biggr \} , 
\nonumber \\
&=&\,  {{G_M^{*}(0)} \over {\kappa_V}} \; 
\biggl \{ F_2^p(Q^2) - F_2^n(Q^2) \biggr \} \, ,  
\label{eq:gmsumrule} 
\end{eqnarray}
where $F_2^p - F_2^n$ is the isovector combination of the 
proton (p) - neutron (n) Pauli form factors. 
Because the sum rule of Eq.~(\ref{eq:gmsumrule}) is independent of $\xi$, 
we only need to constrain the GPD $E^q$ for $\xi = 0$ in order to 
evaluate $G_M^\ast$. 
The sum rule (\ref{eq:gmsumrule}) was 
used in Ref.~\cite{Stoler:2002im}, using a  model 
\cite{Stoler:2001xa}  in which
the Gaussian ansatz  for GPDs is  modified at large $Q^2$ by  terms having 
a power-law behavior. Refs.~\cite{Diehl:2004cx,guidal} used parametrizations 
which are motivated from the expected Regge behavior of the 
GPDs at small $x$ and $Q^2$.
Guidal {\it et al.}~\cite{guidal} parametrized 
the function $E^q(x, 0, Q^2)$ at low $Q^2$ 
through a Regge-type form (denoted by model $R1$) as: 
\begin{eqnarray}
E^q_{R1}  (x,0,Q^2) = e^q (x)\,  x^{\alpha' Q^2}.
\label{eq:gpde_r1}
\end{eqnarray}
The forward magnetic densities  $e^q (x)$ - unlike the usual forward 
parton densities $q(x)$ - are unfortunately not known from 
experiment at present. 
The simplest idea is to take them proportional to the valence 
up-quark ($u_v(x)$) and down-quark ($d_v(x)$) densities as:
\begin{eqnarray}
e^u (x) = \frac{\kappa_u}{2} u_v(x)  
 \qquad {\rm and}  \qquad e^d (x) = \kappa_d d_v(x) \  ,
\label{eq:er1}
\end{eqnarray}
which satisfy the normalization constraint of 
Eq.~(\ref{eq:ffsumrulehe}) at $Q^2 = 0$:
\begin{eqnarray}
\kappa_q = \int d x \, e^q(x). 
\label{eq:normeo}
\end{eqnarray}
where $\kappa_u$ and $\kappa_d$ are defined in Eq.~(\ref{eq:kappaud}). 
One thus sees that the functions $e^q(x)$ encode the quark distribution 
information giving rise to the nucleon anomalous magnetic moments.  
\newline
\indent
As shown in Ref.~\cite{guidal}, the Regge model $R1$ fits the nucleon Dirac  
($F_1$) and Pauli ($F_2$) form factor data for small momentum transfers 
$Q^2 \lesssim 0.5$\,GeV$^2$.
However, at larger $Q^2$ the $R1$ model gives too strong suppression, 
and it consequently falls considerably short of the data for 
$Q^2 > 1$\,GeV$^2$.
\newline
\indent
Experimentally,  the proton helicity flip  form factor $F_2(Q^2)$ has a faster 
power fall-off  
at large $Q^2$ than $F_1(Q^2)$.  
This  means that the  $x\sim 1$ behavior of
the functions $e(x)$ and $q(x)$ should be different.
To produce a faster decrease  with $Q^2$, the $x\sim 1 $ limit of the density 
 $e^q(x)$ should have extra powers of $1-x$ compared to that  of 
 $q(x)$.
Aiming to avoid introducing too many free
parameters, Guidal {\it et al.}~\cite{guidal} tried 
the next simplest ansatz for $e^q(x)$ by just multiplying the 
valence quark distributions by an additional
factor $(1 - x)^{\eta_q}$, i.e. by taking: 
\begin{eqnarray}
e^u (x) = \frac{\kappa_u}{N_u} (1-x)^{\eta_u} u_v(x)  
 \qquad {\rm and}  \qquad e^d (x) = \frac{\kappa_d}{N_d} 
 (1-x)^{\eta_d} d_v(x) \  ,
\label{eq:e2}
\end{eqnarray}
where  the normalization factors $N_u$ and $N_d$  
\begin{eqnarray}
N_u \,=\,  \int _{0}^{1}dx \; (1 - x)^{\eta_u} \, u_v(x) \, ,
\quad \quad \quad \quad 
N_d \,=\,  \int _{0}^{1}dx \; (1 - x)^{\eta_d} \, d_v(x) \, ,
\label{eq:nd}  
\end{eqnarray} 
guarantee the condition of Eq.~(\ref{eq:normeo}). 
In such modified Regge parametrization (denoted by $R2$), the GPD  
$E^q$ entering the sum rule Eq.~(\ref{eq:gmsumrule}) for $G_M^\ast$ 
was parametrized in Ref.~\cite{guidal} as:
\begin{eqnarray}
E^q_{R2}(x, 0, Q^2) \,&=&\, {\kappa_q  \over {N_q}} 
\, (1 - x)^{\eta_q} \, q_v(x) \, 
{{x^{\alpha' \, (1 - x) \, Q^2}}} \, , 
\label{eq:gpde_r2}
\end{eqnarray}
with $q_v$ the valence quark distribution ($q = u, d$), and  
where the powers $\eta_u$ and $\eta_d$ have been determined from a fit 
to the nucleon form factor data as   
$\alpha^{'} = 1.105$\, GeV$^{-2}$, $\eta_u$ = 1.713 and $\eta_d$ = 0.566. 
Note that a value $\eta_q = 2$ 
corresponds to a $ 1/Q^2$ asymptotic behavior of the ratio
$F_2^q(Q^2)/F_1^q(Q^2)$ at large $Q^2$. 
\begin{figure}
\centerline{  \epsfxsize=10cm%
  \epsffile{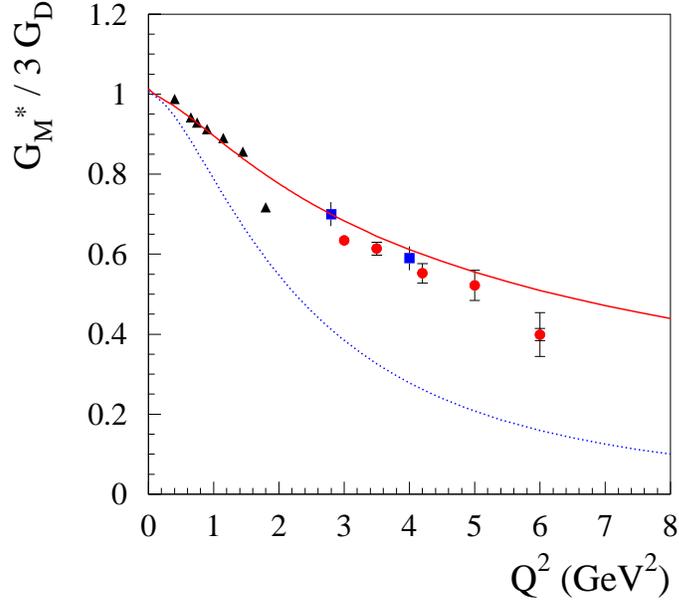} 
}
\caption{
The $N \to \Delta$ magnetic 
transition (Jones-Scadron) form factor $G_M^*$, relative to the 
dipole form (multiplied by a factor 3). 
The curves are calculated from a parametrization of the 
$N \to \Delta$ magnetic GPD $H_M$ of Ref.~\cite{guidal}. Blue dotted curve : 
Regge GPD parametrization according to Eq.~(\ref{eq:gpde_r1}).  
Red solid curve : modified Regge GPD parametrization according to 
Eq.~(\ref{eq:gpde_r2}).
The data points are from JLab Hall C~\cite{Frolov:1998pw} (blue squares), and  
JLab CLAS : Refs.~\cite{Joo:2001tw} (black triangles) 
and~\cite{Ungaro:2006df} (red circles). 
The data of Refs.~\cite{Joo:2001tw,Ungaro:2006df} have been analyzed 
using the unitary isobar model of Ref.~\cite{Aznauryan:2002gd}.}
\figlab{fig:gmdel}
\end{figure}
\newline
\indent
In the following estimates, the unpolarized valence quark distributions 
are taken at input scale $\mu^2$ = 1 GeV$^2$ from the 
MRST2002 global NNLO fit~\cite{Martin:2002dr} as:  
\begin{eqnarray}
u_v &=& 0.262 \, x^{-0.69} (1 - x)^{3.50} 
\left( 1 + 3.83 \, x^{0.5} + 37.65 \, x \right),  \\
d_v &=& 0.061 \, x^{-0.65} (1 - x)^{4.03} 
\left( 1 + 49.05 \, x^{0.5} + 8.65 \, x \right) . 
\end{eqnarray}
For the one-parameter Regge form $R1$ of Eq.~(\ref{eq:gpde_r1})  
the same parameter value $\alpha^{'}= 1.105 \,$ GeV$^{-2}$ 
was chosen as in the model $R2$, which gives a good description of 
the proton charge radius. In \Figref{fig:gmdel}, we show the 
results for the sum rule predictions for $G_M^\ast$ using the 
GPD parametrizations of Eqs.~(\ref{eq:gpde_r1}) and (\ref{eq:gpde_r2}). 
It is seen that both the Regge and modified Regge 
parametrizations yield a magnetic $N \to \Delta$ form factor which decreases 
faster than a dipole, in qualitative agreement with the data. 
The $R1$ Regge parametrization though gives too large suppression 
at larger $Q^2$, 
as was also observed for the nucleon elastic form 
factors in Ref.~\cite{guidal}. For the modified Regge parametrization $R2$, 
it is seen that the sum rule prediction based on the large $N_c$ estimate of 
Eq.~(\ref{eq:gmsumrule}) gives a good quantitative 
description of the data over the whole $Q^2$ range without adjusting 
any parameters beyond the three parameters $\alpha^\prime$, $\eta_u$, 
and $\eta_d$ which were determined from a fit to the nucleon elastic form 
factor data. 

\subsubsection{Model for the electric quadrupole $N \to \Delta$ GPD}
\label{sec:gpdhemodel}

In this section, we make a very first attempt to model the electric quadrupole 
GPD, $H_E$. As in our modeling of $H_M$, we will 
also be guided by the large $N_c$ limit. In the large $N_c$ limit, 
Eq.~(\ref{eq:qpdelrnlargenc}) provides a relation between the $N \to \Delta$ 
quadrupole moment and the neutron charge radius $r_n^2$, 
which for $N_c = 3$ reduces to:
\begin{eqnarray} 
Q_{p \to \Delta^+} \,=\, \frac{1}{\sqrt{2}} \, r_n^2 .
\label{eq:qpdelrnlargenc2}
\end{eqnarray}
Using Eq.~(\ref{eq:qndel}), we can express Eq.~(\ref{eq:qpdelrnlargenc2}) 
in a relation for $G_E^*(0)$, which reads to leading accuracy in the 
$1/N_c$-expansion as:
\begin{eqnarray}
G_E^\ast(0) \,=\, - \frac{1}{6} \, r_n^2 \, \frac{1}{\sqrt{2}} \, 
\frac{(M_\Delta^2 - M_N^2)}{2} .
\label{eq:ge0largenc}
\end{eqnarray}
For small values of $Q^2$, we can express the neutron electric form factor 
as $G_E^n(Q^2) \approx - r_n^2 \, Q^2 / 6$. Therefore, a natural extension 
of the large $N_c$ relation of Eq.~(\ref{eq:ge0largenc}) to finite $Q^2$ 
is given by:
\begin{eqnarray}
G_E^\ast(Q^2) \,=\, \frac{1}{\sqrt{2}} \, \frac{(M_\Delta^2 - M_N^2)}{2} \, 
\frac{G_E^n(Q^2)}{Q^2} .
\label{eq:geqlargenc}
\end{eqnarray}
\begin{figure}
\centerline{  \epsfxsize=10cm%
  \epsffile{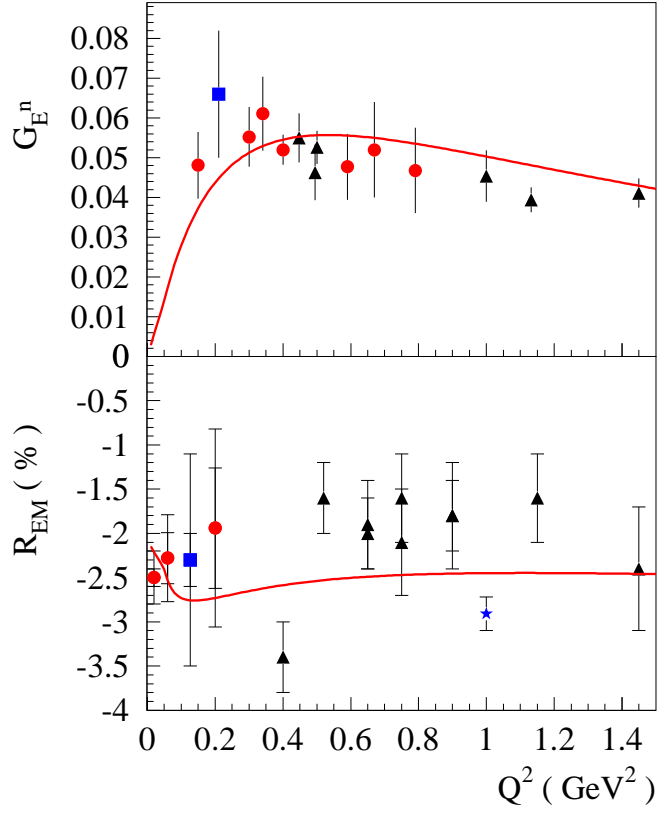} 
}
\caption{GPD calculations for the $Q^2$ dependence of the
neutron electric form factor $G_E^n$ (upper panel) in comparison with  
the $N \to \Delta$ $R_{EM}$ ratio (lower panel). 
For the neutron electric form factor, the modified Regge GPD parametrization 
$R2$ of Ref.~\cite{guidal} is used.  
For the $R_{EM}$ ratio, the large $N_c$ relations 
Eq.~(\ref{eq:geqlargenc}) for $G_E^\ast$, and 
Eq.~(\ref{eq:gmsumrule}) for $G_M^\ast$ are used, which express the 
$N \to \Delta$ form factors in 
terms of nucleon elastic form factors, which in turn 
are calculated using the modified Regge GPD parametrization $R2$.  
The data points for $G_E^n$ are from 
MAMI~\cite{Herberg:ud,Ostrick:xa,Becker:tw,Rohe:sh} (red circles), 
NIKHEF~\cite{Passchier:1999cj} (blue square), 
and JLab~\cite{Zhu:2001md,Warren:2003ma,Madey:2003av} (black triangles). 
The data points for $R_{EM}$ 
are from BATES at $Q^2=0.127$~\cite{Sparveris:2004jn} (blue square);   
MAMI (red circles): $Q^2=0$~\cite{Beck:1999ge},
$Q^2=0.06$~\cite{Stave:2006ea},  
$Q^2=0.2$~\cite{Sparveris:2006}; 
JLab CLAS~\cite{Joo:2001tw} (black triangles); 
and JLab HallA~\cite{Kelly05} (blue star). 
}
\figlab{fig:gengendel}
\end{figure}
\newline
\indent
The prediction which follows from the large $N_c$ motived 
expression of Eq.~(\ref{eq:geqlargenc}) is 
tested in \Figref{fig:gengendel} by comparing the 
$Q^2$ dependence of the neutron electric form factor $G_E^n$ 
and the $N \to \Delta$ $R_{EM}$ ratio. 
For $G_E^n$ we use the modified Regge parametrization $R2$ of 
\cite{guidal}, which is seen to give a fairly good description of the 
available double polarization data. The $R_{EM}$ ratio is calculated using 
the large $N_c$ relations Eq.~(\ref{eq:geqlargenc}) for $G_E^\ast$, and 
Eq.~(\ref{eq:gmsumrule}) for $G_M^\ast$, as discussed 
in Sect.~\ref{sec:gpdmodel}. 
These relations express the 
$N \to \Delta$ form factors in terms of nucleon elastic form factors.  
By using the three 
parameter $R2$ Regge form for the nucleon elastic form factors, 
we obtain in this way a prediction for $R_{EM}$ without adjusting any 
additional parameter. One sees that this yields a $R_{EM}$ ratio 
which has both the right size and displays 
a relatively flat $Q^2$ behavior, up to a $Q^2$ value of about  
1.5~GeV$^2$, in surprisingly good agreement with the data. 
\newline
\indent
The form factor $G_E^\ast$ is obtained from the first moment of 
the electric quadrupole $N \to \Delta$ GPD $H_E$ through 
the sum rule of Eq.~(\ref{eq:gmgegcsumrule}). We can therefore use 
Eq.~(\ref{eq:geqlargenc}) to propose a relation between the 
$N \to \Delta$ GPD $H_E$ and the neutron electric GPD combination, 
which is consistent with this form factor sum rule, as: 
\begin{eqnarray}
H_E(x, 0, Q^2) \,=\, \frac{1}{\sqrt{2}} \, 
\frac{(M_\Delta^2 - M_N^2)}{Q^2} \, 
\left\{ H^{(n)}(x,0,Q^2) - \frac{Q^2}{4 M_N^2} E^{(n)}(x,0,Q^2) \right\} , 
\label{eq:helargenc}
\end{eqnarray}
where the neutron GPDs are obtained in terms of the $u$- and $d$-quark 
flavor GPDs as: 
$H^{(n)} = -1/3 H^u + 2/3 H^d$, and $E^{(n)} = -1/3 E^u + 2/3 E^d$.
\newline
\indent
An interesting topic for future work is 
to perform different model calculations for 
$H_M$, $H_E$, and $H_C$, as well as provide lattice QCD predictions for 
its moments, in order to cross-check the above estimates for $H_M$ and $H_E$.

\subsubsection{GPDs and transverse structure of hadrons}

\begin{figure}
\centerline{  \epsfxsize=15cm%
  \epsffile{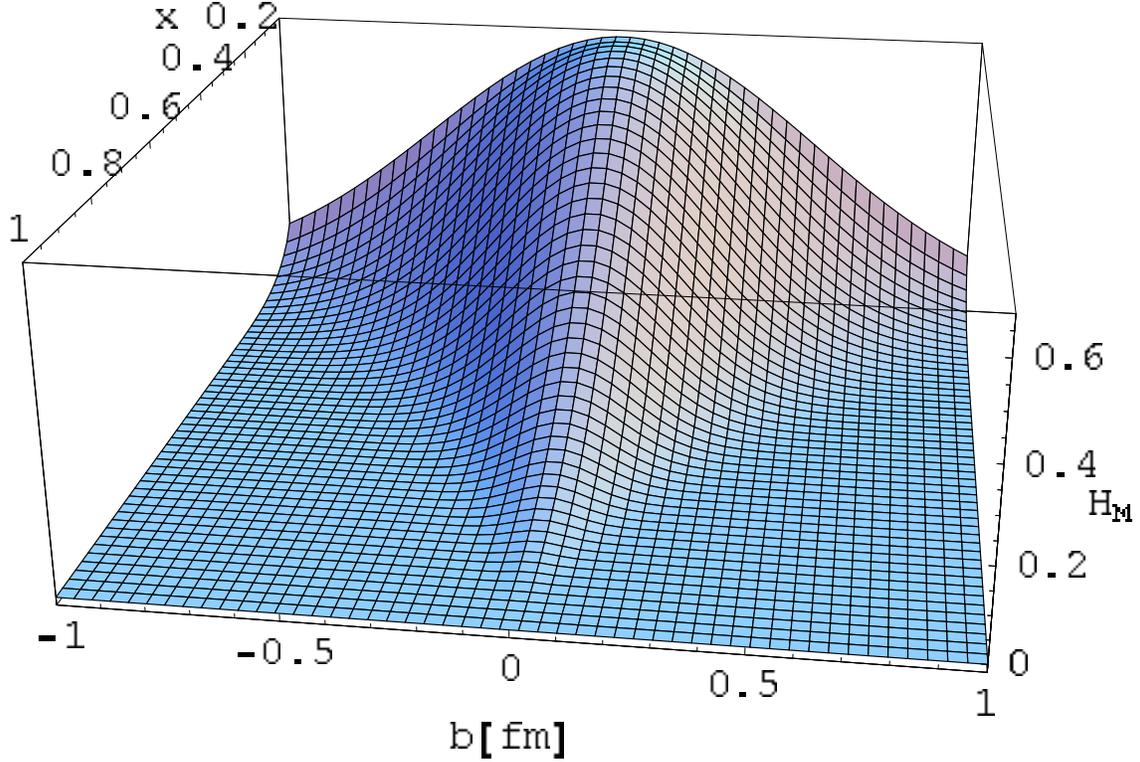} 
}
\caption{
The magnetic dipole $N \to \Delta$ GPD $H_M$ 
as function of the quark momentum fraction $x$ and 
the quark position $b$ in the transverse plane (perpendicular to the 
average direction of the fast moving baryons), 
where $b$ stands for either 
$({\bf b_\perp})_x$ or $({\bf b_\perp})_y$.
The calculation is based on the modified Regge parametrization $R2$ 
of Eq.~(\ref{eq:gpde_r2}). 
}
\figlab{fig:gpdb_gmdel}
\end{figure}

The interplay between the $x$ and $Q^2$-dependence of the GPDs 
contains new nucleon structure information beyond the information encoded 
in forward parton distributions depending only on $x$, or form factors 
depending only on $Q^2$. 
It has been shown that by a Fourier transform of the
$Q^2$-dependence of GPDs, it is conceivable to access the
distributions of parton in the transverse plane, 
see Ref.~\cite{Burkardt:2000za}, 
and to provide a 3-dimensional picture of the nucleon~\cite{Belitsky:2003nz}.  
\newline
\indent
For $\xi = 0$, one can define 
the  impact parameter versions of GPDs 
which are obtained through a Fourier integral 
in transverse momentum $q_\perp$. For the GPD $E^q$, this reads as:
\begin{eqnarray}
E^q(x, {\bf b_\perp}) \,& \equiv &\, 
\int \frac{d^2 {\bf q_\perp}}{(2 \pi)^2} \, 
e^{i {\bf b_\perp \cdot q_\perp}} \;
E^q (x, 0,- {\bf q_\perp^2}) ,  
\end{eqnarray}
and analogous definitions for the other GPDs. 
These impact parameter GPDs have the physical meaning of measuring the 
probability to find a quark which carries longitudinal 
momentum fraction $x$ at a transverse position 
${\bf b_\perp}$ (relative to the transverse center-of-momentum) in a nucleon, 
see Refs.~\cite{Burkardt:2000za,Burkardt:2002hr}.
\newline
\indent
When translating  the GPD parametrization $R2$ of Eq.~(\ref{eq:gpde_r2}), 
into the  impact parameter space, we obtain for the GPD $E$:
\begin{eqnarray}
E^q (x, {\bf b_\perp}) \,&=&\, \frac{\kappa_q}{N_q} \, (1 - x)^{\eta_q} 
\, q_v(x) \, 
\frac{e^{- {\bf b_\perp}^2 \,/\, [ - 4 \, \alpha' \,
 (1-x) \ln x ]}}{4 \pi \, \left[-\alpha' (1-x) \ln x  \right]}  . 
\label{eq:her2b}
\end{eqnarray}
Using the large $N_c$ relation Eq.~(\ref{eq:hmparam}) we can then express 
the impact parameter version of the magnetic GPD $H_M$ as:
\begin{eqnarray}
H_{M}(x,{\bf b_\perp}) & = & 2  \frac{G_M^*(0)}{\kappa_V}  
 \left\{ E^{u}(x, {\bf b_\perp}) - E^{d}(x, {\bf b_\perp}) \right\} .
\label{eq:hmparamb} 
\end{eqnarray}
\newline
\indent
In \Figref{fig:gpdb_gmdel}, we display the impact parameter magnetic 
GPD $H_M(x, {\bf b_\perp})$ using the Regge parametrization $R2$ of 
Eq.~(\ref{eq:her2b}) for $E^q(x, {\bf b_\perp})$.  
It is clearly seen from this image that for large values of $x$, the 
quark distributions are concentrated at small values of ${\bf b_\perp}$, 
reflecting the distribution of valence quarks in the core of the $N$ and 
$\Delta$. On the other hand, at small values of $x$, the distribution in 
transverse position extends much further out.

\subsection{Perturbative QCD (pQCD)}
\label{sec2_pqcd}

\subsubsection{pQCD predictions for helicity amplitudes and form factors}

\begin{figure}
\centerline{  \epsfxsize=9cm%
  \epsffile{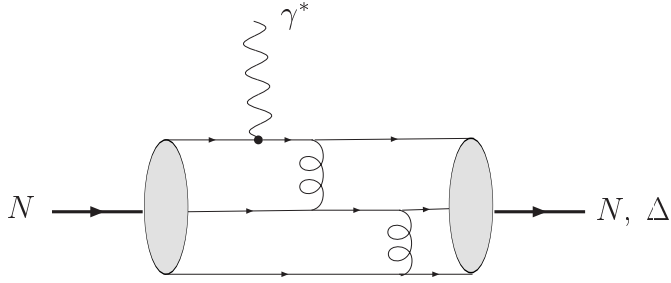} 
}
\caption{Perturbative QCD picture for the nucleon elastic and $N \to \Delta$ 
transition electromagnetic form factors. The highly virtual photon resolves 
the leading three-quark Fock states of $N$ and $\Delta$, described by 
a distribution amplitude. The large momentum is transferred between the quarks 
through two successive gluon exchanges (only one of several possible 
lowest-order diagrams is shown). }
\figlab{fig:ff_pqcd}
\end{figure}
\indent
The electro-excitation of the $\Delta$ 
provides a famous test for perturbative QCD, 
where scaling and selection rules for dominant helicity amplitudes 
were derived and are expected to be valid at sufficiently high momentum 
transfers $Q^2$~\cite{Lepage:1980fj}. A photon of sufficient high 
virtuality will 
see a nucleon (or $\Delta$) consisting of three massless quarks moving 
collinear with the nucleon. 
When measuring an elastic nucleon form factor or a 
$N \to \Delta$ transition form factor, the final state consists again of 
three massless collinear quarks. In order for this (unlikely process) to 
happen, the large momentum of the virtual photon has to be transferred 
among the three quarks through two hard gluon exchanges as illustrated in 
\Figref{fig:ff_pqcd}. This hard scattering mechanism is generated by 
valence quark configurations with small transverse size and finite 
light cone momentum fractions of the total hadron momentum carried by each 
valence quark. The hard amplitude can be written in a factorized 
form~\cite{Chernyak:1977as,Chernyak:1977fk,Efremov:1979qk,Lepage:1980fj}, 
as a product of a perturbatively calculable hard scattering amplitude and 
two distribution amplitudes describing how the large longitudinal momentum 
of the initial and final hadrons is shared between their constituents. 
Because each gluon in such hard scattering process carries a virtuality 
proportional to $Q^2$, this leads to the pQCD prediction that the helicity 
conserving nucleon Dirac form factor $F_1$ should fall as $1/Q^4$ 
(modulo $\ln Q^2$ factors) at sufficiently high $Q^2$. 
Processes such as in 
\Figref{fig:ff_pqcd}, where the interactions among the quarks proceed 
via gluon or photon exchange, both of which are vector interactions, 
conserve the quark helicity in the limit when the quark masses or off-shell 
effects can be neglected. 
In contrast to the helicity conserving form factor $F_1$, the nucleon 
Pauli form factor $F_2$ involves a helicity flip between the initial and 
final nucleons. Hence it requires one helicity flip at the quark level, which 
is suppressed at large $Q^2$. Therefore, for collinear quarks, i.e. 
moving in a light-cone wave function state with 
orbital angular momentum projection 
$l_z = 0$ (along the direction of the fast moving hadron), 
the asymptotic prediction for $F_2$ leads to a 
$1/Q^6$ fall-off at high $Q^2$. 
\begin{figure}
\centerline{  \epsfxsize=9cm%
  \epsffile{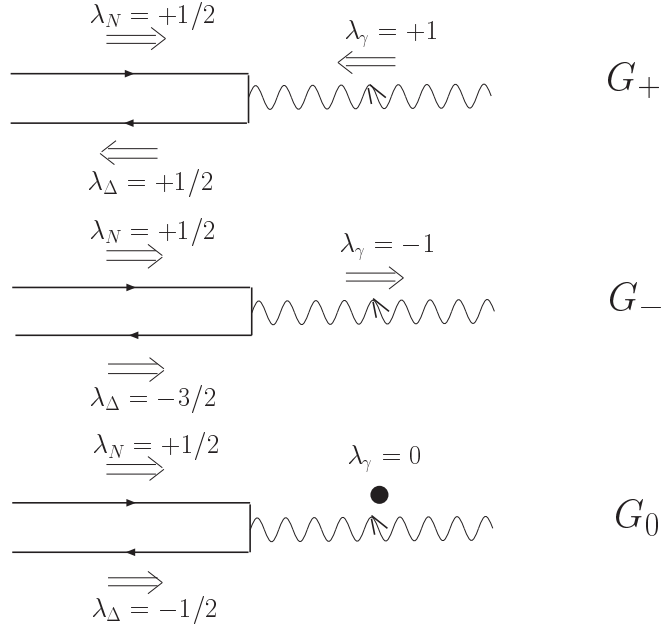} 
}
\caption{$\gamma^\ast N \to \Delta$ helicity amplitudes in the Breit frame, 
where $N$, $\Delta$, and $\gamma^\ast$ momenta are collinear with the 
$N$ momentum, and where the $N$ and $\Delta$ three-momenta have the same 
magnitude but opposite sign. The $\gamma$, $N$, and $\Delta$ helicities 
($\lambda_\gamma$, $\lambda_N$, and $\lambda_\Delta$ respectively) are 
indicated on the figure. The corresponding helicity form factors in the 
notation of Ref.~\cite{Carlson:1985mm} are indicated on the right. }
\figlab{fig:ndel_helff}
\end{figure}
\newline
\indent
To derive the analogous relations for the $\gamma^* N \Delta$ transition 
form factors, it is most convenient to work in a Breit frame, in which 
the photon, $N$, and $\Delta$ momenta are collinear with the incoming $N$. 
The outgoing $\Delta$ momentum has opposite direction but the same magnitude 
three-momentum. In this way, three independent $N \to \Delta$ helicity form 
factors were introduced in Ref.~\cite{Carlson:1985mm} as illustrated in 
\Figref{fig:ndel_helff}. 
They are defined as (for $m = +, -, 0$):
\begin{eqnarray}
G_m \,&\equiv&\, \frac{1}{2 M_N}  
\langle \; \Delta, \, \lambda_\Delta = \lambda_\gamma - \frac{1}{2} 
\,|\, \epsilon_\mu^{(m)} \cdot J^\mu \,|\, N, \, \lambda_N = +\frac{1}{2} 
\;\rangle, 
\end{eqnarray}
where $J^\mu$ is the electromagnetic current operator, and the factor 
$1/(2 M_N)$ is chosen to make $G_m$ dimensionless. 
The $\gamma$, $N$, and $\Delta$ helicities are denoted by 
$\lambda_\gamma$, $\lambda_N$, and $\lambda_\Delta$ respectively. 
The transverse polarization vectors (for a photon moving in the $z$-direction) 
are $\epsilon^{\pm} = (0, \mp 1, -i, 0) /\sqrt{2}$, 
and the longitudinal polarization vector satisfies 
$\epsilon_\mu^{(0)} \cdot \epsilon^{(0) \mu} = 1$, 
$\epsilon_\mu^{(0)} \cdot \epsilon^{(\pm) \mu} = 0$, and 
$\epsilon_\mu^{(0)} \cdot q^\mu = 0$.
One can express the $N \to \Delta$ helicity form factors 
$G_+$, $G_-$, and $G_0$ in terms of the Jones-Scadron form factors 
$G_M^\ast$, $G_E^\ast$, and $G_C^\ast$ as:
\begin{eqnarray}
G_- \, &\equiv& \, - \frac{\sqrt{3}}{2 \, \sqrt{2}} \, 
\frac{Q_- \, (M_\Delta + M_N)}{2 \, M_N^2} \, 
\left\{ G^*_M \,+\, G^*_E \right\} , 
\nonumber \\
G_+ \, &\equiv& \, \frac{1}{2 \, \sqrt{2}} \,
\frac{Q_- \, (M_\Delta + M_N)}{2 \, M_N^2} \, 
\left\{ G^*_M \,-\,3\, G^*_E \right\} , 
\nonumber \\
G_0 \, &\equiv& \, \frac{1}{2} \, 
\frac{Q_- \, (M_\Delta + M_N)}{2 \, M_N^2} \, \frac{Q}{M_\Delta} \,
\left\{ \,-\, G^*_C \right\} .
\label{eq:jshel}
\end{eqnarray}
\newline
\indent
Using the hadron helicity-conserving property of QCD at high $Q^2$, it is 
then easy to derive the asymptotic behavior of the $\gamma^\ast N \Delta$ 
transition form factors. Since $G_+$ is the only helicity amplitude with 
the same helicity between initial $N$ and final $\Delta$, it will be 
the leading amplitude. 
At large $Q^2$, two-gluon exchange between collinear quarks yields a 
$1/Q^3$ behavior for $G_+$ \cite{Carlson:1985mm}. 
The amplitudes $G_0$ and $G_-$, requiring 
helicity flips, are asymptotically zero relative to $G_+$. 
For each helicity flip, one expects one additional $(m/Q)$ power suppression 
in pQCD, where $m$ is some quark mass scale. In this way, 
$G_0$ ($G_-$), requiring one (two) helicity flips, yield a large $Q^2$ 
behavior as $1/Q^4$ ($1/Q^5$) respectively~\cite{Carlson:1985mm}.
\newline
\indent 
Using the relations of Eq.~(\ref{eq:jshel}), one can also derive the large 
$Q^2$ behavior of the Jones Scadron form factors as well as the $R_{EM}$ 
and $R_{SM}$ ratios. For the magnetic $\gamma^\ast N \Delta$ 
form factor $G_M^\ast$, one obtains a $1/Q^4$ behavior for  
$Q^2 \to \infty$. 
Furthermore, because $G_-$ is asymptotically zero relative to $G_+$, 
Eq.~(\ref{eq:jshel}) yields $G_E^\ast \to - G_M^\ast$ in the 
limit $Q^2 \to \infty$. This yields equivalently $R_{EM} \to +1$ for 
$Q^2 \to \infty$. Likewise, the $1/Q^4$ behavior for $G_0$ yields 
$R_{SM} \to$ {\it constant} for $Q^2 \to \infty$~\cite{Carlson:1985mm}. 
\begin{figure}
\centerline{  \epsfxsize=10cm%
  \epsffile{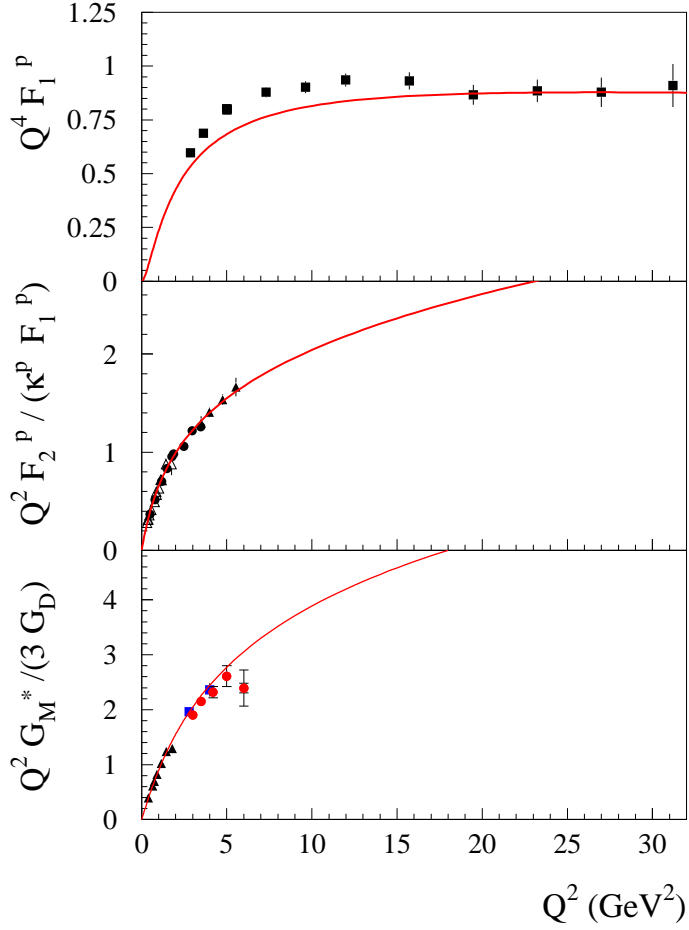} 
}
\caption{Test of baryon form factor scaling predictions. 
Top panel : proton Dirac form factor multiplied by $Q^4$;  
middle panel : ratio of Pauli to Dirac proton form factors 
multiplied by $Q^2$; 
lower panel : ratio of the $N \to \Delta$ magnetic 
transition (Jones-Scadron) form factor $G_M^*$ multiplied by $Q^2$ to the 
dipole form factor $G_D$. 
The data for $F_1^p$ are from \cite{Sill:1992qw} (solid squares).  
Data for the ratio $F_2^p / F_1^p$ are 
from \cite{Jones:1999rz,Punjabi:2005wq} (solid circles), 
\cite{Gayou:2001qt} (open triangles), 
and \cite{Gayou:2001qd} (solid triangles). 
Data for $G_M^*$ as in \Figref{fig:gmdel}.
The red curve is the calculation based on the  
three parameter modified Regge GPD parametrization $R2$ of Ref.~\cite{guidal}.}
\figlab{fig:scaling}
\end{figure}
\newline
\indent
We can test how well the above baryon form factor pQCD scaling predictions 
are satisfied at the available momentum transfers, see 
\Figref{fig:scaling}. One firstly sees from \Figref{fig:scaling} 
that the proton Dirac form factor, which has been measured up to about 
30 GeV$^2$, displays an approximate $1/Q^4$ scaling above 10 GeV$^2$. 
For the proton ratio $F_2^p/F_1^p$, the data up to 5.6 GeV$^2$ show 
no sign of a $1/Q^2$ behavior as predicted by pQCD. 
Instead, the data show that the ratio $F_2^p/F_1^p$ falls less 
fast than $1/Q^2$ with increasing $Q^2$.
In Ref.~\cite{Belitsky:2002kj}, the assumption of quarks moving 
collinearly, underlying the pQCD prediction, has been investigated. 
It has been shown in Ref.~\cite{Belitsky:2002kj} that by 
including components in the nucleon light-cone wave functions with quark 
orbital angular momentum projection $l_z = 1$, one obtains 
the behavior $F_2/F_1 \to \ln^2 (Q^2 / \Lambda^2)/ Q^2$ at large $Q^2$, 
with $\Lambda$ a non-perturbative mass scale. Choosing $\Lambda$ in the range 
$0.2 - 0.4$~GeV, Ref.~~\cite{Belitsky:2002kj} found that the 
data for $F_2^p/F_1^p$ support such double-logarithmic enhancement. 
A same analysis, including states of orbital angular momentum 
projection $l_z = 1$ in the $N$ or $\Delta$ light-cone wave functions, 
which find their physical origin in the transverse momentum 
of the quarks in the $N$ and $\Delta$, was also performed 
for the $\gamma^\ast N \Delta$ Coulomb transition form factor 
$G_C^\ast$~\cite{Idilbi:2003wj}. 
It was shown in Ref.~\cite{Idilbi:2003wj} that due to the orbital motion 
of the partons, 
$R_{SM}$ acquires a double-logarithmic correction 
$\ln^2 (Q^2 / \Lambda^2)$ at large $Q^2$ 
compared with the standard scaling analysis, 
according to which $R_{SM} \to$ {\it constant} for $Q^2 \to \infty$. 
The arguments of Refs.~\cite{Belitsky:2002kj,Idilbi:2003wj} 
still rely on pQCD and it remains to be seen by 
forthcoming data at higher $Q^2$ if this prediction already sets in in the 
few GeV$^2$ region. 
\newline
\indent
The test of the scaling behavior of the magnetic $\gamma^\ast N \Delta$ 
form factor $G_M^\ast$ is also shown on \Figref{fig:scaling}. As pQCD 
predicts a $1/Q^4$ asymptotic scaling behavior, the ratio $G_M^\ast/G_D$ 
(with $G_D$ the dipole form factor) should approach a constant. 
The data for $G_M^\ast$ up to about 6 GeV$^2$ again do not support this 
scaling behavior. One sees instead from \Figref{fig:scaling} 
that the data for $G_M^\ast$ seem to support a similar $Q^2$ behavior 
as for $F_2$. The pQCD prediction for the $R_{EM}$ ratio which should 
approach +100\% at high $Q^2$ also fails dramatically. The $R_{EM}$ ratio 
is measured to be minus a few percent out to 
$Q^2 \simeq 6$~GeV$^2$~\cite{Ungaro:2006df}, 
with no clear indication of a zero crossing. 
Also the $R_{SM}$ ratio which has also been measured up to 
$Q^2 \simeq 6$~GeV$^2$~\cite{Ungaro:2006df},  
does not seem to settle to a constant at large $Q^2$. 

\subsubsection{The road to ``asymptopia''}

Although at high enough $Q^2$, the pQCD scaling predictions should set in, 
the available data for the nucleon and $N \to \Delta$ electromagnetic 
form factors show that one is still far away from this regime. 
This has been further investigated in several theoretical approaches. 
\newline
\indent
In Refs.\cite{Belyaev:1995uw,Belyaev:1995ya}, it has been argued that the 
above described hard scattering mechanism is suppressed at accessible 
momentum transfers relative to the Feynman mechanism~\cite{feynman}, 
also called soft mechanism. The soft mechanism involves only one active 
quark, and the form factor is obtained as an overlap of initial and final 
hadron wave functions. 
The hard scattering mechanism on the other hand, 
involving three active quarks, requires the exchange of two gluons each of 
which brings in a suppression factor $\alpha_s / \pi \sim 0.1$. One therefore 
expects the hard scattering mechanism for $F_1^p$ or $G_M^\ast$  
to be numerically suppressed by a factor 1/100 compared to the soft term. 
Even though the soft mechanism is suppressed asymptotically by a power of 
$1/Q^2$ relative to the hard scattering mechanism, 
it may well dominate at accessible values of $Q^2$.
In Refs.\cite{Belyaev:1995uw,Belyaev:1995ya}, the soft contribution 
to the $\gamma^\ast N \Delta$ form factors has 
been estimated using a model based on local quark-hadron duality. 
In this approach it was found that the $\gamma N \Delta$ transition is 
dominated by the magnetic form factor $G_M^\ast$, while the electric 
quadrupole $G_E^\ast$ and Coulomb quadrupole $G_C^\ast$ form factors are small 
at accessible momentum transfers, in qualitative agreement with the data. 
\newline
\indent
In a more recent work~\cite{Braun:2005be}, the soft contribution 
to the $\gamma^\ast N \Delta$ form factors was 
evaluated within the light-cone sum rule approach. 
In this approach, one analyzes a matrix element, in which the $\Delta$ is 
represented by an interpolating field $\eta_\mu$ 
of Ref.~\cite{Ioffe:1981kw}. More specifically, one
computes the correlation function 
of this interpolating field and the electromagnetic current operator 
$J_\nu$ given by the matrix element:
\begin{eqnarray}
T_{\mu \nu}(P, q) = i \int d^4 y \, e^{i q \cdot y} \langle 0 \,|\, 
T\{ \eta_\mu(0) J_\nu(y) \} \,|\, N(P) \rangle,
\end{eqnarray}
between the vacuum and a single-nucleon state $| N(P)\rangle$. 
It was found in Ref.~\cite{Braun:2005be} that the sum rule for $G_M^\ast$ is 
dominated by contributions of subleading twist-4. 
They involve quark configurations  
with a minus light-cone projection of one of the quark field operators, 
which can be interpreted as the importance of orbital angular momentum.  
The calculations of Ref.~\cite{Braun:2005be} are in agreement with the 
experimental observations that the $R_{EM}$ and $R_{SM}$ ratios are small, 
although 
in the region of low $Q^2 < 2$~GeV$^2$, the result for $G_M^\ast$ is a factor 
of two below the data. 
\newline
\indent
In Sect.~\ref{sec2_gpd}, we have shown that the nucleon elastic and 
$N \to \Delta$ transition form factors can be obtained from model independent 
GPD sum rules. These GPDs, represented by the lower blob in 
\Figref{fig:ndelta_dvcs}, are non-perturbative objects which include 
higher Fock components in the $N$ and $\Delta$ wave functions. One 
can use a GPD parametrization to provide an estimate of the soft 
contributions, and expects this non-perturbative approach 
to be relevant in the low and intermediate $Q^2$ region for the form factors.  
This is shown in \Figref{fig:scaling} 
(solid curves) from which one sees that the GPD Regge parametrization $R2$, 
discussed in Sect.~\ref{sec:gpdmodel} is able to explain 
at the same time an approximate $1/Q^4$ 
behavior for $F_1^p$ and a behavior for $F_2^p/F_1^p$ which falls less steep 
than $1/Q^2$. For $G_M^\ast$, one sees from \Figref{fig:scaling} that 
the GPD sum rule evaluation based on the large $N_c$ relation of 
Eq.~(\ref{eq:gmsumrule}) is supported by the available data up to about 
6 GeV$^2$. 
Forthcoming experiments at the Jefferson Lab 12 GeV facility will extend the 
data for $F_2^p/F_1^p$ and $G_M^\ast$ to $Q^2$ values around 
15 GeV$^2$. Such measurements will allow to quantify in detail the higher 
Fock components in the $N$ and $\Delta$ wave functions versus 
the simple three-quark Fock component, and pave the road to ``asymptopia''. 

%% file: chap3_delta.tex
\section{Phenomenology of pion photo- and electroproduction}
\label{sec3}

We will first review the dynamical model calculations
which have been carried out by various groups and then turn to
MAID, which is a variation of the effective Lagrangian method and
has been very successful in describing the data. Lastly, we give a
brief description on the current status of the dispersion-relation
approach.

\subsection{Introductory remarks}
The electromagnetic production of pions is the main source
of information
about the electromagnetic properties of nucleon
resonances, such as the $\Delta$(1232). Certainly the experimental measurements
alone are not enough to obtain a quantitative insight, a theoretical interpretation
of the resonance-excitation mechanism is necessary to extract, {\it e.g.}, the strength of the
$\ga N\De$ transition. One can distinguish three
major theoretical tools which were developed
in the last millennium for this purpose:
\begin{itemize}
\item[(i)] {\it dispersion theory}, first proposed by
Chew {\it et al.}~\cite{CGLN56} for photoproduction, and by Fubini {\it et al.}
\cite{Fubini61} for electroproduction, has been successfully
applied in the first multipole analyses~\cite{Donnachie67,Schwela67}
as well as
in the modern ones~\cite{Hanstein:1997tp,Azn:2003,Kamalov02}.
The dispersion approach is based on general principles,
such unitarity,
analyticity, crossing symmetry, as well as relies on a phenomenological input from
the $\pi N$ scattering. A more detailed description
of this approach is given in Subsect.~\ref{sec3_dr}.
\item[(ii)] {\it effective Lagrangian approach}, where both $\pi N$
scattering and pion production are calculated based on the same
effective Lagrangian in terms of hadron fields. First attempts
to develop this picture were made by Peccei~\cite{Peccei69}, Olsson and
Osypowsky~\cite{Olsson75}. More recently, this approach was extensively
developed at RPI~\cite{Davidson91},
Madrid~\cite{SpainModel}, Gent~\cite{Gent},
KVI~\cite{KVItheory},
and Giessen~\cite{Feuster:1998cj}.
It is important to emphasize that in this approach the
effective Lagrangian is used only at tree level (nowadays
usually by a unitarization procedure). This can be considered as a
weakness of this approach since, as is now known from $\chi$EFT calculations,
the chiral loop corrections give rise to interesting and appreciable effects.
\item[(iii)] {\it dynamical models}, where the tree-level effective Lagrangian is treated
as a hadron-exchange potential of a quantum-mechanical
scattering problem.
First such models were developed by Tanabe and
Ohta \cite{Tanabe85} and Yang \cite{Yang85}. Many more were developed
over the past two decades, see {it e.g.},
\cite{NBL90,Lee91,SL,SG96,KY99,Chen03,Pascal04}.
Three of these will be detailed below, where also the pros and cons
of this approach in general will be addressed.
\end{itemize}

In addition, the phenomenological multipole solutions SAID~\cite{GWU}
and MAID~\cite{MAID}
have proven to be useful in interpreting the experiment, as well as in
obtaining the empirical information about individual amplitudes.
Without these tools, it is extreamly difficult to extract the
various amplitudes from experiment. In the rest of this subsection
we discuss how it can be done in principle.

\subsubsection{Measurement of pion photoproduction amplitudes}

Naively, without considering the discrete ambiguities, one would
conclude that seven measurements are needed to determine the four
pion photoproduction helicity amplitudes of $H_i's, (i=1,4)$ (four
magnitudes plus three phases) up to an arbitrary overall phase.
The discrete ambiguities arise because the observables are all
bilinear product of helicity amplitudes. To determine the
amplitudes, mathematically speaking, amounts to solve a set of
nonlinear equations. The solutions are not necessarily unique and
the ambiguities could arise if the number of observables measured
are not enough. According to Ref. \cite{Barker75}, nine
measurements are required if all amplitudes would be determined
without discrete ambiguities. Careful analysis by Chiang and
Tabakin \cite{Chiang97}, however, reveals that only eight
measurements would be sufficient to resolve all ambiguities,
namely, four appropriately chosen double-spin observables, along
with three single-spin observables and the unpolarized
differential cross section.
\newline
\indent In pion photoproduction, there are 15 polarization
observables (for detail see \cite{DT92}). Among them three are
three single polarization observables, namely, the polarized
photon asymmetry $\Sigma(\theta)$,  polarized target asymmetry
$T(\theta)$ and the recoil nucleon polarization $P(\theta)$. The
rest are the 12 double polarization observables  which describe
reactions with polarized beam-polarized target, polarized
beam-recoil nucleon polarization, and polarized target-recoil
nucleon polarization. They are commonly denoted as $E, F, G, H,
Cx', Cz', Ox', Oz', Tx', Tz', Lx',$ and $Lz'$. For definition see,
e.g., Refs.~\cite{Barker75,Knochlein95}.
\newline
\indent We illustrate here only the differential cross section and
one of the single polarization observables, the polarized photon
asymmetry, which has been extensively investigated experimentally
in the study of $E2/M1$ mixing ratio for the $N\ra\De$ transition.
\newline
\indent The unpolarized differential cross section for the pion
photoproduction can be expressed in terms of the helicity
amplitudes $H_i$'s (see Appendix) as follows:
 \bea
\frac{d\sigma}{d\Omega}=\frac{|\bk|}{2|\bq|} \sum^4_{i=1}\mid
H_i\mid^2. \label{eq:photoxsec}
\eea The {\it polarized photon
asymmetry} is defined as,
\begin{eqnarray}
\Sigma=\frac{d\sigma_{\perp} -
d\sigma_{\parallel}}{d\sigma_{\perp} + d\sigma_{\parallel}},
\end{eqnarray}
where $d\sigma_{\perp}\,( d\sigma_{\parallel})$ is the
differential cross section for a linearly polarized photon with
polarization vector perpendicular (parallel) to the reaction
plane. In terms of the helicity amplitudes, $d\sigma_{\perp}\,(
d\sigma_{\parallel})$ are given as
\begin{eqnarray}
\frac{d\sigma_{\perp}}{d\Omega} &=&  \frac{|\bk|}{2|\bq|}(\mid
H_1+H_4\mid^2 + \mid H_2-H_3\mid^2)\,,
\nonumber \\
\frac{d\sigma_{\parallel}}{d\Omega} &=&  \frac{|\bk|}{2|\bq|}(\mid
H_1-H_4\mid^2 + \mid H_2+H_3\mid^2)\,.
\end{eqnarray}
\newline
\indent
The expressions given above contain contributions from all
partial waves. However, in the case of $\pi^0$ photoproduction in
the $\Delta(1232)$ resonance region where most of the current
experimental information are derived from, the dominant
contributions come from $s$- and $p$-wave multipoles. Therefore, a
truncated multipole approximation is often used in the analysis of
the data for this reaction and only $E_{0+},E_{1+},M_{1+}$ and
$M_{1-}$ multipoles are kept. Such an approximation greatly
simplifies the analysis and allows one to express the differential
cross section in the following simple form~:
\begin{eqnarray}
\frac{d\sigma_j}{d\Omega}=\frac{k}{q}[ A_j + B_j\cos{\theta} +
C_j\cos^2{\theta}] \label{eq:ABC}
\end{eqnarray}
where the coefficients $A_j,\,B_j$ and $C_j$ are bilinear
functions of the $s$- and $p$- wave multipoles and $j$ indicates
the parallel $(\parallel)$, perpendicular $(\perp)$, and
unpolarized $(0)$ components. These coefficients are quadratic
functions of the $s$- and $p$-wave amplitudes. The parametrization
\Eqref{ABC} can be used separately for $d\sigma_{\bot}$ and
$d\sigma_{\|}$ in the analysis of the photon asymmetry. In
Refs.~\cite{Beck:1997ew,LEGS97}, it was found that $d\sigma_{\|}$
is very sensitive to the small $E_{1+}$ multipole due to the
interference with large $M_{1+}$. We can see this from the
following expressions~:
\begin{eqnarray}
A_{\|} & = & |E_{0+}|^2 + \mid 3E_{1+} + M_{1+} - M_{1-}\mid^2\,,
\nonumber \\
B_{\|} & = & 2 Re \,[E_{0+} (3E_{1+} + M_{1+} - M_{1-})^*]\,,
\\
C_{\|} & = & 12 Re \,[E_{1+} (M_{1+} - M_{1-})^*]\,. \nonumber
\label{ABCpar}
\end{eqnarray}
In Ref.~\cite{Beck:1997ew,LEGS97} this sensitivity was used for
the experimental determination of the $R_{EM}$ ratio of
Eq.~(\ref{eq:remhel}), because at the $\Delta$ resonance position
one has,
\begin{eqnarray}
 R_{EM} \simeq \frac{1}{12}\frac{C_{\|}}{A_{\|}}.
\end{eqnarray}

\subsubsection{Measurement of pion electroproduction amplitudes}

Consider now the pion
electroproduction\footnote{Until Sect.~\ref{sec6}
we assume that the electroproduction process proceeds via
the one-photon exchange. In this case,  electroproduction
differs from photoproduction only in that the incoming photon is
now virtual, and hence in additional to
transverse polarizations
can have the longitudinal or scalar ones.} which kinematically is illustrated
in \Figref{kinematics}. In this diagram the
 four-momenta of the initial and final electron are defined as,
$l_i=(e_i,\,\bq_{iL}),\,
l_f=(e_f,\,\bq_{fL})$, while the momenta of the other particles
are given as,
\begin{itemize}
\item in the {\it lab} frame:  $q=(\omega_{\bq_L},\,\bq_L)$, $p=(M_N,\,{\bf
0})$, $p^\prime=(p^\prime_{0L},\,{\bf p^\prime_L})$, and $
k=(\omega_{\bk_L},\,\bk_L)$;
\item in the center-of-mass ({\it c.m.}) of the $\pi N$ system:
$q=(\omega_\bq,\,\bq)$, $ k=(\omega_\bk,\,\bk)$,  $p=(E,\,-\bq)$, and $
p^\prime=(E^\prime,\,-\bk)$.
\end{itemize}
More kinematical details can be found in the Appendix.
\begin{figure}[t,b,h]
\begin{center}
\epsfig{file=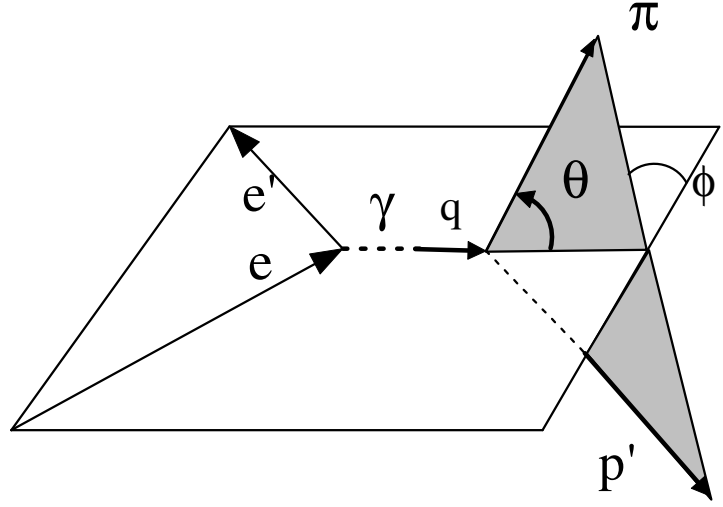,width=10cm}
\end{center}
\caption{Kinematics for pion the electroproduction on the nucleon.}
\figlab{kinematics}
\end{figure}
\newline
\indent
The 5-fold differential cross
section, in the case of the unpolarized target and beam, can be
written as:
\begin{equation}
\frac{d\sigma}{d\Omega_{f}\,de_{f}\,d\Omega_{\pi}} = \Gamma\,
\frac{d\sigma_v}{d\Omega_{\pi}}\,, \label{eq:diff1}
\end{equation}
which defines the virtual photon cross section,
\begin{eqnarray}
\frac{d\sigma_v}{d\Omega_{\pi}}  &=&
\frac{d\sigma_T}{d\Omega_{\pi}} +  \varepsilon
\,\frac{d\sigma_L}{d\Omega_{\pi}} +
\varepsilon\,\frac{d\sigma_{TT}}{d\Omega_{\pi}}\,
\cos{2\phi_{\pi}} + \sqrt{2\varepsilon
(1+\varepsilon)}\,\,\frac{d\sigma_{TL}}{d\Omega_{\pi}}\,
  \cos{\phi_{\pi}}
\nonumber\\
& + & h \sqrt{2\varepsilon
(1-\varepsilon)}\,\,\frac{d\sigma_{TL'}}
{d\Omega_{\pi}}\,\sin{\phi_{\pi}} ,
\label{eq:diff2}
\end{eqnarray}
where
\begin{equation}
\varepsilon=\left[ 1 + 2\frac{{\bf q}^2_L}{Q^2}\tan^2
\frac{\theta_e}{2} \right]^{-1}\,,\qquad \Gamma =
\frac{\alpha_{em}}{2\pi^2}\ \frac{e_f}{e_i}\,\frac{K}{Q^2}\
\frac{1}{1-\varepsilon}\,. \label{eq:flux}
\end{equation}
$\varepsilon, h, \theta_e, \bq_L, e_i,$ and $e_f$ denote the
degree of transverse polarization of the virtual photon, the
helicity of the incoming electron, the angle between the incident
and final electrons, the three-momentum of the virtual photon,
energy of the initial and final electron in the {\it lab} frame,
respectively. $\Gamma$ is the flux of the virtual photon
and $K=(s-M_N^2)/2M_N$, often
called the `equivalent photon energy'.
\newline
\indent
In \Eqref{diff1},  $e_f, \Omega_f$ denote the
energy and solid-angle of the scattered electron in the
laboratory, respectively, and $\phi$ is the tilt angle between the
electron scattering plane and the reaction plane, as shown in
\Figref{kinematics}. $d\Omega_\pi$ is the pion solid-angle
differential measured in the {\it c.m.} frame of the final pion
and nucleon. Since $\Gamma$ can be interpreted as the number of
virtual photons per electron scattered into $de_f$ and
$d\Omega_f,$, the quantity on the r.h.s. of \Eqref{diff1},
$d\sigma_v/d\Omega_{\pi}$ then represents the {\it c.m.}
differential cross section  for pion photoproduction by virtual
photons.
\newline
\indent
Also let us note  that \Eqref{diff2} differs
slightly from Eq. (20) of Ref.~\cite{DT92} often used
in the literature, where the degree of longitudinal
polarization of the virtual photon
$\varepsilon_L=(Q^2/\omega_{\bq_L}^2)\varepsilon$ appears in some
of the coefficients in \Eqref{diff2}. The difference can be viewed
as a different definition for the longitudinal cross sections
which simply modifies their relations with the nuclear response
functions \cite{Raskin89} by a kinematical constant, as will be
specified below.
\newline
\indent The first two terms in  \Eqref{diff2} are the transverse
(T) and longitudinal (L) cross sections. They do not depend on the
pion azimuthal angle $\phi$. The fourth and fifth terms describe
the transverse-longitudinal interferences (TL and TL'). They
contain an explicit factor $\sin\theta $ \cite{Akerlof67} and
therefore vanish along the axis of the momentum transfer. The same
is true for the third term, a transverse-transverse interference
(TT), proportional to $\sin^2\Theta $.
\newline
\indent In the pion-nucleon {\it c.m.} frame,  it is useful to
express these cross sections in terms of five response functions
which depend only on three independent variables, i.e.
$R=R(Q^2,W,\Theta)$. The corresponding expressions are the
following
\begin{eqnarray}
\frac{d\sigma_T}{d\Omega_{\pi}}=\frac{|\bk|}{q_W}\,R_T\,,
\hspace{1.2cm}
&\frac{d\sigma_{TT}}{d\Omega_{\pi}}=\frac{|\bk|}{q_W}\,R_{TT}\,,
\nonumber\\
\frac{d\sigma_L}{d\Omega_{\pi}}=\frac{|\bk|}{q_W}
\frac{Q^2}{\omega_{\bq}^2}\,R_L\,,\hspace{0.6cm}
&\frac{d\sigma_{TL}}{d\Omega_{\pi}}=\frac{|\bk| }{q_W}\,
\frac{Q}{\omega_{\bq}}\,R_{TL}\,,\hspace{0.8cm}
&\frac{d\sigma_{TL'}}{d\Omega_{\pi}}=\frac{|\bk| }{q_W}\,
\frac{Q}{\omega_{\bq}}\,R_{TL'}\,. \label{eq:resp}
\end{eqnarray}
with $q_W=(W^2-M_N^2)/2W$ the "photon equivalent {\it c.m.}
energy", and $\omega_{\bq}=(W^2-Q^2-M_N^2)/2W$ the virtual photon
energy in the {\it c.m.} frame. The five response functions may be
expressed in terms of the six independent CGLN amplitudes
$F_1,...,F_6$. Explicit expressions can be found in
Ref.~\cite{DT92}. In terms of the helicity amplitudes we can get
simple expressions for the response functions:
\begin{eqnarray}
R_T &=& \frac{1}{2}(\mid H_1 \mid^2+\mid H_2 \mid^2+\mid H_3
\mid^2+\mid H_4 \mid^2)\,,
\nonumber \\
R_L &=&(\mid H_5 \mid^2+\mid H_6 \mid^2)\,,
\nonumber \\
R_{TT} &=& Re\, (H_3 H_2^*-H_4 H_1^*)\, ,
\nonumber \\
R_{TL} &=& \frac{1}{\sqrt{2}}\,Re[H_5^*(H_1-H_4)+H_6^*(H_2+H_3)],
\nonumber \\
R_{TL'} &=& -\frac{1}{\sqrt{2}}\,Im\,
[(H_1-H_4)H_5^*+(H_2+H_3)H_6^*]\, .
\end{eqnarray}
\indent Many pion electroproduction experiments in the $\Delta$
resonance region have been performed to study the electromagnetic
$N\ra\Delta$ transition. Most of them are on $\pi^0$ production
reaction, with either unpolarized electron beam $(h=0)$ or
longitudinal electron beam $(h\pm1)$. By measuring coincident
cross sections at different azimuthal angles $\phi$ ($\phi =0$, in
plane and $\phi\neq 0$ out of plane) and different polar angles
$\theta$, it has been possible to obtain data on $\sigma_{TT},
\sigma_{LT}, \sigma_{LT'}$ and unpolarized cross section
$\sigma_0=\sigma_T+\varepsilon\sigma_L$. Values for the $M_{1+}$,
$R_{EM}$, and $R_{SM}$ have been inferred from the models which
provide an overall agreement with the data. This will be discussed
in more detail in Sect.~\ref{sec5}.
\newline
\indent We mention in passing that, again naively, without
considering the discrete ambiguities, one would expect that eleven
measurements would be required to determine the six pion
electroproduction helicity amplitudes of $H_i's, (i=1,...,6)$ (six
magnitudes plus five phases) up to an arbitrary overall phase.
However, in contrast to the case of photoproduction, the question
of how many measurements are needed to determine all amplitudes
without discrete ambiguities has not been properly addressed yet.

\subsection{Dynamical models}

Here we turn to the discussion of the dynamical models
and as an example consider three of the more recent ones:
the model of Sato and Lee (SL) \cite{SL}, DMT (Dubna-Mainz-Taipei) model
\cite{KY99},
and the DUO (dynamical Utrecht-Ohio) model~\cite{Pascal04}.

\subsubsection{General framework: unitarity, relativity, gauge invariance}

The dynamical model formulation for the pion electromagnetic
production reaction starts from the following scattering equation
\beq \bmat t_{\pi\pi} \,&t_{\pi\ga}\\
t_{\ga\pi} \,&t_{\ga\ga}\emat =
  \bmat v_{\pi\pi}\,& v_{\pi\ga}\\
v_{\ga\pi} \,& v_{\ga\ga}\emat +
 \bmat v_{\pi\pi} \,& v_{\pi\ga}\\
v_{\ga\pi} \,& v_{\ga\ga}\emat \bmat G_\pi \,& 0\\
0 \,& G_\ga \emat\\
\bmat t_{\pi\pi} \,& t_{\pi\ga}\\
t_{\ga\pi} \,& t_{\ga\ga}\emat, \label{coupledchannelt}\eeq for
the following four processes \bea &\pi N \rightarrow \pi N,
\hspace{1.0cm} &\pi N
\rightarrow \ga^* N,\nonumber\\
   &\ga^* N \rightarrow \pi N, \hspace{1.0cm} &\ga^* N \rightarrow
\ga^* N, \eea where the transition matrices $t$'s are related to
the elements of the $T$-matrix by
a kinematical factor and $v$'s are the driving potentials of the
$\pi N$ scattering $(\pi\pi)$, pion electromagnetic absorption
$(\pi\ga)$, production $(\ga\pi)$, and the nucleon Compton
scattering $(\ga\ga)$. $G_\pi$ and $G_\ga$ are, respectively, the
pion-nucleon and photon-nucleon two-particle propagators.
\newline
\indent To first order in $e$, the transition matrix element for
$\gamma^*N\rightarrow\pi N$, in a dynamical model is then given as
\bea t_{\ga\pi}=v_{\ga\pi}+v_{\ga\pi} G_0 t_{\pi N},
\label{tgampi} \eea with \bea t_{\pi N}=v_{\pi N}+v_{\pi N} G_0
t_{\pi N}, \label{tpiN} \eea where we have replaced the subscript
$\pi\pi$ used in \eqref{coupledchannelt} by $\pi N$ and introduced
$G_0\equiv G_\pi$ which will be used hereafter. In this
approximation, only the integral equation Eq.~(\ref{tpiN}) for
$\pi N$ scattering has to be solved and the rest is determined in
a one-loop calculation.
\newline
\indent In a Bethe-Salpeter (BS) formulation, Eqs.~(\ref{tgampi})
and (\ref{tpiN}) are four-dimensional equations and the driving
terms $v's$ represent the sums of all two-particle irreducible
amplitudes. Two approximations are commonly made to simplify the
solution of the BS equation \cite{Hung01}. The first is to
approximate the driving potentials by the tree diagrams of an
effective chiral Lagrangian and the other is to replace $G_0$ by a
propagator $g_0$ which would reduce the dimensionality of the
integral equations from four to three. $g_0$ is chosen to have an
invariant form so that the covariance of the original equation is
preserved. The reduction in the dimensionality is achieved with a
$\delta-$function in $g_0$ which imposes a constraint on the time
component of the relative momentum variable, i.e., the relative
energy. In addition, this new propagator must be chosen such that
the resulting  scattering amplitude has a correct elastic cut from
the elastic threshold $(m_\pi+M_N)^2$ to $\infty$ in the complex
$s$ plane, as required by the unitarity  condition.
\newline
\indent It is well known for example, see Ref. \cite{Klein74},
that the choice of $g_0$ is rather arbitrary and can be done in
infinitely many different ways. In Ref. \cite{Hung01}, a class of
three dimensional equations of the following form,
\begin{eqnarray}
\hat G_0(k;P) &= & \frac{1}{(2\pi)^3}\int \frac{ds'}{
s-s'+i\varepsilon} f(s,s') [\alpha
(s,s') P \hspace{-0.10in} / +k \hspace{-0.08in} / +M_N]  \nonumber  \\
 & \times& \delta^{(+)}([\eta_N(s')P' + k]^2
- m_{N}^2)  \delta^{(+)} ([\eta_\pi (s')P'- k]^2 -  m_{\pi}^2).
\label{eq:2GH}
\end{eqnarray}
were employed to investigate the $\pi N$ scattering.  In
\Eqref{2GH}, $P'=\sqrt{\frac{s'}{s}}P$ defines the "off-shellness"
of the intermediate states. The superscript (+) associated with
$\delta$-functions means that only the positive energy part is
kept in defining the  nucleon propagator. $k  =
\eta_\pi(s)p-\eta_N(s)q$ is the relative momentum $k$ with
$\eta's$ any function of $s$ constrained by the condition
$\eta_\pi(s)+\eta_N(s)=1$. To have a correct $\pi N$ elastic cut,
the arbitrary functions $f(s,s')$ and $\alpha(s,s')$ must satisfy
the conditions
\begin{eqnarray}
 f(s,s)   =   1, \hspace{1.0cm}
\alpha(s,s) = \eta_N(s).\label{eq:f-alpha}
\end{eqnarray}
It is easy to verify that for $(m_\pi+M_N)^2 \leq s \leq \infty$,
Eqs. (\ref{eq:2GH}) and (\ref{eq:f-alpha}) give the correct
discontinuity of the propagator $\hat{G}_0$
\begin{eqnarray}
Disc[\hat{G}_0(k;P)] & =&\frac{-i}{(2\pi)^2} (\eta_N(s)
\hspace{-0.10in} /  + k \hspace{-0.08in} / +M_N)
\delta^{(+)}([\eta_N(s) P + k]^2-M_N^2) \nonumber \\
       &\times &
          \delta^{(+)}([\eta_\pi (s) P -k]^2 - m_\pi^2).
          \label{eq:disc}
\end{eqnarray}
\indent The class of propagators given in  \Eqref{2GH} has the
feature that both particles in the intermediate states are put
equally off-mass-shell such that the relative energy dependence in
the interaction is removed. Several three dimensional formulations
developed in the literature, including those developed by
Blankenbecler and Sugar \cite{bs}, Kadyshevsky  \cite{kady},
Thompson \cite{thom}, and Cooper and Jennings  \cite{CJ89}, can be
derived from using Eqs.~(\ref{eq:2GH}) and (\ref{eq:f-alpha}) and
were studied in detail in \cite{Hung1} for $\pi N$ scattering.
All of these schemes set
$\eta_N(s)=\varepsilon_N(s)/(\varepsilon_N(s)+\varepsilon_\pi(s))$
and $\eta_\pi(s) =
\varepsilon_\pi(s)/(\varepsilon_N(s)+\varepsilon_\pi(s)),$ where
$\varepsilon_N(s)=(s+M_N^2-m_\pi^2)/2\sqrt{s}$ and
$\varepsilon_\pi(s)=(s-M_N^2+m_\pi^2)/2\sqrt{s}$ are the center of
mass ({\it c.m.}) energies of nucleon and pion, respectively. The
resultant $\pi N$ interaction obtained with Cooper-Jennings
reduction scheme  has been extensively used in the DMT model
calculation for the pion electromagnetic production reactions. The
functions $\alpha(s,s')$ and $f(s,s')$ specific to the
Cooper-Jennings reduction are \bea \alpha(s,s')&=&\eta_N(s), \\
f(s,s') &=&
 \frac{4\sqrt{ss'}\varepsilon_N(s')\varepsilon_\pi(s')}
 {ss'-(m_N^2-m_\pi^2)^2}.\label{eq:alpha-f}\eea
\newline
\indent
 The resulting three-dimensional scattering equations then
take the Lipmann-Schwinger form \bea
t_{\ga\pi}=v_{\ga\pi}+v_{\ga\pi} g_0 t_{\pi N}, \hspace{1.0cm}
t_{\pi N}=v_{\pi N}+v_{\pi N} g_0 t_{\pi N}. \label{eq:LSeq}\eea
\indent One of the most important features of the dynamical model
approach is that it provides a unified theoretical framework to
describe $\pi N$ scattering and pion production in a consistent
way. In most cases, $v_{\pi N}$ is derived from a tree
approximation to an effective chiral Lagrangian which involves
field operators of the nucleon, $\Delta$, pion,  rho meson, and a
fictitious scalar meson $\sigma$ \cite{Hung01}. Pseudovector $\pi
NN$ coupling is used as it is consistent with the leading order of
chiral perturbation theory. It leads to a driving term which
includes the direct and crossed $N$ and $\Delta$ terms, and the
$t$-channel $\sigma$- and $\rho$-exchange terms.

Furthermore in the dynamical models, one needs to specify
the $\gamma N \Delta$ vertex, shown in \Figref{treevertex}. It is
described by \Eqref{diagndel1} with $g_M$, $g_E$, and $g_C$, the
magnetic dipole $M1$, electric quadrupole $E2$, and Coulomb
quadrupole $C2$ excitation strength of the $\Delta$, respectively.
They are then only three new parameters  in the dynamical model,
besides those determined by the $\pi N$ scattering and
$V\ra\pi\ga$ reactions, to be determined from the electromagnetic
pion production data.
\newline
\indent Since multiple pion rescattering in the final state is
treated explicitly, unitarity in inherent in \Eqref{LSeq}. In
fact, if we take \bea g_0=\frac {1} {E-H_0+i\varepsilon},
\label{eq:LSpropagator}\eea with $H_0$ the free Hamiltonian of the
pion-nucleon system, then a multipole decomposition of
\Eqref{LSeq} gives \cite{Yang85}
\begin{eqnarray}
t^{\alpha}_{\gamma\pi}(k_E,q;E+i\varepsilon)&=&e^{ i\delta_{\alpha}}\, \cos{\delta_{\alpha}}\nonumber\\
&\times&\left[v^{\alpha}_{\gamma\pi}(k_E,q) + P\int_0^{\infty} dk'
\frac{k'^2\,R^{\alpha}_{\pi
N}(k_E,k')\,v^{\alpha}_{\gamma\pi}(k',q)}{E-E_{\pi N}(k')}\right],
\label{eq:tmult}
\end{eqnarray}
where $\delta_{\alpha}$ and $R^{\alpha}_{\pi N}$ are the $\pi N$
scattering phase shift and reaction matrix, in channel $\alpha$,
respectively; $k_E$ is the pion on-shell momentum and $q=\mid {\bf
q}\mid$ is the photon momentum. In the energy region where only
two channels are open: $\pi N$ elastic scattering and single pion
photo- or electroproduction, $v_{\gamma\pi},\, v_{\pi N}$ and the
reaction matrix $R_{\pi N}$ are real numbers. In this case we see
explicitly from  \Eqref{tmult} that  the phase of the
$t_{\gamma\pi}^{\alpha}$ is equal to the $\pi N$ scattering phase
in the corresponding channel, e.g.,
\begin{eqnarray}
t^{\alpha}_{\gamma\pi}(k_E,q;E)= \mid
t^{\alpha}_{\gamma\pi}(k_E,q;E)\mid e^{i\delta_{\alpha}(E)}.
\label{eq:FW}
\end{eqnarray}
This is the well known Fermi-Watson theorem~\cite{Watson54}. This
theorem, which is a consequence the unitarity of the S-matrix, and
time-reversal invariance, imposes a phase condition on the
individual multipole amplitudes below the two-pion production
threshold. It simplifies the analysis of the pion photo- and
electroproduction processes in the $\Delta$ resonance region
connecting the real and imaginary parts of the reaction
amplitudes.
\newline
\indent Gauge invariance of the electromagnetic interaction
requires the following current conservation condition, \bea
t_{\ga\pi}(\varepsilon \rightarrow
q_\mu)=0,\label{eq:current-conserv}\eea as $t_{\ga\pi}$ is
proportional to $T_{fi}=\varepsilon_\mu J^\mu$ of
\Eqref{T-current}. The first term $v_{\ga\pi}$ in $t_{\ga\pi}$ of
\Eqref{LSeq} clearly satisfies the current conservation condition
of \Eqref{current-conserv} for on-mass-shell incoming and outgoing
particles since  it is obtained from the tree diagrams of a
Lagrangian resulting from gauging a chiral effective Lagrangian
for $\pi N$ scattering. However, there is a fundamental difficulty
to impose the current conservation condition on the second term
$t_{\ga\pi}$ of \Eqref{LSeq}, which represents the sum of the
following ladder series, \bea v_{\ga\pi} g_0 t_{\pi N}= v_{\ga\pi}
g_0 v_{\pi N}+ v_{\ga\pi} g_0v_{\pi N} g_0 v_{\pi N} + \cdots.
\label{eq:ladder}\eea It is well-known that the sum of a set of
diagrams is gauge invariant if photon is hooked to every line
which carries charge. Consequently, $v_{\ga\pi} g_0 t_{\pi N}$ is
not a gauge invariant quantity. It is inherent in the
approximation scheme which leads to \Eqref{LSeq}. There are many
recipes proposed \cite{Yang88,NBL90,KY99,Pascal04} to make it to
satisfy the current conservation condition of
\Eqref{current-conserv}. However, they are all {\it ad hoc} in
nature and hence not unique.
\newline
\indent The dynamical model approach as summarized in \Eqref{LSeq}
hence contains four theoretical ingredients. The first one is the
choice of $v_{\ga\pi}$. It is commonly chosen to consist of tree
diagrams of a chiral effective Lagrangian \cite{Olsson75}, which
include the Born terms in pseudoscalar coupling, contribution from
$t-$channel $(\rho,\omega)$ vector meson exchanges, and $s-$ and
$u-$channel $\Delta-$exchanges. Different dynamical calculations
differ mostly in the $\rho NN$ and $\omega NN$ coupling constants
used. The second ingredient concerns the choice of the
three-dimensional propagator $g_0$. Another input is the model
chosen for $t_{\pi N}$ since experimental $\pi N$ phase shifts
constrain only the on-shell behavior while the physical pion
production multipole amplitude $t_{\gamma \pi}$ depends on the
half-off-shell matrix elements of $t_{\pi N}$. The off-shell $\pi
N$ rescattering effects have been shown  \cite{Kamalov01} to play
an important role to explain the threshold $\pi^0$ data. The last
theoretical ingredient is the recipe employed to satisfy the
current conservation condition of \Eqref{current-conserv}.

\subsubsection{Sato-Lee model}

The Sato-Lee (SL) model \cite{SL} made a strong effort to derive
the driving terms of $v_{\ga\pi}$ and $v_{\pi N}$ of \Eqref{LSeq}
in a consistent manner.
\newline
\indent SL started from a model Lagrangian with $N, \De, \pi,$ and
$ \rho$ fields which would generate the tree diagrams of ChPT. The
strong interaction Lagrangian is then extended to include the
$\ga$ field with "minimal substitution". The Lagrangians which
describe the $\ga\pi V$ interaction of \Eqref{Vgapi} with
$V=(\rho,\omega)$ and the $\ga N\De$ vertex of \Figref{treevertex}
are then added. The unique feature of SL's calculation is that
they then use a unitary transformation method, called the SKO
method \cite{SKO92}, to derive from the above-mentioned Lagrangian
an energy-independent effective Hamiltonian. The essence of the
SKO method is to systematically eliminate the virtual processes
from the considered Hamiltonian by using unitary transformation.
The so-called "virtual processes" are the processes like $N
\leftrightarrow N\pi$, $N \leftrightarrow N\rho$, $N
\leftrightarrow \pi\De$, and $\pi \leftrightarrow \pi\rho$, which
can not take place in the free space because of the
energy-momentum conservation. The effects of the virtual processes
are included as effective operators in the resultant Hamiltonian.
Another advantage of such a scheme is that it does not have to
perform the renormalization for the nucleon as the $N
\leftrightarrow N\pi$ vertex has been transformed away since it is
a "virtual" process.
\newline
\indent The final effective energy-independent Hamiltonian of SL
model then takes the form \bea H_{eff}=H_0+v^B_{\pi
N}+v^B_{\ga\pi}+(h^{(0)}_{\pi N\De}+h^{(0)}_{\ga
N\De}+h.c.),\label{eq:Heff}\eea where $v^B_{\pi N}$ is the
background $\pi N$ potential, and $v^B_{\ga\pi}$ describes the
background $\ga N\ra\pi N$ transition. $v^B_{\pi N}$ contains Born
terms and $t-$channel $\rho$ exchange, while $v^B_{\ga\pi}$ is
consisted of Born terms in PV coupling and $t-$channel
$(\rho,\omega)$ vector-meson exchange. $h^{(0)}_{\pi N\De}$ and
$h^{(0)}_{\ga N\De}$ denote the $\pi N\ra\De^{(0)}$ and
$\ga^*N\ra\De^{(0)}$ excitations of a bare $\De^{(0)}$,
respectively. The matrix elements of $h^{(0)}_{\pi N\De}$ and
$h^{(0)}_{\ga N\De}$ take the familiar forms, cfr.
Sect.~\ref{sec2}. \bea <\De^0\mid h^{(0)}_{\pi N\De}\mid\bk a>&=&
-\frac{f^0_{\pi
N\De}}{m_\pi}\frac{i}{\sqrt{(2\pi)^3}}\frac{1}{\sqrt{2\omega(\bk)}}
\sqrt{\frac{E_N(\bk)+M_N}{2E_N(\bk)}}({\bf S\cdot\bk})T_a,\\
<\De^0\mid h^{(0)}_{\ga N\De}\mid\bq >&=&
-\frac{1}{\sqrt{(2\pi)^3}}\frac{1}{\sqrt{2\omega(\bq)}}
 \sqrt{\frac{E_N(\bq)+M_N}{2E_N(\bq)}}\frac{3(M_\De+M_N)}
 {4M_N(E_N(\bq)+M_N)}T_3\nonumber\\
 &\times&\{iG^*_M(Q^2){\bf S}\times\bq\cdot
 \bvare+G^*_E(Q^2)({\bf S}\cdot\bvare\bsig\cdot\bq+
 {\bf S}\cdot\bq\bsig\cdot\bvare)\nonumber\\
 &-&\frac{G^*_C(Q^2)}{M_\De}
{\bf S}\cdot\bq\bsig\cdot\bq\varepsilon_0)\}, \eea where $T$ and
${\bf S}$ are the isospin and spin $\frac 12\ra \frac32$
transition operators, respectively.
\newline
\indent
 The driving terms
$v's$ in \Eqref{LSeq} in the SL model are then given by \bea
v_{\pi N}=v^B_{\pi N}+v^\De_{\pi N},\hspace{1.5cm}
v_{\ga\pi}=v^B_{\ga\pi}+v^\De_{\ga\pi},\label{eq:Vs1} \eea where
\bea v^\De_{\pi N}=\frac{h^{(0)^\dagger}_{\pi N\De}h^{(0)}_{\pi
N\De}}{E-M^{(0)}_\De},\hspace{1.5cm}
v^\De_{\ga\pi}=\frac{h^{(0)^\dagger}_{\pi N\De}h^{(0)}_{\ga
N\De}}{E-M^{(0)}_\De}.\label{eq:Vs2}\eea The energy denominators $
(E-M^{(0)}_\De)$ in \Eqref{Vs2} arise because SL have chosen $g_0$
in \Eqref{LSeq} to be the Schroedinger propagator, i.e, $g_0=
1/(E-H_0)$.
\newline
\indent Following Ref.~\cite{Tanabe85}, SL decomposed,
with the use of two-potential formula, the resulting $t$-matrix as
follows \cite{Hsiao98},
\begin{eqnarray}
t_{\gamma\pi}(E)=\tilde{t}_{\gamma\pi}^B(E) +
\tilde{t}_{\gamma\pi}^{\Delta}(E), \hspace{2.3cm} t_{\pi
N}(E)=\tilde{t}_{\pi N}^B(E) + \tilde{t}_{\pi
N}^{\Delta}(E),\label{eq:ts1}
\end{eqnarray}
where the background contributions $\tilde{t}_{\gamma\pi}^B(E)$
and $\tilde{t}_{\pi N}^B(E)$ are given by,
\begin{eqnarray}
\tilde{t}_{\gamma\pi}^B(E)=v_{\gamma\pi}^B+v_{\gamma\pi}^B\,g_0(E)
\,\tilde{t}_{\pi N}^B(E), \hspace{1.0cm}\tilde{t}_{\pi
N}^B(E)=v_{\pi N}^B+v_{\pi N}^B\,g_0(E) \,\tilde{t}_{\pi N}^B(E).
\label{eq:tB}
\end{eqnarray}
This kind of background terms are further denoted as
``non-resonant" background because they do not contain any
resonance contributions from $v_{\gamma\pi}^{\Delta}$ or $v_{\pi
N}^{\Delta}$ of \Eqref{Vs2}. Note that in  partial channel
$\alpha$, the on-shell matrix elements of both
$\tilde{t}_{\gamma\pi}^B(k_E,q;E+i\varepsilon)$ and
$\tilde{t}_{\pi N}^B(E)(k_E,k_E;E+i\varepsilon) $ have the same
phase $\delta^B_\alpha$  as can be proven in the same way as
\Eqref{tmult}.
\newline
\indent The $\De$ contribution to $\pi N$ scattering $\tilde{t}_{
\pi N}^{\Delta}(E)$ in \Eqref{ts1}   takes the form \bea
\tilde{t}_{\pi N}^{\Delta}(E)= \bar h_{\pi
N\Delta}^{\dagger}(E)g_\De h_{\pi N\De}(E),
\label{eq:tpiNR}
\eea
where $h_{\pi N\Delta}(E)$ describes the dressed $\pi NN$ vertex
\begin{eqnarray}
h_{\pi N\Delta}(E)=h_{\pi N\Delta}^{(0)}+h_{\pi
N\Delta}^{(0)}\,g_0(E) \,\tilde{t}_{\pi N}^B(E),\nonumber\\
\bar h_{\pi N\Delta}^{\dagger}(E)=h_{\pi
N\Delta}^{(0)\dagger}+\tilde{t}_{\pi N}^B(E)\,g_0(E) \,h_{\pi
N\Delta}^{(0)\dagger}.
 \label{hpiN_Rpisl1}
\end{eqnarray}
It can be easily seen that the physical matrix elements of both
$h_{\pi N\Delta}(E)$ and $\bar h_{\pi N\Delta}^{\dagger}(E)$,
i.e., $h_{\pi N\Delta}(E+i\varepsilon,\bk_E)$ and $\bar h_{\pi
N\Delta}^{\dagger}(E+\varepsilon,\bk_E)$, have the phase
$\delta_\alpha$. Note that $\bar h_{\pi N\Delta}^{\dagger}(E)\neq
h_{\pi N\Delta}^{\dagger}(E)$. $g_\De(E)$ is the dressed $\De$
propagator, \bea g^{-1}_{\Delta} = g^{-1}_0 -
\Sigma_{\Delta}(E),\eea where the $\Delta$ self-energy
$\Sigma_{\Delta}(E)$  is given by \bea \Sigma_{\Delta}(E) = h_{\pi
N\Delta}^{(0)}g_0  \bar h_{\pi N\Delta}^{\dagger}(E).  \eea The
matrix element of $\Sigma_{\Delta}(E)$ is related to the dressed
mass $M_\De(E)$ and the width $\Gamma_\De(E)$ of the physical
$\De$ by \bea
<\De^{(0)}\mid\Sigma_{\Delta}(E+i\varepsilon)\mid\De^{(0)}>
=M_\De(E)-i\frac{\Gamma_\De(E)}{2}. \label{eq:SigmaDe}\eea
\newline
\indent Similarly, the $\De$ contribution to $\ga^*N$ reaction
$\tilde{t}_{\gamma\pi}^{\Delta}(E)$ can be expressed as \bea
\tilde{t}_{\gamma\pi}^{\Delta}(E)= \bar h_{\pi
N\Delta}^{\dagger}(E)g_\De h_{\ga N\Delta}(E), \label{eq:tpiNR2}
\eea where
\begin{eqnarray}
h_{\ga N\Delta}(E)&=&h_{\ga N\Delta}^{(0)}+h^{(0)}_{\pi
N\De}\,g_0(E) \,\tilde{t}_{\ga \pi}^B(E)\nonumber\\
&=&h_{\ga N\Delta}^{(0)}+h_{\pi N\De}(E)\,g_0(E) \,v_{\ga \pi}^B.
 \label{eq:hgaNDe}
\end{eqnarray}
\indent To solve the integral equation for $\tilde t^B_{\pi N}$ in
\Eqref{tB}, SL introduced dipole form factors for $v^B_{\pi N}$
and the $\pi N\De$ vertex. The form factors they introduced for
the different pieces in $v^B_{\pi N}$ are different. The
off-energy-shell matrix elements of $v^B_{\ga\pi}$ are needed to
calculate $\tilde t^B_{\ga\pi}$ of \Eqref{tB} and $h_{\ga
N\Delta}(E)$ of \Eqref{hgaNDe}. This is dictated by the unitary
transformation they used as it requires that the time components
time component of the momentum in the propagator of any $\gamma^*
N \rightarrow \pi N$ amplitude is evaluated by the external
momenta associated with the strong interaction vertices.
\newline
\indent
For the electroproduction, Sato and Lee used the following
substitution for the current operator \bea J^\mu \ra J^\mu - \frac
{q\cdot J}{n\cdot q}n^\mu, \label{eq:SLGI} \eea with $n=(0,0,0,1)$
for $q=(\omega,0,0,q)$ to preserve   gauge invariance in their
calculation. This amounts to adding interaction currents $J_z
=J_z(SL) + J_z(int)$ such that $J_z= q^0/|\vec{q}| J_0(SL)$,
instead of $J_z(SL)$, is used in the calculations of longitudinal
cross sections. This simple prescription is identical to what has
been commonly used in many nuclear physics calculations
\cite{deForest66}.
\newline
\indent In short, Sato and Lee derived an effective Hamiltonian
consisting of bare $\De\leftrightarrow\pi N, \ga N$ vertex
interactions and energy-independent meson-exchange $\pi N, \ga
N\ra\pi N$ transition potential operators, by applying a unitary
transformation to a model Lagrangian of $N, \De, \pi, \rho,
\omega,$ and $\ga$ fields in order to achieve consistency in
describing $\ga^*\pi$ reactions and $\pi N$ scatterings. Form
factors are added to the resultant effective Hamiltonian and with
the use of a Schroedinger propagator, the parameters are adjusted
to give a good description of the existing data.


\subsubsection{Dubna-Mainz-Taipei model}

Dubna-Mainz-Taipei (DMT) model was developed in two stages. When the dynamical approach
was first proposed in Ref. \cite{Yang85}, a phenomenological
separable form for $v_{\pi N}$ and the Schroedinger propagator
$(E-H_0)^{-1}$, together with a $v_{\ga\pi}$ derived from a chiral
effective Lagrangian, were employed in \Eqref{LSeq}. Efforts were
then made to construct a meson-exchange model  for the $\pi N$
scattering \cite{Lee91,Hung01} (called Taipei-Argonne $\pi N$
model hereafter), in order to achieve consistency in the treatment
of $\pi N$ scattering and $\ga\pi$ reactions. However, the
resultant $\pi N$ interaction was used in \Eqref{LSeq} only to
describe the pion threshold photoproduction in \cite{Lee91}.
Putting together the meson-exchange models of $v_{\ga\pi}$ and
$v_{\pi N}$ and apply them to the photo- and electroproduction of
pion in the resonance region was performed only recently in Ref.
\cite{KY99}.
\newline
\indent In the TA $\pi N$ model, several three-dimensional
reduction schemes of the Bethe-Salpeter equation for a model
Lagrangian involving $N, \De, \pi, \sigma,$ and $\rho$ fields were
investigated. It was found that all of the resultant
meson-exchange $\pi N$ models can give similar good description of
the $\pi N$ phase shifts up to $400$~MeV. However, they have
significant differences in describing the $\pi NN$ and $\pi N\De$
form factors and $\pi N$ off-shell $t-$matrix elements.
\newline
\indent The TA model contains almost identical physics input as
the SL $\pi N$ model. Namely, the Lagrangian they considered both
consist of $\pi, \rho, N, \De$ fields except that the TA model
also includes the $t-$channel $\sigma$ exchange. The relevant
interaction Lagrangian used
in the TA model is given as follows:
\begin{eqnarray}
{\mathcal L}_I & = & \frac{f^{(0)}_{\pi NN}}{m_\pi} \bar N
\gamma_5 \gamma_\mu
 \vec{\tau} \cdot \partial^\mu \vec{\pi} N
    -g^{(s)}_{\sigma\pi\pi}m_{\pi} \sigma (\vec{\pi} \cdot \vec{\pi})
    -\frac{g^{(v)}_{\sigma\pi\pi}}{2m_\pi}\sigma \partial^{\mu}\vec{\pi}\cdot\partial_{\mu}\vec{\pi}
     \nonumber \\
    & &  -g_{\sigma NN}\bar{N}\sigma N
   -g_{\rho NN} \bar{N} \{ \gamma_\mu \vec{\rho}\,{}^\mu +
\frac{\kappa_V^\rho}{4M_N} \sigma_{\mu\nu} (\partial ^\mu
\vec{\rho}\,{}^\nu -
\partial^\nu \vec{\rho}\,{}^\mu) \} \cdot \frac{1}{2}\vec{\tau} N  \nonumber \\
& & -g_{\rho\pi\pi} \vec{\rho}\,{}^{\mu} \cdot (\vec{\pi} \times
\partial_{\mu}
 \vec{\pi}) -
\frac{g_{\rho\pi\pi}}{4m_{\rho}^2}(\delta - 1)(\partial^\mu
\vec{\rho}\,{}^\nu -
\partial^\nu \vec{\rho}\,{}^\mu) \cdot (\partial_\mu \vec{\pi} \times
\partial_\nu \vec{\pi})  \nonumber \\
& & + \{\frac{g^{(0)}_{\pi N\Delta}}{m_\pi}  \bar\Delta_\mu
[g^{\mu\nu} -(Z+\frac{1}{2})\gamma^\mu\gamma^\nu] \vec{T^\dagger}
N \cdot
 \partial_{\nu} \vec \pi + h.c.\}, \label{largrang}
\end{eqnarray}
where $\Delta_{\mu}$ is the Rarita-Schwinger field operator. The
parameters, namely, all the coupling constants, the bare nucleon
and $\De$ masses $m^{(0)}_N$ and $m^{(0)}_\De$, and the
"off-mass-shell coupling parameters" Z for the $\pi N\De$ vertex
\cite{Olsson75}, are adjusted such that the predictions of $t_{\pi
N}$ agree well with the experimental data, including nucleon mass,
renormalized $\pi NN$ couplings constant $f^2_{\pi NN}/4\pi=0.079$
and the $s-$ and $p-$wave $\pi N$ phase shifts.
\newline
\indent The effective chiral Lagrangian for the strong interaction
is then extended to include the electromagnetic interaction in a
gauge invariant way, i.e., with the minimal substitution of
$\pa_\mu\ra\pa_\mu-ieA_\mu$. In addition, the $t-$channel
vector-meson $(\rho,\omega)$ exchange is known to make
non-negligible contribution. The relevant effective Lagrangian is
given by
\begin{eqnarray} \label{eq:LagV}
\mathcal{L}_{V\pi\gamma} &=&
  \frac{e \lambda_V}{m_\pi}\,
  \varepsilon_{\mu \nu \rho \sigma}\,(\partial^{\mu}A^{\nu})\,
  \pi_i\, \partial^{\rho}(\omega^{\sigma} \delta_{i3} + \rho_i^{\sigma}) \,,
  \label{eq:Vgapi} \\
\mathcal{L}_{VNN} &=&
  g_{VNN} {\bar N} \left( \gamma_{\mu} V^{\mu} -
  \frac{\kappa_V}{2 M_N} \sigma_{\mu \nu} \partial^{\nu} V^{\mu}
  \right)  N \,, \label{eq:VNN}
\end{eqnarray}
where $N$ and $V\,(=\rho,\omega)$ denote the nucleon and  vector
meson field, respectively. $\lambda_V$ is the radiative coupling
determined by $V\ra\pi\ga$ decay. The interaction described above
would give rise to a contribution to $v_{\ga\pi}$ of \Eqref{LSeq}
which contains the Born terms in pseudoscalar coupling and  the
$t-$channel $(\rho,\omega)$ vector-meson exchanges and is normally
called the background transition potential and denoted by
$v^B_{\ga\pi}$.
\newline
\indent
However,
different theoretical treatments  lead to difference in $v_{\pi
N}$ even for the same set of diagrams considered. For example, for
the $u-$channel nucleon exchange diagram, TA model has (omitting the isospin
indeces)\bea
v^{u-ex}_{\pi N}(p',k';p,k)= \left(\frac{f_{\pi NN}}{m_\pi}\right)^2
\,\gamma_5\,
k\pslash / \, S_N(p'- k)\,\gamma_5 \, k'\pslash / ,\label{eq:uexchTA}\eea
where $S_N(p) =(p \hspace{-0.08in} / - M_N+i\varepsilon)^{-1}$  is the
nucleon propagator. In SL model, however, with the use of the
unitary transformation, \Eqref{uexchTA} becomes, \bea
v^{u-ex}_{\pi N}(p',k';p,k)= (\frac{f_{\pi NN}}{m_\pi})^2 \gamma_5
k\pslash /\frac 12 [\,S_N(p'- k) + S_N(p-k')\gamma_5 k'\pslash
/\,\,] \gamma_5 k'\pslash / ,\eea instead. In addition, SL used
$g_0=1/(E-H_0)$ in solving the equation  $t_{\pi N}=v_{\pi
N}+v_{\pi N}g_0t_{\pi N}$, while the TA model considered many
different three-dimensional propagators.
\newline
\indent In the DMT model calculations of \cite{KY99} for
electromagnetic pion production, only the meson-exchange $\pi N$
model for $v_{\pi N}$, obtained with the Cooper-Jennings
\cite{CJ89} three-dimensional reduction scheme, has been employed.
It was chosen because Cooper and Jennings \cite{CJ89} argued that
the reduction scheme they developed has the advantage of
preserving the chiral symmetry contained in the original
Lagrangian. However, two remarks are in order here. First, even
though the $t_{\pi N}$ matrix elements used in the DMT
calculations are obtained within the Cooper-Jennings' reduction
scheme, namely, the time component of the relative momentum in
$v_{\pi N}$ is fixed according to the Cooper-Jennings
prescription, all the particles are put on mass-shell  in the
derivation of $v_{\ga\pi}$ of DMT. In addition, the
Cooper-Jennings propagator was used in solving
 $t_{\pi N}=v_{\pi N}+v_{\pi N} g_0 t_{\pi N}$ but the Schroedinger
 propagator is used for $g_0$ instead in the evaluation of
 $t_{\ga\pi}=v_{\ga\pi}+v_{\ga\pi} g_0
t_{\pi N}.$ With the differences in $v_{\pi N}$ and $g_0$ as
expounded above, it will not be a surprise if the off-shell
behaviors of the $t_{\pi N}$ obtained in SL and TA would differ
substantially in some kinematical regions.
\newline
\indent The prescription used to preserve gauge invariance in the
DMT model goes as follows. The aim is to continue $k$ in
$v_{\ga\pi}^\alpha(k,q)$, which is originally derived only for the
on-energy-shell values $k_E$, to the off-energy-shell region as
called for in \Eqref{tmult}. It starts from the expressions for
the CGLN amplitudes $F_i's$ derived from the tree approximation to
the effective chiral Lagrangian used for the model. The $F_i's$
can be expressed as linear combination of the the invariant
amplitudes $A_i's$ of \Eqref{invariant} as given in Eqs. (Ia-b) in
the Appendix of Ref. \cite{Dennery61}, where the coefficients are
functions of $Q^2$, the energies of photon, pion, initial and
final nucleon, $q_0, k_0, E,$ and $E'$, respectively, and the
total energy $W$. These energy variables are then replaced with
expressions in three-momentum $|\bf q|$ and $|\bf k|$, e.g., $,
q_0=\sqrt {|\bq|^2-Q^2}, k_0=\sqrt{m_\pi^2+\bk^2}$ and
$E=\sqrt{M_N^2+\bq^2}$ etc, while $W$ is always kept as the
initial energy, namely, $W=|\bq|+\sqrt{M_N^2+\bq^2}$. Such
substitutions have the property that the threshold behaviors of
the $F_i's$ are retained, i.e., \bea F_1\propto const.,
\hspace{1.0cm} &F_2\propto |\bq||\bk|, \hspace{1.0cm} &F_3 \propto
|\bq||\bk|,\nonumber \\
 F_4\propto |\bk|^2,\hspace{1.4cm}  & F_7\propto
const.,\hspace{1.0cm}
 & F_8\propto |\bk|. \label{eq:Fthresh}\eea
In the invariant amplitudes, the Mandelstam invariants $t$ and $u$
appearing in the $t$- and $u$-channel exchange diagrams, are
expressed in terms of the three-momentum $|{\bf q}|$ and $|{\bf
k}|$ by~: \bea t &=&
(p-p')^2=2M_N^2-2\sqrt{M_N^2+\bq^2}\sqrt{M_N^2+\bk^2}+2\bq\cdot\bk,
\label{eq:t} \\
u &=& (p-k)^2=M_N^2+m_\pi^2-2
\sqrt{M_N^2+\bq^2}\sqrt{m_\pi^2+\bk^2} -2\bq\cdot\bk. \label{eq:u}
\eea The resulting expressions of $F_i's$, which are functions of
$|{\bf q}|, |{\bf k}|$ and $x=cos\theta$ are then used to obtain,
according to Eq. (If) in Ref. \cite{Dennery61}, the multipole
amplitudes $v^\alpha_{\ga\pi}(k,q)$ where $k$ can now be
off-energy-shell. A dipole form factor $[
(\alpha^2+\bk_E^2)/(\alpha^2+\bk^2)]^2$ with $\alpha=440$ MeV is
then multiplied to all the resultant multipoles to ensure the
convergence in the integral in \Eqref{tmult}.
\newline
\indent There is another difference between SL and DMT dynamical
model calculations. It lies in the scheme in the separation of
background and resonance contributions. In the SL model, the full
$t_{\ga\pi}$ is decomposed as in Eqs.
(\ref{eq:ts1}-\ref{eq:hgaNDe}). In the DMT model, $t_{\ga\pi}$ is
instead decomposed as follows,
\begin{eqnarray}
t_{\gamma\pi}(E)=t_{\gamma\pi}^B(E) +
t_{\gamma\pi}^{\Delta}(E),\label{eq:tgammapi33}
\end{eqnarray}
where
\begin{eqnarray}
t_{\gamma\pi}^B(E)&=v_{\gamma\pi}^B+v_{\gamma\pi}^B\,g_0(E)\,t_{\pi
N}(E)\,,
\label{eq:DMT-bcr}\\
t_{\gamma\pi}^{\Delta}(E)&=v_{\gamma\pi}^\De+v_{\gamma\pi}^\De\,g_0(E)
\,t_{\pi N}(E). \label{eq:DMT-res}
\end{eqnarray}
Here $t_{\gamma\pi}^B$, in contrast to ${\tilde t}_{\gamma\pi}^B$
of \Eqref{tB}, includes the contributions not only from  the
nonresonant mechanisms  but also some of the contributions from
the $\Delta$ excitation as contained in the full $t_{\pi N} $which
lead to the renormalization of the vertex $\gamma^*N\Delta$.
However, all processes which start with the electromagnetic
excitation of a bare resonance are summed up in
$t_{\gamma\pi}^{\Delta}$. One feature of such a decomposition is
that each term in \Eqref{tgammapi33} would fulfill the condition
imposed by Fermi-Watson theorem, i.e., their respective multipole
amplitude in channel $\alpha$ would have the same $\pi N$ phase
shift $\delta_\alpha$ as in \Eqref{FW}.
\begin{figure}[h]
\begin{center}
\epsfig{file=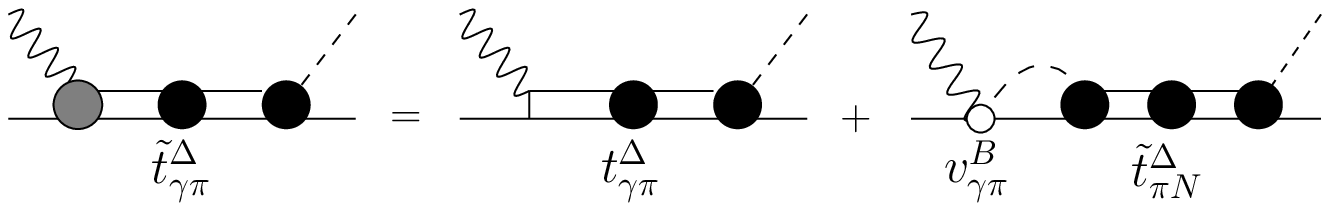,width=12cm}
\end{center}
\caption{Relation between dressed and bare $\gamma N\Delta$
vertices of \Eqref{tR-DMT1}. The shaded and solid circles
represent the dressed $\ga N\De$ and $\pi N\De$ vertices,
respectively.} \figlab{dressed_res}
\end{figure}
\newline
\indent Multipole amplitude of $t_{\gamma\pi}^B$ can be evaluated
in a straightforward manner as in \Eqref{tmult} once $t_{\pi N}$
is given. For the resonance contribution $t_{\gamma\pi}^{\Delta}$,
it is easy to see that it takes the form of,
\begin{eqnarray}
t_{\gamma\pi}^{\Delta}(E)= \bar h_{\pi
N\Delta}^{\dagger}(E)g_\De(E) h_{\gamma N\Delta}^{(0)}.
\label{eq:tRDMT2}
\end{eqnarray}
The relation between $t_{\gamma\pi}^{\Delta}$ and $\tilde
t_{\gamma\pi}^{\Delta}$ of \Eqref{tpiNR}, the $\Delta$ resonance
contribution with dressed and bare e.m. vertices as depicted in
\Figref{dressed_res}, is given by
\begin{eqnarray}
{\tilde t}_{\gamma\pi}^{\Delta}(E)=t_{\gamma\pi}^{\Delta}(E)+
v_{\gamma\pi}^B\,g_0(E)\,{\tilde t}_{\pi N}^{\Delta}(E)\, .
\label{eq:tR-DMT1}
\end{eqnarray}
\indent Note that $g_\De(E)$, which depends on the physical mass
$M_\De$ and total width $\Gamma_\De$, appears in ${\tilde
t}_{\gamma\pi}^{\Delta}(E), t_{\gamma\pi}^{\Delta}$ and ${\tilde
t}_{\pi N}^{\Delta}(E)$, so the resonance position and the total
width can be extracted from $\pi N$ scattering. The DMT model
calculation  takes advantage of this fact and  writes the
resonance multipole amplitude of \Eqref{tRDMT2} in the following
form~:
\begin{equation}
t_{\gamma\pi}^{\Delta,\alpha}(W,Q^2)\,=\,
A_{\alpha}^{\Delta}(Q^2)\,
\frac{f_{\gamma\Delta}(W)\Gamma_{\Delta}(W)\,M_{\Delta}\,f_{\pi\Delta}(W)}
{M_{\Delta}^2-W^2-iM_{\Delta}\Gamma_{\Delta}(W)}
\,e^{i\phi_{\alpha}(W)}, \label{eq:BWDMT}
\end{equation}
where $f_{\pi\Delta}$ is the usual Breit-Wigner factor describing
the decay of a resonance with total width $\Gamma_{\Delta}(W)$ and
physical mass $M_{\Delta}$. Namely, the well known Breit-Wigner
form for the resonance contribution is assumed, in close analogy
to the standard way of analysis of the experimental data as done
in \cite{MAID98}.
 The expressions for the form factors
$f_{\gamma\Delta}, \, f_{\pi \Delta}$ and total width
$\Gamma_{\Delta}$ will be considered in the next subsection. The
energy dependence in $f_{\gamma\Delta}(W)$ is introduced to
account for the effect of possible excitation of the $\De$ via
inelastic channels like $\pi \De$. The phase $\phi(W)$ in Eq.
(\ref{eq:BWDMT}) is used to adjust the phase of the resonance
contribution to be equal to the corresponding $\pi N$ phase shift
$\delta_{\pi N}$ in the (3,3) channel as required by the
Fermi-Watson theorem. At $W=M_{\Delta}=1232$ MeV $\phi=0$ for any
$Q^2$. Therefore, this phase does not affect the $Q^2$ dependence
of the $\gamma N\Delta$ vertex.
\begin{figure}[tbp]
\centerline{\epsfxsize=12cm%
\epsffile{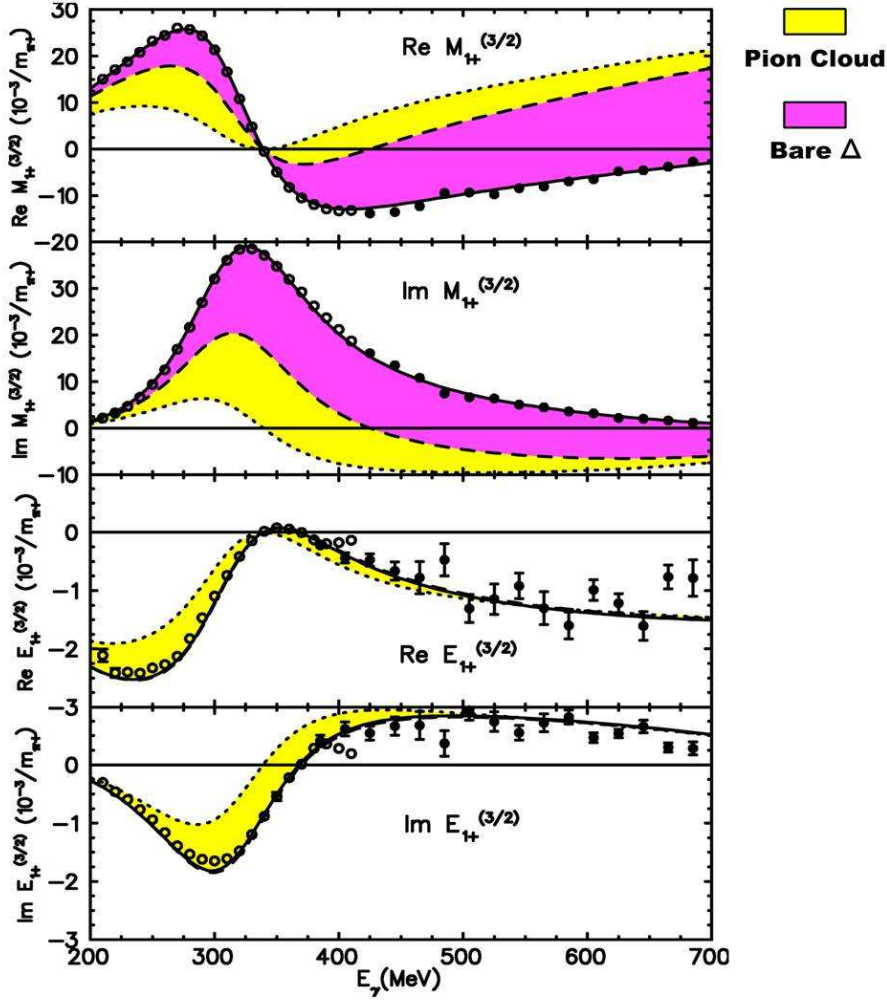}
}
\caption{ Real and imaginary parts of the $M_{1+}^{(3/2)}$ and
$E_{1+}^{(3/2)}$ multipoles. The dashed (dotted) curves are the
results obtained within the DMT model
for $t_{\gamma\pi}^B$ including (excluding) the
principal value integral contribution in \Eqref{tmult}.
The solid curves are the full DMT results including also the
bare $\Delta$ excitation.
For the $E_{1+}$ multipole, the dashed
and solid curves practically coincide
due to the small value of the bare ${\bar{\mathcal E}}^{\Delta}$.
The open circles are the results from the Mainz
dispersion relation analysis~\cite{HDT}, whereas the solid circles
are obtained from the VPI analysis~\cite{VPI97}.
}
\figlab{DMT_M1E1}
\end{figure}
\newline
\indent In the SL and DMT models, the $Q^2$ dependence for the
$\gamma N\Delta$ vertex is parametrized in the same way as in Ref.
\cite{MAID98}, i.e.
\begin{eqnarray}
 A_{\alpha}^{\Delta}(Q^2)= A_{\alpha}^{\Delta}(0)
\frac{
q_\Delta}{q_W}\,(1+\beta_{\alpha}\,Q^2)\,e^{-\gamma_{\alpha}
Q^2}\,G_D(Q^2), \label{barA}
\end{eqnarray}
with $\alpha = M, E, S$ referring to magnetic dipole, electric
quadrupole, or Coulomb quadrupole transitions, where $q_\Delta$ is
defined in Eq.~(\ref{eq:gamomdel}), and where $q_W$ is defined
below Eq.~(\ref{eq:resp}). The form factor
$G_D(Q^2)=1/(1+Q^2/0.71)^2$ is the usual nucleon dipole form
factor. In the case of the magnetic transition the parameters
$\beta$ and $\gamma$ can be determined by fitting
 $A_{M}^{\Delta}(Q^2)$ to the data for the well known $G_M^*$ form
factor, which is obtained from the $M_{1+}$ pion electroproduction
multipole in the (3,3) channel at $W=M_{\Delta}=1232$ MeV, as
given by Eq.~(\ref{eq:multjs}).
\newline
\indent We want to remind the reader that the definition of the
$\Delta$ e.m amplitudes in different models can be different. As
we have seen above in the SL model they can contain contributions
from the excitation of the $\Delta$ resonance via the nonresonant
mechanism and describe the dressed $\gamma N\Delta$ vertex. In the
DMT model such mechanism  is included in the unitarized background
${t}^{B}_{\gamma\pi}$ and the electromagnetic vertices $
A_{\alpha}^\Delta(Q^2)$ describe the bare e.m. vertices.
\newline
\indent In general, in accordance with the considered above
dynamical models, the $G_M^*$ form factor can be decomposed in
three terms
\begin{eqnarray}
 G_M^*(Q^2)= G_M^{bare}(Q^2)+ G_M^{pion\,cloud}(Q^2)+ G_M^{n.r. bcgr.}(Q^2)\,,
\label{GM*3}
\end{eqnarray}
where $G_M^{bare}\,,G_M^{pion\,cloud}$ and $G_M^{n.r. bcgr.}$ are
the contributions from $t_{\gamma\pi}^{\Delta}$, from the second
term in rhs of Eq.~(\ref{eq:tR-DMT1}) and from the
$\tilde{t}_{\gamma\pi}^B$ \Eqref{tB}, respectively. Therefore, different
models, which have a different way for the separation of
background and resonance contributions, the values for the
parameters ${\bar{A}}_{\alpha}^{\Delta}(0)$, $\beta_{\alpha}$ and
$\gamma_{\alpha}$ in Eq. (\ref{barA}) can be different. The same
arguments are also true for the form factors of the electric and
Coulomb transitions.
\newline
\indent
In \Figref{DMT_M1E1}, we show the resonant multipoles $M_{1+}^{(3/2)}$
and $E_{1+}^{(3/2)}$ as obtained in the DMT model.
For $M_{1+}^{(3/2)}$, one sees a large effect of the pion off-shell
rescattering (difference between dotted and dashed curves), which
results from the principal value integral part of  \Eqref{tmult}.
The total pion rescattering (dashed curves) contributes
for half of the $M_{1+}^{(3/2)}$ as seen in
\Figref{DMT_M1E1} for the DMT model, the remaining half originates from the
bare $\gamma N \Delta$ excitation.
Furthermore, one sees that almost all of the $E2$ strength is generated
by the $\pi N$ rescattering.

\subsubsection{Dynamical Utrecht-Ohio model}

The dynamical Utrecht-Ohio (DUO) model has been
developed in Refs.~\cite{tjon00,Pascal04}.
The model is based on a $\pi N$-$\ga N$ coupled-channel
equation which when solved to the first order in the electromagnetic coupling $e$
leads to the electroproduction amplitude, $T_{\pi\ga^\ast}=V_{\pi\ga^\ast}+
T_{\pi\pi} G_\pi V_{\pi\ga^\ast}$, where  $V_{\pi\ga^\ast}$ is
an basic electroproduction potential, $G_\pi$ is the pion-nucleon propagator
and $T_{\pi \pi}$ is the full $\pi N$ amplitude.
Thus,  the pion rescattering effects
are included as the final state interaction.
\begin{figure}[tb]
\centerline{  \epsfxsize=10 cm \epsffile{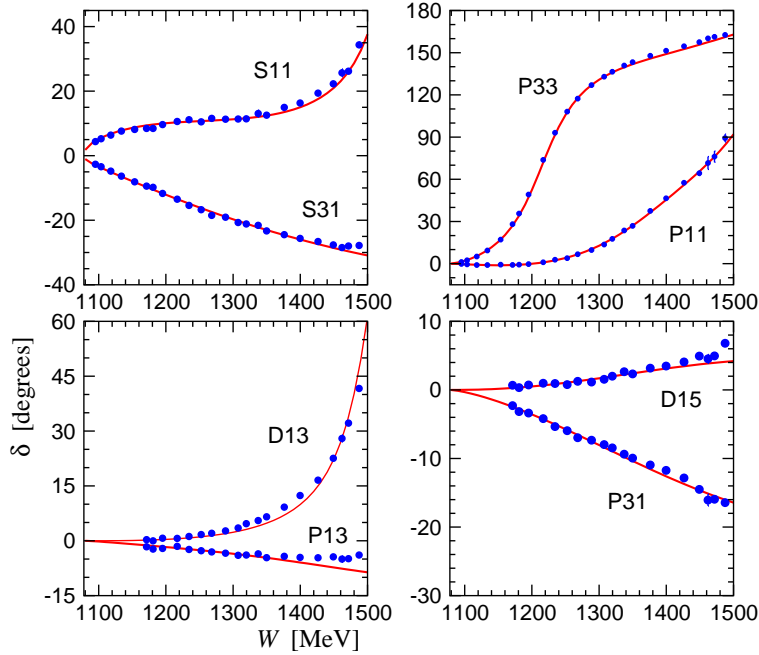}}
\caption{(Color online) The $s$-, $p$-, and some of the $d$-wave
$\pi N$ scattering phase-shifts in the DUO model (red solid curves).
The data points are the results of the SAID
single-energy solution SP06~\cite{GWU}.}
\figlab{pin_fs}
\end{figure}
\newline
\indent
The $\pi N$ amplitude satisfies an integral equation on its own.
The details on constructing $\pi N$ amplitude are presented in~\cite{tjon00}.
The corresponding fit of this model into
the $\pi N$ elastic scattering phase shifts is shown in
\Figref{pin_fs}.
\newline
\indent
In \Figref{gipif} one can see the diagrammatic representation
of the DUO model for the pion electroproduction.
The model potential $V_{\pi\ga}$ includes the Born term
(using the pseudo-vector $\pi NN$ coupling),
the $t$-channel exchange of $\rho-$ and $\w-$ mesons, and
the $\De$-isobar exchange.
\newline
\indent
The $\pi N$ final state interaction dresses the $s$-channel nucleon
and resonance contributions, leading in particular to the mass, field and
coupling constant renormalizations.
Therefore, both $N$- and $\Delta$-pole contributions in $V_{\pi\ga}$
are included using the {\it bare} mass and coupling
parameters obtained from the equation for the $\pi N$ amplitude.
The renormalization conditions together with unitarity demand that the same
propagators and $\pi N$ vertices, including the cutoff functions,
appear in both the $\pi N$ and $\ga N$ potentials. Thus, all these ingredients
are fixed by the analysis of $\pi N$ scattering.
\begin{figure}[tb]
\centerline{  \epsfxsize=10 cm \epsffile{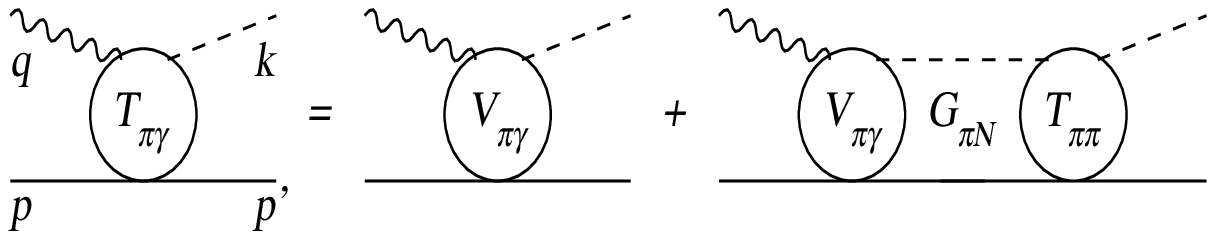}}
\bigskip
\centerline{  \epsfxsize=13 cm \epsffile{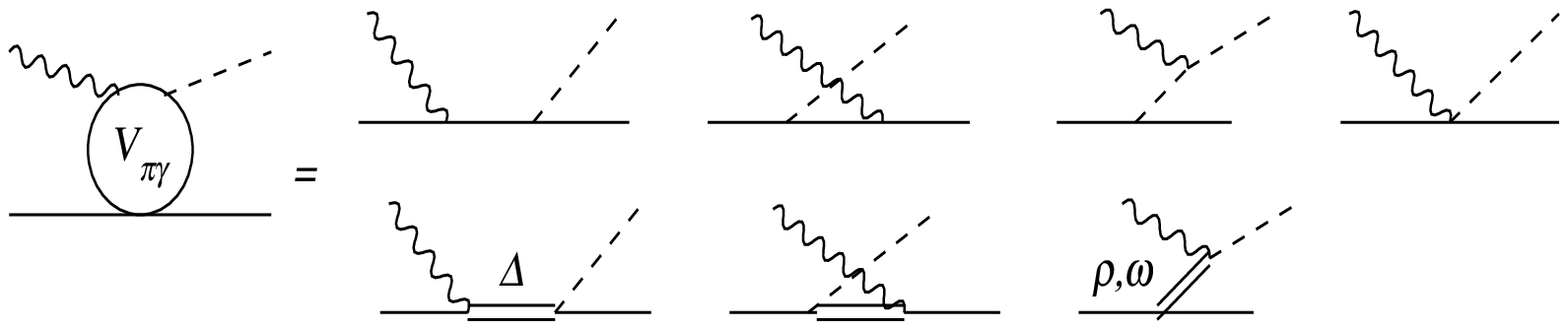}}
\caption{The electromagnetic pion production amplitude
in the DUO model.}
\figlab{gipif}
\end{figure}
\newline
\indent
On the other hand,
the electromagnetic interaction is constrained by the electromagnetic {\it
gauge invariance}.  At this point one is often concerned with the problem
of how to introduce the electromagnetic form factors for nucleon and
pion in a way consistent with gauge invariance.
A common solution to this problem
is to choose all of the electromagnetic form factors
that go into the Born term
 (i.e., nucleon, pion and axial form factors) to be  the same.
This  prescription does enforce the current conservation,
however the Ward-Takahashi (WT)
identities cannot be satisfied in this way. Furthermore, it is clear that
the requirement of gauge invariance should not be able to
restrict the $Q^2$ dependence
of the electromagnetic interaction. Finally, it is known
that these form factors need not to be and {\it are not} the same.
\newline
\indent
The DUO model introduces the procedure where
an arbitrary form factor $F(Q^2)$ can be accommodated by the
following replacement of the current:
\beq
\eqlab{procedure}
J^\mu \rightarrow {J'}^\mu(Q^2)= J^\mu +[F(Q^2)-1] \,O^{\mu\nu} J_\nu\,,
\eeq
where $O ^{\mu \nu}=g^{\mu \nu}-q^{\mu} q^{\nu}/q^{2}$, and $q$
is the photon 4-momentum, $Q^2=-q^2$.
It is easy to see that the resulting
current ${J'}^\mu$ obeys exactly the same WT identities as $J^\mu$.
Thus, as long as gauge-invariance is implemented
at the real-photon point, the inclusion of the form factors via \Eqref{procedure}
will give the gauge-invariant current for $Q^2\neq 0$.
For example, the bare $NN\ga$ and $\pi\pi\ga$ vertex functions and the Kroll-Rudermann term are:
\begin{subequations}
\eqlab{modEMvert}
\begin{eqnarray}
\eqlab{modgaNNvert} \Gamma_{
NN\ga}^{\mu}&=&e\,\gamma^{\mu}+e\,[F_{1}(Q^2)-1] \,O^{\mu \nu}
\gamma_{\nu}+\frac{e\kappa_N}{2m_{N}}\, F_{2}(Q^{2})\,i\sigma ^{\mu \nu}q_{\nu} \,,\\
\eqlab{modgapipivert} \Gamma_{\pi\pi\ga}^{\mu}
&=&e\,(k+k')^{\mu}+e\,[F_{\pi}(Q^2)-1] \,O^{\mu \nu}(k+k')_{\nu}\,,\\
\eqlab{KRmodchannel} J_{KR}^{\mu}&=&\frac{e g_{\pi
N}}{2m_{N}}\left\{
\gamma^{\mu}+[F_{A}(Q^{2})-1]\,O^{\mu \nu}\gamma_{\nu}\right\}\gamma_{5}\,.
\end{eqnarray}
\end{subequations}
This procedure allows one to use the experimentally
determined form factors in the Born terms.

\subsection{MAID - a phenomenological multipole solution}

MAID, developed by the Mainz group, is currently the most
comprehensive unitary isobar model which includes contributions
from all other known resonances up to second and third resonance
regions, in addition to the $\De(1232)$. Its first version was
published in 1999 \cite{MAID98} as commonly referred to as MAID98.
Since then it is updated once every two years to accommodate for
the new data by readjusting the parameters and can be accessed to,
together with DMT, at the website of Ref.~\cite{MAID}.
\newline
\indent The basic equations used for the MAID are similar to those
of the DMT model \cite{KY99}. It also starts from the equation
$t_{\gamma\pi}(E)=v_{\gamma\pi}+v_{\gamma\pi}\,g_0(E)\,t_{\pi
N}(E)$ with $g_0=(E-H_0)^{-1}$. For any resonant channel, the
transition potential $v_{\gamma\pi}$ consists of two terms
\begin{eqnarray}
v_{\gamma\pi}(E)=v_{\gamma\pi}^B
+v_{\gamma\pi}^R(E),\label{eq:vgammapi33}
\end{eqnarray}
where $v_{\gamma\pi}^B$ is the background transition potential and
$v_{\gamma\pi}^R(E)$ corresponds to the contribution of the bare
resonance excitation. The resulting $t$-matrix can be decomposed
into two terms as in Eqs. (\ref{eq:tgammapi33}-\ref{eq:DMT-res}),
\begin{eqnarray}
t_{\gamma\pi}(E)=t_{\gamma\pi}^B(E) +
t_{\gamma\pi}^{R}(E),\label{eq:tgammapiR33}
\end{eqnarray}
where
\begin{eqnarray}
t_{\gamma\pi}^B(E)=v_{\gamma\pi}^B+v_{\gamma\pi}^B\,g_0(E)\,t_{\pi
N}(E), \\
t_{\gamma\pi}^R(E)=v_{\gamma\pi}^R+v_{\gamma\pi}^R\,g_0(E)
\,t_{\pi N}(E).
\end{eqnarray}
\indent In MAID, the background potential
$v^{B,\alpha}_{\gamma\pi}$ is constructed in the same way as in
the DMT model except now the Born terms are calculated with a
hybrid model (HM), i.e., an energy dependent mixing of
pseudovector-pseudoscalar (PV-PS) $\pi NN$ coupling ,
\begin{equation}
{L}_{\pi NN}^{HM}=\frac{\Lambda_m^2}{\Lambda_m^2+{\bf k}_0^2}
{L}_{\pi NN}^{PV}+\frac{{\bf k}_0^2}{\Lambda_m^2+{\bf k}_0^2}
{L}_{\pi NN}^{PS}\,,
\end{equation}
where  ${\bf k}_0$ is the asymptotic pion momentum in the $\pi N$
{\it c.m.} frame which depends only on $W$ and is not an operator
acting on the pion field. From the analysis of the
$M_{1-}^{(3/2)}$ and $E_{0+}^{(3/2)}$ multipoles it was found that
the most appropriate value for the mixing parameter is
$\Lambda_m=450$ MeV \cite{MAID98}. Note that for the pion pole
term, the pion couples with on-shell nucleons only, and HM, PV,
and PS  couplings are all equivalent. As a matter of fact, only
the multipoles $E_{0+},\,M_{1-},\, L_{0+}$, and $L_{1-}$ are
affected by the use of such a mixed coupling scheme.  In all,
 $v^{B,\alpha}_{\gamma\pi}$ depends on 7 parameters:
the PV-PS mixing parameter $\Lambda_m$, 4 coupling constants  and
2 cut-off parameters for the vector-meson exchange contributions.
\newline
\indent In MAID98, the background contribution was given as
$t_{\gamma\pi}^{B,\alpha}({\rm
MAID}98)=v_{\gamma\pi}^{B,\alpha}(W,Q^2)$, a real and smooth
function. The unitarization of the total amplitude was done by
introducing an additional phase $\phi$ in the resonance
contribution which ensures the phase of the total amplitude
(background plus resonance) be equal to the corresponding
pion-nucleon scattering phase $\delta_{\alpha}$. In the new
extended version (MAID2005), all the the background contributions
of $s$-, $p$-, $d$- and $f$-waves
 are complex numbers as  prescribed in
K-matrix approximation, namely,
\bea
 t^{B,\alpha}_{\gamma\pi}(W,Q^2)=v^{B,\alpha}_{\gamma\pi}(W,Q^2)\,[1+it_{\pi N}^{\alpha}(W) ]\, ,
\label{eq:backgr} \eea
where $t^{\alpha}_{\pi N}=[\eta_{\alpha}
\exp(2i\delta_{\alpha})-1]/2i$ is the pion-nucleon elastic
scattering amplitude with the phase shift $\delta_{\alpha}$ and
the inelasticity parameter $\eta_{\alpha}$ taken from the analysis
of the VPI group (SAID program). Below the two-pion threshold
where $\eta_{\alpha}=1$, \Eqref{backgr} becomes
$t^{B,\alpha}_{\gamma\pi}(W,Q^2)=e^{i\delta_\alpha}
cos\delta_\alpha v^{B,\alpha}_{\gamma\pi}(Q^2)$. Such a structure
of the background appears naturally from \Eqref{tmult} if the
principal value integral contribution in the pion rescattering
term is neglected. Part of such a ``pion cloud" contribution
(principal value integral) will be effectively included in the
resonance sector with the use of dressed $\gamma N R$ vertex as
will be explained below.
\newline
\indent As in the DMT model, the resonance contributions
$t_{\gamma\pi}^{R,\alpha}(W,Q^2)$ in MAID is parametrized as in
\Eqref{BWDMT},
\begin{equation}
t_{\gamma\pi}^{R,\alpha}(W,Q^2)\,=\,{\bar{A}}_{\alpha}^R(Q^2)\,
\frac{f_{\gamma R}(W)\Gamma_R\,M_R\,f_{\pi
R}(W)}{M_R^2-W^2-iM_R\Gamma_R} \,e^{i\phi_R}. \label{eq:BW}
\end{equation}
where $\bar A_{\alpha}^R(Q^2)$ should now be interpreted, in
contrast to ${\bar{A}}_{\alpha}^{\Delta}(Q^2)$ of \Eqref{BWDMT},
as corresponding to the excitation of a dressed resonance in order
 to partly account for the neglected principle value integral
 term of \Eqref{tmult} in $t^{B,\alpha}_{\gamma\pi}(W,Q^2)$
 of \Eqref{backgr}.  $f_{\pi N}(W)$ is the usual Breit-Wigner factor
describing the decay of the $N^*$ resonance with total width
$\Gamma_{tot}$, partial $\pi N$-width $\Gamma_{\pi N}$ and spin
$j$,
\begin{equation}
f_{\pi N}(W)=\left[\frac{1}{(2j+1)\pi}\frac{k_W}{\mid {\bf k}\mid}
\frac{M_N}{W_R}\frac{\Gamma_{\pi
N}}{\Gamma_{tot}^2}\right]^{1/2}\,,\qquad
k_W=\frac{W^2-M_N^2}{2W}\,. \label{eq:fpi}
\end{equation}
The factor $C_{\pi N}$ is  $\sqrt{3/2}$ and $-1/\sqrt{3}$ for the
isospin 3/2 and isospin 1/2  multipoles respectively. The energy
dependence of the partial width $\Gamma_{\pi N}$ is given by
\begin{equation}
\Gamma_{\pi N}=\beta_{\pi}\,\Gamma_R\,\left(\frac{\mid{\bf
k}\mid}{k_R} \right)^{2l+1}\,\left(\frac{X^2+k_R^2}{X^2+{\bf
k}^2}\right)^l\, \frac{W_R}{W}\,, \label{eq:gamma}
\end{equation}
where $X$ is a damping parameter, $\Gamma_{tot}(W)$ is the total
width, $k_R$ is the pion {\it c.m.} momentum  at the resonance
peak ($W=W_R$) and $\beta_{\pi}$ is the single-pion branching
ratio. Expressions for the total width can be found in
\cite{MAID98}. The $W$ dependence of the $\gamma NN^*$ vertex is
given by the form factor
\begin{equation}
f_{\gamma N}(W)=\left(\frac{q_W}{q_R}\right)^n\,
\left(\frac{X^2+q_R^2}{X^2+q_W^2}\right)\,, \label{eq:fgam}
\end{equation}
where the damping parameter $X$ is the same as in \Eqref{gamma})
and $q_R=q_W$ at $W=W_R$. The parameter $n$ is defined by the best
fit of the experimental multipoles at the photon point ($Q^2$=0).
At the resonance position $f_{\gamma N}(W_R)=1$. The phase
$\phi(W)$ in \Eqref{BW} is introduced to adjust the phase of the
total multipole to  equal to the corresponding $\pi N$  phase
shift $\delta_{\alpha}$ below the two pion threshold. For the $s$-
and $p$- waves such unitarization procedure is applied up to
$W=1400$ MeV. At higher energies for these waves the phase
$\phi(W)$ is taken as constants. Note that in the case of the
$\Delta(1232)$ excitation, the phase $\phi=0$ at W=1232 MeV. In
this case, the Fermi-Watson theorem is applied up to $W <
1600$~MeV, where the inelasticity parameter $\eta_{\alpha}$ of
pion scattering amplitude is still close to 1. For the resonances
in the $d$- and $f$-waves resonance phase $\phi$ assumed to be
constant and determined from the best fit.
\newline
\indent While in the original version of MAID \cite{MAID98} only
the 7 most important nucleon resonances were included with mostly
only transverse e.m. couplings, in the new extended version MAID05
all four star resonances below $W=2$~GeV are included (with
transverse electric, ${\bar{A}}_{\alpha}^R=\bar{E}_{l\pm}$,
magnetic, ${\bar{ A}}_{\alpha}^R=\bar{M}_{l\pm}$, and coulomb
,${\bar{ A}}_{\alpha}^R=\bar{S}_{l\pm}$) couplings. They are
$P_{33}(1232)$, $P_{11}(1440)$, $D_{13}(1520)$, $S_{11}(1535)$,
$S_{31}(1620)$, $S_{11}(1650)$, $D_{15}(1675)$, $F_{15}(1680)$,
$D_{33}(1700)$, $P_{13}(1720)$, $F_{35}(1905)$, $P_{31}(1910)$ and
$F_{37}(1950)$. For all these 13 resonances the total numbers of
the e.m. couplings is 52 (34 for proton and 18 for neutron
channels) They are independent of the total energy and depend only
on $Q^2$. They can be taken as constants in a single-Q$^2$
analysis, e.g. in photoproduction, where $Q^2=0$ but also at any
fixed $Q^2$, where enough data with W and $\theta$ variation is
available. Alternatively they can also be parametrized as
functions of $Q^2$

\subsection{Dispersion relation approach}
\label{sec3_dr}

The dispersion relations approach, or "S-matrix theory" was
developed mostly in the 50's and 60's of the last century. It was
suggested as an alternative to the field theoretical approach to
the strong interaction to avoid an expansion in the strong
coupling constant. The underlying postulate of this theory is that
the S-matrix is a Lorentz-invariant {\it analytic} function  of
all momentum variables with only those singularities required by
unitarity. It is generally believed that the analyticity arises
from the causality property as in the case of Kramers-Kronig
relations for the dielectric constant.
\newline
\indent The starting point of the dispersion relation approach is
the fixed-$t$ dispersion relations which follow from the assumed
analytic properties, and from crossing symmetry \Eqref{crossing}.
They can be conveniently written in the following matrix notation
as \cite{Donnachie72}: \bea Re \, \tilde A(s,t,Q^2) =
\tilde A^{pole}(s,t,Q^2) +
\frac{P}{\pi}\int^\infty_{s_{thr}}ds'\left[\frac{1}{s'-s}+[\bar\xi]\frac{1}{s'-u}\right]
Im \tilde A(s',t,Q^2), \label{eq:DR}\eea where
$s_{thr}=(m_\pi+M_N)^2$ and $A^{pole}(s,t,Q^2)$ is the Born term
evaluated in pseudoscalar coupling as can be found in
\cite{Hanstein:1997tp,Donnachie72}. It consists of $s$- and $u$-channel
nucleon poles, and $t$-channel pion pole with residues
proportional to the electromagnetic form factors of the nucleon
and pion.
\newline
\indent The next step is to apply the multipole projection to the
dispersion relations \Eqref{DR} which leads to the following set
of coupled integral equations for the CGLN multipoles ${\mathcal
M}_{\alpha}=(E_{l\pm},\,M_{l\pm},\,L_{l\pm}/\omega)$,
\begin{eqnarray}
Re\,{\mathcal M}_{\alpha}(W)&=& {\mathcal M}_{\alpha}^{Pole}(W)+
{\mathcal M}_{\alpha}^{Diag}(W)\nonumber\\  &+&
\frac{1}{\pi}\int^{\infty}_{W_{thr.}}dW'\sum_{\alpha\neq
\beta}\,K_{\alpha\beta}(W,W')\, Im\, {\mathcal M}_{\beta}(W') \,,
\label{dr2a}
\end{eqnarray}
where $\alpha$ and $\beta$ are the set of quantum numbers and
\begin{eqnarray}
{\mathcal
M}_{\alpha}^{Diag}(W)&=&\frac{P}{\pi}\int^{\infty}_{W_{thr.}}dW'
\frac{Im {\mathcal M}_{\alpha}(W')r_{\alpha}(W')}{(W'-W)
r_{\alpha}(W)}\nonumber\\  &+&
\frac{1}{\pi}\int^{\infty}_{W_{thr.}}dW'K_{\alpha\alpha}(W,W')\,
Im {\mathcal  M}_{\alpha}(W'). \label{dr2b}
\end{eqnarray}
The detailed expressions for the kernels $K's$ and the kinematical
factor $r_{\alpha}(W)$ are given in
Refs.~\cite{Hanstein:1997tp,Gehlen68}.
\newline
\indent One of the methods widely used to calculate the dispersion
integrals in Eqs.~(\ref{dr2a})-(\ref{dr2b}) is based on the
Fermi-Watson theorem~\cite{Watson54} such that below the two-pion
threshold, we can use the following relation between the real and
imaginary parts of the amplitude:
\begin{eqnarray}
Im \, {\mathcal M}_{\alpha}(W,Q^2)= Re \,{\mathcal
M}_{\alpha}(W,Q^2)\ \tan{\delta_{\alpha}(W)}\,. \label{Watson}
\end{eqnarray}
If one further makes an assumption about the high-energy behavior
of the multipole phases, we obtain a system of coupled integral
equations for $Re {\mathcal M}_{\alpha}(W)$. This is the standard
method to apply fixed-$t$ dispersion relations to pion
photoproduction at threshold and in the $\Delta(1232)$ resonance
region, which was successfully used by many
authors~\cite{Donnachie67,Schwela67,Hanstein:1997tp,Azn:2003}. The
reliability of this method at low energies ($W<1400$ MeV) is
mainly based on the finding that Eq.~(\ref{Watson}) can be applied
to the important $P_{33}$ multipole, dominated by the
$\Delta(1232)$ resonance contribution, with good accuracy up to
$W=1600$ MeV.
\newline
\indent Another method to calculate the dispersion integrals is
based on isobaric models~\cite{Salin,Loub,Walecka,Crawford} which
allow extending the use of fixed-$t$ DR  to higher energies. With
this approach, the imaginary parts of the pion photo- and
electroproduction multipoles are expressed in terms of background
(${\mathcal M}^B$) and resonance (${\mathcal M}^R$) contributions,
\begin{eqnarray}
 Im \, {\mathcal M}_{\alpha}(W,Q^2)= Im \, {\mathcal M}^B_{\alpha}(W,Q^2)
 + Im \, {\mathcal M}^R_{\alpha}(W,Q^2).
\end{eqnarray}
In a recent work \cite{Kamalov02}, both parts were modeled similar
to MAID and good agreement with the data is found for the neutral
pion photoproduction at threshold.

%% file: chap4_delta.tex
\section{Chiral effective-field theory in the $\De$-resonance region}
\label{sec4}

In this section we review the recent extension of chiral perturbation
theory into the $\De$-resonance region. Such an extension yields
the opportunity to study the pion-production processes in the
$\De$-resonance region within the systematic framework of 
effective field theory (EFT). Some of the applications
that are relevant to the $\ga N\De$ transition will be discussed
here as well.
\newline
\indent
In Sect.~\ref{sec:cl} we remind the reader of some basic facts about the
chiral Lagrangians for the pion and nucleon 
fields.\footnote{More details and insights
can be found in Refs.~\cite{Wei95b,Ecker:1994gg,Kaplan:2005es} and
references therein.}
In Sect.~\ref{sec:rs}, we introduce the spin-3/2 formalism of Rarita
and Schwinger and discuss its consistency with respect to 
the description of a spin-3/2 particle.
In Sect.~\ref{sec:sec4ss3}, the inclusion of
the spin-3/2 $\De$-isobar field in the chiral Lagrangian is discussed
with the emphasis on the consistency of the spin-3/2 theory.
In Sect.~\ref{sec:pc}, we discuss the power counting in 
the presence of the $\De$-resonance, and thus introduce
the ``$\de$-expansion''. In Sect.~\ref{sec:ip}, we 
demonstrate several applications of the $\de$-expansion
in the processes involving the $\De$-resonance
excitation. 

\subsection{Effective chiral Lagrangians}
\label{sec:cl}

The relevant effective Lagrangian of low-energy QCD will in our
case include the pion, the nucleon, and the $\De$-isobar
fields. The color degrees of freedom, as well as heavier mesons
and baryons are assumed to be ``integrated out'', thus setting an
upper limit on the energy range where this theory is valid.
To begin with we need to write down the most general Lagrangian
involving these fields and consistent with the underlying symmetries
of QCD. In particular, the {\it chiral} symmetry is known to govern
the interaction of hadrons at low energies.
\newline
\indent
Let us briefly recall here that
chiral symmetry is a symmetry of massless quark Lagrangian,
\beq
\lag_{\mathrm{quark}} =  i \, \bar q_f \sla{D} \,\, q_f\,,
\eeq
where $q_f$ is the quark field, $D$ is the covariant QCD derivative,
and the summation over the flavor index $f$ is understood.
This Lagrangian is invariant under the following rotations
of quark fields in the flavor space
\beq
\eqlab{chirotation}
q_f \to q_f' = \left[\half (1-\ga_5)\, \exp{(i\th^a_L\tau^a_{ff'})} +
 \half (1+\ga_5)\, \exp{(i\th^a_R \tau^a_{ff'})} \right]\, q_{f'}\,,
\eeq 
where index $a=\overline{1,n_f^2-1}$, with $n_f$ being the number
of flavors, $\th$'s are $2\times (n_f^2-1)$ independent
parameters and $\tau$'s are the SU$(n_f)$ Pauli matrices. Since the rotations
are done independently for the left-handed and right-handed quarks
($\th_L\neq \th_R$), we have a global SU$_L(n_f$)$\times$SU$_R(n_f$)
symmetry, known as chiral symmetry. 
\newline
\indent
In QCD chiral symmetry is broken both spontaneously,
by non-perturbative effects of the QCD vacuum, and explicitly,
by the quark masses. Both mechanisms break chiral symmetry down to
the SU($n_f$) symmetry under rotation \Eqref{chirotation} with 
$\th_L=\th_R$, the isospin symmetry.
The spontaneous chiral-symmetry breaking ($\chi$SB) leads to the generation
of the chiral quark condensate, $\left< \bar q q\right> \simeq
-(230$ MeV$)^3$, and to the appearance of $n_f^2-1$ massless modes ---
Goldstone bosons (GBs). The explicit $\chi$SB mechanism is responsible for
giving the mass to the GBs.
Accordingly, the effective low-energy Lagrangian of QCD 
should contain GB fields, and preserve explicit
chiral symmetry up to the terms vanishing in the limit
of massless GBs masses --- the ``chiral limit''.  
\newline
\indent
In what follows we restrict our consideration to QCD with two flavors
($n_f=2$), 
up and down quarks only. In this case the three Goldstone bosons
are the pions, described by a (pseudo)scalar-isovector
field $\pi^a$. The chiral Lagrangian can conveniently be written 
in terms of the unimodular non-linear
representation of the pion field:
\beq
U(x) = e^{2i\hat \pi(x)/f_\pi} , 
\,\,\,\, \mbox{with} \,\,\,
\hat \pi \equiv  \half \pi^a \tau^a = \frac{1}{2}
\bmat \pi^0  &\sqrt{2}\,\pi^+ \\
\sqrt{2}\,\pi^- &  - \pi^0  \emat,
\eeq
where $\pi^{\pm,0}$ represent the charge eigenstates of the pion.
Under the chiral transformation, $U \to R \,U L^\dagger $, with $L$ and $R$
elements of SU$_L(2)$ and SU$_R(2)$, respectively. Therefore,
the Lagrangian containing an even number of $U$'s is chirally symmetric.
It is also obvious that only derivatives of $U$ can enter
such chirally-symmetric terms. (Terms where $U$ enters without derivatives
reduce trivially to terms with derivatives, e.g., 
$\mbox{Tr}[U^\dagger\,\pa_\mu  U \pa^\mu U^\dagger \,U] = 
\mbox{Tr}[\pa_\mu  U \pa^\mu U^\dagger ]$.) This shows that 
the chirally-symmetric interactions are proportional
to the momentum of Goldstone bosons, and therefore is weak at low energies
---
the fact that plays a crucial role in the construction of the perturbative expansion
in powers of momenta, which is utilized in $\chi$PT.
\newline
\indent
In applications that furthermore will be considered, the
isospin-breaking effects are negligible. We therefore shall
assume exact isospin symmetry, and in particular 
take the masses of $u$ and $d$ 
quarks to be equal ($m_u = m_d\equiv m_q$).
The lowest-order chiral 
Lagrangian of pion fields is then 
given by\footnote{In the notation $\lag^{(i)} $, 
the superscript indicates the order of the chiral Lagrangian,
given here by the number of
derivatives of Goldstone-boson fields and insertions of their mass.}   
\beq
\lag_\pi^{(2)} = \quarter f_\pi^2\, \mbox{Tr}[ \,\pa_\mu  U
\pa^\mu U^\dagger +  2B \,m_q\, ( U + U^\dagger ) ], 
\eeq  
where, at this lowest order, $f_\pi \simeq 92.4$ MeV is the pion decay
constant and $B$ is related to the scalar quark condensate as
$B=-\left< \bar q q\right>/f_\pi^2 $.
The explicit $\chi$SB, linear in $U$, term gives rise to the pion mass.
Expanding in the pion field, 
\beq
\lag_\pi^{(2)} = \half \pa_\mu \pi^a \pa^\mu \pi^a - \half m_\pi^2  
\pi^2  + O(\pi^4), 
\eeq
one can identify $m_\pi^2 = 2B m_q$, which 
is the celebrated Gell-Mann--Oaks--Rennner relation.
\newline
\indent
Next we include the nucleon, described by an isodoublet spinor
field, $N=(p,n)^T$. Its left- and right-handed components, 
$N_{L,R} = \half (1\mp \ga_5 ) N$, transform under  SU$_L(2$)$\times$SU$_R(2$)
as $N_L\to L N_L$ and $N_R\to R N_R$. In this representation it is not easy
to write down a chirally symmetric Lagrangian. In particular, the nucleon
mass term, $M_N\, \bar N N = M_N (\bar N_L N_R + \bar N_R N_L)$, 
breaks the symmetry, while should preserve it exactly
(because $M_N$ does not vanish in the chiral limit).
The trick is to redefine the nucleon field such that the its left- and 
right-handed components transform in the same way, which is achieved
by redefining $N_L' = u\,N_L$ and $N_R' = u^\dagger\,N_R$, with $u=\sqrt{U}$.
Then, $N_{L,R}' \to K\, N_{L,R}'$, where 
\beq
\eqlab{Kisospin}
K=\sqrt{R} \,u \,\sqrt{L} \,u^\dagger 
= \sqrt{L}\,u^\dagger \sqrt{R} \,u 
\eeq
is an SU$_V$(2) matrix which depends on the pion field.
The latter fact demands a little more work for the derivative
of the nucleon field, which now transforms as (omitting from now
on the prime on the redefined nucleon field): 
$\pa_\mu N \to K\pa_\mu N + (\pa_\mu K) N $.
Obviously for the Lagrangian construction 
it is desirable to have a derivative which transforms
in the same way as the field, i.e., $D_\mu N \to K D_\mu N$.
Thus, one is led to consider
\bea
(\pa_\mu K) K^\dagger &=& \sqrt{L}\, (\pa_\mu u^\dagger)\,u \,\sqrt{L^\dagger}
+ K \,u^\dagger \,(\pa_\mu u) \,K^\dagger \nn\\
 &=& \sqrt{R} \,(\pa_\mu u)\, u^\dagger \,\sqrt{R^\dagger}
+ K \,u \,(\pa_\mu u^\dagger)\, K^\dagger\,. 
\eea
Noting that $(\pa_\mu u) u^\dagger + u (\pa_\mu u^\dagger)=0$, 
it is useful
to introduce the SU(2) vector and axial-vector currents,
\begin{subequations}
\bea
v_\mu & \equiv & \half\, \tau^a v_\mu^a(x) = \frac{1}{2i} \left(u \,
\pa_\mu u^\dagger+u^\dagger \pa_\mu  u \right), \\
a_\mu & \equiv & \half \,\tau^a a^{\,a}_\mu(x) =
  \frac{1}{2i} \left(u^\dagger \,
\pa_\mu  u- u \,\pa_\mu  u^\dagger \right) , 
\eea
\eqlab{currents}
\end{subequations}
and observe that, under SU$_L\times$SU$_R$, they transform as 
\begin{subequations}
\bea
v_\mu &\to& K v_\mu K^\dagger + i (\pa_\mu K) K^\dagger, \\
a_\mu &\to& K a_\mu K^\dagger.
\eea
\end{subequations}
Therefore, the ``chiral covariant derivative'' of the nucleon field
can be defined as follows:
\beq 
D_\mu N = \pa_\mu N  + i v_\mu N \to K\, D_\mu N.
\eeq  
The derivative of the axial-vector field can also be defined
in a covariant fashion:
\beq
{\mathcal D}_\mu a_\nu = \pa_\mu a_\nu + i[v_\mu,\,a_\nu] \to
K \, {\mathcal D}_\mu a_\nu \, K^\dagger.
\eeq
We are now in position to write down a chiral Lagrangian
with nucleon fields. Any hermitian Lagrangian built 
from a combination of the nucleon and axial-vector fields, as well as their covariant derivatives,
will be chirally symmetric.  
The lowest-order such Lagrangian is given by:
\beq
\lag^{(1)}_N = \ol N\,( i \sla{D} -{M}_{N} +  g_A \,  
\slaa\,\ga_5 )\, N\,,
\eqlab{Nlagran}
\eeq
where
$g_A\simeq 1.267$ is the nucleon axial-coupling constant.
\newline
\indent
In practice these Lagrangians need to be expanded in the pion field.
To do that conveniently we write
\beq
u = u_1 + i (\hat \pi/f_\pi)\,u_2,
\eeq 
where $u_{1,2}$ are real functions of $\pi^2\equiv 4\hat \pi^2$:
\bea
u_1 & = & \cos\mbox{$\frac{\sqrt{\pi^2}}{2f_\pi}$} = \sum_{n=0}^\infty
\frac{1}{(2n)!} \left( -\frac{\pi^2}{4 f_\pi^2}\right)^n , \\
u_2 &=& \mbox{$\frac{2f_\pi}{\sqrt{\pi^2}} \, 
\sin \frac{\sqrt{\pi^2}}{2f_\pi}$}
= \sum_{n=0}^\infty 
\frac{1}{(2n+1)!} \left( -\frac{\pi^2}{4 f_\pi^2}\right)^n .
\eea 
In terms of these functions, the vector and axial currents are given by
\bea
v_\mu &=& -i\left[u_1\, \pa_\mu u_1 + (1/f_\pi^2)\,\hat\pi\, u_2\, \pa_\mu(\hat\pi\, u_2)\right] 
=  \frac{1}{4f_\pi^2}\, \tau^a\, \veps^{abc} \,
\pi^b\,(\pa_\mu\pi^c)\, u_2^2\nn\\
&=& \frac{1}{4f_\pi^2}\, \tau^a\, \veps^{abc} \,
\pi^b\,(\pa_\mu\pi^c)\,\left(1-\frac{\pi^2}{3! \,2 f_\pi^2}+\frac{\pi^4}{5!\,3 f_\pi^4}
- \frac{\pi^6}{7!\,4 f_\pi^6}+\ldots\right)
 \,,\\
a_\mu &=& u_1 \,\pa_\mu(\hat\pi\, u_2) - (\pa_\mu u_1)\,\hat\pi\,u_2 
=\frac{1}{2f_\pi}\tau^a\pa_\mu\pi^b\,
\left[\de^{ab} \,u_1 u_2 + (\pi^a \pi^b/\pi^2) \,(1 - u_1 u_2)\,\right] \nn\\
&=& \frac{1}{2f_\pi}\tau^a\pa_\mu\pi^b\,\left[ \de^{ab} - 
(\pi^2\de^{ab}-\pi^a\pi^b)\left(\frac{1}{3!f_\pi^2}-\frac{\pi^2}{5! f_\pi^4}+
\frac{\pi^4}{7! f_\pi^6}+ \ldots \right)
\right]\,. 
\eea
The currents, obviously, are of the first order in the 
derivatives of the pion field. 
\newline
\indent
Finally let us note that 
in the presence of the electromagnetic field ($A_\mu$), the electric 
charge
of the pions is accounted for by making the ``minimal substitution'':
$\pa_\mu \pi^a  \to \pa_\mu \pi^a + e  \,\veps^{ab3}A_\mu\pi^b $,
in the above expressions.  Similarly, the proton charge is included
by the minimal substitution in the chiral derivative as: 
$D_\mu N \to D_\mu N - ie \half (1+\tau^3) A_\mu N$.

\subsection{Inclusion of the spin-3/2 fields}
\label{sec:rs}

The $\De(1232)$ is a spin-3/2 resonance. Therefore its spin content can
conveniently be described in terms of a Rarita-Schwinger (RS) field~\cite{Rarita:1941mf}:
$\psi_\mu^{(\si)}$, where $\mu$ is the vector and $\si$ the spinor
index; the latter index is omitted in the following. 
The free Lagrangian of the massive RS field is given by
\beq
\eqlab{RSLag}
\lag_{\mathrm{RS}} = \ol\psi_\mu \left(i\ga^{\mu\nu\al}\,\pa_\al - 
M\,\ga^{\mu\nu} \right) \psi_\nu \,,
\eeq
where  $M$ is the mass, and 
the totally-antisymmetric products
of $\ga$-matrices are defined as: $\ga^{\mu\nu}=\half[\ga^\mu,\ga^\nu]$,
$\ga^{\mu\nu\al}=\half \{ \ga^{\mu\nu}, \ga^\al\}= 
-i\veps^{\mu\nu\al\be}\ga_\be\ga_5$ (using the convention $\veps_{0123}=+1$). The 
corresponding Euler-Lagrange field equations are:
\begin{subequations}
\bea
\eqlab{fieldeq1}
i\ga^{\mu\nu\al}\pa_\al \psi_\nu -
M  \ga^{\mu\nu} \psi_\nu = & \,0, \,& \\
 \pa_\mu (i\ga^{\mu\nu\al}\pa_\al  -
M  \ga^{\mu\nu}) \psi_\nu  = & \,0\, & = \ga^{\mu\nu} \pa_\mu \psi_\nu ,\\
\ga_\mu (i\ga^{\mu\nu\al}\pa_\al  -
M  \ga^{\mu\nu}) \psi_\nu = & \,0\, & = 
-(2i \ga^{\mu\nu} \pa_\mu + 3 M \ga^\nu )\psi_\nu ,
\eea
\end{subequations}
which can equivalently be written as:
\bea
 (i \slad - M )\,\psi_\mu &=& 0, \nn\\
\pa\cdot \psi &=& 0, \\
\ga\cdot\psi &=& 0. \nn
\eea
Thus, the RS field obeys the Dirac equation, supplemented with the
auxiliary conditions, or 
{\it constraints}.\footnote{A canonical method for
determination of constraints is due to Dirac~\cite{Dir64}. 
See Refs.~\cite{Senjanovic:1977vr,Baaklini:1978qa,Yamada:1985vb,Pas98} for applications of Dirac's
method to the spin-3/2 case.}
The constraints ensure that the number of independent components
of the vector-spinor field is reduced to the physical number of
spin degrees of freedom\footnote{In this case the degrees-of-freedom
counting goes as follows. The vector-spinor has 16 components,
the field equations show that there are 8 conditions on them, and the
Dirac equation. The latter halves the number of independent components, 
and thus in total we have: (16-8)/2=4, the number
equal to the number of different spin polarizations of massive
spin-3/2 particle. } (sDOF). 
\newline
\indent
Note that the constraints are built
into the Lagrangian. This is achieved by making the Lagrangian
to be symmetric under a certain local transformation of the RS field.
To exhibit this local symmetry, observe that
the massless RS Lagrangian is, up to a total derivative, symmetric under
\beq
\eqlab{gsym}
\psi_\mu(x) \to \psi_\mu(x) +\pa_\mu \eps(x),
\eeq 
where $\eps$ is a spinor field. This gauge symmetry is a fermionic
analog of the gauge symmetry of the electromagnetic field, 
and just as in that case, it leads to a reduction
of the number of sDOF to 2,
as is required for a massless field with a spin. The mass term
breaks (partially) this gauge symmetry to raise the number of sDOF to 4, as
is required for a massive field with spin 3/2.
\newline
\indent
Clearly, the coupling of the RS field must be compatible
with the free theory construct, in order to preserve the
physical sDOF content of the theory. However this fact has largely been ignored
in the literature on the field-theoretic description of the
$\De$. There are plenty of examples
of the so-called ``inconsistent'' couplings, i.e., couplings
that violate the free-theory constraints.  Besides involving
the unphysical sDOF, such couplings lead to fascinating pathologies,
such as negative-norm states~\cite{Johnson:1960vt,Hagen:1972ea} 
and superluminal (acausal) modes~\cite{Velo:1969bt,Singh:1973gq}.
\newline
\indent
The quest for ``consistent'' spin-3/2 couplings was
raised from time to time, 
see, e.g.,~\cite{Hagen:1982ez,Haberzettl:1998rw,Deser:2000dz,Pilling:2004cu,Wies:2006rv,Napsuciale:2006wr}.
One of the most viable proposals up to date is the one of gauge-invariant
couplings~\cite{Pas98,Pascalutsa:1999zz}, suggesting that the couplings
invariant under the gauge transformation \eref{gsym} are consistent.
Indeed, in the case of gauge-invariant couplings, only the mass term
breaks the gauge symmetry, hence 
changing the sDOF content, and it is known to do that in a correct
way. 
\newline
\indent
A notable feature of the gauge-invariant couplings is 
that the corresponding vertices satisfy a transversality condition:
\beq
\eqlab{transversality}
p_\mu \,\Ga^\mu (p, \ldots) = 0,
\eeq
where $p$ is the four-momentum and $\mu$ is the vector index of
a spin-3/2 leg. The RS propagator, obtained by inverting the
operator in \Eqref{RSLag}, can be written as the following anticommutator:
\beq
\eqlab{RSprop}
   S_{\mu\nu}(p)  = - \third \left\{(\slap-M)^{-1},\,
      (g_{\mu\nu} - \mbox{$\frac{1}{M^2}$} p_\mu p_\nu 
- \half \gamma_{\mu\nu}) \right\}.
\eeq
It apparently contains a spin-1/2 sector, 
that can be made explicit
by writing the propagator in terms of the covariant spin projection operators:
\beq
\eqlab{RSproj}
   S_{\mu\nu}(p)  = - \frac{1}{\slap-M}
     P_{\mu\nu}^{(3/2)}+\frac{2}{3M^2}(\slap+M)
     P^{(1/2)}_{22,\mu\nu} 
      - \frac{1}{\sqrt{3}M} \left(P^{(1/2)}_{12,\mu\nu}
         - P^{(1/2)}_{21,\mu\nu}\right) \ ,
\eeq
where
\begin{equation}
   P^{(3/2)}_{\mu\nu} = g_{\mu\nu} - \frac{1}{3}\gamma_\mu\gamma_\nu
     - \frac{1}{3p^2} (\slap\gamma_\mu p_\nu
        + p_\mu\gamma_\nu \slap)
\end{equation}
projects onto the pure spin-3/2 states, while
\begin{eqnarray}
\label{eq:shalf}
    {P}^{(1/2)}_{22,\mu\nu} & = &
                 p_\mu p_\nu/p^2        , \nonumber \\
    {P}^{(1/2)}_{12,\mu\nu} & = & p^\varrho p_\nu
                 \ga_{\mu\varrho}/(\sqrt{3}\,p^2) , \\ \nonumber
    {P}^{(1/2)}_{21,\mu\nu} & = & p_\mu p^\varrho
                 \ga_{\varrho\nu}/(\sqrt{3}\,p^2) 
\end{eqnarray}
are projection operators onto the spin 1/2 states.
It is easy to see that, in combination with gauge-invariant
couplings satisfying \Eqref{transversality}, 
the spin-1/2 contributions decouple from observables, e.g.,
\beq
\Ga^\mu(p, \ldots) \, S_{\mu\nu}(p) \, \Ga^\mu(p,\ldots) = 
\Ga^\mu(p, \ldots) \, \frac{1}{M-\slap} \,{P}^{(3/2)}_{\mu\nu}(p) \, \Ga^\mu(p,\ldots) 
\eeq
\newline
\indent
Such a decoupling of the lower-spin contribution is of course a desirable
effect and often had been implemented in the literature ``by hand''.
Namely, one would either drop the spin-1/2 terms in the
non-local decomposition of the propagator, \Eqref{RSprop}, 
see, e.g.~\cite{Wil85,Ade86,Bernard:2003xf}), or
use a local decomposition, {\it e.g.}, 
\beq
   S_{\mu\nu}(p)  = - \frac{1}{\slap-M}\frac{p^2}{M^2}
     P_{\mu\nu}^{(3/2)}+\frac{\slap+M}{3M^2}\left(g_{\mu\nu}-\third
\ga_\mu\ga_\nu\right) + \frac{1}{3M^2}  \left(\ga_\mu p_\nu -\ga_\mu p_\nu\right)\ ,
\eeq
and retain only the $P^{(3/2)}$ term therein~\cite{GrS93,Pascalutsa:2003vz,Bernard:2005fy}.
Both methods share a common problem: the {\it ad hoc} deletion
of momentum-dependent terms may destroy the symmetries, such as chiral
and electromagnetic-gauge invariances. For example, while the 
full RS propagator
is guaranteed to obey a Ward-Takahashi identity with an electromagnetic
coupling obtained by `minimal substitution' into \Eqref{RSLag}, the
truncated propagator does not even have an inverse. 
In other words, since the above-mentioned procedures are not based on a 
Lagrangian it is not clear how to implement the symmetries.
In contrast, the spin-3/2 gauge-invariant couplings
ensure the spin-1/2 decoupling automatically. The only question
is then how to implement the spin-3/2 gauge and other symmetries
in the same Lagrangian. Prior to attempting to answer this,
we would like to make one more remark.  
\newline
\indent
While the spin-3/2 gauge-invariant couplings make sense,
perhaps equally consistent seems the idea of having couplings
which break the gauge symmetry in the same way as the mass term.
However, such couplings should then be proportional to the mass,
in order to provide consistency in the massless limit. And the
couplings proportional to the mass can be rewritten in 
a gauge-invariant way by using the free-field equation, \Eqref{fieldeq1},
or equivalently by a field redefinition. To give an example,
consider a coupling of the RS field to a spinor $\mathit{\Psi}$ and a scalar
$\phi$~:
\beq
\lag_{\mathrm{int}} = g \,M \,\mathit{ \ol\Psi} 
\ga^{\mu\nu} \psi_\mu \, \pa_\nu \phi
+ \mathrm{H.c.},
\eeq
where $g$ is a dimension [mass]$^{-2}$ coupling constant. This
coupling is known to affect the constraints in the same
way as the mass term~\cite{Nath:1971wp,Pas98}. Upon the
field redefinition,
\beq
\psi_\mu \to \psi_\mu + g \,\mathit{\Psi}\, \pa_\mu\phi,
\eeq
we have $\lag_{\mathrm{RS}}+\lag_{\mathrm{int}} \to 
\lag_{\mathrm{RS}}+\lag_{\mathrm{int}}'$, with the new coupling
being 
\beq
\lag_{\mathrm{int}} ' = ig \,\mathit{ \ol\Psi} 
\ga^{\mu\nu\al} (\pa_\al \psi_\mu) \, \pa_\nu \phi
+ \mathrm{H.c.},
\eeq
which evidently is a gauge-invariant coupling.
\newline
\indent
One should emphasize here that the above example is quite exceptional,
because in general field redefinitions lead to the appearance of higher-order
couplings, the so-called ``contact terms''. Their appearance, however,
is not troublesome as long as we deal with effective theories, where
all possible higher-order term are present anyway,
unless forbidden by other symmetries. The latter condition 
brings us again to the question of how the spin-3/2 gauge symmetry
will co-exist with other local or global symmetries of the effective theory, 
and in particular the chiral symmetry. 
\newline
\indent
Certainly it is not easy, in many cases impossible, to 
incorporate several different symmetries in a given interaction. 
For example, in the description of electrically charged
RS field, the only possible way to have both the spin-3/2 and 
the electromagnetic gauge symmetries, in a closed form, is 
to allow for general covariance in a de-Sitter geometry,
resulting in an extended supergravity~\cite{Freedman:1976aw}. 
In many other cases, 
e.g., spin-5/2, even such formidable possibilities are unavailable.
Nevertheless,
within the EFT framework one can envision a following method 
for inclusion of the spin-3/2 (and other higher-spin) gauge symmetries.
\newline
\indent
To start with, one may construct the effective Lagrangian disregarding
the higher-spin gauge symmetry. Then, 
in case of a massive spin-3/2 field, the gauge-invariant
couplings can be obtained by the following substitution~\cite{Pas01}:
\beq
\eqlab{sub1}
\psi_\mu \to \psi_\mu'=(M\ga^{\mu\si})^{-1} \ga^{\si\nu\al} 
i\pa_\al \psi_\nu
\eeq
everywhere in the interaction Lagrangian. This step ensures
the spin-3/2 gauge symmetry while providing an on-shell equivalence
of the new and old couplings. Unfortunately in doing so, 
the other symmetries are likely to be violated
since, while the field transforms
covariantly under other symmetries, its derivative does not.
To restore the other symmetries one needs to replace the derivatives
by covariant derivatives (`minimal substitution'): 
\beq
\pa_\mu \psi_\nu  \to D_\mu \psi_\nu, 
\eeq
hence violating again the spin-3/2 gauge symmetry. 
\newline
\indent
To break out
of this loop, we need to note that generically $D_\mu =
\pa_\mu + \Gamma_\mu$, where the ``connection'' $\Gamma_\mu$ is 
 at least one order higher (in the EFT expansion) than $\pa_\mu$.    
Then, if the first substitution restores the spin-3/2 symmetry
to, say, order $n$, the other symmetries will be violated only
at order $n+1$ or higher. So to order $n$ all the symmetries
are satisfied. When going to the next order, we would need to make
the minimal substitution to restore the other symmetries to that order
{\it and} the first substitution to restore the spin-3/2 symmetry.
One can continue this procedure to establish all the symmetries
to any given order in the effective expansion. 
\newline
\indent
We emphasize that substitution \eref{sub1} is certainly 
not unique, other free equations of the RS field can be used, e.g.,
\bea
\psi_\mu &=& (i/M)\,\ga^\nu\, (\pa_\mu \psi_\nu - \pa_\nu \psi_\mu) , \\
\psi^\mu &=& (1/M)\, \veps^{\mu\nu\al\be} \ga_\be \ga_5 \,\pa_\al \psi_\nu,
\eea
all leading to gauge-invariant couplings, equivalent
at the order where the field equations are used. 
Further on we will apply this method
to write down chiral Lagrangians involving the $\De$-isobar field.

\subsection{Chiral Lagrangians with $\De$'s}
\label{sec:sec4ss3}   

The spin-3/2 isospin-3/2 $\De$(1232) can be represented 
by a vector-spinor {\it isoquartet} field,
$\vDe_\mu = (\vDe_\mu^{++},\vDe_\mu^{+}, \vDe_\mu^{0}, \vDe_\mu^{-})^T $,
where the four components correspond to the charge states
of the $\De$-isobar.
Inclusion of an isoquartet into the
chiral Lagrangian can be done by generalizing slightly
the formalism of Subsect.~4.1 for the isodoublet field to the
case of isospin 3/2. It basically amounts to replacing the 
SU(2) generators in the fundamental representation $(\tau^a/2)$ 
by the generators in the isospin-3/2 representation. 
The latter have the following form:  
\begin{subequations}
\bea
\Tau^1 & = & \frac{2}{3} \left(\begin{array}{cccc}
0 & \sqrt{3}/2  & 0 & 0 \\
\sqrt{3}/2 & 0 & 1 & 0 \\
0 & 1 & 0 & \sqrt{3}/2\\
0 & 0 & \sqrt{3}/2 & 0 
\emat, \\
\Tau^2 & = & \frac{2i}{3} \left(\begin{array}{cccc}
0 & -\sqrt{3}/2  & 0 & 0 \\
\sqrt{3}/2 & 0 &-1 & 0 \\
0 & 1 & 0 & -\sqrt{3}/2\\
0 & 0 & \sqrt{3}/2 & 0 
\emat,\\
\Tau^3 & = & \mbox{diag} (1, \third, -\third, -1),
\eea
\end{subequations}
and satisfy $\Tau^a \Tau^a = 5/3$. 
\newline
\indent
The chiral transformation of the isoquartet field is then given as
\beq
\vDe_\mu \to K_4\, \vDe_\mu\,
\eeq  
where $K_4 $ is a four-dimensional SU$_V$(2) matrix. Similarly,
we introduce the isospin-3/2 vector and axial-vector currents:
\begin{subequations}
\bea
v_\mu^{(3/2)} & \equiv & \Tau^a v_\mu^a(x) = \Tau^a 
\,\mbox{Tr}(\tau^a v_\mu), \\
a_\mu^{(3/2)} & \equiv & \Tau^a a^{\,a}_\mu(x) = 
\Tau^a \,\mbox{Tr}(\tau^a a_\mu), 
\eea  
\end{subequations}
and the chiral covariant derivative of the $\De$-field:
\beq 
D_\mu \vDe_\nu = (\pa_\mu + i \,v_\mu^{(3/2)} )\, \vDe_\nu .
\eeq
\newline
\indent
Before writing down the effective 
Lagrangians using these ingredients,
let us remark on the other frequently used representation.
The  so-called {\it isospurion} representation is
based on an isovector-isodoublet field~\cite{JeM91a,TaE96,FeM01}:
$\vDe_\mu^{a}  =  T^a \vDe_\mu$, with $T$ the isospin-1/2-to-3/2 
transition matrices defined as
\begin{subequations}
\bea
T^1 & = & \frac{1}{\sqrt{6}} \left(\begin{array}{cccc}
-\sqrt{3}   & 0 & 1 & 0 \\
0 & -1 & 0 & \sqrt{3}  
\emat, \\
T^2 & = & \frac{-i}{\sqrt{6}} \left(\begin{array}{cccc}
\sqrt{3}  & 0 & 1 & 0 \\
 0 & 1 & 0 & \sqrt{3}
\emat,\\
T^3 & = & \sqrt{\frac{2}{3} } \left(\begin{array}{cccc}
 0 & 1 & 0 & 0\\
 0 & 0 & 1 & 0  
\emat,
\eea
\end{subequations}
and satisfying $T^a T^{b\dagger} = \de^{ab} -\third \tau^a \tau^b$,
$T^{a\dagger} T^a = {1}_4$, $\tau^a T^a=0$.
Under a chiral rotation the isospurion transforms as 
\beq
\vDe^a_\mu \to K^{ab} \,K \,\vDe^b_\mu,    
\eeq
where $K$, an SU$_V$(2) matrix given by \Eqref{Kisospin},
acts on the doublet components, while
$K^{ab} = \half \mbox{Tr}(\tau^a K\tau^b K^\dagger)$
transforms the isovector components. Note that the latter
object has the following important properties:
$K^{ab}\tau^b = K\tau^a K^\dagger$, $K^{ab}K^{bc} = \de^{ac}$,
where $\de$ is the Kronecker symbol.
The transformation properties of the isospin-3/2 field in the
two different representations are related via:
\beq
K_4 = T^{a\dagger} \, K\, T^b \, K^{ab}. 
\eeq
Knowing these relations, one can easily go from
one representation to another. They both, of course, 
are equivalent at the level of observables. 
\newline
\indent
In writing down the chiral 
Lagrangian we adopt the isoquartet representation.
The first-order Lagrangian of the $\De$ is given by:
\beq
\eqlab{deltaLag}
\lag^{(1)}_\De = \ol\vDe_\mu \left(i\ga^{\mu\nu\rho}\,D_\rho - 
M_\De\,\ga^{\mu\nu} \right) \vDe_\nu - 
 \half H_A  \,\ol\vDe_\mu 
\, \slaa^{(3/2)}\, \ga_5\, \vDe^\mu,
\eeq
where  $M_\De$ is the mass of the $\De$-isobar,
$H_A$ is the axial coupling constant of the $\De$ 
given, in the large-$N_c$ limit,
by $H_A=(9/5) g_A$.  Note that the electric
charge of the $\De$ can be accounted for by the following
minimal substitution: $D_\mu \to D_\mu - ie \half(1+3\Tau^3) A_\mu$.
\newline
\indent
The chiral interactions
in \Eqref{deltaLag} obviously do not have the spin-3/2
gauge symmetry, see \Eqref{gsym}. 
We follow the program outlined
in the previous section to incorporate this constraint.
First, we write out the spin-3/2 gauge-invariant 
couplings which are on-shell equivalent to the 
ones in \Eqref{deltaLag}, e.g.,
\bea
\eqlab{delta1Lag}
\lag^{(1)}_\De & = &  \ol\vDe_\mu \left(i\ga^{\mu\nu\rho}\,\pa_\rho - 
M_\De\,\ga^{\mu\nu} \right) \vDe_\nu \nn\\
&-&
\, \frac{1}{M_\De^2}\veps^{\mu\nu\al\be}\,(\pa_\al \ol\vDe_\mu)\,[\,\slash \!\!\! v^{(3/2)} +
 \half H_A \, \slaa^{(3/2)}]\, \ga_5\, (\pa_\be\vDe_\nu).
\eea
The `minimal substitution' into the second term, 
which is required to restore the chiral symmetry, generates
higher-order contributions, such as:
\beq
\eqlab{deltaprime1Lag}
\lag^{(2)}_\De = -\frac{1}{M_\De^2}\veps^{\mu\nu\al\be}\,
(\pa_\al \ol\vDe_\mu)\,[\,\slash \!\!\! v^{(3/2)} \, \ga_5+
 \half H_A \, \slaa^{(3/2)}]\, v^{(3/2)}_\be\,\vDe_\nu+ \mbox{H.c.}
\eeq
The spin-3/2 gauge-symmetry is manifest in the following 
on-shell equivalent Lagrangian:
\beq
\eqlab{deltaprime2Lag}
\lag^{(2)}_\De = -\frac{i}{M_\De^3}\veps^{\mu\nu\al\be}\,
(\pa_\al \ol\vDe_\mu)\,[\,\slash \!\!\! v^{(3/2)} \, \ga_5+
 \half H_A \, \slaa^{(3/2)}]\, v^{(3/2)}_\be \,\ga^\vrho 
(\pa_\vrho \vDe_\nu-\pa_\nu \vDe_\vrho)+ \mbox{H.c.}
\eeq
 This of course is not the complete second-order Lagrangian, 
only the term required by chiral symmetry when the first-order
Lagrangian \eref{delta1Lag} is used.   
\newline
\indent
The terms required by chiral symmetry are not only of higher
order in the derivatives of the pion field, they are are 
of higher order in pion field itself, and hence in many cases
may appear only in multi-loop corrections. Here
we shall focus on a single-pion production with no more than
one-loop contributions, thus it suffices to consider
the couplings with no more than three pion fields.
The relevant terms of the chirally and gauge symmetric
Lagrangian of the $\De$, expanded to third order in the pion field, read
\beq
\lag^{(1)}_{\De\De\pi} = \frac{H_A}{2M_\De f_\pi}  \,
\veps^{\mu\nu\rho\si} \,\ol\De_\mu 
\, \Tau^a \,(\pa_\rho \De_\nu )\,\pa_\si \pi^{a} + O(\pi^3).
\eeq
\newline
\indent
Similarly, we write a few relevant couplings of 
the $N\De$-transition Lagrangian 
in the form which manifests the spin-3/2 gauge symmetry and is 
expanded to leading order in the pion field:
\begin{subequations}
\eqlab{lagran}
\bea
\lag^{(1)}_{N\De} &=&  \frac{i h_A}{2 f_\pi M_\De}
\ol N\, T^a \,\ga^{\mu\nu\la}\, (\pa_\mu \De_\nu)\, \pa_\la \pi^a 
+ \mbox{H.c.}, \\
\lag^{(2)}_{N\De} &=&  \frac{h_1}{2 f_\pi M_\De^2}
\ol N\, T^a \,\ga^{\mu\nu\la}\, (\pa_\la \slad \pi^a) \, (\pa_\mu \De_\nu)
+ \mbox{H.c.}, \\
\lag^{(2)}_{N\De} &=&   \frac{3 i e g_M}{2M_N (M_N + M_\Delta)}\,\ol N\, T^3
\,\pa_{\mu}\De_\nu \, \tilde F^{\mu\nu}  + \mbox{H.c.},\\
\lag^{(3)}_{N\De} &=&  \frac{-3 e}{2M_N (M_N + M_\Delta)} \ol N \, T^3
\ga_5 \left[ g_E (\pa_{\mu}\De_\nu) 
+  \frac{i g_C}{M_\De} \ga^\al  
(\pa_{\al}\De_\nu-\pa_\nu\De_\al) \,\pa_\mu\right] F^{\mu\nu}+ \mbox{H.c.},
\;\;\;\;\; 
\eea
\end{subequations}
where $F^{\mu\nu}$ and $\tilde F^{\mu\nu}$
are the electromagnetic field strength and its dual.
Note that the electric and the Coulomb $\ga N\De$ couplings 
are of one order higher than
the magnetic one, because of the $\ga_5$ which involves 
the ``small components'' of the fermion fields and thus 
introduces an extra power of the 3-momentum.
\newline
\indent
Finally, for the applications below it will also be useful  to
 write out the magnetic moment coupling for the nucleon and
$\De$ fields:
\begin{subequations}
\eqlab{lagran2}
\bea
\lag^{(2)}_N &=&  \frac{i e}{4M_N}\, \ol N \, \half (\kappa^{(S)}_N 
+ \kappa^{(V)}_N\,\tau_3)\,
\si_{\mu\nu}\, N \,F^{\mu\nu}\,,\\
\lag^{(2)}_\De &=&  
\frac{i e}{2M_\De}\, \ol \De_\mu 
\,\half [ 1+\kappa^{(S)}_\De  
+ 3(1+\kappa^{(V)}_\De)\,\Tau_3] \De_\nu\, F^{\mu\nu}\,, 
\eea
\end{subequations}
where $\kappa^{(S)}$, $\kappa^{(V)}$ correspond in both the $N$ and $\De$ case
with the isoscalar and isovector anomalous magnetic moments. Also in both
cases the anomalous magnetic moments are defined as the deviation from the gyromagnetic ratio ($g=\mu/s$) from 2, the natural value for 
an elementary particle of any spin $s$~\cite{Weinberg:1970bu,Ferrara:1992yc,Holstein:2006wi}. 
In this notation the magnetic moment of the $\De$ corresponds with
\bea
\mu_\De = \frac{e}{2M_\De} \left( 3 e_\De + \half \kappa^{(S)}_\De  
+ \thalf \kappa^{(V)}_\De \, \Tau_3 \,\right)\, 
\eea
where $e_\De = (1+3\Tau_3)/2$ is the charge of the $\De$ in units of $e$.
\newline
\indent
We should piont out that such a choice of the anomalous magnetic moment of
the $\De$ is not (yet) widely used. 
One usually defines it as a deviation of the magnetic
moment from the magneton value: $e\,e_\De/(2 M_\De)$. Namely, in
\Eqref{EMmoments}, $F_2^\ast(0)$ corresponds precisely 
with the conventional definition of the anomalous magnetic moment.
The relation between the two conventions is obvious:
$\kappa_\De^{(S,V)} = F_2^{\ast (S,V)}(0) -2\,$.

\subsection{Power counting, renormalization and naturalness}
\label{sec:pc}

The fact that the strength of the chiral interactions 
goes with derivatives of pion fields allows one to organize
a perturbative expansion in powers of pion momentum and mass  --- 
the chiral perturbation theory~\cite{Weinberg:1978kz,Gasser:1983yg}.
The small expansion parameter is $p/\La_{\chi SB}$, where $p$
is the momentum and $\La_{\chi SB}\sim 4\pi f_\pi \approx 1$ GeV
stands for the scale of spontaneous chiral symmetry breaking. 
Based on this expansion,
one should be able to systematically compute
the pion-mass dependence of static quantities, such as nucleon mass,
magnetic moments, as well as the momentum dependence of scattering
processes, such as pion-pion and pion-nucleon scattering. 
It generically is an effective-field theory (EFT) expansion, in this
case a low-energy expansion of QCD. One expects to obtain exactly 
the same answers as from QCD directly, provided the low-energy
constants (LECs) --- the parameters of the effective Lagrangian --- 
are known, either from experiment or from QCD itself. 
\newline
\indent
One of the principal ingredients of an EFT expansion is 
{\it power counting}. The power counting scheme assigns an order
to Feynman graphs arising in loopwise expansion of the amplitudes,
and thus defines which graphs need to be computed to a given order in the
expansion. In a way, it simply is a tool to estimate the size of
different contributions without doing explicit calculations.
Of course, the main requirement on a power-counting scheme
is that it should estimate the relative size of various contributions
correctly. So, if a given graph is power counted to be of subleading
order, but in explicit calculations gives a dominant effect, the
power-counting scheme fails.
\newline
\indent
In $\chi PT$ with pions and nucleons alone, the power counting
for a graph with $L$ loops, $N_\pi$ ($N_N$) internal pion (nucleon)
lines, and $V_k$ vertices from $k$th-order Lagrangian, estimates
its contribution to go as $p^{n_{\chi \mathrm{PT}}}$, with the power
given by~\cite{GSS89}:
\beq
\eqlab{chptindex}
n_{\chi \mathrm{PT}} =  4 L  - 2 N_\pi - N_N + \sum_k k V_k\,.
\eeq
For example, the one-loop correction to the nucleon mass given in
\Figref{nucself}(a) is characterized by 
$L=1$, $N_\pi =1$, $N_N=1$, $V_1=2$, while
the correction to the nucleon electromagnetic vertex
\Figref{nucself}(b)
has $L=1$, $N_\pi =2$, $N_N=1$, $V_1=2$, and $V_2=1$ (
two $\pi NN$ vertices from $\lag^{(1)}$
and the $\ga\pi\pi$ vertex from $\lag^{(2)}$). Therefore,
both of these graphs count as order $p^3$.   
\begin{figure}
\centerline{  \epsfxsize=6cm
  \epsffile{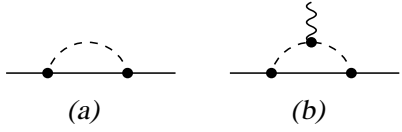} 
}
\caption{Nucleon self-energy and electromagnetic-moment 
corrections to one loop.}
\figlab{nucself}
\end{figure}
\newline
\indent
Another important ingredient is the {\it renormalization} program.
First of all, although the interactions
are non-renormalizeable in the usual sense, one always should be able to 
absorb the infinities into the available LECs. This is ensured
simply by the fact that the effective Lagrangian contains
all the possible terms allowed by symmetries and entering with
arbitrary coefficients.  
Second, the renormalization program should be compatible with 
power counting. This issue is related to the arbitrariness
of renormalization, namely the arbitrariness of the finite part
that remains after the renormalization. In the EFT framework, it is
clear that only the parts which satisfy power counting should remain.
That is, if power counting estimates the graph to be, say, of order
$p^3$, but its unrenormalized contribution contains a piece which goes
as $p^2$, that piece must be completely absorbed into the $p^2$
counter term. Again, symmetries and the generality of the effective
Lagrangian ensure that this can always be done \cite{Gegelia:1999gf}. 
\newline
\indent
In the above example of the one-loop correction to the nucleon mass,
an explicit calculation 
shows~\cite{GSS89} that the loop produces
$O(m_\pi^0)$ and $O(m_\pi^2)$ terms, both of which are large. In fact,
they are ultravioletly divergent, hence are  infinitely large.
However, the appears of such terms is not in
 violation of the power counting, because there are two
LECs: the nucleon mass in the chiral limit, $M^{(0)}$,
and $c_{1N}$, which enter at $O(m_\pi^0)$ and $O(m_\pi^2)$, 
respectively, and renormalize away 
the large contributions coming from the loop.    
The renormalized result (up to $p^4$ terms) 
is given in \Eqref{nucmass},
such that the loop contribution begins
at $O(m_\pi^3)$ in agreement with power counting. 
\newline
\indent
Similarly, in the case of 
the chiral correction to the electromagnetic interaction, 
an explicit calculation will reveal
(see, e.g., \cite{Kubis:2000zd,Holstein:2005db}) 
that the magnetic contribution goes as
\beq
\mbox{\Figref{nucself}(b)}  = 
\frac{e}{2M_N} \veps_\mu \si^{\mu\nu} q_\nu
\, \left(\frac{g_A M_N}{4\pi f_\pi}\right)^2 
\left[ c_0 + c_1 \frac{m_\pi}{M_N} + O(m_\pi^2)\right],
\eeq
where $\veps$ and $q$ are the four-vectors 
of photon polarization and momentum, respectively, 
and where $c_{0,1}$ are 
real coefficients which depend only on isospin. 
Recalling that both the charge $e$ and
the photon momentum count as one power of $p$, we find
that the $c_0$ term is of order $p^2$, hence superficially
the power counting is violated.
However, there is
a LEC coming at this order from $\lag^{(2)}_N$ in \Eqref{lagran2}.
The $c_0$ term, would it be finite or infinite,
can be absorbed by this LEC, thus providing
correct power-counting for the renormalized loop graph.
\newline
\indent
A cornerstone principle of effective field theories in general is
{\it naturalness}, meaning that the (dimensionless) 
LECs must be of ``natural size'', {\it i.e.}, of order of unity.
Any significant fine-tuning of even a single LEC leads, obviously,
to a break-down of the EFT expansion. Therefore, if an EFT
describes well the experimental data, but at the expense of fine-tuned
LECs, the result is negative: 
EFT fails in the description
of those data. Such an EFT can still be useful for getting insights
into the physics beyond the EFT itself. Namely, by looking
at the form of the fine-tuned operators, one might be able to 
deduce which contributions are missing.  
\newline
\indent
For instance, it is well known that the NLO 
$\chi$PT description of the pion-nucleon elastic scattering,
near threshold, requires relatively large values for some of the LECs, 
see {\it e.g.,} Ref.~\cite{BKM},
It is not difficult to see that the operators corresponding 
with those unnatural LECs can be matched to the ``integrated out''
$\De$-resonance contributions. 
The problem is  that the $\De$ is relatively light, its excitation
energy, $\vDe\equiv M_\De-M_N\sim 0.3$ GeV, is still quite
small compared to $\La_{\chi SB} \sim 1$ GeV.
Integrating out the $\De$-isobar degrees of freedom
corresponds to an expansion in powers of $p/\vDe$, with $p\sim m_\pi$,
which certainly is not as good of an expansion as the one 
in the meson sector, in powers of $p/\La_{\chi SB}$.
\newline
\indent
The fine-tuning of the ``Deltaless'' $\chi$PT seems to be 
lifted by the inclusion of an explicit $\De$-isobar.
Also, the limit of applicability of the EFT expansion
is then extended to momenta of order of the resonance excitation
energy, $p\sim \vDe$. Such momenta can still be considered as soft,
as long as $\vDe /\La_{\chi SB}$ can be treated as small.  
The resulting $\chi$PT with pion, nucleon, and $\De$-isobar degrees
of freedom has two distinct light scales: $m_\pi$ and $\De$.
Perhaps the most straightforward way to proceed is to organize a
simultaneous expansion in two different small parameters:
$\eps = m_\pi /\La_{\chi SB}$ and $\de = \vDe /\La_{\chi SB}$. 
However, for power counting purposes, it is certainly more convenient
to have a single small parameter, and thus a relation between
$\eps$ and $\de$ is usually imposed. We emphasize that the relation
is established only at the level of power counting and not in the 
actual calculations of graphs. 
In the literature up to date two such relations between $\eps$ and $\de$
 are used: (i) $\eps \sim \de$, see~\cite{JeM91a,HHK97,TaE96,FeM01,HWGS05}, 
which we will commonly refer to as the
 ``$\eps$-expansion'',  
(ii)  $\eps \sim \de^2$, the 
``$\de$-expansion''\footnote{The same counting
was independently developed by Hanhart 
and Kaiser~\cite{Hanhart:2002bu} in application to the pion production
in nucleon-nucleon collisions.}
of Ref.~\cite{Pascalutsa:2002pi}.
The table below (\Tabref{compare}) summarizes the counting of momenta in the
three expansions: Deltaless  ($\sla{\De}$-$\chi$PT), 
$\eps$-expansion, and $\de$-expansion.
\begin{table}[h]
{\centering
\begin{tabular}{|c||c|c|}
\hline
EFT &  $\quad p\sim m_\pi \quad$ & $\quad p\sim \Delta \quad$\\
\hline
$\sla{\De}$-$\chi$PT & ${\mathcal O}(p)$        & ${\mathcal O}(1)$\\
 $\eps$-expansion  & ${\mathcal O}(\epsilon) $ & ${\mathcal O}(\epsilon) $\\
$\delta$-expansion & ${\mathcal O}(\delta^2) $ & ${\mathcal O}(\delta) $\\
\hline
\end{tabular}  \par }
\caption{The counting of momenta in 
the three different $\chi$EFT expansions discussed in the text.}
\tablab{compare}
\end{table}
\newline
\indent
An unsatisfactory feature of the $\eps$-expansion is 
that the $\De$-resonance 
contributions are always estimated to be of the
same size as the nucleon contributions. 
In reality (revealed by actually computing these contributions),
they are {\it suppressed} at low energies and {\it dominate} 
in the the $\De$-resonance region. 
Thus, apparently the power-counting in the $\eps$-expansion
{\it overestimates} the $\De$-contributions
at lower energies and {\it underestimates} them
at the resonance energies. 
The $\de$-expansion improves on this aspect, as is briefly described
in what follows.
\newline
\indent
In the $\de$-expansion, the power counting depends
on the energy domain, 
since  in the {\it low-energy region} ($p\sim m_\pi$)
and the {\it resonance region} ($p\sim \vDe$), the momentum
counts differently, see \Tabref{compare}. This dependence 
most significantly affects the power counting of the direct resonance
exchanges, the so-called 
one-Delta-reducible (ODR) graphs. Figure~\ref{fig:ODR}
illustrates examples of the ODR graphs for the case of Compton scattering
on the nucleon. These graphs are all characterized by having
a number of ODR propagators, each going as
\beq
S_{ODR}\sim \frac{1}{s-M_\De^2} \sim \frac{1}{2M_\De}\frac{1}{p-\vDe}\, ,
\eeq
where $p$ is the soft momentum, in this case given by the photon energy.
In contrast the nucleon propagator in analogous graphs would go
simply as $S_N\sim 1/p$.
\begin{figure}[t,b,h]
\centerline{  \epsfxsize=11cm
  \epsffile{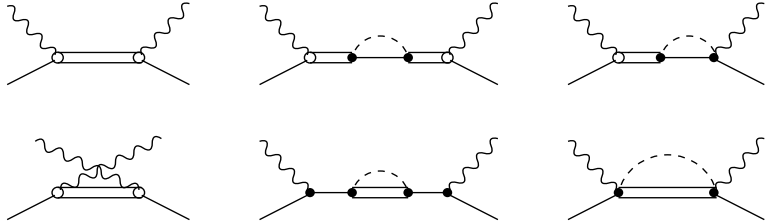} 
}
\caption{Examples of the one-Delta-reducible (1st row) and
 the one-Delta-irreducible (2nd row) graphs in Compton scattering
on the nucleon. }
\figlab{ODR}
\end{figure}
\newline
\indent
Therefore, in the low-energy region, the $\De$ and nucleon 
propagators would
count as ${\mathcal O}(1/\de)$ and ${\mathcal O}(1/\de)$, 
the $\De$ being suppressed by one power of the small parameter
as compared to the nucleon. 
In the resonance region, the ODR graphs obviously 
all become large. Fortunately they all can be subsumed, leading
to ``dressed'' ODR graphs with a definite power-counting index.
Namely, it is not difficult to see that the resummation of
the classes of ODR graphs results 
in ODR graphs with only a single ODR propagator of
the form
\beq
S_{ODR}^\ast = \frac{1}{S_{ODR}^{-1} - \Si }
\sim \frac{1}{p-\vDe-\Si}\,,
\eeq
where $\Si$ is the $\De$ self-energy.
The expansion of the self-energy begins with $p^3$, and hence
in the low-energy region 
does not affect the counting of the $\De$ contributions. However,
in the resonance region the self-energy not only ameliorates
the divergence of the ODR propagator at $s=M_\De^2$ but also
determines power-counting index of the propagator.
Defining the $\De$-resonance region formally as the region of $p$
where
\beq
|p-\vDe | \leq \de^3 \La_{\chi SB}\,,
\eeq
we deduce that an ODR propagator, in this region, counts
as ${\mathcal O}(1/\de^3)$. Note that the nucleon propagator in
this region counts as ${\mathcal O}(1/\de)$, hence is
suppressed by two powers as compared to ODR propagators.
Thus, within the power-counting scheme we have the mechanism for
estimating correctly  the relative size
of the nucleon and $\De$ contributions in the two energy domains.
In \Tabref{counting} we summarize the counting of the nucleon,
ODR, and one-Delta-irreducible (ODI) propagators in 
both the $\eps$- and $\de$-expansion.
\begin{table}[htb]
{\centering \begin{tabular}{||c|c||c|c||}
\hline
&  $\eps$-expansion & 
\multicolumn{2}{|c||}{$\de$-expansion} \\
\cline{2-4} 
  & $p/\La_{\chi SB}\sim \eps$  &  $p\sim m_\pi$ & $p\sim \vDe$ \\
\hline
$\,S_N\,$ &  $1/\eps$ & $1/\de^2$ & $1/\de$\\
$\,S_{ODR}\,$ & $1/\eps$ & $1/\de$ & $1/\de^3$\\
$\,S_{ODI}\,$ & $1/\eps$ & $1/\de$ & $1/\de $ \\
\hline
\end{tabular} \par }
\caption{The counting for the nucleon, one-Delta-reducible (ODR), and
one-Delta-irreducible (ODI) propagators in the two different expansion
schemes. The counting in the $\de$-expansion depends on the energy domain.}
\tablab{counting}
\end{table}
\newline
\indent
We conclude this discussion by giving the general
formula for the power-counting index in the $\de$-expansion.
The power-counting index, $n$, of a given graph
simply tells us that the graph is of the size of ${\mathcal O}(\de^n)$.
For a graph with $L$ loops, $V_k$ vertices of
dimension $k$, $N_\pi$ pion propagators, $N_N$
nucleon propagators, $N_\De$ Delta propagators, $N_{ODR}$
ODR propagators and  $N_{ODI}$ ODI propagators 
(such that $N_\De=N_{ODR}+N_{ODI}$) the index is
\beq
\eqlab{PCindex}
n = \left\{ \begin{array}{cc} 2 n_{\chi \mathrm{PT}} - N_\De\,, 
& p\sim m_\pi ; \nn\\
n_{\chi \mathrm{PT}} - 3N_{ODR} - N_{ODI}\,, & p\sim \De, \end{array}\right.
\eeq
where $ n_{\chi \mathrm{PT}}$, given by \Eqref{chptindex}, 
is the index of the graph in $\chi$PT with no $\De$'s. For further details
on the $\de$ counting we refer to Ref.~\cite{Pascalutsa:2002pi}. 
The rest of this section is devoted to applications of $\chi$EFT to
several processes that are relevant for the $N\to\De$ transition.

\subsection{In practice: the next-to-leading order calculations}
\label{sec:ip}

\subsubsection{Pion-nucleon scattering}

The pion-nucleon ($\pi N$) scattering amplitude at leading order
in the $\de$-expansion in the resonance region, is given 
by the graph $(LO)$ in \Figref{pin2NLO}. This is an example
of an ODR graph and thus the $\De$-propagator counts
as $\de^{-3}$. The leading-order vertices are from $\lag^{(1)}$
and since $p\sim\de$, the whole graph is ${\mathcal O}(\de^{-1})$. 
\begin{figure}
\centerline{  \epsfxsize=9cm
  \epsffile{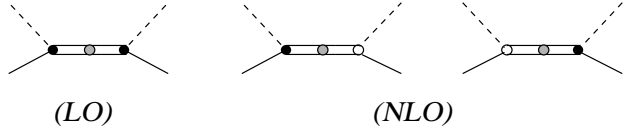} 
}
\caption{The leading and next-to-leading order graphs of the 
$\pi N$-scattering amplitude.}
\figlab{pin2NLO}
\end{figure}
\newline
\indent
At the NLO, the graphs labeled $(NLO)$ \Figref{pin2NLO} begin 
to contribute. The $\pi N\De$ vertices denoted by dots 
stand for the $h_A$ coupling from $\lag^{(2)}_{N\De}$ and the circles
for the $h_1$ coupling from $\lag^{(2)}_{N\De}$, see \Eqref{lagran}.
The NLO graphs are thus ${\mathcal O}(\de^0)$. 
It is not difficult to see that graphs containing the loop correction
to the vertex, as well as the nucleon-exchange graphs, 
begin to contribute at N$^2$LO $[{\mathcal O}(\de)]$.
\newline
\indent
The ODR graphs contribute only to the $P_{33}$ and $D_{33}$
partial waves. The $D_{33}$ contribution is due to the 
``negative-energy states'' contribution and is suppressed by 
$\de^3$, as compared to the ``positive-energy states''
contributions to $P_{33}$. The $D_{33}$ will therefore be omitted
from our considerations.   
\newline
\indent
The $P_{33}$ contribution can
conveniently be  written in terms of the following partial-wave
`$K$-matrix':
\beq
\eqlab{kmat33}
K_{P33}=-\frac{1}{2}\frac{\Gamma(W)}{W-M_\De} \,,
\eeq 
where $W=\sqrt{s}$ is the total energy and $\Gamma$ is an
energy-dependent width, which arises from the $\De$ self-energy.
At this stage it is already taken into account that the
real part of the self-energy will lead to the mass and field renormalization
and otherwise are of N$^2$LO. Thus, only the imaginary part of the
self-energy affects the NLO calculation.
\begin{figure}
\centerline{  \epsfxsize=11cm
  \epsffile{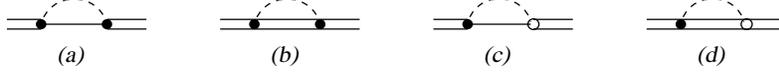} 
}
\caption{The leading and next-to-leading order graphs of the 
$\De$ self-energy.}
\figlab{selfen2NLO}
\end{figure}
\newline
\indent
In the ODR graphs of \Figref{pin2NLO}, 
the $\De$-propagator is dressed by the self-energy
given to NLO by the graphs in \Figref{selfen2NLO}.
The Lorentz-covariant self-energy of the spin-3/2 $\De$ is a 
rank-2 tensor, has the $\ga$-matrix structure and is dependent on
one four-momentum. As such, this object can in general be decomposed onto
a basis of 10 independent scalar functions. The procedure
for computing these functions is cumbersome~\cite{Korpa:2004sh,Kaloshin:2004jh}. 
A consistent procedure
of ``dressing'' the Rarita-Schwinger propagator with such a self-energy
does not exist, because the propagator is derived from the Lagrangian
with only two independent structures: the kinetic and the mass term.
As discussed above, the form of the Rarita-Schwinger Lagrangian
is constrained by the requirement of physical spin degrees-of-freedom 
count, and hence the expression for the self-energy must be constrained
as well. The spin-3/2 gauge-invariant couplings, introduced
in the previous two subsections, automatically ensure 
these constraints.
In that case, 
the most general form of the self-energy can be written as
\beq
\Si_{\al\be} (p) = \Si (\slap) \,{P}^{(3/2)}_{\al\be}(p),
\eeq
where ${P}^{(3/2)}$ is the covariant spin-3/2 projection
operator, and $\Si (\slap)$ has the spin-1/2 Lorentz form, thus
 2 independent Lorentz structures as needed. 
\newline
\indent
The ``dressed'' propagator of the $\De$ to NLO is then given as follows:
\beq
S_{\al\be}(p) = \frac{- Z_2 }{(\slap - M_\De)[1-Z_2\, \Si'(M_\De)] -
(M_\De^{(B)}-M_\De + Z_2\, \Si(M_\De)\,)} 
{P}^{(3/2)}_{\al\be}(p)\,,
\eeq
where $M_\De^{(B)}$ is the ``bare'' mass, $Z_2$ is the field 
renormalization constant, and 
\beq
\Si'(M_\De) = \left.
\frac{\pa}{\pa\slap} \,\Si(\slap)\,\right|_{p\!\!\!/ = M_\De}\,.
\eeq
Our `on-shell renormalization' conditions read:
\bea
Z_2 &=& 1-Z_2 \,\mbox{Re}\Si'(M_\De), \\
0 & = & M_\De^{(B)}-M_\De + Z_2\,\mbox{Re} \Si(M_\De),
\eea
and thus the renormalized NLO propagator is given by 
\beq
S_{\al\be}(p) = \frac{- 1}{(\slap - M_\De)[1- i\,\mbox{Im} \Si'(M_\De)] -
 i\,\mbox{Im}\Si(M_\De)} 
{P}^{(3/2)}_{\al\be}(p)\,.
\eeq
The energy-dependent width in \Eqref{kmat33} is then given by
\beq
\Gamma(W) = -2\, \mbox{Im} \left[\Si(M_\De) + (W-M_\De)\, \Si'(M_\De)\right]\,.
\eeq
and therefore the expression for the K-matrix becomes
\beq
\eqlab{newkmat}
K_{P33} = \frac{\mbox{Im}\Si(M_\De)}{W-M_\De}
+ \mbox{Im}\Si'(M_\De)\ .
\eeq
\newline
\indent
An elementary calculation (see, {\it e.g.}, \cite{PV05}) of the 
graphs in \Figref{selfen2NLO}
yields,  in the region $W \in [ M_N + m_\pi, \, M_\De + m_\pi]$, 
the following result:
\begin{subequations}
\eqlab{NLOwidth}
\bea
{\rm Im}\, \Si(M_\De) &=& - \pi \,\frac{h_A^2 
+ 2 h_A h_1 \frac{\vDe}{M_\De}}{ 24 M_\De^5 \,(8 \pi f_\pi)^2 } \,
[(M_N+M_\De)^2-m_\pi^2]^{5/2} (\vDe^2-m_\pi^2)^{3/2} \,,\\
{\rm Im}\,\Si'(M_\De) &=& - \frac{\pi h_A^2}{ 8 M_\De^6 (8 \pi f_\pi)^2 }
[(M_N+M_\De)^2-m_\pi^2]^{3/2} \sqrt{\vDe^2-m_\pi^2} \nn\\
&\times& \left[ M_\De^4-(M_N^2-m_\pi^2)^2- \third (\vDe^2-m_\pi^2) (M_N^2+M_\De M_N-m_\pi^2)\right] .
\eea
\end{subequations}
\newline
\indent
The $\pi N$ scattering phase-shift is related to the partial-wave
K-matrix simply as
\beq
\eqlab{phase33}
\de_l = \arctan K_l\,,
\eeq
where $l$ stands for the conserved quantum numbers: spin ($J$),
isospin ($I$) and parity ($P$). We emphasize that
the $P_{33}$ phase
(corresponding to $J=3/2=I$, $P=+$) is the only nonvanishing one
at NLO in the resonance region, and is computed by
substituting the NLO expressions for the self-energy [\Eqref{NLOwidth}]
into \Eqref{newkmat}. We can then fix the LECs $h_A$ and $h_1$ 
by fitting the result to the well-established empirical information
about this phase-shift.
\newline
\indent
Thus, in \Figref{pin_phase} the red solid curve shows the NLO description 
of the empirical $P_{33}$ phase-shift represented by the data points. 
The curve is obtained by taking $h_A=2.85$
and $h_1=0$, and is characterized by $\chi^2/\mbox{point} \simeq 1$, 
where we assume $0.5$ degree uncertainty in the empirical values. 
(The best fit is obtained by slightly
decreasing $h_A$ and increasing $h_1$ to about 0.6, however the improvement 
is very small and
we prefer to neglect $h_1$ for simplicity).
The blue dashed line in \Figref{pin_phase} shows the LO result, obtained by 
neglecting $\mbox{Im} \Si'$ and $h_1$, and taking $h_A=2.85$. 
This corresponds with the so-called
``constant width approximation''.
At both
LO and NLO, the resonance width takes the value 
$\Ga(M_\De)\simeq 115$ MeV. 
\begin{figure}[t,h]
\centerline{  \epsfxsize=8cm%
\epsffile{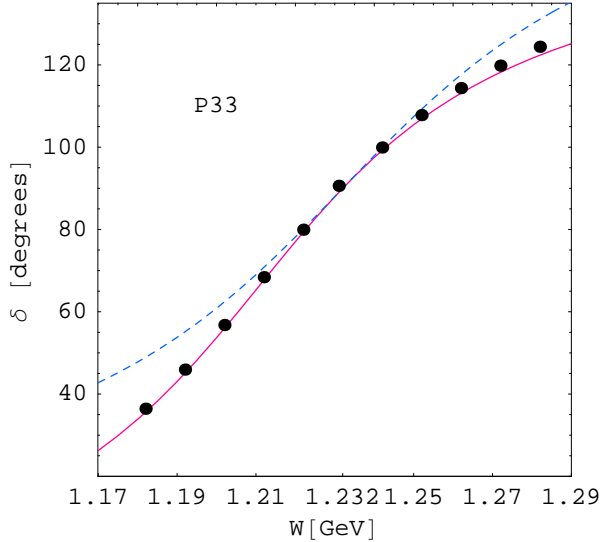}
}
\caption{
(Color online) The energy-dependence of the 
P33 phase-shift of elastic pion-nucleon scattering
in the $\De$-resonance region.
The red solid (blue dashed) curve represents the NLO (LO) result. 
The data points are from the 
SP06 SAID analysis~\protect\cite{Arndt:2002xv}.
}
\figlab{pin_phase}
\end{figure}
\newline
\indent
Of interest is the position of the resonance in the complex plane, 
$\sqrt{s_R} = M_\De^{(pole)} - (i/2) \Ga_\De^{(pole)}$. 
In this NLO calculation,
the ``complex pole'' parameters come out to be
\begin{subequations}
\bea
M_\De^{(pole)} &=& M_\De - \frac{\mbox{Im}\Si\,\mbox{Im}\Si'}{1+(\mbox{Im}\Si')^2} 
\simeq 1.211 \,\mbox{GeV}\,, \\
\Ga_\De^{(pole)} &=& - \frac{2\, \mbox{Im}\Si}{1+(\mbox{Im}\Si')^2} \simeq
 0.097 \,\mbox{GeV} \,.
\eea
\end{subequations}
All these numbers are within the range quoted by the  
Particle Data Group~\cite{PDG2006}.
\newline
\indent
Note that the calculations presented here satisfy (the two-body $\pi N$)
unitarity exactly. Indeed, the partial-wave S-matrix obtained
from the graphs in \Figref{pin2NLO} is given by
\beq
S_l = \frac{1+iK_l}{1-iK_l} = e^{2i\de_l}\,,
\eeq
where $K_l$ and $\de_l$, given respectively 
by Eqs.~\eref{kmat33} and \eref{phase33}, are real numbers, hence
$|S_l | =1$.

\subsubsection{Pion photo- and electroproduction}

We now turn to the analysis of the
pion electroproduction process. Since
we are using the one-photon-exchange approximation (until
Sect.~6), the pion photoproduction can be viewed as
the particular case of electroproduction at $Q^2=0$.
\begin{figure}
\centerline{  \epsfxsize=11cm
  \epsffile{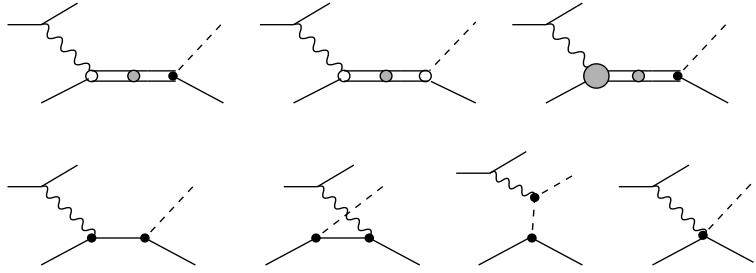} 
}
\caption{Diagrams for the $e N \to e \pi N $ reaction 
to LO and  NLO in the $\delta$-expansion. The dots denote
the vertices from the 1st-order Lagrangian, while the
circles are the vertices from the 2nd order Lagrangian ({\it e.g.}, the 
$\ga N\De$-vertex in the first two graphs is the $g_M$ coupling from $\lag^{(2)}$).}
\figlab{diagrams}
\end{figure}
\begin{figure}
\centerline{  \epsfxsize=11cm
  \epsffile{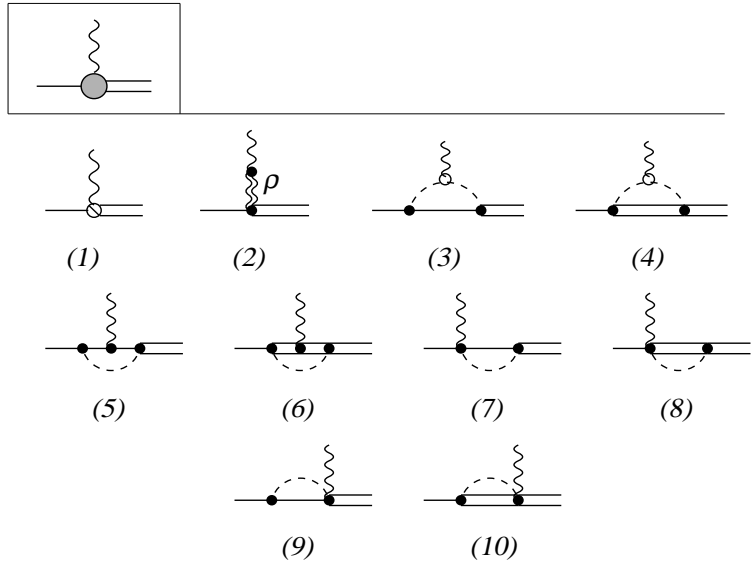} 
}
\caption{The $\ga N\De$ vertex at ${\mathcal O}(\de^3)$.
The sliced vertex (1) stands for the $g_E$ and $g_C$ couplings
from the 3rd order Lagrangian. The wiggly double line in (2) stands
for the vector-meson propagator.}
\figlab{NLOvertex}
\end{figure}
\newline
\indent
The pion electroproduction amplitude to NLO in the $\de$-expansion, in the
resonance region, is given by the graphs in  \Figref{diagrams}, 
where the shaded
blob in the 3rd graph denotes the NLO $\ga N\De$ vertex, given by the graphs
in \Figref{NLOvertex}. The 1st graph in \Figref{diagrams} enters at the LO,
which here is ${\mathcal O}(\de^{-1})$. The $\De$ self-energy in this graph
is included to NLO, see \Figref{selfen2NLO}. 
All the other graphs in \Figref{diagrams}
are of NLO$={\mathcal O}(\de^{0})$. Note that the $\De$-resonance contribution
at NLO is obtained by going to NLO in either the $\pi N\De$ vertex (2nd graph)
or the $\ga N\De$ vertex (3rd graph). Accordingly, 
the $\De$ self-energy in these graphs is included, respectively,
to NLO (\Figref{selfen2NLO}) and to LO [\Figref{selfen2NLO}(a)] in 
the $\pi N\De$ coupling. 
\newline
\indent
The vector-meson diagram,  \Figref{NLOvertex}(2), contributes to 
NLO for $Q^2\sim \La\De$. One includes
it effectively by giving the $g_M$-term a dipole $Q^2$-dependence 
(in analogy to how it is usually done
for the nucleon isovector form factor): 
\beq
g_M\to \frac{g_M}{(1+Q^2/0.71\,\mbox{GeV}^2)^{2}}. 
\eeq
The analogous
effect for the $g_E$ and $g_C$ couplings begins at N$^2$LO. 
\newline
\indent
An important observation is that at $Q^2=0$ 
only the imaginary part (unitarity cut) of the loop graphs in 
\Figref{NLOvertex}
contributes to the NLO amplitude. Their real-part contributions, after the
renormalization of the LECs, begin to contribute at N$^2$LO, for 
$Q^2\ll \vDe\La_{\chi SB}$.
At present we will consider only the NLO calculation where the 
$\pi\De$-loop contributions to the $\ga N\De$-vertex are omitted 
since they do not give the imaginary contributions
in the $\De$-resonance region. We emphasize that such loops might 
become important at this order for $Q^2\sim \vDe\La_{\chi SB}\sim 0.3$ GeV$^2$ 
and should be included for the complete NLO 
result. The present calculation is thus 
restricted to values $Q^2 < 0.3$ GeV$^2$.
\begin{figure}[t,h]
\centerline{
\includegraphics[width=0.45\columnwidth]{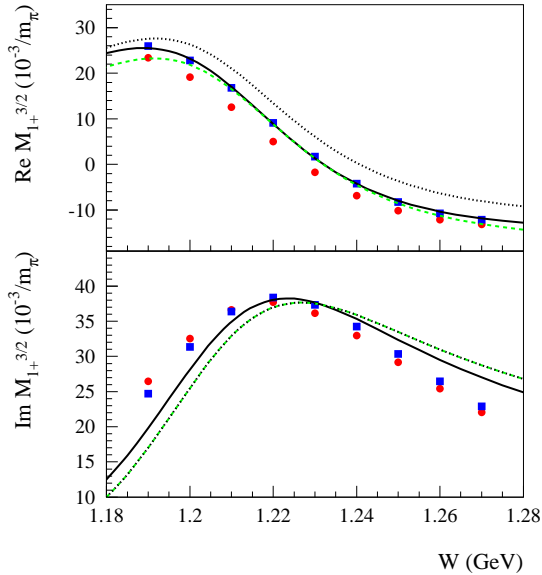}
}
\caption{
(Color online) The 
$M_{1+}^{(3/2)}$ multipole
for pion photoproduction as function of the invariant energy.  
Green dashed curves: 
$\Delta$ contribution without the $\ga N\De$-vertex loop corrections,
[i.e., only the first three graphs in \Figref{diagrams} with
\Figref{NLOvertex}(1) contribution are taken into account].  
Blue dotted curves: adding the Born contributions, 2nd line in 
\Figref{diagrams}, to the dashed curves. 
Black solid curves: the NLO calculation, includes 
all graphs in \Figref{diagrams} as well as the loop corrections. 
The data point are from the 
SAID analysis~(FA04K)~\protect\cite{Arndt:2002xv} (red circles), and from the 
MAID 2003 analysis~\protect\cite{MAID98} (blue squares).
}
\figlab{gap_pin_m1mult}
\end{figure}
\begin{figure}[t,h]
\centerline{\includegraphics[width=0.45\columnwidth]{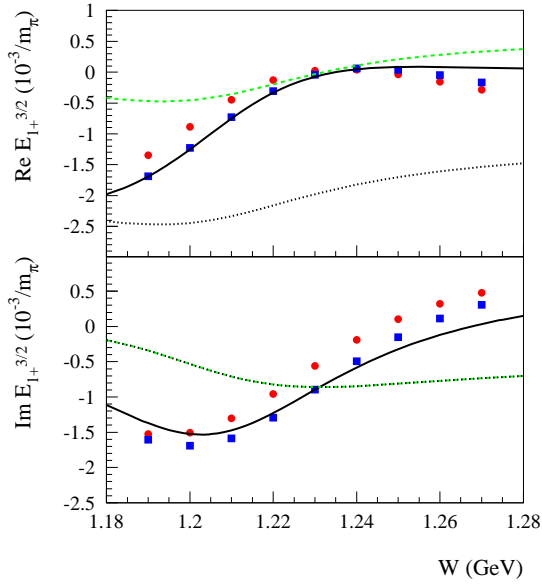}}
\caption{
(Color online) The $E_{1+}^{(3/2)}$ multipole 
for pion photoproduction. Curve conventions and data points
 are the same as in \Figref{gap_pin_m1mult}.  
}
\figlab{gap_pin_e1mult}
\end{figure}
\begin{figure}[t,h]
\centerline{\includegraphics[width=0.45\columnwidth]{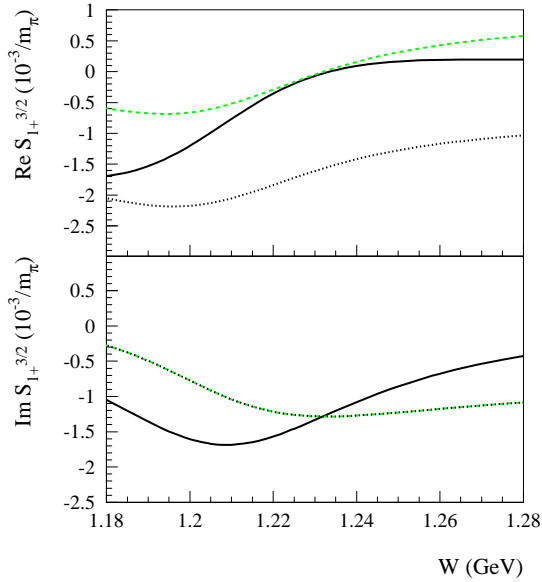}}
\caption{
(Color online) The $S_{1+}^{(3/2)}$ multipole 
at $Q^2=0$ as function of the invariant mass $W$ of 
the $\pi N$ system. Curve conventions 
are the same as in \Figref{gap_pin_m1mult}.  
}
\figlab{gap_pin_s1mult}
\end{figure}
\newline
\indent
In Figs.~\ref{fig:gap_pin_m1mult}--\ref{fig:gap_pin_s1mult} 
we show the results
for the pion electroproduction resonant multipoles 
$M_{1+}^{(3/2)}$, $E_{1+}^{(3/2)}$, and $S_{1+}^{(3/2)}$ as function of the invariant energy $W=\sqrt{s}$
around the resonance position, at $Q^2=0$. 
The $M_{1+}^{(3/2)}$ and $E_{1+}^{(3/2)}$ 
multipoles are well established by the MAID~\cite{MAID98}
and SAID~\cite{Arndt:2002xv} partial-wave solutions, thus allowing one to fit 
two of the three $\ga N\De$ LECs at this order as:~$g_M = 2.97$, $g_E = -1.0$.
The third LEC is adjusted to for a best description of the
pion electroproduction data at low $Q^2$ (see Sect.~\ref{sec5}), yielding
 $g_C=-2.6$.
The latter values translate into $G_M^\ast=3.04$, $G_E^\ast=0.07$, and $G_C^\ast=1.00$
for the Jones-Scadron form-factors at $Q^2=0$.    
As is seen from the figure,
the NLO results (solid curves) give a good description of the energy 
dependence of the resonant multipoles in 
a window of 100 MeV around the $\Delta$-resonance position.
Also, these values yield $R_{EM}= -2.2$ \% and $R_{SM}= -3.4$ \%.
\newline
\indent
The dashed curves in these figures
show the contribution of the $\Delta$-resonant diagram of \Figref{diagrams}
{\it without} the NLO loop corrections in \Figref{NLOvertex}.
For the $M_{1+}$ multipole this is the LO and part of the NLO contributions.
For the $E_{1+}$ and $S_{1+}$ multipole
the LO contribution is absent (recall that $g_E$ and $g_C$ coupling
are of one order higher than the  $g_M$ coupling). 
Hence,  the dashed curve represents
a partial NLO contribution to $E_{1+}$ and $S_{1+}$.
\newline
\indent
Note that such a purely resonant contribution without the loop corrections 
satisfies unitarity in the sense of the Fermi-Watson 
theorem~\cite{Watson54}, which states that the phase of a pion 
electroproduction amplitude ${\mathcal M}_l$ is given by the
corresponding pion-nucleon phase-shift:
${\mathcal M}_l = | {\mathcal M}_l | \, \exp({i\de_l})$.
As a direct consequence of this theorem, 
 the real-part of the resonant multipoles must vanish at the resonance 
position, where the phase-shift crosses $90$ degrees.
\begin{figure}[b]
\centerline{  \epsfxsize=7.8cm%
\epsffile{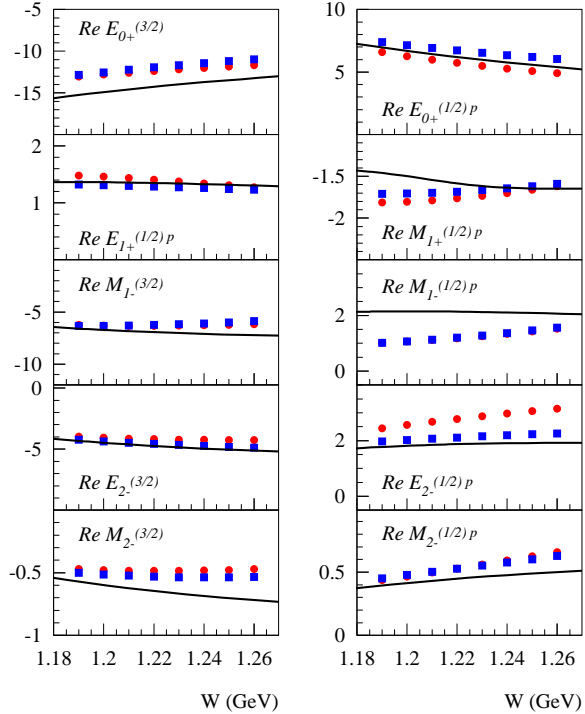}
}
\caption{
(Color online) Non-resonant multipole amplitudes
(in units $10^{-3} / m_\pi$) 
of pion photoproduction as function of the invariant mass $W$ of 
the $\pi N$ system.  
The solid curves result from our NLO calculation. 
The data points are from the 
SAID analysis (FA04K)~\protect\cite{Arndt:2002xv} (red circles), and from the 
MAID 2003 analysis~\protect\cite{MAID98} (blue squares).
}
\figlab{gap_pin_nonresmult}
\end{figure}
\newline
\indent
Upon adding the non-resonant  Born graphs (2nd line in \Figref{diagrams}) 
to the dashed curves, one obtains the dotted curves. 
The non-resonant contributions are purely real at this order and hence 
the imaginary part of the multipoles do not change.
While this is consistent with unitarity for the non-resonant multipoles 
(recall that the non-resonant phase-shifts are zero at NLO), 
the Fermi-Watson theorem in the resonant channels is violated. 
In particular, one sees that the real parts 
of the resonant multipoles  now fail to cross zero at the resonance position.
The complete NLO calculation, shown by the solid curves in the figure
includes in addition the $\pi N$-loop corrections in \Figref{NLOvertex}, 
which obviously restore unitarity. The Fermi-Watson theorem is satisfied 
exactly in this calculation. 
\newline
\indent
Next we examine the results for the non-resonant
multipoles, which all receive contributions of the Born graphs only. 
In Fig.~\ref{fig:gap_pin_nonresmult}, we show the NLO calculations for 
the non-resonant $s$-, $p$- and $d$-wave 
pion photoproduction multipoles in the $\Delta(1232)$ region in 
comparison with the 
two state-of-the-art phenomenological multipole 
solutions MAID and SAID.  
At this order the non-resonant 
multipoles are purely real. The multipole solutions show 
indeed that the imaginary 
parts of non-resonant multipoles, around the $\Delta$ resonance,
are negligible in comparison to their real parts. 
\begin{figure}[t,b]
\centerline{  \epsfxsize=8cm%
\epsffile{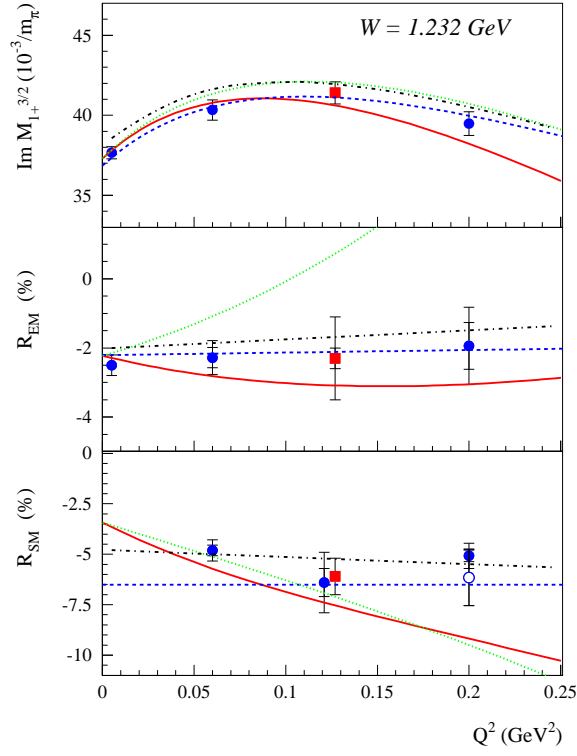}
}
\caption{
(Color online) The resonant multipoles
of pion electroproduction as function $Q^2$ at the $\De$-resonance position.
The green dotted curve is the $\De$-contribution alone.  
The solid curves are results of the NLO calculation of Ref.~\cite{Pascalutsa:2005vq}. 
Also shown are results of the 
SAID analysis (FA04K)~\protect\cite{Arndt:2002xv} (black dashed-dotted curves),
and  the MAID 2003 analysis~\protect\cite{MAID98} (blue dashed curves).
The data points are from BATES at $Q^2=0.127$~\cite{Sparveris:2004jn} and 
MAMI: $Q^2=0$~\cite{Beck:1999ge} ,
$Q^2=0.06$~\cite{Stave:2006ea}, $Q^2=0.121$~\cite{Pospischil:2000ad}, 
$Q^2=0.2$~\cite{Elsner:2005cz}.} 
\figlab{resmult_qsq}
\end{figure}
\newline
\indent
From the figure we conclude that for most of the non-resonant 
multipoles, the parameter-free NLO results, 
agree fairly well with the phenomenological solutions.  
The largest differences are observed for $M_{1-}^{(1/2)p}$ multipole.   
This multipole corresponds with nucleon quantum numbers. The cause 
of the appreciable difference in this channel is largely 
due to the nucleon anomalous magnetic moment
contributions, which are not included in this calculation (since
they appear at N$^2$LO), but which are included in the 
phenomenological solutions. 
\newline
\indent
Finally, in \Figref{resmult_qsq} we show the NLO $\chi$EFT results
for the $Q^2$ dependence of the resonant multipoles, at the resonance
position. The red solid curve is {\it with} and the green dotted 
curve {\it without}
the chiral-loop corrections of \Figref{NLOvertex}. The blue dashed curves
and the black dashed-dotted curves represent the results of MAID and SAID,
respectively. The data point are from the recent MIT-Bates and 
MAMI experiments.
We observe from the figure
that the chiral loops play a crucial role in
the low momentum-transfer dependence of the $R_{EM}$ ratio. 
The effect of the ``pion cloud'' is most
pronounced in  the $E2$ $\ga N\De$ transition.

\subsubsection{Compton scattering}
 
The $\ga N\De$ transition plays an important role in the
$\ga N\to \ga N$ process in the $\De$-resonance region. 
Already well below the resonance the effects of the $\De$ excitation
are appreciable, and mostly due to its significant contributions
to the magnetic and the backward spin polarizabilities 
of the nucleon (cf.~\cite{Pascalutsa:2003zk} and references therein):
\bea
 \be^{(\De)} &=&  \frac{2\,\al_{em} \,g_M^2}{(M_N+M_\De)\,\vDe}\simeq
6 \,\,[\times \mbox{10$^{-4}$ fm$^3$]} \,,\\
\ga_\pi^{(\De)} &=&  \frac{\al_{em}\, g_M^2}{(M_N+M_\De)^3\,\vDe^2}\,
\left(1+ \frac{8\vDe}{M_N}\right)\simeq 9 \,\,[\times\mbox{10$^{-4}$ fm$^4$]}\,.
\eea 
These numbers can for instance be compared with the known empirical values
for the proton case~\cite{Beane:2002wn,Camen:2001st}: 
$\be^{(p-exp)} = 3.2\pm 1.2$ [$\times$ 10$^{-4}$ fm$^3$], 
$\ga_\pi^{(p-exp)} = -38.7\pm 1.8$  [$\times$10$^{-4}$ fm$^4$],
to convince oneself that the effects are significant. 
\newline
\indent
Note also that
the leading chiral-loop contribution to magnetic polarizability is 
$\be^{(p-LO)}\simeq 1.2 $ in the heavy-baryon $\chi$PT~\cite{Bernard:1992qa} 
and
$\be_p^{(p-RLO)}\simeq -2 $ in the manifestly covariant 
$\chi$PT~\cite{Bernard:1991ru,Lvov:1993ex}.
The latter number is certainly preferred, if we are to reconcile the large
$\De$ contribution to the magnetic polarizability with experiment. 
More discussion of this issue can be found in Ref.~\cite{Pascalutsa:2004wm}.
\newline
\indent
The calculations of Compton scattering in $\chi$PT without $\De$'s
are shown to be applicable only to energies not far above the pion-production
threshold, $W=M_N+m_\pi$, see Refs.~\cite{McGovern:2001dd,Beane:2004ra}. 
Traditionally
calculations of this reaction in the resonance region are done
using isobar-type of models~\cite{KVItheory,Feuster:1998cj,Kondratyuk:2001qu} 
or dispersion-relation 
approaches~\cite{L'vov:1996xd,Drechsel:2002ar}. Only relatively
recently, first attempts to compute the $\De$-resonance region in a $\chi$EFT 
framework have been completed to
some degree of 
success~\cite{Pascalutsa:2002pi,Hildebrandt:2003fm,Hildebrandt:2005ix}.
\newline
\indent
In \Figref{fig2} we show the results of Ref.~\cite{Pascalutsa:2002pi} 
for the Compton-scattering
observables, namely, differential cross-section and the linear beam asymmetry.
The results are shown as a function of the photon lab energy, 
$\w = (s-M_N^2)/(2M_N)$,
and the scattering angle in the center-of-mass system $\th_{c.m.}$.
The complete calculation (red solid curves) represents the NLO result in the
$\de$-expansion over both the low-energy and the resonance energy region.
It is instructive to compare it to the NLO calculation in 
the Deltaless  HB$\chi$PT (blue dashed lines). The latter clearly breaks 
down at energies
above the pion-production threshold, $\w_{\mathrm{thr}} = m_\pi$.
These results also demonstrate that the such computed $\De$ 
contributions (inferred
by the difference between the red solid and blue dashed curves),
while being small at low energies, are dominating the resonance region 
($\w$ around 340 MeV), in qualitative agreement with the power-counting
of the $\de$-expansion scheme.
\begin{figure}[tb]
\centering
\includegraphics[width=10.5cm]{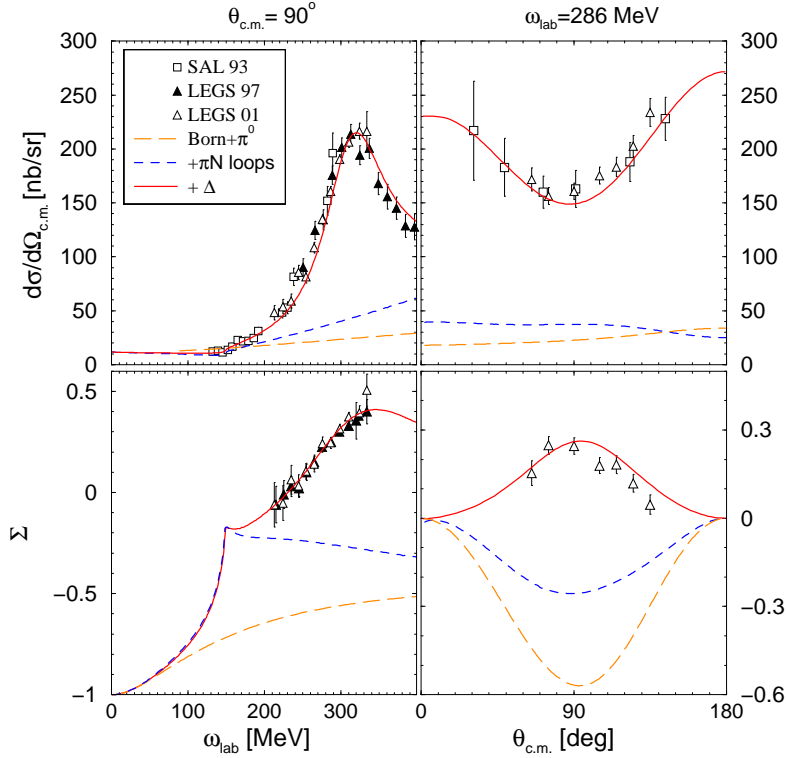}
\caption{\small 
The \ceft\ calculations of the $\ga$p$\to\ga$p
differential cross section ($d\si/d\Omega_{c.m.}$)
and the photon beam asymmetry ($\Sigma$) compared with recent experimental
data: SAL93~\cite{Hal93}, LEGS97~\cite{LEGS97CS}, 
LEGS01~\cite{Blanpied:2001ae}. 
Left panel: energy dependence at a fixed scattering angle. Right panel: angular
dependence at a fixed energy. The long-dashed orange line represents the 
sum of nucleon and pion Born graphs, 
the blue dashed line gives the NLO $\sla{\De}$-HB$\chi$PT prediction, and
the red solid line is the full result at NLO in the 
$\delta$-expansion~\cite{Pascalutsa:2002pi}.}
\figlab{fig2}
\end{figure}
\newline
\indent
In all fairness one should note that the NLO results shown in 
\Figref{fig2} are obtained using $g_M=2.6$ and $g_E=-6.0$, the values of LECs
which are inconsistent with the photoproduction analysis in the
same expansion scheme~\cite{Pascalutsa:2005vq} 
(see also the previous subsection).
The source of the discrepancy is most likely to be the use of the heavy-baryon
expansion in the Compton-scattering work~\cite{Pascalutsa:2002pi}. 
Relativistic effects are expected
to be important in the resonance region. As mentioned above, they 
already make a significant impact on such low-energy quantities as 
polarizabilities. 
A future study of Compton scattering in a manifestly covariant $\chi$EFT 
framework is called for to clarify this issue.

\subsubsection{Radiative pion photoproduction}

The radiative pion photoproduction ($\gamma N \to \pi N \gamma^\prime$) in the
$\De$-resonance region is used to access experimentally 
the magnetic dipole  moment (MDM) of the $\De$~\cite{Kotulla:2002cg}.
On the theory side this reaction had extensively been studied 
within the isobar type of 
models ({\it e.g.},~\cite{Machavariani:1999fr,Drechsel:2000um,Drechsel:2001qu,Chiang:2004pw}). 
Here, however, we only discuss the more recent
study~\cite{PV05} performed within the $\chi$EFT framework.
\begin{figure}
  \epsfxsize=8.5cm
\centerline{  \epsffile{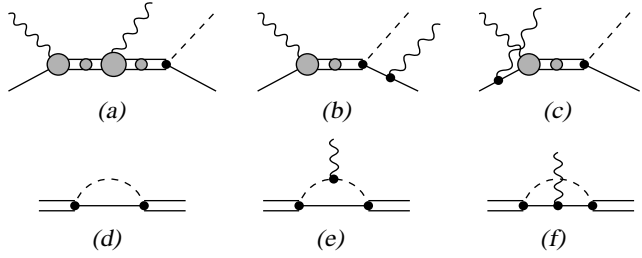} }
\caption{Diagrams for the $\gamma p \to \pi^0 p \gamma^\prime$ reaction 
at NLO in the $\delta$-expansion.}
\figlab{MDMdiagrams}
\end{figure}
\newline
\indent
Consider the amplitude for the $\gamma p \to \pi^0 p \gamma^\prime$ reaction. 
The optimal sensitivity to the MDM term is achieved when the incident 
photon energy is in the vicinity of $\De$, 
while the outgoing photon energy is  of order of $m_\pi$.
In this case the  $\gamma p \to \pi^0 p \gamma^\prime$ amplitude to 
NLO in the $\de$-expansion is given by the diagrams 
\Figref{MDMdiagrams}(a),( b), and (c),
where the shaded blobs, in addition to the couplings from the chiral 
Lagrangian, contain the one-loop corrections 
shown in \Figref{MDMdiagrams}(e), (f).
\newline
\indent
The general form of the $\ga \De\De$ vertex is given in \Eqref{gadeldeltree}. 
To NLO it suffice to keep only the $F_1^\ast$ and $F_2^\ast$ terms, 
which both receive corrections from chiral loops 
\Figref{MDMdiagrams}(e), (f). It is important to emphasize
that the Ward-Takahashi identity, 
\beq
q_\mu  \Gamma^{\mu\al\be} (p',p) = 
e\left[(S^{-1})^{\al\be}(p')-(S^{-1})^{\al\be}(p)\right],
\eeq
to NLO leads to the relation $F_1^\ast(0) =1-\Si'(M_\De)$, 
where $\Si'$ is given in \Eqref{NLOwidth}. 
This condition is verified exactly in the NLO calculation~\cite{PV05}.
\newline
\indent
Figure~\ref{chiral} shows the pion mass dependence of real and
imaginary  parts of the $\Delta^+$ and $\Delta^{++}$ MDMs, according to 
the calculation of Ref.~\cite{PV05}. Each of the two solid curves has a free 
parameter, a counterterm $\kappa_\De$ from $\lag^{(2)}_\De$,
adjusted to agree with the lattice data at larger values of $m_\pi$.
As can be seen from 
Fig.~\ref{chiral}, the $\Delta$ MDM develops an 
imaginary part when $m_\pi<\vDe$, 
whereas the real part has a pronounced cusp at $m_\pi = \vDe$. 
The dashed-dotted curve in Fig.~\ref{chiral} 
shows the result~\cite{Pascalutsa:2004ga}
for the magnetic moment of the proton. One can see that 
$\mu_{\Delta^+}$ and $\mu_p$, while having very distinct behavior
for small $m_\pi$, are approximately equal for larger values of $m_\pi$. 
\begin{figure}[t,b,h]
\centerline{\includegraphics[width=11cm]{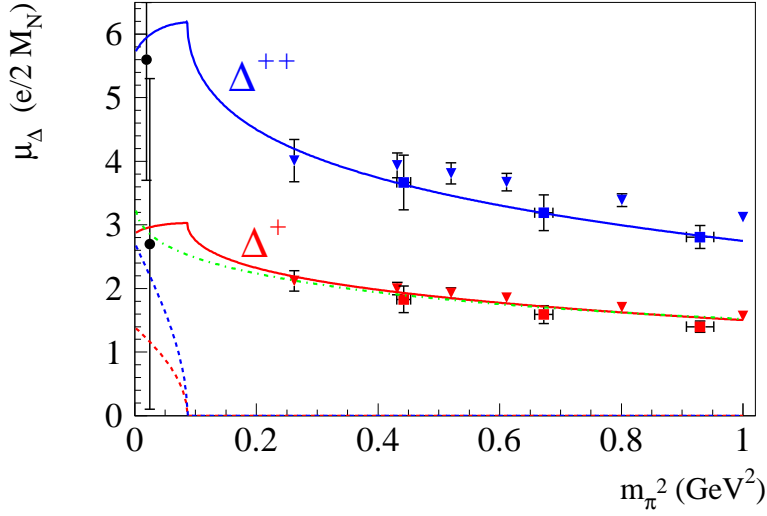}}
\caption{Pion mass dependence of the real (solid curves) and imaginary
(dashed curves) parts of $\Delta^{++}$ and $\Delta^{+}$ 
MDMs [in nuclear magnetons].
Dashed-dotted curve is the result for the proton magnetic 
moment from Ref.~\cite{Pascalutsa:2004ga}. 
The experimental data points for $\De^{++}$ and $\De^+$ (circles) 
are the values quoted by the PDG~\cite{PDG2006}. 
Quenched lattice data are from Refs.~\cite{Lein91} (squares)  
and  from Ref.~\cite{Lee:2005ds} (triangles).  }
\label{chiral}
\end{figure}
\newline
\indent
We next consider the NLO results for the $\gamma p \to \pi^0 p \gamma^\prime$ 
observables. 
The NLO calculation outlined above,  
completely fixes the imaginary  part of the 
$\gamma \Delta \Delta$ vertex\footnote{For an alternative recent calculation
of the imaginary part of the $\De$ MDM see Ref.~\cite{Hacker:2006gu}}. 
The expansion for the real part of the $\gamma \Delta \Delta$ begins
with LECs from $\lag^{(2)}$ which represent the isoscalar and isovector
MDM couplings: $\kappa_\De^{(S)}$ and $\kappa_\De^{(V)}$ in \Eqref{lagran2}. 
A linear combination of these parameters, 
$\mu_{\De^+} =[3+(\kappa_\De^{(S)}+\kappa_\De^{(V)})/2](e/2M_\De)$,  
is to be extracted from the $\gamma p \to \pi^0 p \gamma^\prime$ observables. 
Several such observables
 are shown in Fig.~\ref{fig:cross} for an incoming photon 
energy $E_\gamma^{lab} = 400$~MeV as function of the emitted photon 
energy $E_\gamma^{\prime \, c.m.}$. 
\newline
\indent
In the soft-photon limit 
($E_\gamma^{\prime \, c.m.} \to 0$), the 
$\gamma p \to \pi^0 p \gamma^\prime$ 
reaction is completely determined from the bremsstrahlung off 
the initial and final protons.  
The deviations of the $\gamma p \to \pi^0 p \gamma^\prime$ observables, 
away from the soft-photon limit, will then allow to study the 
sensitivity to $\mu_{\Delta^+}$. It is therefore very useful to 
introduce the ratio~\cite{Chiang:2004pw}:
\begin{eqnarray}
\label{eq:R1}
  R \,\equiv \, \frac{1}{\sigma_\pi} \cdot
  E^\prime_\gamma
  \frac{d\sigma}{dE^\prime_\gamma} ,
\end{eqnarray}
where $d\sigma / dE^\prime_\gamma$ is the 
$\gamma p \to \pi^0 p \gamma^\prime$ 
cross section integrated over the pion and photon angles, and 
$\sigma_\pi$ is the angular integrated cross section 
for the $\gamma p \to \pi^0 p$ process weighted with the bremsstrahlung 
factor, as detailed in~\cite{Chiang:2004pw}. 
\newline
\indent
This ratio $R$ has the property that in the soft-photon limit, the 
low energy theorem predicts that $R \to 1$. One firstly sees from 
Fig.~\ref{fig:cross} that the EFT calculation exactly satisfies this 
low-energy theorem. Furthermore, the EFT result for $R$ shows clear deviations 
from unity at higher outgoing photon energies, which are in good 
agreement with the first data for this process~\cite{Kotulla:2002cg}. 
The sensitivity of the EFT calculation to the $\mu_\Delta$ is a very 
promising setting for the dedicated second-generation experiment  
by the Crystal Ball Coll.\ at MAMI, which is currently 
under analysis (for first results, see Ref.~\cite{CB}). 
It improves upon the statistics of the first experiment (Fig.~\ref{fig:cross})
by at least two orders of magnitude and will allow 
for a reliable extraction of 
$\mu_{\Delta^+}$ using the EFT calculation presented here. 
\begin{figure}
\centerline{
  \epsfxsize=8cm
  \epsffile{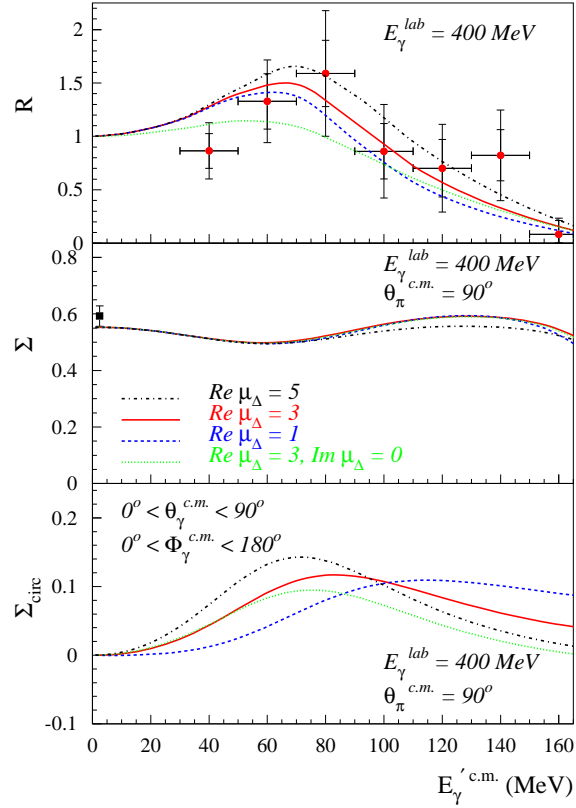}
}
\caption{The outgoing photon energy dependence of the 
$\gamma p \to \pi^0 p\ga'$ observables for different values of 
$\mu_{\Delta^+}$ (in units $e/2 M_\Delta$). 
Top panel: the ratio of $\gamma p \to \pi^0 p \gamma^\prime$ to 
$\gamma p \to \pi^0 p$ cross-sections Eq.~(\ref{eq:R1}). 
Data points are from~\cite{Kotulla:2002cg}. 
Middle panel: the linear-polarization photon asymmetry 
of the $\gamma p \to \pi^0 p \gamma^\prime$  
cross-sections differential 
w.r.t.\ the outgoing photon energy and pion c.m. angle. 
The data point at $E^\prime_\gamma = 0$ corresponds with the 
$\gamma p \to \pi^0 p$ photon asymmetry from~\cite{Beck:1999ge}.
Lower panel: the circular-polarization photon 
asymmetry (as defined in~\cite{Chiang:2004pw}), 
where the outgoing photon angles have been integrated over the 
indicated range. }
\label{fig:cross}
\end{figure}
\newline
\indent
Besides the cross section, the  
asymmetries for linearly and circularly polarized incident photons
are also displayed in \Figref{cross}.  
The photon asymmetry for linearly polarized photons,  $\Sigma$, at $E^\prime_\gamma = 0$
exactly reduces  to the 
$\gamma p \to \pi^0 p$ asymmetry. From Fig.~\ref{fig:cross} one sees
that in the soft-photon limit, 
the NLO calculation is in good agreement with the experimental data point,
and predicts a nearly constant energy-dependence. It also predicts a
weak sensitivity on the MDM value. 
The linear beam asymmetry is therefore an excellent
observable for a consistency check of the EFT calculation. 
\newline
\indent
The asymmetry for circularly polarized photons, 
$\Sigma_{circ}$, (which is exactly zero for a 
two body process due to reflection symmetry w.r.t.\ the reaction plane) 
has been proposed~\cite{Chiang:2004pw} as a unique observable 
to enhance the sensitivity to $\mu_\Delta$. 
Indeed, in the soft-photon limit, where the 
$\gamma p \to \pi^0 p \gamma^\prime$ process reduces to a two-body process, 
$\Sigma_{circ}$ is exactly zero. 
Therefore, its value at higher outgoing photon energies 
is directly proportional to $\mu_\Delta$. 
One sees from Fig.~\ref{fig:cross} (lower panel) 
that the EFT calculation 
supports this observation, and shows sizably different asymmetries 
for different values of $\mu_\Delta$. 
A combined fit of all three observables shown in Fig.~\ref{fig:cross} 
will therefore allow for a very stringent test 
of the EFT calculation, which can then be used 
to extract the $\Delta^+$ MDM.

\subsubsection{Errors due to neglect of higher-order effects}

\begin{figure}[t,h]
\centerline{  \epsfxsize=13cm
  \epsffile{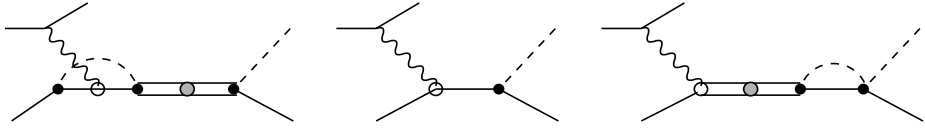} 
}
\caption{Examples of N$^2$LO contributions to
the $e N \to e \pi N $ reaction.
Open circles denote the \EM\ coupling to the
anomalous magnetic moment of the nucleon. }
\figlab{NNLOgraphs}
\end{figure}

An advantange of the \ceft\ framework over, say, dynamical models,
is that one can make a resonable estimate of the theoretical uncertainty
of the calculations due to the neglect of higher-order effects. 
Consider the pion electroproduction case which we have computed above to NLO. 
Some of the neglected next-next-to-leading order (N$^2$LO) 
contributions are shown in \Figref{NNLOgraphs}. Of course,
there is no substitute for an actual calculation of those
effects, but at present we only know that they must be suppressed
by at least one power of $\de$ ($=\De/\La_{\chi SB}$) 
as compared to the NLO and two powers of $\de$  
as compared to the LO contributions. Therefore, we can {\it estimate}
the size of the N$^2$LO contribution to an amplitude $A$ as:
$A_{\mathrm{NLO}}\, \de $, or $A_{\mathrm{LO}}\, \de^2 $. The theoretical
uncertainty of a calculation up to and including NLO can thus be
estimated as:
\begin{subequations}
\eqlab{naive}
\beq
A_{\mathrm err} = | A_{\mathrm{LO}} + A_{\mathrm{NLO}}| \, \de^2.
\eeq
In cases where the amplitude does not receive any LO contributions,
we have simply:
\beq
A_{\mathrm err} = | A_{\mathrm{NLO}}| \, \de.
\eeq
\end{subequations}
\newline
\indent
There are a few caveats in the
implementation of such a naive estimate. First of all, as we increase the
light scales, the expansion parameter must increase too. And this
does not concern $\De$ alone but, in our case, also $m_\pi$ and $Q^2$.
Therefore, it is more realistic to estimate the error using an
expansion parameter such as: 
\beq
\tilde \de = \frac{1}{3}
 \left[\frac{\De}{M_N} + 
\left(\frac{m_\pi}{M_N}\right)^{1/2} 
+ \left(\frac{Q^2}{M_N^2}\right)^{1/2} \right],
\eeq
where all the light scales are treated on equal footing and 
hence are averaged over. 
\newline
\indent
Secondly, what if the amplitude happens to vanish at some
kinematical point. According to \Eqref{naive} the theoretical calculation
at that point would be perfect, which is of course  unlikely to be true
in reality. So, when considering dependencies on kinematical variable(s),
we shall take an {\it average of the error} over some appropriate
region of that variable.  
\newline
\indent
Given these two points, we are led to the following 
formula for the theoretical uncertainty of the NLO
calculation for an amplitude $A$,
\beq
\eqlab{Aerror}
A_{err} = \left\{ \begin{array}{cc}
|A|_{av}  \, \tilde\de^2, & \mbox{LO}\neq 0 \\
|A|_{av}  \, \tilde\de, & \mbox{LO} = 0,\,
\end{array}\right.
\eeq
and the subscript ``{\it av}'' indicates
that the appropriate averaging is performed.
\newline
\indent
The theoretical uncertainty of the NLO
calculation
of an observable $O$ is:
\beq
\eqlab{Oerror}
O_{err} = \left\{ \begin{array}{cc}
2 |O|_{av}  \, \tilde\de^2, & \mbox{LO}\neq 0 \\
2 |O|_{av}  \, \tilde\de, & \mbox{LO} = 0,\,
\end{array}\right.
\eeq
where the factor of 2 takes into account that an observable
is a product of two amplitudes.

%% file: chap5_delta.tex
\section{Results for observables and discussion}
\label{sec5}

In this Section, we review the current status of pion photo- and 
electroproduction data in the $\Delta$-resonance region, 
and discuss the results of the theoretical 
approaches described above, namely dynamical models and \ceft . 
An extensive review of meson productions from the nucleons,
including single pion, two pions and $\eta$ up to second resonance
region, has been published recently \cite{Burkert:2004sk}. Another
recent review \cite{Krusche:2003ik} on meson productions also goes
up to the second resonance region, but is restricted only to
photoproduction.
\newline
\indent
After some introductory remarks in Sect.~\ref{sec51}, we shall consider 
pion photoproduction in Sect.~\ref{sec52}. 
In Sect.~\ref{sec53}, the focus will be on the electroproduction case 
for both the low and larger $Q^2$ regions.

\subsection{Introductory remarks}
\label{sec51}

The first experimental work on meson photoproduction dates as
early as 1949. However, most of the present precision data became 
available thanks to the progress made in accelerator and detector
technology during the last two decades. At the present time, the
new generation of electron accelerators at JLab, ELSA, and MAMI
are dedicated to the study of nuclear and hadronic structure and
are equipped with state-of-the-art detector systems. The high intensity
electron beams, besides being used for electron induced reactions,
also provide good photon sources for photonuclear reactions via
bremsstrahlung. Another way to produce high energy photon beams is
through Compton backscattering.  Laser backscattering systems
have been installed at BNL (LEGS), ESRF (GRAAL), and at SPring8
(LEPS) to produce photon beams to investigate nucleon resonances
and meson photoproduction.
\newline
\indent 
Turning to the theory, let us recall the key 
ingredients which enter into the
dynamical models and the \ceft \ calculations. 
Chiral perturbation theory ($\chi$PT) is now widely accepted as the
"basic" theory to describe the low energy interactions of
Goldstone bosons among themselves and with other hadrons. 
There is generally good agreement between the $\chi$PT predictions and
experiments \cite{chiral96}. For example, state-of-the-art $\chi$PT
calculation on $\ga p\ra\pi^0 p$ at threshold
\cite{Bernard91,Bernard}, which has gone up to order $p^4$ and
found large contributions from one-loop charge-exchange
rescattering, agree beautifully with the new precise measurements
\cite{Fuchs96,Bergstrom,NIKHEF,Distler,Bernstein97,Schmidt}.
\newline
\indent
The \ceft \ $\delta$-expansion presented in Sect.~\ref{sec4} 
provides an extension of $\chi$PT into the $\Delta$-resonance region. 
In such scheme, for reactions such as $\pi N$ scattering and
electromagnetic pion production, diagrams with a dressed $\De$ in
the intermediate states (the ODR graphs) as depicted in \Figref{pin2NLO}a 
and \Figref{diagrams} (upper left diagram) are the leading 
diagrams\footnote{It is interesting to note that 
the LO \ceft \ calculation provides the field 
theoretic justification for the previously employed 
$\De-$saturation model (see {\it e.g.} \cite{Sauer86,Yang85}).}.
In the NLO, $\pi N$ scattering is still dominated by the 
$\De-$excitation diagrams as depicted in \Figref{pin2NLO}. 
For pion electromagnetic production, only the Born terms with electric $\ga
NN$ coupling, in addition to the loop correction to the $\ga N\De$
vertex as given by graphs in \Figref{NLOvertex} , 
enter in the NLO. The only question left then is 
whether the expansion converges. Namely, whether NNLO terms
are smaller than NLO. At NNLO, for example, the background
$\pi N$ interaction corresponding to the crossed Born diagram
would begin to contribute to the $\De$ self-energy in the two-loop
diagrams which is the main driving mechanism in generating the
$\De-$resonance in the Chew-Low theory. Convergence problem aside, EFT
in the resonance region does provide a consistent expansion scheme
which contains the fundamental features of QCD, namely chiral
symmetry, gauge invariance, covariance, and crossing symmetry, as
in the case for $\chi$PT for low-energy hadron phenomena.
\newline
\indent
As in $\chi$PT, dynamical models also start from an effective
chiral Lagrangian. However, the calculation of the 
$S$-matrix in dynamical models is not 
based on field theory. 
Rather, the effective Lagrangian is used
to construct a potential used in a Lippmann-Schwinger type of 
scattering equation. The solution of the scattering equation 
includes the rescattering
effects to all orders and thereby unitarity is ensured while the
crossing and the other symmetries are not. 
In dynamical models applied to pion electromagnetic production,  
considered in Sect.~\ref{sec3}, gauge invariance 
is problematic. The only hope is that the
{\it ad hoc} schemes which have been proposed provide good
approximations. The success of DMT for the $\ga p\ra\pi^0 p$ at
threshold \cite{KY99} offers a hope in this direction.
\newline
\indent
Dynamical models 
\cite{pearce,Hung01,GrS93,SL,afnan99,tjon00} have been able to
provide a good description of $\pi N$ scattering lengths and the
phase shifts in $S, P,$ and $D$ waves up to $W\simeq 1.5$ GeV,
see for instance \Figref{pin_fs}. It should be pointed out here that within
the dynamical models, the $\De$ resonance is a coherent contribution of an
elementary $\De$ and Chew-Low graphs as obtained in the cloudy bag
model~\cite{Cloudy80}. One advantage of dynamical 
model calculations over \ceft \ is that
they provide a unitary framework up to higher energies because they 
do not rely on a perturbative expansion in small energy scales.

\subsection{Pion photoproduction}
\label{sec52}

The majority of the photopion production data comes from  
 measurements of the differential cross section by unpolarized
 photons on unpolarized nucleons. All other photoproduction
 experiments involve the use of polarized photon beams, polarized
 nucleon targets, a combination of both polarization or the
 measurement of the polarization of the recoil nucleon.
\newline
\indent
 The earliest of the polarization experiments used unpolarized
 photon beams and unpolarized targets and measured the recoil
 polarization of the nucleon, $P(\theta)$, by a scattering
 of this recoiling nucleon on a secondary target. 
 High-energy polarized photon beams have been produced either by
 laser backscattering or coherent bremsstrahlung in a crystal.
 They have been used to measure the beam asymmetry,
 $\Sigma(\theta)$, from an unpolarized target.
\newline
\indent
 The first data on the target asymmetry, $T(\theta)$, was obtained in
 1972 for the $\ga + p\ra\pi^+ +n$ process. Since then an impressive set of
 measurements of $d\sigma/d\Omega, P(\theta), T(\theta)$, and
 $\Sigma(\theta)$ bas been assembled on the final states $\pi^+ n,
 \pi^0 p$, and $\pi^- p$. The data coverage on the first two
 channels spans the whole resonance region with good angular
 coverage. The $\pi^- p$ channel data, on the other hand, is more
 scarce since it involves the neutron (deuterium) target.
\newline
\indent
 Double polarization observables have also been measured. They fall
 into three categories: beam-target, beam-recoil, and
 target-recoil. Polarized beam-polarized target experiments yield
 $G(\theta)$ and $H(\theta)$ and have been measured at Kharkov
 \cite{Belyaev:1985sp,Belyaev85} and at 
MAMI \cite{Ahrens:2000bc,Ahrens:2004pf,Ahrens:2005zq}.
 Beam-recoil and target-recoil
 measurements are experimentally more difficult as they involve
 a second scattering process to analyze the recoil nucleon polarization.
 No data have been reported for them.
 The most precise determination to date of the $R_{EM}$ ratio in the
 photo-excitation of the $\Delta(1232)$ has come from the
 simultaneous measurements of $p(\vec \ga,p)\pi^0$ and $p(\vec \ga,n)\pi^+$
 with polarized photon beams \cite{Beck:1997ew,LEGS97}.
\newline
\indent
 A comprehensive account of the status of photoproduction
 experiments can be found in \cite{Arndt:2002xv} and the full database
 can be accessed at the SAID website \cite{GWU}.
%
\begin{table}[tb]
\centering
\begin{tabular}{||c|c|c|c|c|c|}
\hline \hline
       &  $A_{1/2}$   &  $A_{3/2}$    & $Q_{p\rightarrow\Delta^+}$ & $\mu_{p\rightarrow\Delta^+}$ &$R_{EM}$ \\
&$(10^{-3}$~GeV$^{-\frac12})$&$(10^{-3}$~GeV$^{-\frac12})$&(fm$^2$)&$(\mu_N)$&(\%)\\
\hline
 PDG   & -135$\pm$6       &  -250$\pm$ 8   &   -0.0846$\pm$0.0033        &      3.46$\pm$0.03    &   -2.5$\pm$0.5\\
\hline
 MAMI  & -131       &  -251    &   -0.082       &    3.45       & -2.5\\
\hline
 LEGS  & -135.7       &  -266.9    &   -0.108        &      3.642     & -3.07 \\
\hline
 DMT    & -134  &  -256  &   -0.081 &  3.52 & -2.4 \\
&(-80)&(-136)&(0.009)&(1.922)&(0.47)\\
\hline
 SL    & -118  &  -228  &   -0.081 &   3.13 & -2.7 \\
&(-84)&(-153)&(-0.027)&(2.13)&(-1.3)\\
\hline
 DUO    & -131.5   & -255    &  -0.091   &  3.49  & -2.7  \\
\hline
 \ceft \ & -133   & -253    &  -0.077   &  3.48  & -2.2   \\
\hline \hline
\end{tabular}
\caption{Comparison of the values for
 the helicity amplitudes, $Q_{N\rightarrow\Delta}$, 
$\mu_{N\rightarrow\Delta}$, and $R_{EM}$ extracted from experiment, 
with their values in dynamical models 
(DMT~\cite{KY99}, SL~\cite{SL}, DUO~\cite{Pascal04}) 
and \ceft \ calculations~\cite{Pascalutsa:2005ts,Pascalutsa:2005vq}. 
The numbers within the parenthesis, in the cases of DMT and SL,
 correspond to the bare values.}
\label{table_helicity-amp}
\end{table}
\newline
\indent
Let us turn to the results of the theoretical approaches considered above.
In the \ceft \ \cite{Pascalutsa:2005ts,Pascalutsa:2005vq}
and in the dynamical model calculations of 
SL~\cite{SL}, DMT~\cite{KY99}, and DUO~\cite{Pascal04}, the
$\ga N\De$ coupling constants are all obtained by fitting to the
experiments. We present in Table~\ref{table_helicity-amp} the
resulting values for the helicity amplitudes $A_{1/2}, A_{3/2}$,
transition electric quadrupole moment $Q_{p\rightarrow\Delta^+}$,
transition magnetic dipole moment, $\mu_{p\rightarrow\Delta^+}$,
and $R_{EM}$, together with the PDG \cite{PDG2006} values and
experimental results measured at 
MAMI~\cite{Beck:1997ew,Beck:1999ge} 
and LEGS~\cite{LEGS97,Blanpied:2001ae}. 
The values for $Q_{p\rightarrow\Delta^+}$ and
transition magnetic dipole moment $\mu_{p\rightarrow\Delta^+}$
listed in the same row with the PDG values are the estimates 
of Eqs.~(\ref{eq:qndelexp}) and (\ref{eq:mundelexp}) respectively.
%
%
\begin{figure}[htbp]
\centerline{ \epsfxsize=10cm%
\epsffile{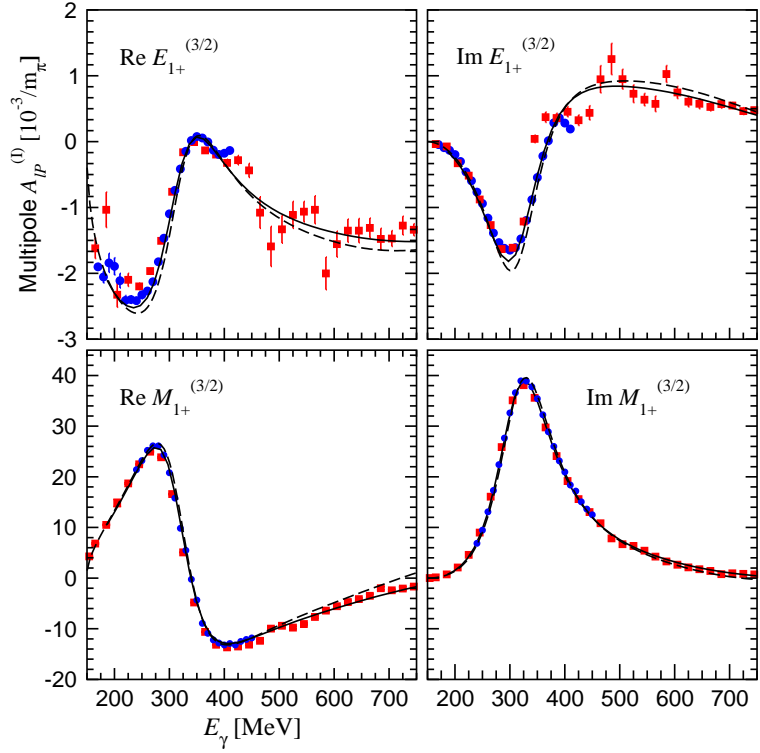}
}
\caption{ Real and imaginary parts of the $M_{1+}^{(3/2)}$ and
$E_{1+}^{(3/2)}$ multipoles. 
The solid curves are the full DMT results, whereas the dashed curves 
are the results of the DUO model. 
The blue circles are the results from the Mainz
dispersion relation analysis~\cite{Hanstein:1997tp}, whereas the red squares 
are obtained from the VPI analysis~\cite{VPI97}. 
}
\figlab{figM1E1}
\end{figure}
\begin{figure}[t]
\centerline{ \epsfxsize=13cm%
\epsffile{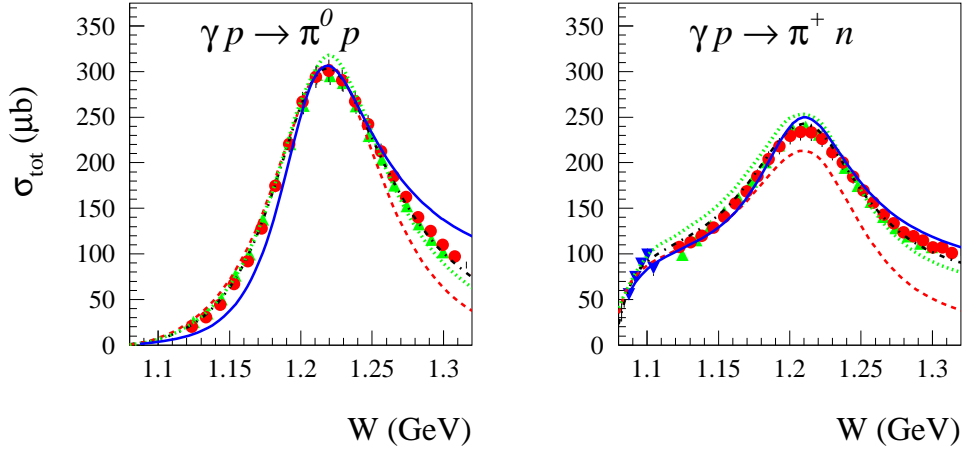}
}
\caption{%
The total cross sections for
$\gamma p \to \pi^0 p$ (left panel) and
$\gamma p \to \pi^+ n$ (right panel)
through the $\Delta(1232)$ resonance region as function of the total 
{\it c.m.} energy $W$. 
Dashed-dotted (black) curves : DMT01 model.
Dashed (red) curves : SL model.
Dotted (green) curves : DUO model.
Solid (blue) curves : NLO \ceft \ calculation.
The data points are from
Refs.~\cite{McPherson64} (inverted blue triangles),
\cite{MacCormick:1996jz} (green triangles),
and \cite{Ahrens:2000bc} (red circles).
}
\label{fig:tot}
\end{figure}
\begin{figure}[bt]
\centerline{ \epsfxsize=12cm%
\epsffile{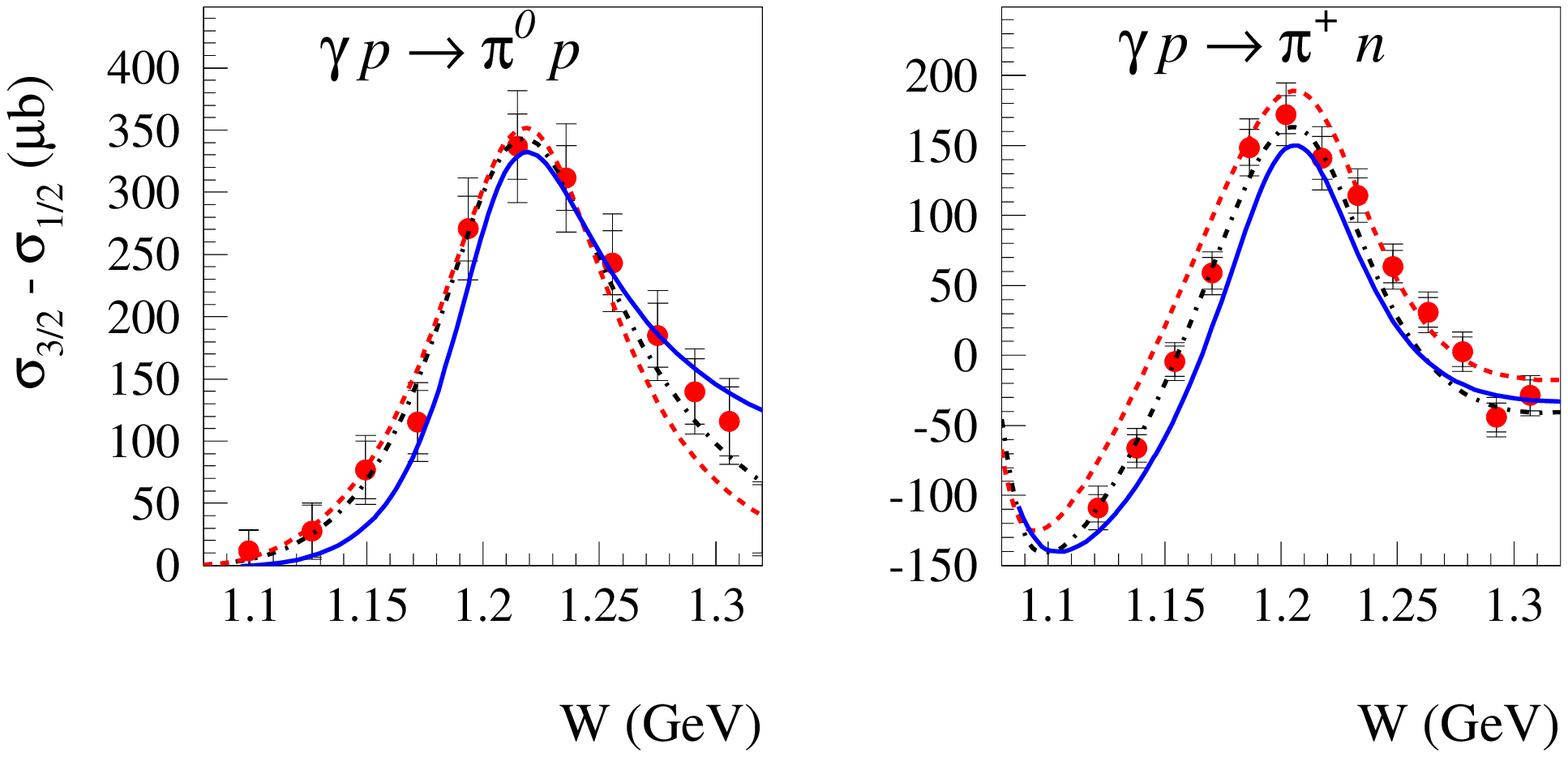}
}
\caption{%
The helicity difference total cross section
$\sigma_{3/2} - \sigma_{1/2}$ for
$\gamma p \to \pi^0 p$ (left panels) and
$\gamma p \to \pi^+ n$ (right panels)
through the $\Delta(1232)$ resonance region.
Curve conventions as in~\Figref{tot}.
The data points are from
MAMI~\cite{Ahrens:2000bc} (red circles).
}
\label{fig:hel}
\end{figure}
\begin{figure}[ht]
\centerline{ \epsfxsize=11cm%
\epsffile{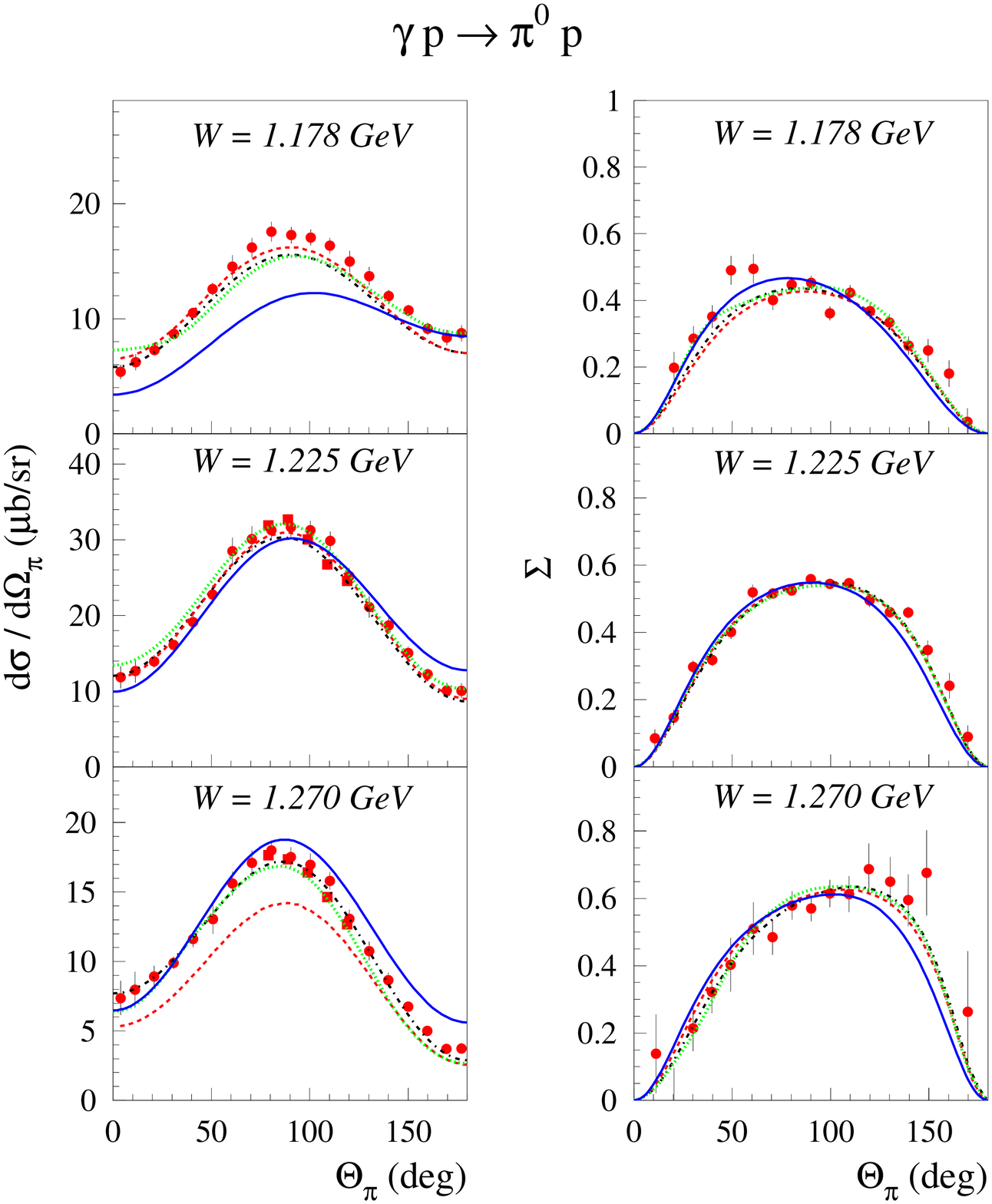}
}
\caption{%
The differential cross section (left panels) and
photon asymmetry (right panels) for $\gamma p \to \pi^0 p$
through the $\Delta(1232)$ resonance region.
Curve conventions as in~\Figref{tot}.
The data points are from
MAMI : Refs.~\cite{Beck:1999ge,Leukel01} (red circles)
and~\cite{Ahrens:2004pf} (red squares).}
\label{fig:pio}
\end{figure}
\begin{figure}[ht]
\centerline{ \epsfxsize=11cm%
\epsffile{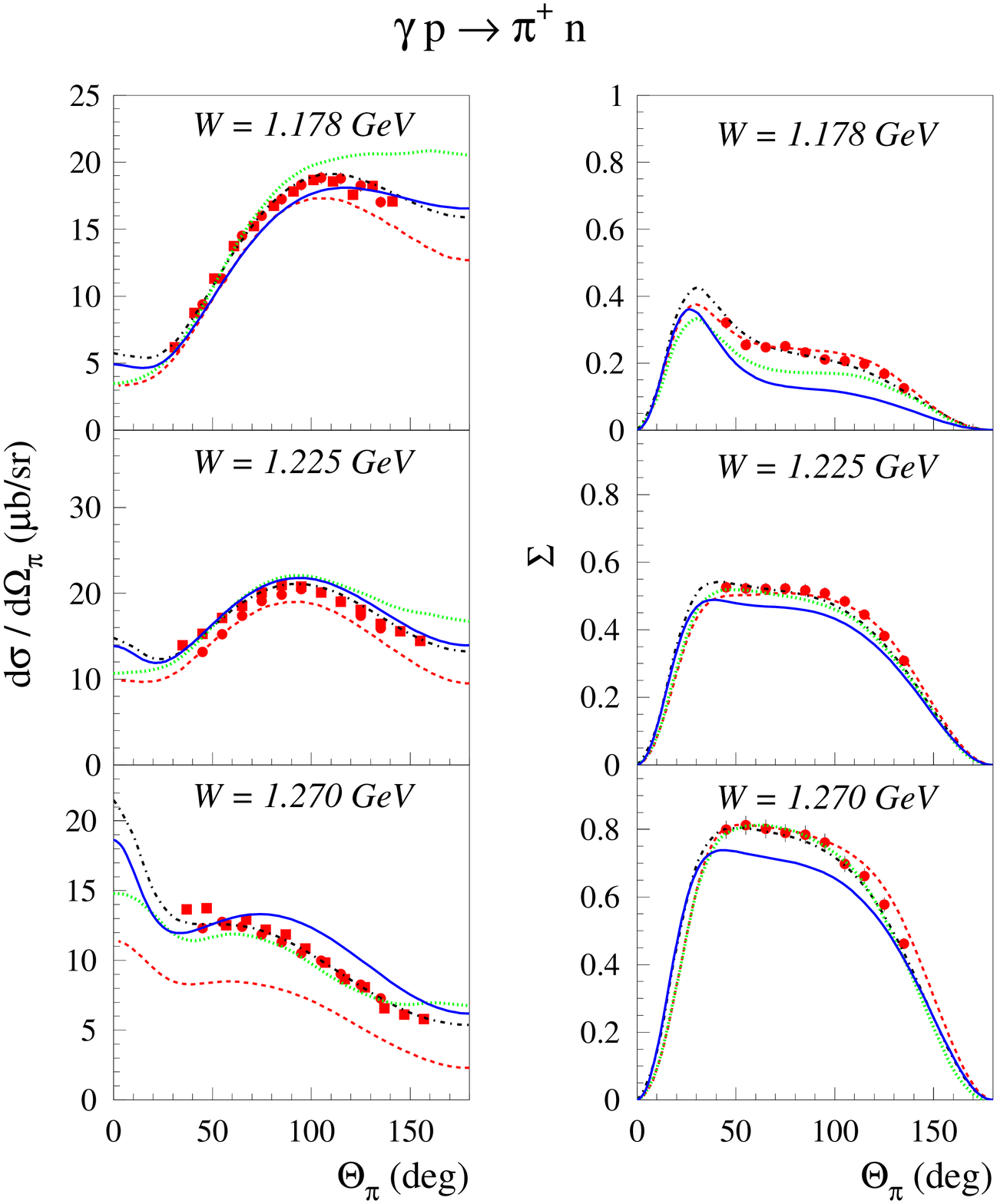}
}
\caption{%
The differential cross section (left panels) and
photon asymmetry (right panels) for $\gamma p \to \pi^+ n$
through the $\Delta(1232)$ resonance region.
Curve conventions as in~\Figref{tot}.
The data points are from
MAMI : Refs.~\cite{Beck:1999ge} (red circles)
and~\cite{Ahrens:2004pf} (red squares).}
\label{fig:pip}
\end{figure}
\begin{figure}[ht]
\centerline{ \epsfxsize=11cm%
\epsffile{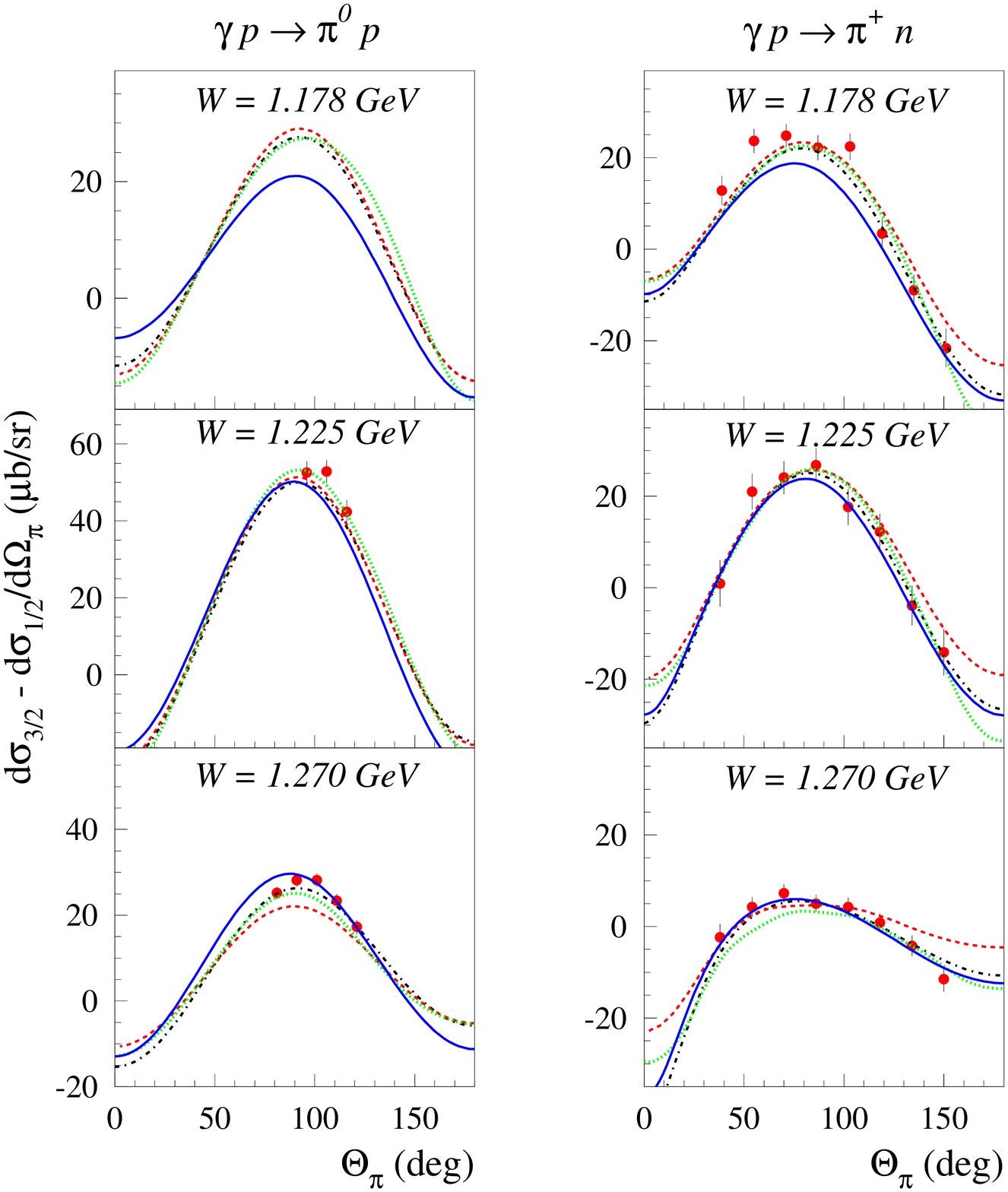}
}
\caption{%
The helicity difference differential cross section
$d \sigma_{3/2} - d \sigma_{1/2}$ for
$\gamma p \to \pi^0 p$ (left panels) and
$\gamma p \to \pi^+ n$ (right panels)
through the $\Delta(1232)$ resonance region.
Curve conventions as in~\Figref{tot}.
The data points are from
MAMI~\cite{Ahrens:2000bc} (red circles).
}
\label{fig:heldiff}
\end{figure}
\begin{figure}[ht]
\centerline{ \epsfxsize=11cm%
\epsffile{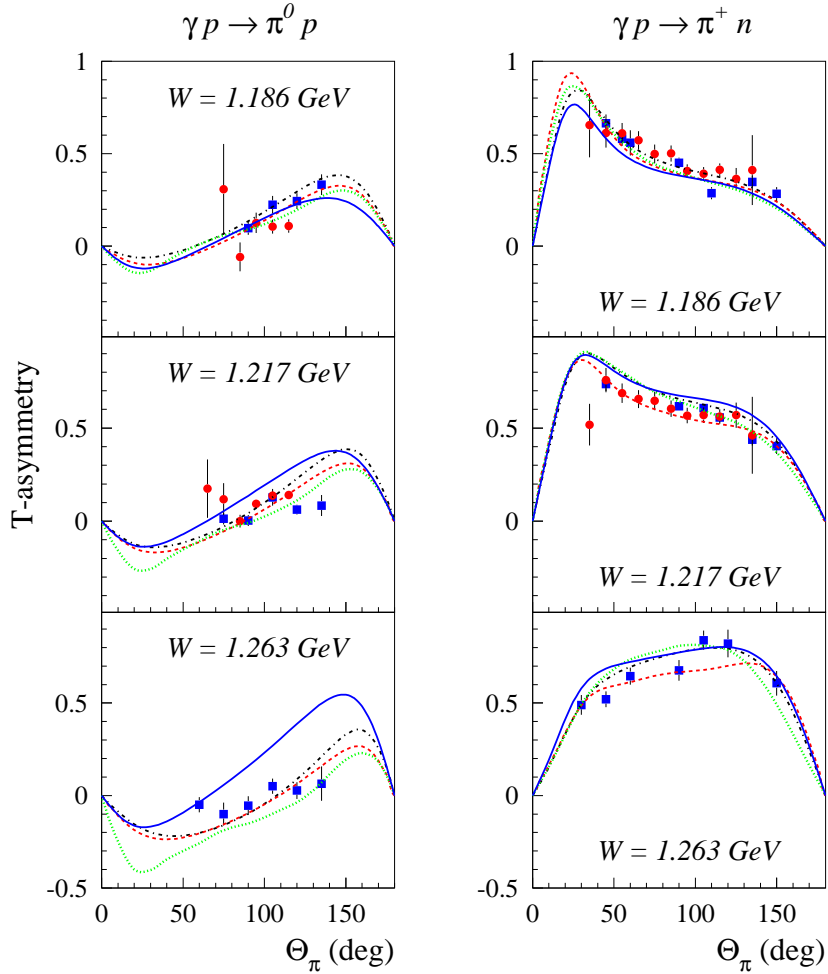}
}
\caption{%
The target asymmetry $T$ for
$\gamma p \to \pi^0 p$ (left panels) and
$\gamma p \to \pi^+ n$ (right panels)
through the $\Delta(1232)$ resonance region.
Curve conventions as in~\Figref{tot}.
The data points for $\gamma p \to \pi^0 p$ are from
Refs.~\cite{Belyaev:1983xf} (blue squares)
and \cite{Bock:1998rk} (red circles).
The data points for $\gamma p \to \pi^+ n$ are from
Refs.~\cite{Getman:1981qt} (blue squares)
and \cite{Dutz:1996uc} (red circles).
}
\label{fig:target}
\end{figure}
\begin{figure}[ht]
\centerline{ \epsfxsize=11cm%
\epsffile{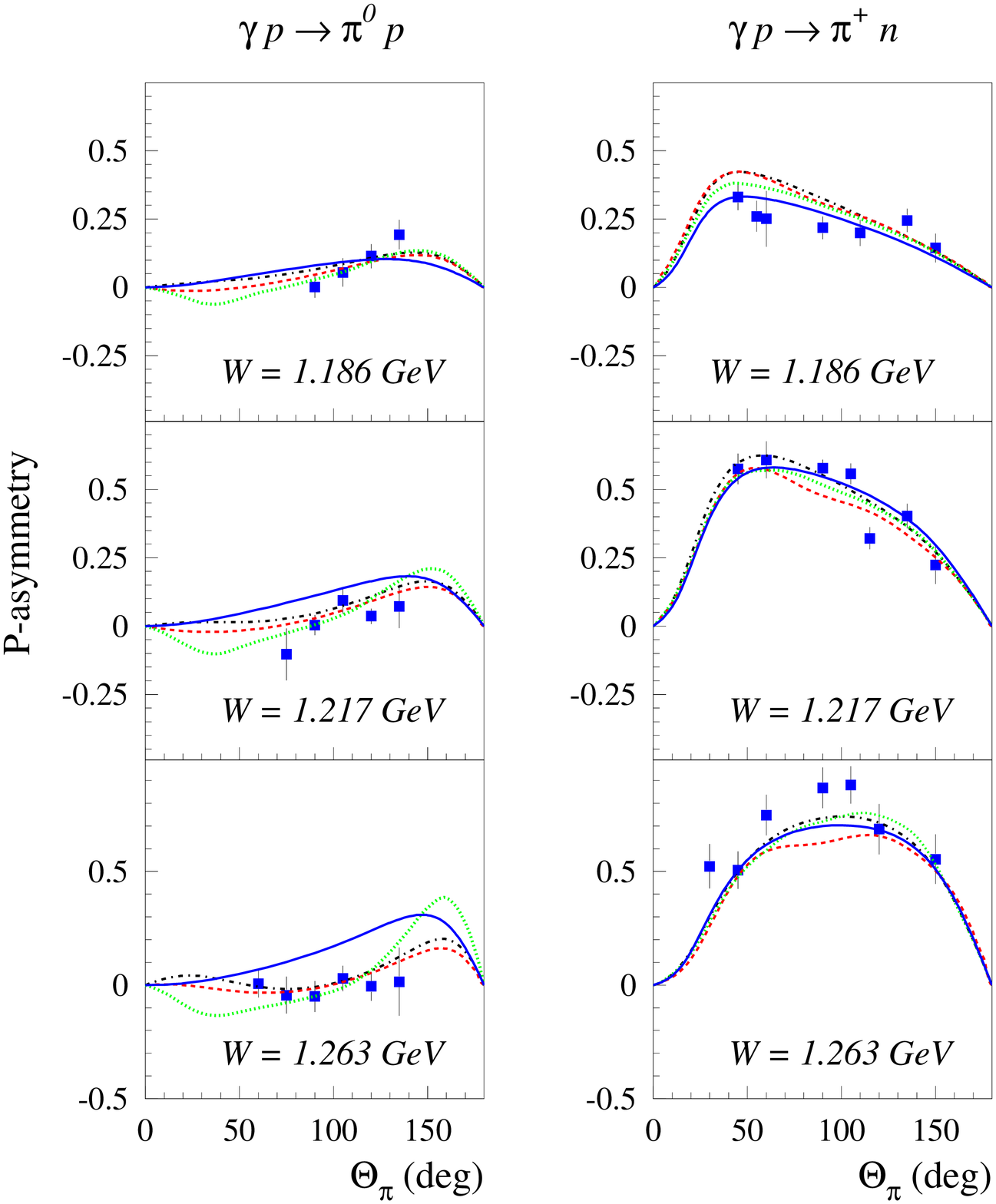}
}
\caption{%
The recoil asymmetry $P$ for
$\gamma p \to \pi^0 p$ (left panels) and
$\gamma p \to \pi^+ n$ (right panels)
through the $\Delta(1232)$ resonance region.
Curve conventions as in~\Figref{tot}.
The data points for $\gamma p \to \pi^0 p$ are from
Ref.~\cite{Belyaev:1983xf}.
The data points for $\gamma p \to \pi^+ n$ are from
Ref.~\cite{Getman:1981qt}.
}
\label{fig:recoil}
\end{figure}
\begin{figure}[ht]
\centerline{ \epsfxsize=12cm%
\epsffile{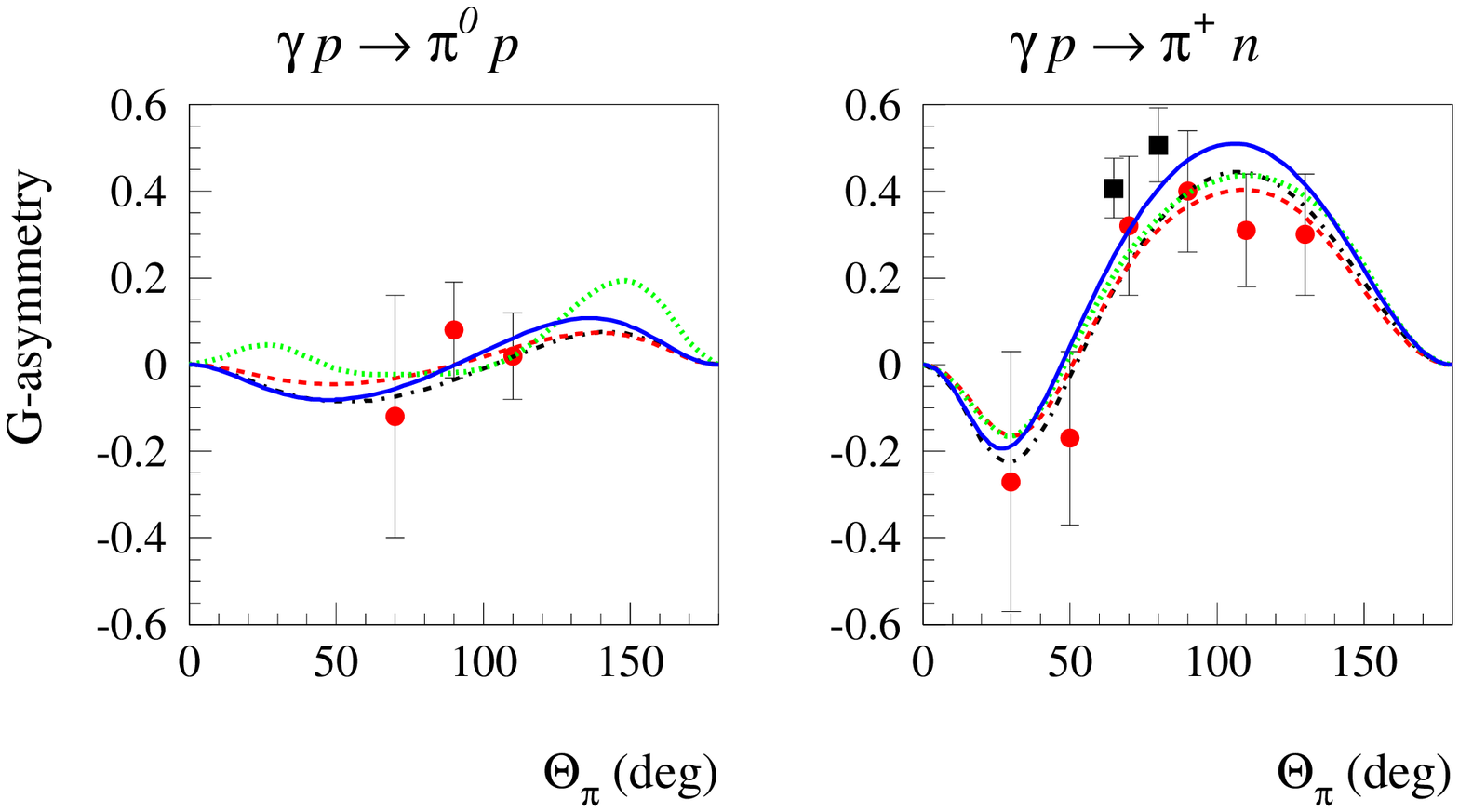}
}
\caption{%
The double polarization asymmetry $G$ for
$\gamma p \to \pi^0 p$ (left panel) and
$\gamma p \to \pi^+ n$ (right panel) at $W = 1.232$~GeV.
Curve conventions as in~\Figref{tot}.
The data points are from
Kharkov~\cite{Belyaev:1985sp} (black squares) and from
MAMI~\cite{Ahrens:2005zq} (red circles) .
}
\label{fig:gasymm}
\end{figure}
\begin{figure}[ht]
\centerline{ \epsfxsize=12cm%
\epsffile{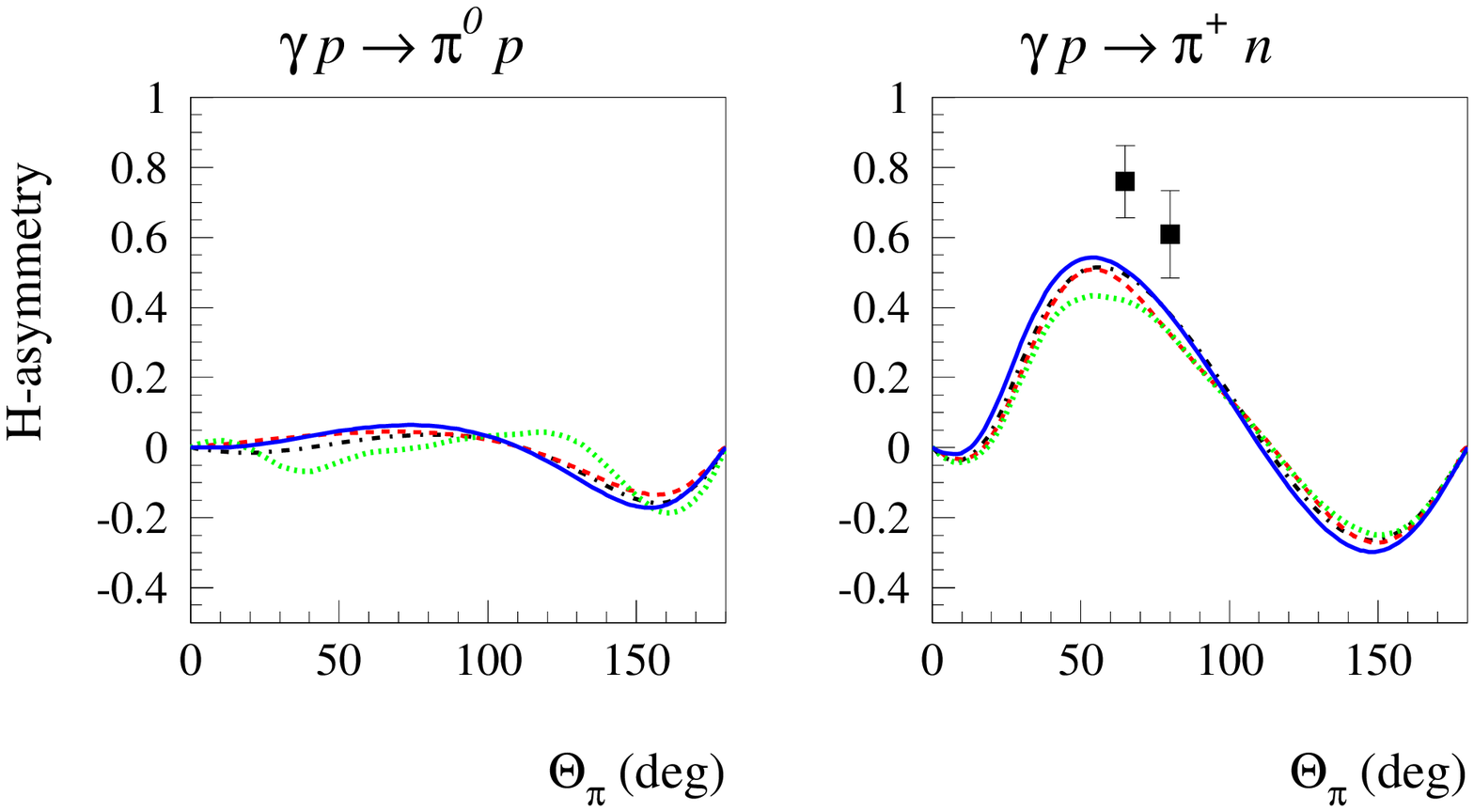}
}
\caption{%
The double polarization asymmetry $H$ for
$\gamma p \to \pi^0 p$ (left panel) and
$\gamma p \to \pi^+ n$ (right panel) at $W = 1.232$~GeV.
Curve conventions as in~\Figref{tot}.
The data points are from
Kharkov~\cite{Belyaev:1985sp} (black squares).
}
\label{fig:hasymm}
\end{figure}
\newline
\indent
In the DMT and DUO calculations, the 
best fit to the multipoles obtained in the recent analyses of
the Mainz~\cite{Hanstein:1997tp} and VPI groups~\cite{VPI97} is shown in
\Figref{figM1E1}. 
One sees that both models are able to describe the energy dependence of 
the resonant multipoles very accurately throughout the 
$\Delta$-resonance region. 
\newline
\indent
Interpreting the results of the dynamical models, one finds that 
a sizeable part of the $M1$ $\gamma N \Delta$ strength and 
almost all of the $E2$ strength is generated 
by the $\pi N$ rescattering.  
One sees from Table~\ref{table_helicity-amp} that the bare values
for the helicity amplitudes obtained in both SL and DMT, as given
within the parentheses, amount to only about $60\%$ of the
corresponding dressed values, and are close to the predictions of the
constituent quark model, as was first pointed out in Ref.~\cite{SL}. 
The large reduction of the helicity amplitudes from the
dressed to the bare ones results from the fact that the principal
value integral part of  \Eqref{tmult}, which represents the
effects of the off-shell pion rescattering, contributes
approximately for half of the $M_{1+}^{(3/2)}$, as seen in 
\Figref{DMT_M1E1} for the DMT model. 
The large pion rescattering effects 
obtained in the dynamical models are generally interpreted as an
important manifestation of the presence of the pion cloud in the
nucleon. 
In \ceft, the size of the
$\pi N$ contribution to the $M1$ and $E2$ transitions is quite different
from the dynamical models, due to the 
renormalization of the LECs $g_M$ and $g_E$ at the resonance 
position for $Q^2 = 0$.
The renormalized $\pi N$ loop contributions then 
satisfy power counting and as a consequence contribute substantially less than 
in dynamical models.
\newline
\indent
The resonant multipoles $M_{1+}^{(3/2)}$ 
and $E_{1+}^{(3/2)}$ obtained in the \ceft \ calculation at 
NLO are shown in
\Figref{gap_pin_m1mult} and \Figref{gap_pin_e1mult}, respectively.
The imaginary part of these multipoles
arise from purely the $\De$ contibutions, see the upper graphs in 
\Figref{diagrams}.
 The Born graphs (2nd line of \Figref{diagrams})
contribute to the real part of the resonant multipoles, 
as well as to the non-resonant multipoles as shown in
\Figref{gap_pin_nonresmult}. Up to NLO in the $\delta$-expansion, 
the non-resonant multipoles are thus purely real, whereas
the corresponding $\pi N$ phase shifts are zero, in accordance
 with the Fermi-Watson.
\newline
\indent
The predictions of \ceft\ and the dynamical model
calculations of SL, DMT, and DUO for the total and differential cross
sections, single polarizations including beam and target asymmetry
and recoil polarization, and double polarizations $G$ and $H$ for
both $\ga p\ra\pi^0 p$ and $\ga p\ra\pi^+ n$ are shown in 
Figs.~\ref{fig:tot}-\ref{fig:hasymm}.
\newline
\indent
For the $\gamma p \to \pi^0 p$ and $\gamma p \to \pi^+ n$ processes, 
one sees from Figs.~\ref{fig:tot}-\ref{fig:hasymm} that below the resonance 
the dynamical models are in overall agreement with the available 
data. Above the resonance, the SL model starts to fall 
increasingly below the $\pi^0 p$ and in particular the 
$\pi^+ n$ cross section data, see 
Figs.~\ref{fig:tot}, \ref{fig:pio}, and \ref{fig:pip}. 
For the DUO model at higher energies, one observes more structure 
in the $\pi^0 p$ $P, T, G$, and $H$ observables for forward and backward 
angles, where however the data base is scarce. 
The DMT model shows good agreement with all $\pi^0 p$ and $\pi^+ n$ 
observables throughout the $\Delta$-energy range. 
\newline
\indent
The presented NLO \ceft \ calculations (solid blue curves) are done in the 
scheme designed to work in a small energy domain around the resonance, 
as discussed in Sect.~\ref{sec4}. 
It is therefore of interest to see how far away from the resonance such 
calculations actually work. 
One sees from Figs.~\ref{fig:tot}-\ref{fig:hasymm} 
that the NLO calculation reproduces well the pion angular dependencies 
of the $\pi^0 p$ and $\pi^+ n$ observables around resonance. 
Below the resonance, 
the \ceft \ results for the $\pi^0 p$ cross sections, and to 
lesser extent for the $\pi^+ n$ cross sections, start to fall 
below the data. It was checked that this is mainly due to the absence  
of the anomalous magnetic couplings in the Born diagrams and in the 
$\pi N$ loop corrections to the $\gamma N \Delta$ vertex. 
Such effects are coming 
in at NNLO in the $\delta$-expansion. For energies about 50 MeV or higher 
above resonance, the NLO begins to overestimate the $\pi^0 p$ 
cross sections, and to a lesser extent the $\pi^+ n$ cross sections, 
which also shows up in larger deviations from the data 
for $\Sigma$ (for $\pi^+n$) and $T$ and $P$ (for $\pi^0 p$). 
Nevertheless, given the simplicity of the NLO calculations in the 
$\delta$-expansion and its potential for systematic improvement, one 
can be pleased with the results. 
Overall, the \ceft \ to NLO in the $\delta$-expansion provides a 
simple and elegant description of pion photoproduction observables in a 
100 MeV window around the $\Delta$-resonance. 
The precision of the available data can be exploited to study the 
higher order effects in the $\gamma N \Delta$ transition 
using calculations beyond NLO.

\subsection{Pion electroproduction}
\label{sec53}

The pion production by electrons was first observed by Panofsky and
his collaborators at Stanford in 1955. Even though it was first
attempted at Cornell in 1965, it was not until in the beginning of
1970's that double-arm or coincidence experiments, in which the
scattered electron and one of the recoiling hadrons are measured
in coincidence, were actively pursued. A summary of the
coincidence experimental results on pion electroproduction, up
until early 1980's can be found in the 
extensive review of Ref.~\cite{Foster83}.
\newline
\indent
Extensive sets of high statistics data of pion electroproduction
have been collected at BATES, MAMI, and JLab in the last several
years. The majority of them are on the $ep\ra e'p\pi^0$ reaction.
They cover a large range of invariant mass $W$, from threshold up to
$2.5$~GeV and a wide range in momentum-transfer squared
$Q^2=0.05 - 6$~GeV$^2$, and the full coverage of pion polar and azimuthal
angles. Single and double polarization
measurements have also been performed. Single polarization
experiments are carried out with polarized electron beams to
measure the beam helicity asymmetry. Double polarization experiments
are performed with a polarized electron beam together with a polarized
target or with recoil nucleon polarization measurements. Data on the
threshold $\pi^0$ production are measured mostly at low $Q^2$
region in order to test the chiral perturbation theory. They have been 
obtained prominently at MAMI, see  Ref.~\cite{Merkel02} and references
contained therein.
\newline
\indent
Data on the $\pi^+ n$ channel were much less abundant in the past. Most
of them are in the mass range of the first and second resonances and
at forward pion {\it c.m.} angles, with some at backward angles. 
As in the case of photoproduction, data on the $\pi^- p$ channel from 
a neutron (deuterium) target is even more scarce. Currently, CLAS at JLab
has an active programs to measure cross sections for the $\pi^+ n$
\cite{DeVita02} and $\pi^- p$ \cite{Klimenkao06} channels.
\newline
\indent
Measurements made before the early 1980's are summarized in Tables 5-7
in Ref.~\cite{Foster83}. The new electroproduction data in the
resonance region, measured recently at BATES, ELSA, JLab, and
MAMI, up until 2003 are given in Table 2 in Ref.~\cite{Burkert:2004sk}.
\newline
\indent
Electroproduction of neutral pions on a proton target has been employed most
extensively for the purpose of determining the $Q^2$ evolution of
the $R_{EM}$ and $R_{SM}$ ratios. 
Many experiments have been performed at BATES, MAMI,
and JLab. At BATES, the experiments have been set up for 
central invariant mass $W = 1.232$~GeV and 
$Q^2 = 0.127$~GeV$^2$. With the use of either
unpolarized or polarized electron beams and choice of specific
kinematics, quantities like induced polarization, and $\sigma_0,
\sigma_{LT}, \sigma_{TL'},$ and $\sigma_{TT}$ have been measured
\cite{Warren98,Mertz:1999hp,Kunz:2003we,Sparveris:2004jn}. Both
unpolarized and polarized experiments are done at MAMI
\cite{Pospischil:2000ad,Bartsch02,Elsner:2005cz,Stave:2006ea} with
$Q^2$ ranges from a low value of $0.06$~GeV$^2$ \cite{Stave:2006ea}
to $0.2$~GeV$^2$. In \cite{Pospischil:2000ad}, the recoil proton
polarization was measured in the double-polarized $p(\vec e,e'\vec
p)\pi^0$ reaction and all three proton polarization components
were measured simultaneously. All three halls at JLab, Hall A
\cite{Laveissiere04,Kelly05}, Hall B (CLAS)
\cite{Joo:2001tw,Joo02b,Ungaro:2006df}, and Hall C
\cite{Frolov:1998pw}, have programs on $\pi^0$ electroproduction
with either unpolarized or polarized electrons and a large variety
of observables have been measured.  The experiment of 
Ref.~\cite{Ungaro:2006df} measured $Q^2$ values as high as $6.0$~GeV$^2$. 
The aim of such large $Q^2$ measurements is to map out the transition 
towards the pQCD regime. The large amount of
precision data have been used to extract the $N\ra\Delta(1232)$
transition form factors from $Q^2 =0$ to 6~GeV$^2$. They also
provide very stringent tests for theoretical models.
\begin{figure}
\centerline{  \epsfxsize=11cm
  \epsffile{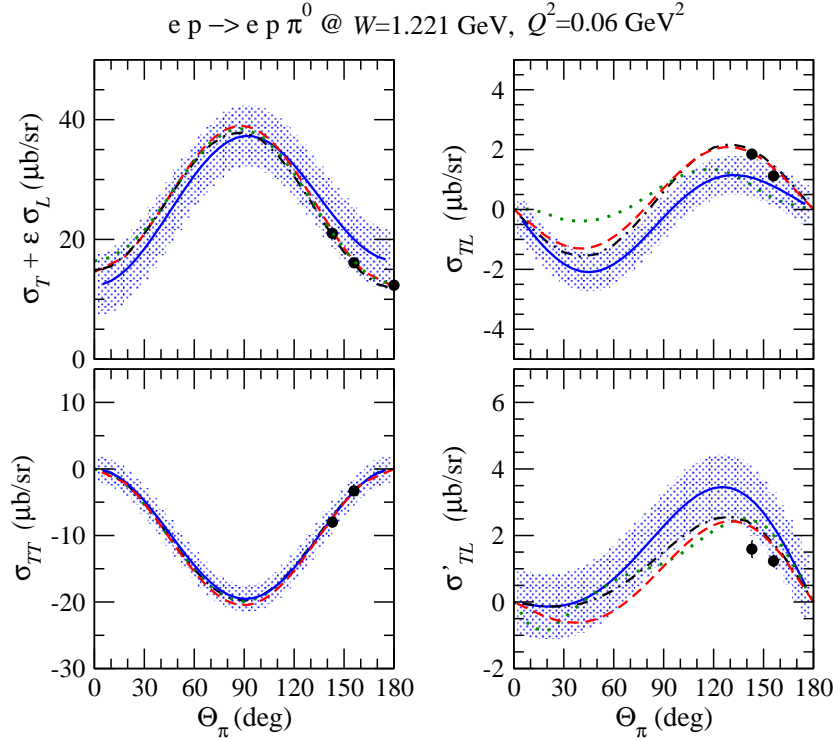} 
}
\caption{The pion angular dependence 
of the $\ga^\ast p \to \pi^0 p$ cross sections at 
$W = 1.221$~GeV and $Q^2$ = 0.06~GeV$^2$. 
Dashed-dotted (black) curves : DMT01 model.
Dashed (red) curves : SL model.
Dotted (green) curves : DUO model.
Solid (blue) curves : \ceft \ results of 
Refs.~\cite{Pascalutsa:2005ts,Pascalutsa:2005vq}. 
The bands provide an estimate of the theoretical error for the \ceft \ 
calculations. 
Data points are from the MAMI experiment of 
Ref.~\cite{Stave:2006ea}.
}
\label{fig:epio_cross1}
\end{figure}
\begin{figure}
\centerline{  \epsfxsize=11cm
  \epsffile{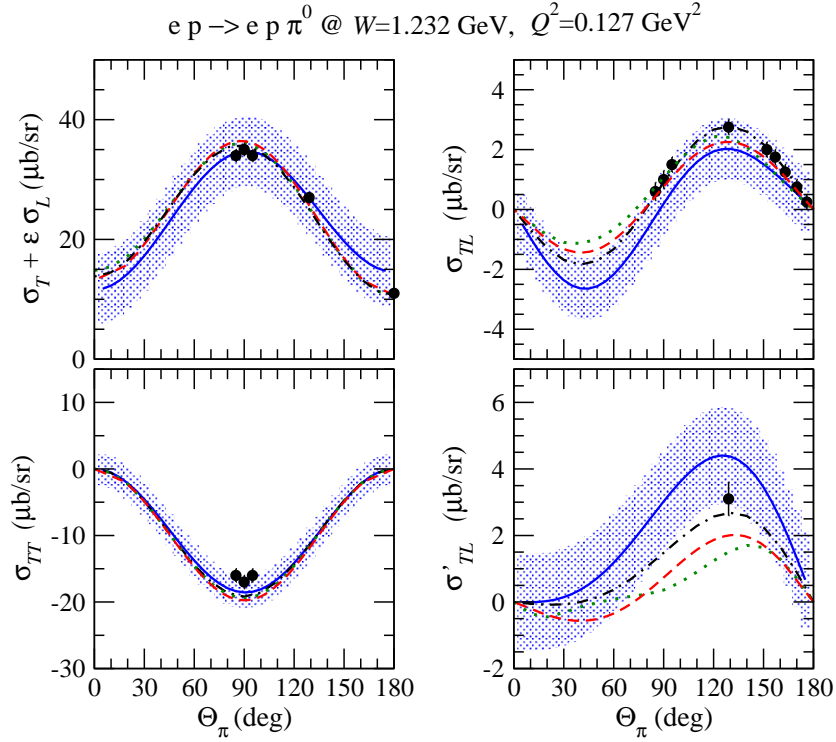} 
}
\caption{The pion angular dependence 
of the $\ga^\ast p \to \pi^0 p$ cross sections at 
$W = 1.232$~GeV and $Q^2$ = 0.127~GeV$^2$. 
Curve conventions as in \Figref{epio_cross1}. 
The bands provide an estimate of the theoretical error for the \ceft \ 
calculations. 
Data points are from BATES 
experiments~\cite{Mertz:1999hp,Kunz:2003we,Sparveris:2004jn}.
}
\label{fig:epio_cross2}
\end{figure}
\begin{figure}
\centerline{  \epsfxsize=11cm
  \epsffile{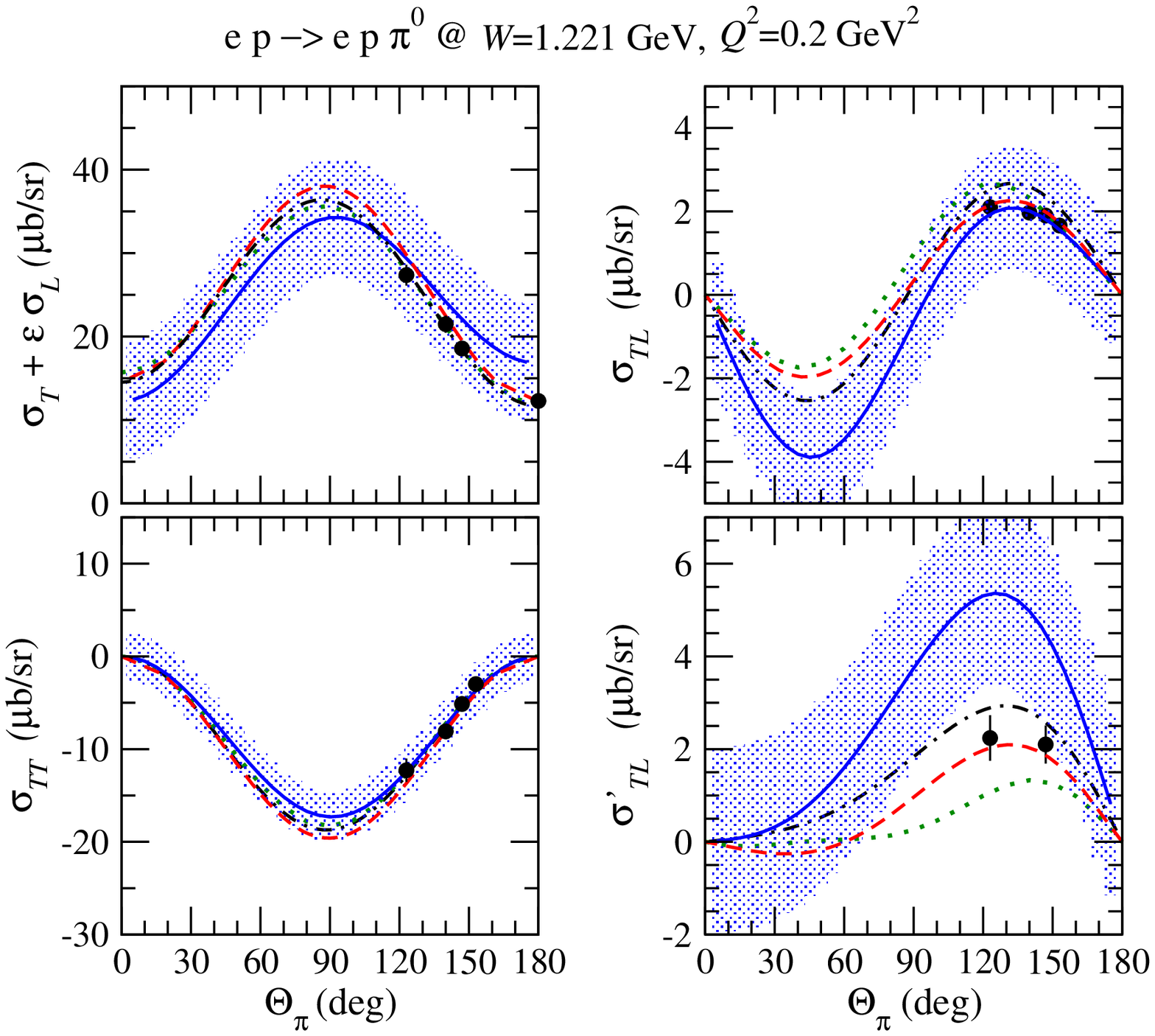} 
}
\caption{The pion angular dependence 
of the $\ga^\ast p \to \pi^0 p$ cross sections at 
$W = 1.221$~GeV and $Q^2$ = 0.20~GeV$^2$. 
Curve conventions as in \Figref{epio_cross1}. 
The bands provide an estimate of the theoretical error for the \ceft \ 
calculations. 
Data points are from the MAMI experiment of 
Ref.~\cite{Sparveris:2006}.
}
\label{fig:epio_cross3}
\end{figure}
\newline
\indent
In Figs.~\ref{fig:epio_cross1}-\ref{fig:epio_cross3}, 
the different virtual photon absorption cross sections 
around the resonance position are displayed 
at three different $Q^2$ values: $Q^2 = 0.06, 0.127, 0.20$~GeV$^2$, 
where recent precision data are available. 
We compare these data with the predictions made within the \ceft \ 
framework~\cite{Pascalutsa:2005ts,Pascalutsa:2005vq} and using the 
SL, DMT, and DUO dynamical models. 
\newline
\indent
In the \ceft \ calculations, 
the low-energy constants $g_M$ and $g_E$, were fixed 
from the resonant pion photoproduction multipoles.  
Therefore, the only other low-energy constant from the chiral Lagrangian  
entering the NLO calculation is $g_C$. 
The main sensitivity on $g_C$ enters in $\sigma_{TL}$. 
A best description of the $\si_{TL}$ data at low $Q^2$ 
is obtained by choosing $g_C = -2.6$. 
In Figs.~\ref{fig:epio_cross1}-\ref{fig:epio_cross3}, 
we also include as an illustration the theoretical uncertainty 
of the NLO \ceft \ result, estimated using \Eqref{Oerror}, where 
the average is taken over the range of  $\Theta_\pi$ 
\footnote{Note that $\si_{TL}$ and $\si_{TL'}$ do not receive any LO
contributions and therefore the LO$=0$ case in \Eqref{Oerror}
must be applied in the estimate.}. 
From Figs.~\ref{fig:epio_cross1}-\ref{fig:epio_cross3}, one 
sees that the NLO \ceft\ calculation, within its accuracy,
is consistent with the experimental data for these observables at 
low $Q^2$. Only for $\sigma_{TL'}$, the \ceft\ calculation overpredicts the 
data with increasing $Q^2$. This is because this observable is sensitive to 
the background multipoles $E_{0+}$ and $S_{0+}$ which are not very accurately 
reproduced in the NLO \ceft\ calculation, see \Figref{gap_pin_nonresmult}.   
\newline
\indent
The dynamical models are in basic agreement with each other and the data 
for the transverse cross sections. Differences between the models do show 
up in the $\sigma_{TL}$ and $\sigma_{TL'}$ cross sections which involve 
the longitudinal amplitude. In particular for $\sigma_{TL'}$ the differences 
reflect to large extent how the non-resonant $S_{0+}$ multipole is 
described in the models. 
\begin{figure}[tbp]
\centerline{ \epsfxsize=11cm%
\epsffile{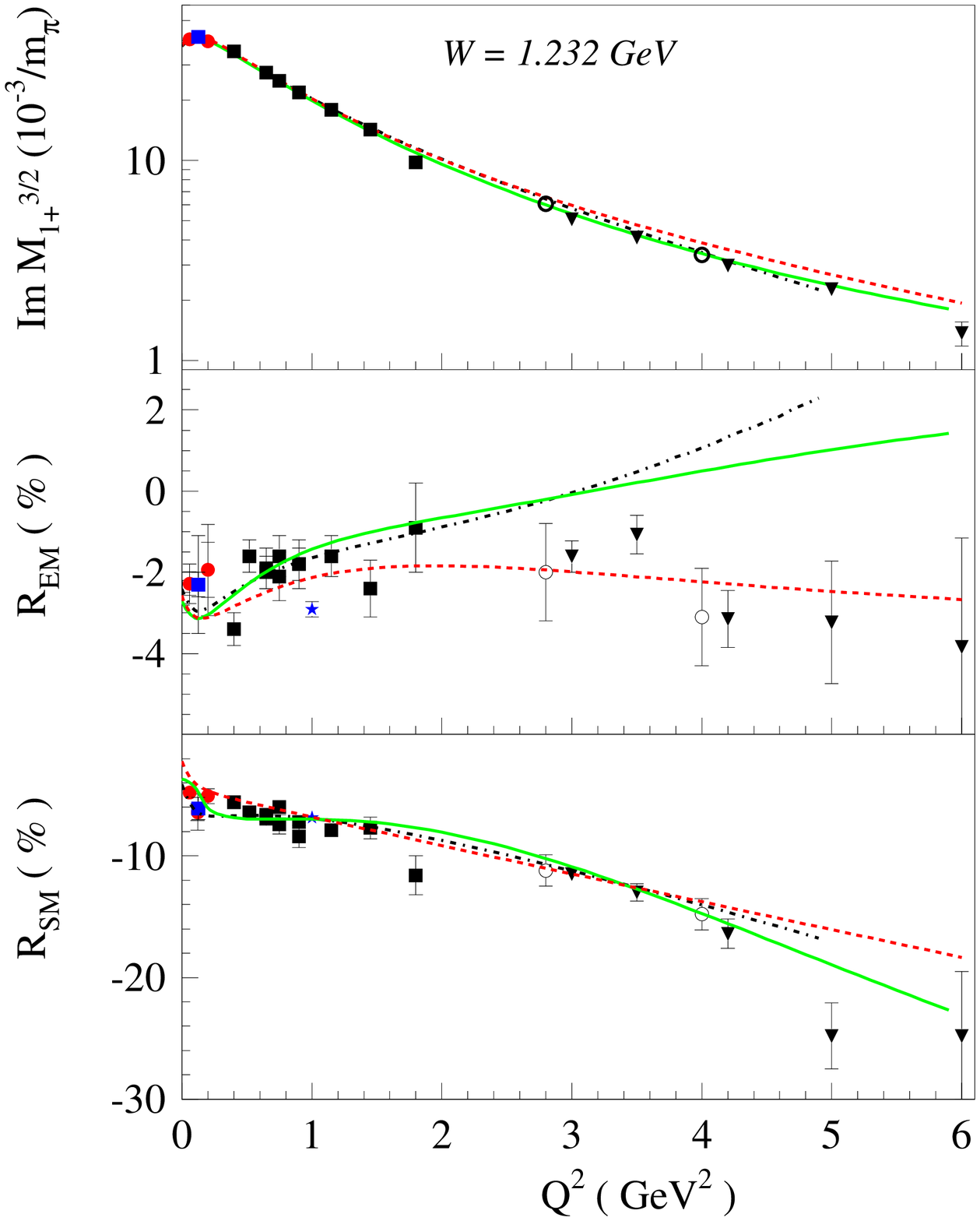}
}
\caption{$Q^2$ dependence of the resonant multipoles at $W = 1.232$~GeV. 
Upper panel : imaginary part of $M_{1+}^{(3/2)}$;  
middle panel : $R_{EM}$; 
lower panel : $R_{SM}$. 
The curves are the results obtained in different dynamical models~:
DMT01 (dashed-dotted black curves); 
SL (dashed red curves); 
DUO (solid green curves).
The data are from 
BATES~\cite{Sparveris:2004jn} (blue squares),    
MAMI~\cite{Beck:1999ge,Stave:2006ea,Sparveris:2006} (red circles),  
JLab/Hall C~\cite{Frolov:1998pw} (open circles), 
JLab/CLAS~\cite{Joo:2001tw} (black squares) and \cite{Ungaro:2006df} 
(inverted triangles),  
and JLab/Hall A~\cite{Kelly05} (blue stars).
}
\figlab{mult_largeq}
\end{figure}
\newline
\indent
In \Figref{mult_largeq}, we compare the data for the resonant multipoles 
up to the largest measured values of $Q^2$ with the 
results obtained for the three dynamical models 
(SL, DMT, DUO) reviewed in this work. 
One sees that all three models give a very accurate account of the 
dominant $M_{1+}^{(3/2)}$ multipole. They are also in good agreement with 
the $R_{SM}$ ratio, which grows to a value of around $- 25$ \% around 
$Q^2 \simeq 6$~GeV$^2$. 
Only for the $R_{EM}$ ratio, the three dynamical models 
show clear differences. While in the DMT and DUO models the 
$R_{EM}$ ratio shows a zero crossing around $Q^2 \simeq 3$~GeV$^2$, 
the SL result is in good agreement with the data, which give small negative 
values for $R_{EM}$ (around $-4$ \%) at the largest measured values 
of $Q^2 \simeq 6$~GeV$^2$. As remarked in Sect.~\ref{sec2_pqcd}, the data are 
very far from the pQCD prediction 
for $R_{EM}$ which should approach $+100$~\% at asymptotically large values 
of $Q^2$. The DMT calculation~\cite{KY99} 
finds nearly as large values for the 
subdominant (according to pQCD) helicity amplitude $A_{3/2}$ as for 
the pQCD dominant amplitude $A_{1/2}$, at accessible values 
of $Q^2$. In both the calculations and the data, one therefore  
sees no onset of hadron helicity conservation in the $\gamma N \Delta$ 
transition up to $Q^2 \simeq 6$~GeV$^2$.  

%% file: chap6_delta.tex
\section{Beyond the one-photon exchange approximation}
\label{sec6}

In all the up-to-date studies 
of the $\Delta$ excitation by electrons ($e N \to e \Delta$), 
as discussed in the previous sections, 
the electromagnetic interaction between the electron and the nucleon 
is assumed to be mediated by a single photon exchange (\OPE). 
In this Section, we discuss effects beyond this approximation, 
in particular the effects due to two-photon exchange (\TPE). 
The spectacular discrepancy between the 
polarization-transfer~\cite{Jones:1999rz,Gayou:2001qt,Gayou:2001qd}
and Rosenbluth-separation measurements~\cite{Arrington:2003df} 
of the nucleon elastic form factors is largely due to  \TPE-exchange 
effects~\cite{Guichon:2003qm,BMT03,YCC04}.  
Subsequently, the \TPE \ corrections to the 
$e N \to e \Delta$ process were studied in Ref.~\cite{Pascalutsa:2005es},  
where these \TPE \ corrections 
were evaluated in a parton model. 
Recently, the \TPE \ corrections to $e N \to e \Delta$ were studied 
in a hadronic model~\cite{Kondratyuk:2006ig}. 
In what follows, we will review the general formalism for the 
\TPE \ corrections to $e N \to e \Delta$ and discuss their effect 
within the partonic calculation of Ref.~\cite{Pascalutsa:2005es}.

\subsection{General formalism}

When describing the $e N \to e \Delta$ process, 
the total number of helicity amplitudes is 32, and is reduced to 16 by 
using parity invariance. 
Furthermore, in a gauge theory lepton helicity 
is conserved to all orders in perturbation 
theory when the lepton mass is zero. Neglecting the lepton mass,    
reduces the number of helicity amplitudes to 8. 
Recall that in the \OPE\ approximation, there are only three independent
helicity amplitudes which can be expressed in terms of 
three form factors of the $\ga^* N\De$ transition as discussed in 
Sect.~\ref{sec2_def}. 
\newline
\indent
The $\gamma^* N \Delta$ transition form factors are  
usually studied in the $e N \to e \pi N$ process in the
$\De$-resonance region. In the \OPE\ approximation, 
the $\gamma^* N \to \pi N$ virtual photon absorption cross section 
$d \sigma_v / d \Omega_\pi$ was presented in Sect.~\ref{sec3}. 
Beyond the \OPE\ approximation, one can still formally define a 
cross section $d \sigma_v / d \Omega_\pi$ from the experimentally 
measured $e N \to e \pi N$ five-fold differential cross sections 
using Eq.~(\ref{eq:diff1}), with momentum transfer $q$ defined as 
$q \equiv k - k^\prime$.  
However, beyond the \OPE \ approximation 
$\Gamma$ does not have the physical interpretation any more of a 
virtual photon flux factor. 
The expression of the thus defined cross section 
$d \sigma_v / d \Omega_\pi$ will then also be modified from 
the \OPE\ expression of Eq.~(\ref{eq:diff2}). 
For unpolarized nucleons, the cross section at $W = M_\Delta$ can, 
in general, be parametrized as (using the notations introduced in 
Sect.~\ref{sec3})~:
\begin{eqnarray}
\frac{d \sigma_v}{d \Omega_\pi}&=& \sigma_{0}    
+ \varepsilon \, \cos (2 \phi_\pi) \, \sigma_{TT} 
+\, h \, \varepsilon \, \sin (2 \phi_\pi) \, \sigma^\prime_{TT} \nonumber \\
&+&\sqrt{2 \varepsilon}\,\veps_+ \cos \phi_\pi \, \sigma_{TL}
+ h  \sqrt{2 \varepsilon }\, \veps_- \sin \phi_\pi \, \sigma^\prime_{TL} , \ \ 
\label{eq:crossv6}
\end{eqnarray}
with $h = \pm 1$ the lepton helicity and 
$\varepsilon_\pm \equiv  \sqrt{1 \,\pm\, \varepsilon}$.
\newline
\indent
It is convenient to multipole expand Eq.~(\ref{eq:crossv6}) 
for the $e p \to e \Delta^+ \to e \pi N$ process ({\it i.e.} keeping only 
the $\Delta$ multipoles) as~:
\begin{eqnarray}
\sigma_0
&=& A_0 \,    
+ \half (3 \cos^2 \theta_\pi - 1) \, A_2 ,
 \nn \\
\sigma_{TT}
&=& \sin^2 \theta_\pi \, C_0 , \quad \quad \quad \quad
\sigma_{TL}
= \half  \sin(2 \theta_\pi) \, D_1,
\nn \\
\sigma^\prime_{TT}
&=& \sin^2 \theta_\pi \, C^\prime_{0} , \quad \quad \quad \quad
\sigma^\prime_{TL} = \half   \sin(2 \theta_\pi) \, D^\prime_{1},
\label{eq:crossv7}
\end{eqnarray}
where $A_0$ can 
be written as: 
\begin{eqnarray}
A_{0 } = {\mathcal I} \, \frac{e^2}{4 \pi} \frac{Q_-^2}{4 \, M_N^2} 
\frac{(M_\Delta + M_N)}{(M_\Delta - M_N)} \frac{1}{M_\Delta \, \Gamma_\Delta} 
(G_M^*)^2\, \sigma_R,
\label{eq:aom}
\end{eqnarray} 
where ${\mathcal I}$ denotes the isospin factor which depends on the 
final state in the $\Delta^+ \to \pi N$ decay as~: 
${\mathcal I}(\pi^0 p) = 2/3$ and ${\mathcal I}(\pi^+ n) = 1/3$, 
and where $\sigma_R$ denotes the reduced cross section $\sigma_R$,  
including \TPE\ corrections.
\newline
\indent
We are now in position to 
discuss the corrections to $R_{EM}$ and $R_{SM}$ 
as extracted from $\sigma_{TT}$ and $\sigma_{TL}$. 
Experimentally, these ratios have been extracted 
at $W = M_\Delta$ using~:
\begin{eqnarray}
R_{EM}^{exp,I} &=& \frac{3 A_2 - 2 C_0}{12 A_0}
	\stackrel{1\gamma}{=}  
	R_{EM}  + \varepsilon \, \frac{4 M_\Delta^2 Q^2}{Q_+^2\, Q_-^2} R_{SM}^2
	+ \ldots
	\label{eq:rem1a} \nonumber \\
R_{SM}^{exp} &=& \frac{Q_+ \, Q_-}{Q \, M_\Delta} \frac{D_1}{6 A_0}
	\stackrel{1\gamma}{=} R_{SM} - R_{SM} R_{EM} + \ldots 
\label{eq:rsm1}
\end{eqnarray}
where the omitted terms involve cubic products of $R_{EM}$ and $R_{SM}$.  
These formulas are usually applied~\cite{Joo:2001tw} 
by neglecting the smaller quantities 
$R_{SM}^2$ and $R_{EM} \cdot R_{SM}$. 
Ref.~\cite{Pascalutsa:2005es}, 
however, kept these quadratic terms, 
and proposed a second method of extracting $R_{EM}$: 
\begin{eqnarray}
R_{EM}^{exp,II} = \frac{-(A_0 - A_2) - 2 \, C_0}{ 3(A_0 - A_2) - 2 \, C_0 }
&\stackrel{1\ga}{=}  R_{EM} . 
\label{eq:rem1b} 
\end{eqnarray}
This method apparently avoids corrections at the one-photon level,
which may prove to be significant at larger momentum transfer
due to an appreciable contribution of the $R_{SM}^2$ term.
\newline
\indent
We denote the corrections to $R_{EM}$ and $R_{SM}$  by~:
\begin{eqnarray}
R \,&\simeq&\, R^{exp} + \delta R^{1 \gamma} 
+ \delta R^{2 \gamma} ,
\end{eqnarray}
where $\delta R^{1 \gamma}$ 
denotes the corrections due to the quadratic terms 
in Eqs.~(\ref{eq:rem1a}),(\ref{eq:rem1b}), which are~: 
\begin{eqnarray}
\delta R_{EM}^{1 \gamma, I} &=& 
- \varepsilon \, \frac{4 \, M_\Delta^2 \, Q^2}{Q_+^2 \, Q_-^2} \,
R_{SM}^2 ,
\nonumber \\
\delta R_{EM}^{1 \gamma, II} &=& 0 , \\
\delta R_{SM}^{1 \gamma} &=& R_{EM} \cdot R_{SM} .\nn
\end{eqnarray}
\begin{figure}[t]
\centerline{ \epsfxsize=10cm  
\epsffile{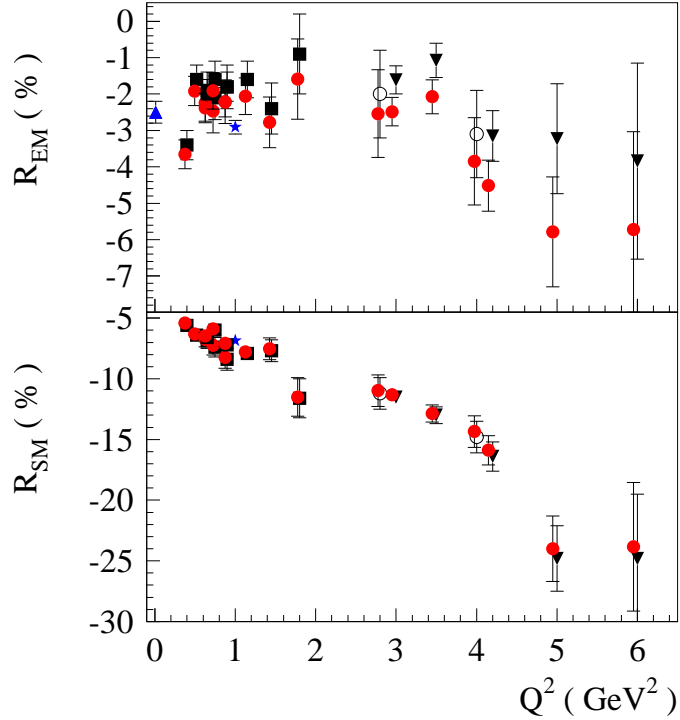} 
}
\caption{
\label{fig:remrsm_qsqr1ga}
$Q^2$ dependence of the \OPE\ corrections, $\delta R^{1 \gamma}$, to 
$R_{EM}$ (upper panel) and $R_{SM}$ (lower panel) at $W = 1.232$~GeV. 
The uncorrected data are from JLab/Hall C~\cite{Frolov:1998pw} (open circles), 
JLab/CLAS~\cite{Joo:2001tw} (solid squares) and \cite{Ungaro:2006df} 
(inverted triangles).  
The error bars reflect the statistical error only.   
The red circles include the corrections $\delta R_{EM}^{1 \gamma, I}$.  
The triangle at $Q^2 = 0$ is the real photon point from~\cite{Beck:1999ge}, 
whereas the star at $Q^2 = 1$~GeV$^2$ is the extraction from the 
JLab/Hall A recoil polarization experiment~\cite{Kelly05}, which 
provides an independent way to extract $R_{EM}$ and $R_{SM}$.
}
\end{figure}
\newline
\indent
In Fig.~\ref{fig:remrsm_qsqr1ga}, we show the effect of the \OPE\ 
corrections on $R_{EM}$ and $R_{SM}$. We see that its  
effect on $R_{SM}$ is negligible, whereas it yields a systematic 
downward shift of the $R_{EM}$ result. This shift becomes more pronounced 
at larger $\varepsilon$, and it is found that at the highest 
$Q^2$ values 
it decreases $R_{EM}$ by around 2 \%. To avoid such a correction, it calls 
for extracting $R_{EM}$ according to the procedure $II$ as we outlined above. 
Alternatively, double polarization experiments provide a very useful 
cross check on the extraction of the $R_{EM}$ and $R_{SM}$ ratios. 
In Ref.~\cite{Kelly05}, the angular distributions of 14 recoil polarization 
response functions and two Rosenbluth combinations have been measured 
for $e p \to e p \pi^0$ at $Q^2 = 1$~GeV$^2$. The analysis 
of Ref.~\cite{Kelly05} avoids making  
a multipole truncation. In particular for $R_{EM}$, one 
sees from Fig.~\ref{fig:remrsm_qsqr1ga} that 
it yields a value which is lower than the one of Ref.~\cite{Joo:2001tw} based 
on a multipole truncation, in agreement with the trend observed when applying 
the correction of Eq.~(\ref{eq:rem1a}).
\newline
\indent
In the following, we briefly describe the partonic estimate 
of Ref.~\cite{Pascalutsa:2005es} for  
the \TPE \ contribution to the $N \to \Delta$ 
electroproduction amplitudes at large $Q^2$.

\subsection{Partonic calculation of two-photon exchange effects}

\begin{figure}[t]
\centerline{  \epsfxsize=6cm
  \epsffile{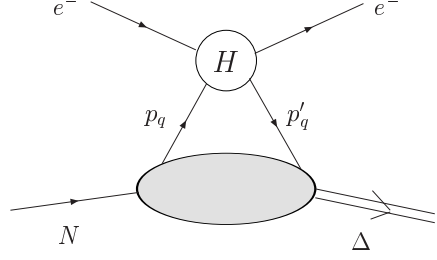} 
}
\caption{Handbag approximation for the $e N \to e \Delta$ process  
 at large momentum transfers. In the partonic scattering process 
($H$), the lepton scatters from quarks with momenta $p_q$ and $p'_q$. 
The lower blob represents the $N \to \Delta$ GPDs.}
\label{fig:handbag}
\end{figure}

In the partonic framework, illustrated in Fig.~\ref{fig:handbag},  
the \TPE\ amplitude is related 
to the $N \to \Delta$ GPDs, as discussed in Sect.~\ref{sec2_gpd}. 
This calculation  involves a hard scattering subprocess on a quark, 
which is embedded in the proton by means of
the $N \to \Delta$ GPDs. 
Besides the dominant vector GPD $H_M$, corresponding with the magnetic 
dipole $\gamma N \Delta$ transition considered in Sect.~\ref{sec2_gpd}, 
it also involves the leading axial-vector GPD, denoted by $C_1$,  
which are the leading GPDs in the large $N_c$ 
limit~\cite{Frankfurt:1999xe} contributing to this transition.  
Analogously to the relation of Eq.~(\ref{eq:gmgegcsumrule}) relating the 
GPD $H_M$ to the $\gamma^* N \Delta$ magnetic dipole form factor $G_M^\ast$, 
one can relate the GPD $C_1$ to the $\gamma^* N \Delta$ axial transition 
form factor $C_{5}^{A}$, introduced by Adler~\cite{Adl75}.
\newline
\indent
The $ e N \to e \Delta$ handbag amplitude can then be specified in
terms of the following two characteristic integrals~\cite{Pascalutsa:2005es}~:
\begin{eqnarray}
A^* &=& \int_{-1}^1 \frac{dx}{x}  \    
	\left[ \frac{\hat s - \hat u}{Q^2} g_M^{\,hard} + g_A^{(2\gamma)}\right]
	\sqrt{\frac{2}{3}}   \,\frac{1}{6} \, H_M	\,, 
			 \\
C^* &=& \int_{-1}^1 \frac{dx}{x} \ \left[ \frac{\hat s - \hat u}{Q^2} g_A^{(2\gamma)} + g_M^{\,hard}\right]
	\, \mathrm{sgm}(x) \,\frac{1}{6}\,  C_1 ,	\ \ \  
\label{eq:ac}
\end{eqnarray}
where all hard scattering quantities in the square brackets are given in 
Ref.~\cite{YCC04}. 
\newline
\indent
In terms of the above integrals $A^*$ and $C^*$, the reduced cross section 
of Eq.~(\ref{eq:aom}) including \TPE \ corrections, can be expressed 
as~\cite{Pascalutsa:2005es}~:
\begin{eqnarray}
\sigma_R &=& 1 + 3 \, R_{EM}^2 + \varepsilon \, 
\frac{16 \, M_\Delta^2 \, Q^2}{Q_+^2 \, Q_-^2} \, R_{SM}^2 
\nn \\
&+&\, \frac{1}{G_M^*}  \sqrt{\frac{2}{3}}  
\left[ \frac{A^*}{2}   
	\frac{Q^2 \varepsilon_+ \varepsilon_- }{Q_+ \, Q_-}  
+ 2 \, C^*  \frac{Q^2}{Q_-^2}  \varepsilon_-^2 \, 
\frac{M_N}{M_N + M_\Delta} \right].
\label{eq:crossred} 
\end{eqnarray}
Furthermore, the \TPE \ exchange corrections to $R_{EM}$ 
and $R_{SM}$, are~:
\begin{eqnarray}
\delta R_{EM}^{2 \gamma, I} &=& - 
\frac{1}{8} \sqrt{\frac{3}{2}} \frac{Q^2}{Q_+ \, Q_-} \, 
\frac{\varepsilon_-^3 \, \varepsilon_+}{\varepsilon} \, 
\frac{1}{G_M^*} \, A^*  
+ \frac{1}{4} \sqrt{\frac{2}{3}} \, \frac{Q^2}{Q_-^2} \, 
\frac{\varepsilon_-^2 \, \varepsilon_+^2}{\varepsilon} \, 
\frac{M_N}{(M_N + M_\Delta)} \, \frac{1}{G_M^*} \, C^* \, ,
\nonumber \\
\delta R_{EM}^{2 \gamma, II} &=& 2 \;
\delta R_{EM}^{2 \gamma, I} ,
\nonumber \\
\delta R_{SM}^{2 \gamma} &=& 
\sqrt{\frac{2}{3}} \, \frac{(Q^2 - M_\Delta^2 + M_N^2)}{4 \, M_\Delta^2} \, 
\frac{Q_+}{Q_-} \, \frac{1}{\sqrt{2 \, \varepsilon}}
\frac{\varepsilon_-^2}{\varepsilon_+} \, \frac{M_N}{(M_N + M_\Delta)} 
\, \frac{1}{G_M^*} \, C^* \, .
\end{eqnarray}
\indent
To provide numerical estimates for the $2 \gamma$ corrections, 
one needs a model for the $N \to \Delta$ GPDs
which appear in the integrals $A^*$ and $C^*$.   
In Ref.~\cite{Pascalutsa:2005es}, the large
\( N_{c} \) relations discussed in Sect.~\ref{sec2_gpd} were used.  
For the GPD $H_M$, the model of Sect.~\ref{sec:gpdmodel} is used, whereas  
the axial GPD $C_1$ is expressed through the isovector combination 
$C_{1}(x, 0 , Q^2) =  
\sqrt{3} [ \tilde{H}^{u} - \tilde{H}^{d} ](x, 0 , Q^2)$ 
of the nucleon axial GPDs, see Ref.~\cite{Frankfurt:1999xe}.
\begin{figure}[t]
\centerline{ \epsfxsize=8.5cm  
\epsffile{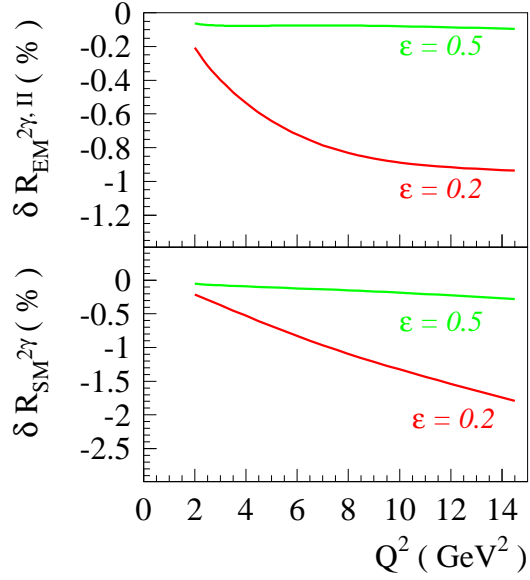} 
}
\caption{\label{fig:remrsm_qsqr2ga}  
$Q^2$ dependence of the \TPE\ corrections  
to $R_{EM}$ (upper panel) and $R_{SM}$ (lower panel) 
at $W = 1.232$~GeV and for different values of $\varepsilon$. 
Figure from Ref.~\cite{Pascalutsa:2005es}. 
}
\end{figure}
\newline
\indent
In Fig.~\ref{fig:remrsm_qsqr2ga} we show the \TPE \ corrections to 
$R_{EM}$ and $R_{SM}$  
estimated using the modified Regge GPD model. 
We see that the \TPE \ effects are mainly pronounced at small $\varepsilon$ 
and larger $Q^2$. For $R_{EM}$ they are well below 1 \%, whereas they yield 
a negative correction to $R_{SM}$ by around 1 \%, when $R_{SM}$ is extracted 
from $\sigma_{TL}$ according to Eq.~(\ref{eq:rsm1}). 
\begin{figure}[t]
\centerline{  \epsfxsize=10cm
  \epsffile{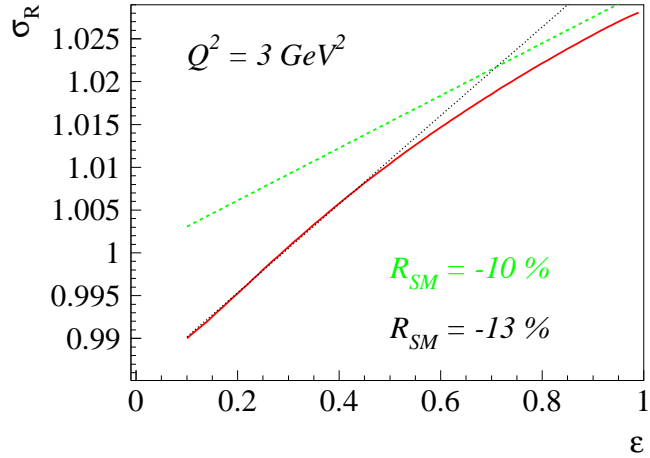} 
}
\vspace{-0.5cm}
\caption{\label{fig:remrsm3} 
Rosenbluth plot of the reduced cross section $\sigma_R$ 
of Eq.~(\ref{eq:crossred}) for the $e N \to e \Delta$ 
reaction at $Q^2 = 3$~GeV$^2$.
The dashed curve corresponding with $R_{SM} = -10 \%$  
is the \OPE \ result, 
whereas the solid curve represents the result including 
\TPE \ corrections for the same values of $R_{SM}$. 
The dotted curve corresponding with  $R_{SM} = -13 \%$ 
corresponds with a linear fit to the total result in an intermediate 
$\varepsilon$ range. Figure from Ref.~\cite{Pascalutsa:2005es}.  
}
\end{figure}
\newline
\indent
Another way to obtain $R_{SM}$ is from a Rosenbluth-like analysis of
the cross section $\sigma_0$ for the $e N \to e \Delta$ reaction.  In
Fig.~\ref{fig:remrsm3}, we show the reduced cross section from
Eq.~(\ref{eq:crossred}). One sees how a different extraction of
$R_{SM}$ can yield a significantly different result.  Starting from a
value of $R_{SM}$ (e.g., --10 \%) extracted from $\sigma_{TL}$, adding
in the 2$\gamma$-exchange corrections gives a sizable change in the
slope of the Rosenbluth plot.  When fitting the total result by a
straight line in an intermediate $\varepsilon$ range, one extracts a
value of $R_{SM}$ around  3 percentage units lower than  its value as
extracted from $\sigma_{TL}$. This is sizable, as it corresponds with
a 30\% correction on the absolute value of $R_{SM}$. 
This is similar using the Rosenbluth method to extract the 
proton elastic form factor ratio $G_E^p/G_M^p$,  
where the 2$\gamma$-corrections were found to be large~\cite{Guichon:2003qm}.
\newline
\indent
Note that obtaining $R_{SM}$ from the $D_1/A_0$ ratio depends on
isolating the $\cos \phi_\pi$ term in the cross section.  This term
implicitly uses the decay $\Delta \to N\pi$ to give us some information
about the polarization of the $\Delta$.  The ``$D_1/A_0$'' method is
thus the analog of the polarization transfer measurement of $G_E/G_M$
in the elastic case.  Like the elastic case, the calculations of 
Ref.~\cite{Pascalutsa:2005es} show that there are
sizeable corrections to the Rosenbluth-like determination of $R_{SM}$, and
significantly smaller corrections if $R_{SM}$ is obtained from
$D_1/A_0$.   It will therefore 
be interesting to confront the $D_1/A_0$ extraction of $R_{SM}$ with the new 
Rosenbluth separation data in the $\Delta$ region up to $Q^2 \simeq
5$~GeV$^2$ which are presently under analysis~\cite{Tvaskis:2005ex,Tvaskis}. 

%% file: chap7_delta.tex
\section{Conclusions and outlook}
\label{sec7}

In this work we have been reviewing the recent progress 
in understanding the nature of the $\Delta$-resonance
and of its electromagnetic excitation.
\newline
\indent
On the experimental side, 
high-precision data for low-$Q^2$ pion electroproduction 
in the $\Delta$-resonance region have become available 
from BATES, MAMI and JLab. As a result of these measurements, a 
consistent picture has emerged for the small quadrupole ratios 
$R_{EM}$ and $R_{SM}$. The electric quadrupole over magnetic dipole 
ratio $R_{EM}$ is around $-2$~\%. 
The Coulomb quadrupole over magnetic dipole ratio $R_{SM}$ is around 
$-6$~\% in the $Q^2$ range up to about 1~GeV$^2$, and becomes more negative 
with increasing values of $Q^2$. 
The measurements at large $Q^2$ show no trend towards the pQCD predictions. 
Up to the largest measured $Q^2$ values 
around 6 GeV$^2$, the dominant magnetic dipole $\gamma^* N \Delta$ 
form factor decreases faster than the pQCD scaling prediction $\sim 1/Q^4$, 
and the $R_{EM}$ ratio shows no trend towards a zero crossing 
to reach the pQCD prediction of +100 \%. 
\newline
\indent
The small value of the $R_{EM}$ ratio has been 
understood in the large $N_c$ limit where the $R_{EM}$ ratio is of 
order $\sim 1/N_c^2$. The large $N_c$ limit leads to
the spin-flavor symmetry which is used in many quark models. 
In such quark models the small non-zero values of $R_{EM}$ and 
$R_{SM}$ are interpreted as indicators of hadron deformation, because they 
require $d$-wave components in the $N$ and/or $\Delta$-wave functions. 
The constituent quark models yield however too small (absolute) 
values for the $R_{EM}$ and $R_{SM}$ ratios, and underpredict the 
$M1$ $N \to \Delta$ transition by about 25 \%. A natural way to 
improve on such quark models is to include pion degrees of freedom. 
A variety of models have shown that pionic effects 
to the $\gamma N \Delta$ transition are indeed sizeable.  
In particular the strong tensor interaction due to one-pion exchange 
between quarks may yield to a simultaneous spin flip of two quarks in the 
nucleon and yield non-zero $E2$ and $C2$ amplitudes even if 
the quark wave functions have no $d$-state admixture.  
\newline
\indent
The effect of pionic degrees of freedom in the $\ga N\De$ transition
has been discussed within the context of dynamical models. We have
reviewed in detail three examples of such models and compared their
results. These dynamical models yield a good description of experimental
data and provide a model-dependent framework for extracting the
$\ga N\De$ form factors.
\newline
\indent
A systematic framework to account for the pion loop effects in the 
$N \to \Delta$ transition has emerged in recent years in terms of
the chiral effective field theory. 
Namely, the extensions of chiral perturbation theory to the
$\Delta$(1232)-resonance energy region have been developed.  
The $\de$-expansion is based on 
the small parameter $\de$ equal to the excitation
energy of the resonance over the chiral symmetry-breaking scale.
The other low-energy scale of the theory, the pion mass,
counts as $\de^2$, which is crucial for an adequate counting
of the $\De$-resonance contributions in both the low-energy
and the resonance energy regions.  
This framework has been applied to the process of
pion electroproduction. 
A complete calculation of this process in the resonance region up to,
and including, next-to-leading order in the  $\de$-expansion 
satisfies  gauge and chiral symmetries perturbatively, and
Lorentz-covariance, analyticity, unitarity (Watson's theorem) exactly.
The low-energy constants (LECs) entering at this order
are the $\ga N\De$ couplings $g_M$, $g_E$, $g_C$ characterizing
the $M1$, $E2$,  $C2$ transitions, respectively. 
They are extracted from a fit to the pion photoproduction data, 
and for $g_C$ to the pion electroproduction at low $Q^2$. 
Once these LECs are fitted, both the energy ($W$) and $Q^2$ 
dependencies follow as predictions. The NLO 
results were found to give a good description of both the 
energy-dependence around $\Delta$-resonance 
and the low $Q^2$ dependence of the pion electroproduction data. 
\newline
\indent
We have also reviewed the recent progress achieved in the 
lattice QCD simulations of the $\gamma N \Delta$ transition. 
They now clearly establish a non-zero negative value for the 
$R_{EM}$ and $R_{SM}$ ratios. However the present lattice 
calculations have been obtained for pion masses well above 300 MeV. 
A naive linear extrapolation in the quark mass to its physical value, 
yields a value for $R_{SM}$ which is however in disagreement with 
experiment. It has been reviewed that 
the \ceft\ framework  plays a  dual role
in that it allows for
an extraction of resonance parameters from observables and 
predicts their pion-mass dependence. 
It was seen that the opening of the 
$\De\to \pi N$ decay channel at $m_\pi = M_\De-M_N$
induces a pronounced  non-analytic
behavior of the $R_{EM}$ and $R_{SM}$ ratios. 
While the linearly-extrapolated lattice QCD results 
for $R_{SM}$ are in disagreement with experimental data, the 
\ceft\ prediction of the non-analytic dependencies suggests that
these results are in fact consistent with experiment.
\newline
\indent
The quark structure of the 
electromagnetic $N \to \Delta$ transition can be accessed in 
hard scattering processes such as the deeply 
virtual Compton scattering processes $\gamma^* N \to \gamma \Delta$, 
where the virtual photon $\gamma^*$ carries a large virtuality. 
The non-perturbative information in this process can be 
parametrized in terms of $N \to \Delta$ generalized parton distributions. 
The $N \to \Delta$ transition form factors are obtained as first moments 
of these GPDs. In the large $N_c$ limit, the dominant magnetic GPD is 
related to an isovector combination of nucleon GPDs. 
Using this large $N_c$ relation combined with 
a Regge parametrization for the nucleon GPDs, 
fitted to the elastic nucleon form factors, 
the magnetic dipole GPD was calculated. Its first moment yields a prediction 
for the $Q^2$ dependence for the $M1$ $\gamma^* N \Delta$ form factor. 
This prediction is very well in agreement with the 
recent data up to the highest measured $Q^2$ value of around 6~GeV$^2$.   
This calculation also shows that the $M1$ $\gamma^* N \Delta$ form factor  
follows approximately 
the same fall-off as the nucleon (isovector) Pauli form factor, 
yielding a faster fall-off at large $Q^2$ compared with the $1/Q^4$ 
pQCD prediction.  Likewise, we have used a large $N_c$ relation 
between the low $Q^2$ $R_{EM}$ ratio and the neutron electric form factor 
to calculate the electric quadrupole $N \to \Delta$ GPD. 
The resulting prediction for $R_{EM}$ is in good agreement with the data 
up to around 1.5~GeV$^2$. 
\newline
\indent
At large momentum transfer, we discussed  
the \TPE \ contribution to the 
$e N \to e \Delta(1232) \to e \pi N$ process.  
The \TPE \ amplitude has been related in a partonic model 
to the $N \to \Delta$ GPDs. 
For $R_{EM}$, the \TPE \ corrections were found to be small, 
below the 1 \% level. However, the neglect of a quadratic term  
$R_{SM}^2$ in a truncated multipole expansion to extract $R_{EM}$ 
yields corrections at the 1-2 \% level.    
The \TPE \ corrections to $R_{SM}$ on the other hand are 
substantially different when extracting this quantity 
from an interference cross section or from unpolarized cross sections, 
as has been observed before for the elastic $e N \to e N$ process.   
They turn out to be important when extracting $R_{SM}$ at large $Q^2$ 
from the unpolarized longitudinal cross section using a 
Rosenbluth separation. 
\newline
\indent
We end this review by spelling out a few open problems and challenges 
(both theoretical and experimental) in this field~:

\begin{enumerate}

\item {\it Issues in $\chi$EFT:}
\begin{itemize}
\item[a)]
To interpret the precise data on pion photo- and 
electroproduction at low $Q^2$ with minimal model assumptions, 
requires \ceft\ calculations to go beyond making predictions of 
$\gamma^* N \Delta$ form factors 
and directly perform the comparison at the level of  the observables. 
The \ceft\ to NLO in the $\delta$-expansion power counting 
scheme have shown to yield an overall satisfying description 
of a large number of pion photo- and electroproduction observables. 

\item[b)]
The \ceft \ calculations in the $\delta$-expansion have 
been obtained at NLO for the $\gamma^* N \to \pi N$ process. 
The present theoretical uncertainties of these calculations are larger  
than the precision of present data. 
The next step for the theory is to go beyond a NLO calculation. 

\item[c)]
The \ceft \ calculation also allows to make predictions for Compton 
and virtual Compton scattering processes in the $\Delta$ region. 
A comparison of the NLO predictions for $\gamma^* N \to \pi N$, real 
and virtual Compton scattering may yield a powerful cross-check 
of the theory.

\item[d)]
In the low $Q^2$ region, it will be very worthwhile 
to measure the complementary channels 
in particular virtual Compton scattering in the $\Delta$ region.

\item[e)]
The M1 excitation of the $\Delta$ gives a large (para)magnetic contribution 
to the magnetic polarizability of the nucleon 
( $\approx 6 \cdot 10^{-4}$~fm$^3$ ). 
The challenge is to reconcile 
this large contribution with the small value observed in experiment. 

\item[f)]
At present, \ceft \ calculations are performed in different power counting 
schemes. A challenge it to combine the low-energy chiral expansions around 
threshold with the $\delta$-expansion around $\Delta$-resonance. 
 \end{itemize}

\vspace{.5cm} 
\item {\it Resonances on the lattice}:
\begin{itemize}
\item[]
Lattice QCD calculations with dynamical quarks have just started for the 
electromagnetic $N \to \Delta$ form factors. 
They are currently performed for pion masses above 300 MeV where the 
$\Delta$ is stable. 
The \ceft \ shows a strong non-analytic behavior in the quark mass 
for the $\gamma^* N \Delta$ form factors when the 
pion mass becomes smaller than the $N - \Delta$ mass difference.  
It will be a challenge to implement the unstable nature of the 
$\Delta$ on the lattice and to extend these 
calculations to pion mass values sizeably below 300 MeV. 
\end{itemize}

\vspace{.5cm}
\item {\it Development of dynamical models:}
\begin{itemize}
\item[]
Building in symmetries in a consistent way is a non-trivial challenge for 
the dynamical models. Future challenges in this field are 
to find ways to systematically improve on 
approximations, identify the limitations, and quantify the 
theoretical uncertainties.
\end{itemize}

\vspace{.5cm}
\item {\it The $N \to \De$ GPDs:}
\begin{itemize}
\item[a)]
The parametrization of the $N \to \Delta$ GPD for the C2 transition  
has not been addressed until now. Model calculations of $N \to \Delta$ GPDs 
as well as lattice simulations of its moments will be very useful in order 
to make a comparison with the quark distributions in a nucleon, 
as parametrized through the nucleon GPDs. 
\item[b)] 
Of high interest is the measurement of the  
deeply virtual Compton scattering $\gamma^* N \to \gamma \Delta$ process 
to directly access the $N \to \Delta$ GPDs.
\end{itemize}

\vspace{.5cm}
\item {\it Large momentum transfer regime}:
\begin{itemize}

\item[a)] 
At presently accessible momentum transfers ($Q^2 \le 6$ GeV$^2$),
the pQCD prediction ($R_{EM}\to 1$, $R_{SM}\to$ const) seems 
to be nowhere in sight.
A challenge is to map out the transition from \ceft \ at low $Q^2$ to pQCD  
by measuring the $\gamma^* N \Delta$ form factors to larger $Q^2$ 
(e.g. in the near future at the planned JLab 12 GeV facility).

\item[b)]
At larger $Q^2$, constrain the model dependence in the extraction of 
multipoles. In this respect, double polarization observables have 
already been shown to be a very valuable cross-check and should be 
further investigated.

\item[c)]
Check the importance of $2 \gamma$ exchange corrections by 
Rosenbluth measurements (via the $\sigma_L$ cross section) of $R_{SM}$.

\end{itemize}

\end{enumerate}

In retrospect, the impressive level of precision obtained in the 
measurements of the electromagnetic excitation of the $\Delta$-resonance 
during the past few years has challenged the theory to 
arrive at a QCD based description of the $\Delta$-resonance. Several new 
theoretical developments are under way and have shown 
promise and potential. We hope that 
the present work will stimulate further efforts in this direction 
and contribute to the very fruitful interplay between theory and 
experiment in this field.

%% file: appendix.tex
\section{General analysis of the electromagnetic pion production on the nucleon}

In this Appendix we present the general
properties of 
the pion photo- and electroproduction reactions: 
\bea
\gamma(q)+N(p)& \rightarrow & N(p')+\pi(k) 
\label{eq:gamapi}\,,\\
e(l_i)+N(p) & \rightarrow & e^\prime(l_f)+N(p^\prime)+\pi(k)\,.
\eea
Namely, we consider the kinematics, various decompositions
into invariant amplitudes, the
multipole expansion, and the isospin structure.

\subsection{Kinematical details}
In both cases, 
the invariant $T$-matrix can be expressed
as \bea T_{fi}= \epsilon_\mu J^\mu, \label{eq:T-current}\eea where
$J^\mu$ is the nucleon electromagnetic current matrix element \bea
J^\mu=<p^\prime k|j^\mu|p>, \label{eq:current}\eea and
$\epsilon_\mu$ is the photon polarization vector for
photoproduction and \bea \epsilon_\mu =- \frac {e\bar
u(q_f)\gamma_\mu u(q_i)}{Q^2},\eea for electroproduction, with
$Q^2=-q^2=-(l_f-l_i)^2$.
\newline
\indent 
The T-matrix  can be expanded in terms of
Lorentz-covariant operators as~:
\begin{eqnarray}
T_{fi}=\bar{u}(p')\left[\sum_{i=1}^n
A_i(s,t,Q^2)M_i\right]u(p),\label{eq:invariant}
\end{eqnarray}
where $s=(p+q)^2$, $t=(p'-p)^2$ are the Mandelstam variables, 
$n=4 (6)$ for
photoproduction (electroproduction), and $A$'s are scalar amplitudes. 
The expressions for the operators $M_i$ can be found in
Refs.~\cite{CGLN56,Fubini61,Dennery61}.

\subsection{CGLN decomposition}
Traditionally the Lorentz structure of the photopion amplitude
is expressed in terms of the Pauli matrices and
(two-component) spinors~\cite{CGLN56}:
\begin{eqnarray}
T_{fi}(\ga^\ast N\to \pi N) =\frac{4\pi
W}{M_N}\chi^{\dagger}_f\,{\mathcal F}\,\chi_i,\label{eq:CGLN1}
\end{eqnarray}
with~${\mathcal F}$ given by~\cite{DT92,Knochlein95}:
\bea
{\mathcal F}&=& i{\tilde\bsig \cdot {\boldsymbol \epsilon}}\,F_{1}
+{\bsig }\cdot\hat{\bf k}\,{\bsig }\cdot(\hat{\bf
q}\times{\boldsymbol \epsilon})\,F_{2} +i{\bsig }\cdot\hat{\bf
q}\,\tilde\bk\cdot{\boldsymbol \epsilon}\,F_{3} + i{\bsig }\cdot
\hat {\bf k}\,\tilde\bk\cdot{\boldsymbol \epsilon}\,F_{4}
\nonumber\\
&&+i{\bsig }\cdot\hat{\bf q}\, \hat{\bf q}\cdot{\boldsymbol
\epsilon}\,F_{5} +i{\bsig }\cdot \hat {\bf
k}\,\hat\bq\cdot{\boldsymbol \epsilon}\,F_{6}-i{\bsig
}\cdot\hat{\bf k}\epsilon_0\,F_{7} -i{\bsig }\cdot\hat{\bf q}
\epsilon_0\,F_{8}, \label{eq:CGLN} \eea
 where
${\bsig}$ is spin operator of the nucleon and $\hat {\bf k}={\bf
k}/{\bf |k|}, \,\hat {\bf q}={\bf q}/{\bf |q|}$. Note that with
the use of $\tilde\bsig=\bsig-(\bsig\cdot\hat\bq)\hat\bq$  and
$\tilde\bk=\hat\bk-(\hat\bk\cdot\hat\bq)\hat\bq$ in the 
first term of \Eqref{CGLN}, the $F_5$ and $F_6$ terms
differ from the ones often used in the
literature, {\it e.g.}, \cite{Dennery61,Donnachie67}. 
\newline
\indent 
Obviously, the first four terms in \Eqref{CGLN} 
arise due the transverse polarizations of the photon, 
while amplitudes $F_5$ and $F_6$ come from the longitudinal polarization
of the virtual photon. 
The longitudinal component of the
current is related to the scalar component via current
conservation and hence:
 \bea |\bq| F_5= q_0 F_8,
\hspace{0.8cm} |\bq| F_6= q_0 F_7.
\eea 
\newline
\indent 
Note also that the scalar amplitudes in \Eqref{CGLN} 
(commonly knows the `CGLN amplitudes')
are complex functions of three independent kinematical variables, {\it e.g.},
$F_i=F_i(W,\Th_\pi,Q^2)$, with the
total energy $W$, the
pion scattering angle $\Theta_\pi$, and the photon virtuality $Q^2$.

\subsection{Helicity amplitudes}
Cross sections and polarization observables of pion
photo- and electroproduction are also often expressed in terms
of the helicity amplitudes (see, {\it e.g.}, \cite{Donnachie72}):
 \bea
f_{\mu',\lambda\mu}=<\mu'|\mathcal
F|\lambda\mu>, 
\label{eq:helicity-f}
\eea 
where $\mu$ ($\mu'$) denotes the helicity of the intitial (final) nucleon
and $\la$ the photon helicity. 
\newline
\indent 
There are 6 (4) independent helicity amplitudes
for the case pion electro- (photo-) production.
They are defined Table~\ref{tab:helicit-f} provided $\mu'=\frac 12$
and the phase convention of Jacob and Wick \cite{JW59} is used.
\begin{table}[h]
\centering
\begin{tabular}{|c|ccc|}
\hline\hline
\backslashbox{\(\mu\)}{\(\lambda\)}& \hspace{1.0cm}$-1$    &\hspace{1.0cm}1     &\hspace{1.0cm}0    \hspace{1.0cm}\\
\hline $ \frac 12$                 & \hspace{1.0cm}$H_1$ &\hspace{1.0cm}$H_4$ &\hspace{1.0cm}$H_5$\hspace{1.0cm}\\
        $  -\frac 12$              & \hspace{1.0cm}$ H_2$&\hspace{1.0cm}$-H_3$&\hspace{1.0cm}$H_6$\hspace{1.0cm}\\
\hline\hline
        \end{tabular}
\label{tab:helicit-f}
\caption{Helicity amplitudes $i f_{\mu'=\frac
12,\lambda\mu}$.}
\end{table}
\newline
\indent 
The six helicity amplitudes 
$H_1,...,H_6$ can be expressed as the
following linear combination of the six independent CGLN
amplitudes $F_1,...,F_6$:
\begin{eqnarray} H_1 &=& -\sqrt{\half}\sin\theta
\,\cos(\theta/2) \, (F_3+F_4)\,,
\nonumber \\
H_2 &=&  \sqrt{2}\cos(\theta/2) \, [(F_2-F_1)+
\half (1-\cos\theta)\,(F_3-F_4)]\,,
\nonumber \\
H_3 &=& \sqrt{\half} \sin\theta \,\sin(\theta/2) \,
(F_3-F_4)\,,
\\
H_4 &=&  \sqrt{2}\sin(\theta/2) \, [(F_1+F_2)+
\half(1+\cos\theta)\,(F_3+F_4)]\,,
\nonumber \\
H_5 &=& \cos(\theta/2) \,(F_5+F_6)\,,\qquad
\nonumber \\
H_6 &=& \sin(\theta/2) \,(F_6-F_5)\,. \nonumber
\end{eqnarray}

\subsection{Multipole decomposition}
 Multipole amplitudes are obtained with the use of
eigenstates of parity and angular momentum, instead of plane wave
states. For electroproduction, there are six types of transitions
possible to a $\pi N$ final state with angular momentum $l$ and
parity $P$, which are classified according to the character of the
photon, transverse or scalar (or alternatively, longitudinal) and
the total angular momentum $J=l\pm\frac 12$ of the final state. In
addition, the transverse photon states can either be electric,
with $P=(-1)^L$, or magnetic, with $P=(-1)^{L+1}$, where $L$ is
the total orbital angular momentum of the photon. 

The expansion of
the CGLN amplitudes in terms of the multipole amplitudes
is given as follows~\cite{Dennery61}:
\begin{eqnarray} \label{cm}
F_{1} & = & \sum_{l\geq0}\{(lM_{l+}+E_{l+})P_{l+1}^{\prime}
+[(l+1)M_{l-}+E_{l-}]P_{l-1}^{\prime}\}, \nonumber\\
F_{2} & = & \sum_{l\geq1}[(l+1)M_{l+}+lM_{l-}]P_{l}^{\prime}, \nonumber\\
F_{3} & = & \sum_{l\geq1}[(E_{l+}-M_{l+})P_{l+1}^{\prime\prime}
+(E_{l-}+M_{l-})P_{l-1}^{\prime\prime}],   \nonumber\\
F_{4} & = & \sum_{l\geq2}(M_{l+}-E_{l+}-M_{l-}-E_{l-})P_{l}^{\prime\prime}, \\
F_{5} & = & \sum_{l\geq0}[(l+1)L_{l+}P_{l+1}^{\prime}
-lL_{l-}P_{l-1}^{\prime}],\nonumber \\F_{6} & = &
\sum_{l\geq1}[lL_{l-}-(l+1)L_{l+}]P_{l}^{\prime}\,\nn
 \label{eq:CGLN2}
\end{eqnarray}
where $P_{l}^{\prime}s$ are the derivatives of the Legendre
polynomials of argument $x=cos\theta $. Note that in the
literature the scalar transitions are sometimes employed and
described by $S_{l\pm}$ multipoles, which correspond to the
multipole decomposition of the amplitudes $F_7$ and $F_8$. They
are connected with the longitudinal ones by $L_{l\pm}=
(\omega_\bq/{\mid {\bf q}\mid}) S_{l\pm}.$

\subsection{Isospin decomposition}

Assuming isospin conservation, there are three independent
isospin amplitudes for the single pion production off a nucleon by a photon. 
These are: $A^{(0)}$ for the isoscalar photon,
and $A^{(1/2)}$ and $A^{(3/2)}$ for the isovector photon. The latter two
correspond to the $\pi N$ system with total isospin $I=1/2$ and
$I=3/2$, respectively. The isospin structure of any photopion 
amplitude ({\it e.g.}, the multipoles) can be written as:
\begin{eqnarray}
A=A^{(+)}\delta_{\alpha 3}+A^{(-)}\frac
12[\tau_\alpha,\tau_3]+A^{(0)}\tau_\alpha,\label{eq:isospin}
\end{eqnarray}
where $\alpha$ is the pion isospin index and $\tau_\alpha$ the
nucleon isospin matrices. The isospin amplitudes $A^{(1/2)}$, and
$A^{(3/2)}$ are related to the amplitudes of
Eq.~(\ref{eq:isospin}) as~: \bea A^{(3/2)} = A^{(+)} - A^{(-)},
\qquad A^{(1/2)} = A^{(+)} + 2 A^{(-)}. \eea It will also be
useful to define  the proton $_pA^{(1/2)}$ and neutron
$_nA^{(1/2)}$ amplitudes with total isospin 1/2,
\begin{eqnarray} \label{pnam}
_pA^{(1/2)} =A^{(0)}+\frac{1}{3}A^{(1/2)}\,,\qquad _nA^{(1/2)}
=A^{(0)}-\frac{1}{3}A^{(1/2)}\,.
\end{eqnarray}
With this convention the physical amplitudes for the four physical
pion photo- and electroproduction processes are
\begin{eqnarray}
A(\gamma^*p\rightarrow n\pi^+) &  = &
\sqrt{2}\,\left[_pA^{(1/2)}-\frac{1}{3}A^{(3/2)}\right]
\,, \nonumber\\
A(\gamma^*p\rightarrow p\pi^0)  & =  &
_pA^{(1/2)}+\frac{2}{3}A^{(3/2)}\,, \nonumber\\
A(\gamma^*n\rightarrow p\pi^-) &  = &
\sqrt{2}\,\left[_nA^{(1/2)}+\frac{1}{3}A^{(3/2)}\right]
\,, \\
A(\gamma^*n\rightarrow n\pi^0) &  = &
-_nA^{(1/2)}+\frac{2}{3}A^{(3/2)}\,\nn. \label{physam}
\end{eqnarray}
\newline
\indent 
Under the interchange of $s$ and $u$ (crossing), the
functions $A_i^{(\pm,0)}$ are either even or odd, i.e., and this
crossing symmetry can readily be derived from the assumption of
charge conjugation C invariance for photo- and electroproduction.
These crossing symmetry properties can be compactly expressed in
matrix form with a six-vector $\tilde A$ with elements
$A_1,..,A_6$, \bea \tilde A(s,t,u,Q^2) = [\bar\xi] \tilde
A(u,t,s,Q^2), \label{eq:crossing} \eea where $[\xi]$ is a $6\times
6$ diagonal matrix \bea [\bar\xi] = \xi\,\, diag \{1, \, 1, \, -1,
\, 1, \, -1, \, -1\},\eea and the parameter $\xi$ is defined by
$\xi=1$ for isospin index $(+,0),$ and $\xi=-1$ for isospin index
$(-)$.

%% file: ackno.tex
\section*{\center{Acknowledgements}}

The work  of V.~P and M.~V. is supported in part by DOE grant no.\
DE-FG02-04ER41302 and contract DE-AC05-06OR23177 under
which Jefferson Science Associates operates the Jefferson Laboratory.  
Furthermore, the work of S.N.Y. is supported in part by the National 
Science Council of ROC under grant No. NSC94-2112-M002-025.
\newline
\indent
We like to thank 
C. Alexandrou, 
I. Aznauryan, 
A. Bernstein, 
D. Drechsel, 
S.S. Kamalov, 
C. W. Kao, 
T.S.H. Lee, 
D.H. Lu, 
C. Papanicolas,
C. Smith, 
N. Sparveris, 
S. Stave, 
L. Tiator, 
and R. Young,
for useful discussions and correspondence during the course of this work. 
\newline
\indent
We also like to thank A. Bernstein and C. Papanicolas 
for organizing a very stimulating workshop 
on ``The Shape of Hadrons'' (Athens, April 27-29, 2006), 
where many of the topics reviewed in this work were discussed.

%% file: review_delta.bbl
\begin{thebibliography}{99}



\bibitem{Anderson:1952nw}
 H.~L.~Anderson, E.~Fermi, E.~A.~Long and D.~E.~Nagle,
 Phys.\ Rev.\  {\bf 85}, 936 (1952).

\bibitem{Brown:1975di}
 G.~E.~Brown and W.~Weise,
 Phys.\ Rept.\  {\bf 22}, 279 (1975).

\bibitem{Cattapan:2002rx}
 G.~Cattapan and L.~S.~Ferreira,
 Phys.\ Rept.\  {\bf 362}, 303 (2002).

\bibitem{Hanhart:2003pg}
  C.~Hanhart,
  Phys.\ Rept.\  {\bf 397}, 155 (2004).

\bibitem{GZK}
K. Greisen, Phys. Rev. Lett. {\bf 16}, 748 (1966);
G. T. Zatsepin and V. A. Kuzmin, JETP Lett. {\bf 4}, 78 (1966).

\bibitem{Mucke:1998mk}
  A.~Mucke, J.~P.~Rachen, R.~Engel, R.~J.~Protheroe and T.~Stanev,
  Publ.\ Astron.\ Soc.\ Austral.\  {\bf 16}, 160 (1999). 

\bibitem{Weinberg:1978kz}
  S.~Weinberg,
  Physica A {\bf 96}, 327 (1979).

\bibitem{Gasser:1983yg}
  J.~Gasser and H.~Leutwyler,
  Annals Phys.\  {\bf 158}, 142 (1984); 
  Nucl.\ Phys.\ B {\bf 250}, 465 (1985).

\bibitem{BKM}
V.~Bernard, N.~Kaiser, and U.~G.~Mei{\ss}ner,
Int.\ J.\ Mod.\ Phys.\ E {\bf 4}, 193 (1995).


\bibitem{DT92}
D. Drechsel and L. Tiator,
J. Phys. G {\bf 18}, 449 (1992).

\bibitem{Krusche:2003ik}
  B.~Krusche and S.~Schadmand,
  Prog.\ Part.\ Nucl.\ Phys.\  {\bf 51}, 399 (2003).

\bibitem{Burkert:2004sk}
  V.~D.~Burkert and T.~S.~H.~Lee,
  Int.\ J.\ Mod.\ Phys.\ E {\bf 13}, 1035 (2004).



\bibitem{Jones:1972ky}
H.~F.~Jones and M.~D.~Scadron,
Ann.~Phys.\  {\bf 81}, 1 (1973).

\bibitem{SL}
T. Sato and T.-S.H. Lee,
Phys.\ Rev.\ C {\bf 54}, 2660 (1996);
{\it ibid.}\ {\bf 63}, 055201 (2001).

\bibitem{PDG2006}
  W.~M.~Yao {\it et al.}  [Particle Data Group],
  J.\ Phys.\ G {\bf 33}, 1 (2006).

\bibitem{Berends:1974zp}
  F.~A.~Berends and A.~Donnachie,
  Nucl.\ Phys.\ B {\bf 84}, 342 (1975).

\bibitem{Arndt:1990ej}
  R.~A.~Arndt, R.~L.~Workman, Z.~Li and L.~D.~Roper,
  Phys.\ Rev.\ C {\bf 42}, 1864 (1990).

\bibitem{Tiator:2003xr}
  L.~Tiator, D.~Drechsel, S.~S.~Kamalov and S.~N.~Yang,
  Eur.\ Phys.\ J.\ A {\bf 17}, 357 (2003).



\bibitem{Ash67}
W.~W.~Ash {\it et al.}, Phys. Lett. {\bf 24B}, 165 (1967).

\bibitem{Weber:1978dh}
  H.~J.~Weber and H.~Arenhovel,
  Phys.\ Rept.\  {\bf 36}, 277 (1978).

\bibitem{Noz90}
S. Nozawa and D.B. Leinweber,
Phys. Rev. D {\bf 42}, 3567 (1990).

\bibitem{Beck:1997ew}
  R.~Beck {\it et al.},
  Phys.\ Rev.\ Lett.\  {\bf 78}, 606 (1997).

\bibitem{Beck:1999ge}
  R.~Beck {\it et al.},
  Phys.\ Rev.\ C {\bf 61}, 035204 (2000).

\bibitem{LEGS97}
G.~Blanpied  {\em et al.},  Phys.\ Rev.\ Lett.\  {\bf 79}, 4337 (1997).

\bibitem{Blanpied:2001ae}
  G.~Blanpied {\it et al.},
  Phys.\ Rev.\ C {\bf 64}, 025203 (2001).

\bibitem{Tiator:2000iy}
  L.~Tiator, D.~Drechsel, O.~Hanstein, S.~S.~Kamalov and S.~N.~Yang,
  Nucl.\ Phys.\ A {\bf 689}, 205 (2001).

\bibitem{Isgur}
  N.~Isgur and G.~Karl,
  Phys.\ Rev.\ D {\bf 18}, 4187 (1978);
  {\it ibid.}\ D {\bf 19}, 2653 (1979)
  [Erratum-ibid.\ D {\bf 23}, 817 (1981)];
   {\it ibid.}\ D {\bf 20}, 1191 (1979).


\bibitem{DeRujula:1975ge}
  A.~De Rujula, H.~Georgi and S.~L.~Glashow,
  Phys.\ Rev.\ D {\bf 12}, 147 (1975).

\bibitem{Koniuk:1979vy}
  R.~Koniuk and N.~Isgur,
  Phys.\ Rev.\ D {\bf 21}, 1868 (1980)
  [Erratum-ibid.\ D {\bf 23}, 818 (1981)].

\bibitem{Isgur:1981yz}
  N.~Isgur, G.~Karl and R.~Koniuk,
  Phys.\ Rev.\ D {\bf 25}, 2394 (1982).

\bibitem{Becchi:1965}
C.~M.~Becchi and G.~Morpurgo,
  Phys.\ Lett. {\bf 17}, 352 (1965).

\bibitem{Donoghue:1975yg}
  J.~F.~Donoghue, E.~Golowich and B.~R.~Holstein,
  Phys.\ Rev.\ D {\bf 12}, 2875 (1975).


\bibitem{Glashow79}
S.~L.~Glashow, Physica {\bf 96A}, 27 (1979).

\bibitem{Gershtein:1981zf}
  S.~S.~Gershtein and G.~V.~Jikia,
  Sov.\ J.\ Nucl.\ Phys.\  {\bf 34}, 870 (1981)
  [Yad.\ Fiz.\  {\bf 34}, 1566 (1981)].

\bibitem{Bourdeau:1987ih}
  M.~Bourdeau and N.~C.~Mukhopadhyay,
  Phys.\ Rev.\ Lett.\  {\bf 58}, 976 (1987).

\bibitem{Gogilidze87}
S.~A.~Gogilidze, Yu.~S.~Surovtsev and F.~G.~Tkebuchava,
Sov.~J.~Nucl.~Phys. {\bf 45}, 674 (1987)
[Yad. Piz. {\bf 45}, 1085 (1987)].

\bibitem{Drechsel:1984ie}
  D.~Drechsel and M.~M.~Giannini,
  Phys.\ Lett.\ B {\bf 143}, 329 (1984).

\bibitem{Giannini:1990pc}
  M.~M.~Giannini,
  Rept.\ Prog.\ Phys.\  {\bf 54}, 453 (1990).

\bibitem{Capstick:1989ck}
  S.~Capstick and G.~Karl,
  Phys.\ Rev.\ D {\bf 41}, 2767 (1990).

\bibitem{Capstick:1992uc}
  S.~Capstick,
in
  Phys.\ Rev.\ D {\bf 46}, 2864 (1992).


\bibitem{Kaelbermann:1983zb}
  G.~Kaelbermann and J.~M.~Eisenberg,
  Phys.\ Rev.\ D {\bf 28}, 71 (1983).

\bibitem{Thomas:1982kv}
  A.~W.~Thomas,
  Adv.\ Nucl.\ Phys.\  {\bf 13}, 1 (1984).

\bibitem{Bermuth:1988ms}
  K.~Bermuth, D.~Drechsel, L.~Tiator and J.~B.~Seaborn,
  Phys.\ Rev.\ D {\bf 37}, 89 (1988).

\bibitem{Lu:1996rj}
  D.~H.~Lu, A.~W.~Thomas and A.~G.~Williams,
  Phys.\ Rev.\ C {\bf 55}, 3108 (1997).

\bibitem{Luprivate}
D.~H.~Lu, private communication.

\bibitem{Fiolhais:1996bp}
  M.~Fiolhais, B.~Golli and S.~Sirca,
  Phys.\ Lett.\ B {\bf 373}, 229 (1996).

\bibitem{Wirzba:1986sc}
  A.~Wirzba and W.~Weise,
  Phys.\ Lett.\ B {\bf 188}, 6 (1987).

\bibitem{Abada:1995db}
  A.~Abada, H.~Weigel and H.~Reinhardt,
  Phys.\ Lett.\ B {\bf 366}, 26 (1996).

\bibitem{Walliser:1996ps}
  H.~Walliser and G.~Holzwarth,
  Z.\ Phys.\ A {\bf 357}, 317 (1997).


\bibitem{Dia86}
D.~I.~Diakonov and V.~Petrov,
Nucl. Phys. {\bf B272}, 457 (1986).

\bibitem{Dia88}
D.~I.~Diakonov, V. Petrov and P. Pobylitsa,
Nucl. Phys. {\bf B306}, 809 (1988).

\bibitem{Chr96}
Chr.V. Christov, A. Blotz, H.-C. Kim, P. Pobylitsa, T. Watabe,
Th. Meissner, E. Ruiz Arriola and K. Goeke,
Prog. Part. Nucl. Phys. {\bf 37}, 91 (1996).

\bibitem{Watabe:1995xy}
  T.~Watabe, C.~V.~Christov and K.~Goeke,
  Phys.\ Lett.\ B {\bf 349}, 197 (1995).

\bibitem{Silva:1999nz}
  A.~Silva, D.~Urbano, T.~Watabe, M.~Fiolhais and K.~Goeke,
  Nucl.\ Phys.\ A {\bf 675}, 637 (2000).


\bibitem{Buchmann:1996bd}
  A.~J.~Buchmann, E.~Hernandez and A.~Faessler,
  Phys.\ Rev.\ C {\bf 55}, 448 (1997).

\bibitem{Grabmayr:2001up}
  P.~Grabmayr and A.~J.~Buchmann,
  Phys.\ Rev.\ Lett.\  {\bf 86}, 2237 (2001).

\bibitem{Buchmann:2004ia}
  A.~J.~Buchmann,
Phys.\ Rev.\ Lett.\  {\bf 93}, 212301 (2004).

\bibitem{Dillon:1998zm}
  G.~Dillon and G.~Morpurgo,
  Phys.\ Lett.\ B {\bf 448}, 107 (1999).

\bibitem{Buchmann:2001gj}
  A.~J.~Buchmann and E.~M.~Henley,
  Phys.\ Rev.\ C {\bf 63}, 015202 (2001).

\bibitem{Buchmann:2002xq}
  A.~J.~Buchmann and E.~M.~Henley,
  Phys.\ Rev.\ D {\bf 65}, 073017 (2002).

\bibitem{BM75}
A.~Bohr and B.~Mottelson,
{\it Nuclear Structure II} (Benjamin, Reading, MA, 1975).

\bibitem{Faessler:2006ky}
  A.~Faessler, T.~Gutsche, B.~R.~Holstein, V.~E.~Lyubovitskij, D.~Nicmorus and K.~Pumsa-ard,
  arXiv:hep-ph/0608015.


\bibitem{'tHooft:1973jz}
  G.~'t Hooft,
  Nucl.\ Phys.\ B {\bf 72}, 461 (1974).

\bibitem{Witten:1979kh}
  E.~Witten,
  Nucl.\ Phys.\ B {\bf 160}, 57 (1979).

\bibitem{Jenkins:1998wy}
  E.~Jenkins,
  Ann.\ Rev.\ Nucl.\ Part.\ Sci.\  {\bf 48}, 81 (1998).

\bibitem{Lebed:1998st}
  R.~F.~Lebed,
  Czech.\ J.\ Phys.\  {\bf 49}, 1273 (1999).

\bibitem{Jenkins:1994md}
  E.~Jenkins and A.~V.~Manohar,
  Phys.\ Lett.\ B {\bf 335}, 452 (1994).

\bibitem{Jenkins:2002rj}
  E.~Jenkins, X.~d.~Ji and A.~V.~Manohar,
  Phys.\ Rev.\ Lett.\  {\bf 89}, 242001 (2002).

\bibitem{Buchmann:2002mm}
  A.~J.~Buchmann, J.~A.~Hester and R.~F.~Lebed,
  Phys.\ Rev.\ D {\bf 66}, 056002 (2002).


\bibitem{Cohen:2002sd}
  T.~D.~Cohen,
  Phys.\ Lett.\ B {\bf 554}, 28 (2003).

\bibitem{Cohen:2006up}
  T.~D.~Cohen and R.~F.~Lebed,
 Phys.\ Rev.\ D {\bf 74}, 056006 (2006).


\bibitem{Pascalutsa:2003zk}
  V.~Pascalutsa and D.~R.~Phillips,
  Phys.\ Rev.\ C {\bf 68}, 055205 (2003).

\bibitem{Leinweber:1992pv}
  D.~B.~Leinweber, T.~Draper and R.~M.~Woloshyn,
  Phys.\ Rev.\ D {\bf 48}, 2230 (1993).

\bibitem{Alexandrou:2002nn}
  C.~Alexandrou, P.~de Forcrand and A.~Tsapalis,
  Phys.\ Rev.\ D {\bf 66}, 094503 (2002).


\bibitem{Alexandrou:2003ea}
  C.~Alexandrou {\it et al.},
  Phys.\ Rev.\ D {\bf 69}, 114506 (2004).

\bibitem{Cohen:1992kk}
  T.~D.~Cohen and D.~B.~Leinweber,
  Comments Nucl.\ Part.\ Phys.\  {\bf 21}, 137 (1993).

\bibitem{Elsner:2005cz}
  D.~Elsner {\it et al.},
  Eur.\ Phys.\ J.\ A {\bf 27}, 91 (2006).

\bibitem{Pospischil:2000ad}
  T.~Pospischil {\it et al.},
  Phys.\ Rev.\ Lett.\  {\bf 86}, 2959 (2001).

\bibitem{Stave:2006ea}
  S.~Stave {\it et al.},
  arXiv:nucl-ex/0604013.

\bibitem{Sparveris:2004jn}
  N.~F.~Sparveris {\it et al.}  [OOPS Collaboration],
  Phys.\ Rev.\ Lett.\  {\bf 94}, 022003 (2005).

\bibitem{Joo:2001tw}
  K.~Joo {\it et al.}  [CLAS Collaboration],
  Phys.\ Rev.\ Lett.\  {\bf 88}, 122001 (2002).

\bibitem{Alexandrou:2004xn}
  C.~Alexandrou, P.~de Forcrand, H.~Neff, J.~W.~Negele, W.~Schroers and
A.~Tsapalis,
  Phys.\ Rev.\ Lett.\  {\bf 94}, 021601 (2005).


\bibitem{Butler:1993ht}
  M.~N.~Butler, M.~J.~Savage and R.~P.~Springer,
  Phys.\ Lett.\ B {\bf 304}, 353 (1993).

\bibitem{JeM91a}
E.~Jenkins and A.~V.~Manohar,
Phys.\ Lett.\ B {\bf 255}, 558 (1991).


\bibitem{Gellas:1998wx}
  G.~C.~Gellas, T.~R.~Hemmert, C.~N.~Ktorides and G.~I.~Poulis,
  Phys.\ Rev.\ D {\bf 60}, 054022 (1999).

\bibitem{HHK97}
T.~Hemmert, B.~R.~Holstein and J.~Kambor,
Phys.\ Lett.\ B {\bf 395}, 89 (1997);
  J.\ Phys.\ G {\bf 24}, 1831 (1998).

\bibitem{Gail:2005gz}
  T.~A.~Gail and T.~R.~Hemmert,
  arXiv:nucl-th/0512082.

\bibitem{Pascalutsa:2005ts}
  V.~Pascalutsa and M.~Vanderhaeghen,
Phys. Rev. Lett. {\bf 95}, 232001 (2005).

\bibitem{Pascalutsa:2005vq}
  V.~Pascalutsa and M.~Vanderhaeghen,
  Phys.\ Rev.\ D {\bf 73}, 034003 (2006).

\bibitem{Pascalutsa:2002pi}
  V.~Pascalutsa and D.~R.~Phillips,
  Phys.\ Rev.\ C {\bf 67}, 055202 (2003).


\bibitem{KY99}
S.~S. Kamalov and S.~N. Yang, Phys.\ Rev.\ Lett. {\bf 83}, 4494 (1999);
 
S.~S. Kamalov, S.~N. Yang, D. Drechsel, O. Hanstein, and L. Tiator, 
Phys.\ Rev.\ C {\bf 64}, 032201(R) (2001).

\bibitem{MAID98}
  D.~Drechsel, O.~Hanstein, S.~S.~Kamalov and L.~Tiator,
  Nucl.\ Phys.\ A {\bf 645}, 145 (1999).

\bibitem{Arndt:2002xv}
  R.~A.~Arndt, W.~J.~Briscoe, I.~I.~Strakovsky and R.~L.~Workman,
  Phys.\ Rev.\ C {\bf 66}, 055213 (2002).

\bibitem{DeSanctis:2005vq}
  M.~De Sanctis, M.~M.~Giannini, E.~Santopinto and A.~Vassallo,
  Nucl.\ Phys.\ A {\bf 755}, 294 (2005).


\bibitem{Mertz:1999hp}
  C.~Mertz {\it et al.},
  Phys.\ Rev.\ Lett.\  {\bf 86}, 2963 (2001).



\bibitem{Leinweber:2001ui}
  D.~B.~Leinweber, A.~W.~Thomas and R.~D.~Young,
  Phys.\ Rev.\ Lett.\  {\bf 86}, 5011 (2001);

 W.~Detmold, W.~Melnitchouk, J.~W.~Negele, D.~B.~Renner and A.~W.~Thomas,
  {\it ibid.}\  {\bf 87}, 172001 (2001).

\bibitem{Hemmert:2003cb}
  T.~R.~Hemmert, M.~Procura and W.~Weise,
  Phys.\ Rev.\ D {\bf 68}, 075009 (2003).

\bibitem{Banerjee:1994bk}
  M.~K.~Banerjee and J.~Milana,
  Phys.\ Rev.\ D {\bf 52}, 6451 (1995).

\bibitem{Leinweber:1999ig}
  D.~B.~Leinweber, A.~W.~Thomas, K.~Tsushima and S.~V.~Wright,
  Phys.\ Rev.\ D {\bf 61}, 074502 (2000).

\bibitem{Ross}
  R.~D.~Young, D.~B.~Leinweber, A.~W.~Thomas and S.~V.~Wright,
   Phys.\ Rev.\ D {\bf 66}, 094507 (2002);

  R.~D.~Young, D.~B.~Leinweber and A.~W.~Thomas,
  Prog.\ Part.\ Nucl.\ Phys.\  {\bf 50}, 399 (2003);

  D.~B.~Leinweber, A.~W.~Thomas and R.~D.~Young,
  Phys.\ Rev.\ Lett.\  {\bf 92}, 242002 (2004).

\bibitem{Bernard:2003xf}
  V.~Bernard, T.~R.~Hemmert and U.~G.~Meissner,
  Phys.\ Lett.\ B {\bf 565}, 137 (2003).

\bibitem{Frink:2005ru}
  M.~Frink, U.~G.~Meissner and I.~Scheller,
  Eur.\ Phys.\ J.\ A {\bf 24}, 395 (2005). 

\bibitem{Bernard:2005fy}
  V.~Bernard, T.~R.~Hemmert and U.~G.~Meissner,
  Phys.\ Lett.\ B {\bf 622}, 141 (2005).

\bibitem{HWGS05}
C.~Hacker, N.~Wies, J.~Gegelia and S.~Scherer,
  Phys.\ Rev.\ C {\bf 72}, 055203 (2005).

\bibitem{Pascalutsa:2005nd}
  V.~Pascalutsa and M.~Vanderhaeghen,
  Phys.\ Lett.\ B {\bf 636}, 31 (2006).

\bibitem{Labrenz:1996jy}
  J.~N.~Labrenz and S.~R.~Sharpe,
  Phys.\ Rev.\ D {\bf 54}, 4595 (1996).


\bibitem{Bernard:2001av}
  C.~W.~Bernard {\it et al.},
  Phys.\ Rev.\ D {\bf 64}, 054506 (2001).

\bibitem{Cloet03}
  R.~D.~Young, D.~B.~Leinweber and A.~W.~Thomas,
  Nucl.\ Phys.\ Proc.\ Suppl.\  {\bf 129}, 290 (2004).

\bibitem{PV05}
  V.~Pascalutsa and M.~Vanderhaeghen,
  Phys.\ Rev.\ Lett.\  {\bf 94}, 102003 (2005).

\bibitem{Alexandrou:2005em}
  C.~Alexandrou {\it et al.},
  PoS {\bf LAT2005}, 091 (2006)
  [arXiv:hep-lat/0509140].


\bibitem{Ji98a}
X. Ji and J. Osborne,
Phys. Rev. D {\bf 58}, 094018 (1998).

\bibitem{Col99}
J.~C.~Collins and A.~Freund,
Phys. Rev. D {\bf 59}, 074009 (1999).

\bibitem{Rad98}
A.~V.~Radyushkin,
Phys. Rev. D {\bf 58}, 114008 (1998).

\bibitem{Muller:1998fv}
D.~Muller, D.~Robaschik, B.~Geyer, F.~M.~Dittes and J.~Horejsi,
Fortsch.\ Phys.\  \textbf{42}, 101 (1994).

\bibitem{Ji:1996ek}
X.~D.~Ji,
Phys.\ Rev.\ Lett.\  \textbf{78}, 610 (1997);
Phys.\ Rev.\ D \textbf{55}, 7114 (1997).

\bibitem{Radyushkin:1996nd}
  A.~V.~Radyushkin,
  Phys.\ Lett.\ B {\bf 380}, 417 (1996).

\bibitem{Ji:1998pc}
  X.~D.~Ji,
  J.\ Phys.\ G {\bf 24}, 1181 (1998).

\bibitem{Goeke:2001tz}
K.~Goeke, M.~V.~Polyakov and M.~Vanderhaeghen,
Prog.\ Part.\ Nucl.\ Phys.\  {\bf 47}, 401 (2001).

\bibitem{Diehl:2003ny}
  M.~Diehl,
  Phys.\ Rept.\  {\bf 388}, 41 (2003).

\bibitem{Belitsky:2005qn}
  A.~V.~Belitsky and A.~V.~Radyushkin,
  Phys.\ Rept.\  {\bf 418}, 1 (2005).


\bibitem{Frankfurt:1999xe}
L.~L.~Frankfurt, M.~V.~Polyakov, M.~Strikman and M.~Vanderhaeghen,
Phys.\ Rev.\ Lett.\  {\bf 84}, 2589 (2000).

\bibitem{GMV03}
  P.~A.~M.~Guichon, L.~Mosse and M.~Vanderhaeghen,
  Phys.\ Rev.\ D {\bf 68}, 034018 (2003).

\bibitem{Guidal:2003ji}
M.~Guidal, S.~Bouchigny, J.~P.~Didelez, C.~Hadjidakis, E.~Hourany and M.~Vanderhaeghen,
Nucl.\ Phys.\ A {\bf 721}, 327 (2003).

\bibitem{Stoler:2002im}
  P.~Stoler,
  Phys.\ Rev.\ Lett.\  {\bf 91}, 172303 (2003).

\bibitem{Burkardt:2004bv}
M.~Burkardt,
Phys.\ Lett.\ B \textbf{595}, 245 (2004).

\bibitem{Stoler:2001xa}
P.~Stoler,
Phys.\ Rev.\ D {\bf 65}, 053013 (2002).

\bibitem{Diehl:2004cx}
  M.~Diehl, T.~Feldmann, R.~Jakob and P.~Kroll,
  Eur.\ Phys.\ J.\ C {\bf 39}, 1 (2005).

\bibitem{guidal}
  M.~Guidal, M.~V.~Polyakov, A.~V.~Radyushkin and M.~Vanderhaeghen,
  Phys.\ Rev.\ D {\bf 72}, 054013 (2005).

\bibitem{Martin:2002dr}
A.~D.~Martin, R.~G.~Roberts, W.~J.~Stirling and R.~S.~Thorne,
Phys.\ Lett.\ B {\bf 531}, 216 (2002).


\bibitem{Frolov:1998pw}
V.~V.~Frolov {\it et al.},
Phys.\ Rev.\ Lett.\  {\bf 82}, 45 (1999).

\bibitem{Ungaro:2006df}
  M.~Ungaro, P.~Stoler, I.~Aznauryan, V.~D.~Burkert, K.~Joo and L.~C.~Smith
                  [CLAS Collaboration],
  arXiv:hep-ex/0606042.

\bibitem{Aznauryan:2002gd}
  I.~G.~Aznauryan,
  Phys.\ Rev.\ C {\bf 67}, 015209 (2003).


\bibitem{Herberg:ud}
C.~Herberg {\it et al.},
Eur.\ Phys.\ J.\ A {\bf 5}, 131 (1999).

\bibitem{Ostrick:xa}
M.~Ostrick {\it et al.},
Phys.\ Rev.\ Lett.\  {\bf 83}, 276 (1999).


\bibitem{Becker:tw}
J.~Becker {\it et al.},
Eur.\ Phys.\ J.\ A {\bf 6}, 329 (1999).

\bibitem{Rohe:sh}
D.~Rohe {\it et al.},
Phys.\ Rev.\ Lett.\  {\bf 83}, 4257 (1999).

\bibitem{Passchier:1999cj}
I.~Passchier {\it et al.},
Phys.\ Rev.\ Lett.\  {\bf 82}, 4988 (1999).

\bibitem{Zhu:2001md}
H.~Zhu {\it et al.}  [E93026 Collaboration],
Phys.\ Rev.\ Lett.\  {\bf 87}, 081801 (2001).


\bibitem{Warren:2003ma}
G.~Warren {\it et al.}  [Jefferson Lab E93-026 Collaboration],
Phys.\ Rev.\ Lett.\  {\bf 92}, 042301 (2004).


\bibitem{Madey:2003av}
R.~Madey {\it et al.}  [E93-038 Collaboration],
Phys.\ Rev.\ Lett.\  {\bf 91}, 122002 (2003).


\bibitem{Sparveris:2006}
N.~Sparveris, 
to appear in Proceedings of the Workshop ``Shape of Hadrons'', 
Athens, 2006. Eds. C.N. Papanicolas and A.M. Bernstein, AIP (2006).  

\bibitem{Kelly05}
J.J.~Kelly {\it et al.}, 
Phys.\ Rev.\ Lett.\ {\bf 95}, 102001 (2005).


\bibitem{Burkardt:2000za}
  M.~Burkardt,
  Phys.\ Rev.\ D {\bf 62}, 071503 (2000)
  [Erratum-ibid.\ D {\bf 66}, 119903 (2002)].

\bibitem{Belitsky:2003nz}
  A.~V.~Belitsky, X.~d.~Ji and F.~Yuan,
  Phys.\ Rev.\ D {\bf 69}, 074014 (2004).

\bibitem{Burkardt:2002hr}
M.~Burkardt,
Int.\ J.\ Mod.\ Phys.\ A {\bf 18}, 173 (2003).


\bibitem{Lepage:1980fj}
  G.~P.~Lepage and S.~J.~Brodsky,
  Phys.\ Rev.\ D {\bf 22}, 2157 (1980).

\bibitem{Chernyak:1977as}
  V.~L.~Chernyak and A.~R.~Zhitnitsky,
  JETP Lett.\  {\bf 25}, 510 (1977)
  [Pisma Zh.\ Eksp.\ Teor.\ Fiz.\  {\bf 25}, 544 (1977)].

\bibitem{Chernyak:1977fk}
  V.~L.~Chernyak, A.~R.~Zhitnitsky and V.~G.~Serbo,
  JETP Lett.\  {\bf 26}, 594 (1977)
  [Pisma Zh.\ Eksp.\ Teor.\ Fiz.\  {\bf 26}, 760 (1977)].

\bibitem{Efremov:1979qk}
  A.~V.~Efremov and A.~V.~Radyushkin,
  Phys.\ Lett.\ B {\bf 94}, 245 (1980).

\bibitem{Carlson:1985mm}
  C.~E.~Carlson,
  Phys.\ Rev.\ D {\bf 34}, 2704 (1986).

\bibitem{Sill:1992qw}
A.~F.~Sill {\it et al.},
Phys.\ Rev.\ D {\bf 48}, 29 (1993).

\bibitem{Jones:1999rz}
M.~K.~Jones {\it et al.}  [Jefferson Lab Hall A Collaboration],
Phys.\ Rev.\ Lett.\  {\bf 84}, 1398 (2000).

\bibitem{Punjabi:2005wq}
V.~Punjabi {\it et al.},
  Phys.\ Rev.\ C {\bf 71}, 055202 (2005)
  [Erratum-ibid.\ C {\bf 71}, 069902 (2005)].

\bibitem{Gayou:2001qt}
O.~Gayou {\it et al.},
Phys.\ Rev.\ C {\bf 64}, 038202 (2001).

\bibitem{Gayou:2001qd}
O.~Gayou {\it et al.}  [Jefferson Lab Hall A Collaboration],
Phys.\ Rev.\ Lett.\  {\bf 88}, 092301 (2002).

\bibitem{Belitsky:2002kj}
  A.~V.~Belitsky, X.~d.~Ji and F.~Yuan,
  Phys.\ Rev.\ Lett.\  {\bf 91}, 092003 (2003).

\bibitem{Idilbi:2003wj}
  A.~Idilbi, X.~d.~Ji and J.~P.~Ma,
  Phys.\ Rev.\ D {\bf 69}, 014006 (2004).

\bibitem{Belyaev:1995uw}
  V.~M.~Belyaev and A.~V.~Radyushkin,
  Phys.\ Lett.\ B {\bf 359}, 194 (1995).

\bibitem{Belyaev:1995ya}
  V.~M.~Belyaev and A.~V.~Radyushkin,
  Phys.\ Rev.\ D {\bf 53}, 6509 (1996).

\bibitem{feynman}
R.~P.~Feynman, {\it Photon-Hadron Interactions}
(Benjamin, Reading, MA, 1972).

\bibitem{Braun:2005be}
V.~M.~Braun, A.~Lenz, G.~Peters and A.~V.~Radyushkin,
  Phys.\ Rev.\ D {\bf 73}, 034020 (2006).

\bibitem{Ioffe:1981kw}
  B.~L.~Ioffe,
  Nucl.\ Phys.\ B {\bf 188}, 317 (1981)
  [Erratum-ibid.\ B {\bf 191}, 591 (1981)].





\bibitem{CGLN56}
G.F. Chew, M.L. Goldberger, F.E. Low, and Y. Nambu, Phys.\ Rev.\
{\bf 106}, 1337 (1956).

\bibitem{Fubini61}
S. Fubini, Y. Nambu, and V. Wataghin, Phys.\ Rev.\ {\bf 111}, 329 (1958).


\bibitem{Donnachie67}
F.~A.~Berends, A.~Donnachie, and D.~L.~Weaver,
Nucl. Phys. B {\bf 4}, 1 (1967); {\it ibid.} {\bf 4}, 54 (1967);
{\it ibid.} {\bf 4}, 103 (1967).

\bibitem{Schwela67}
D. Schwela, H. Rollnik, R. Weizel, and W.
Korth, Z. Phys. {\bf 202}, 452 (1967); 

D. Schwela and R. Weizel,
Z. Phys. {\bf 221}, 71 (1969).

\bibitem{Hanstein:1997tp}
  O.~Hanstein, D.~Drechsel and L.~Tiator,
  Nucl.\ Phys.\ A {\bf 632}, 561 (1998).


\bibitem{Azn:2003}
I.~G.~Aznauryan,
  Phys.\ Rev.\ C {\bf 67}, 015209 (2003).

\bibitem{Kamalov02}
S.~S. Kamalov {\it et al.},
Phys. Rev. C {\bf 66}, 065206 (2002).


\bibitem{Peccei69}
R.D. Peccei, Phys.\ Rev.\ {\bf 181}, 1902 (1969).

\bibitem{Olsson75}
M.G. Olsson and E.T. Osypowski, Nucl. Phys. B
{\bf 87}, 399 (1975); Phys.\ Rev.\ D {\bf 17}, 174 (1978).

\bibitem{Davidson91}
R.M. Davidson, N.C. Mukhopadhyay, and R.S.
Wittman, Phys.\ Rev.\ D {\bf 43}, 71 (1991).

\bibitem{SpainModel}
  H.~Garcilazo and E.~Moya de Guerra,
  Nucl.\ Phys.\ A {\bf 562}, 521 (1993);

C.~Fernandez-Ramirez, E.~Moya de Guerra and J.~M.~Udias,
  Annals Phys.\  {\bf 321}, 1408 (2006);
  Phys.\ Rev.\ C {\bf 73}, 042201 (2006).


\bibitem{Gent}
  M.~Vanderhaeghen, K.~Heyde, J.~Ryckebusch and M.~Waroquier,
  Nucl.\ Phys.\ A {\bf 595}, 219 (1995).

\bibitem{KVItheory}
  V.~Pascalutsa and O.~Scholten,
  Nucl.\ Phys.\ A {\bf 591}, 658 (1995);

O.~Scholten, A.~Y.~Korchin, V.~Pascalutsa and D.~Van Neck,
Phys.\ Lett.\ B {\bf 384}, 13 (1996);

  A.~Y.~Korchin, O.~Scholten and R.~G.~E.~Timmermans,
 Phys.\ Lett.\ B {\bf 438}, 1 (1998);

  A.~Usov and O.~Scholten,
  Phys.\ Rev.\ C {\bf 72}, 025205 (2005); {\it ibid.} {\bf 74}, 015205 (2006).

\bibitem{Feuster:1998cj}
  T.~Feuster and U.~Mosel,
  Phys.\ Rev.\ C {\bf 59}, 460 (1999);

  G.~Penner and U.~Mosel,
  Phys.\ Rev.\ C {\bf 66}, 055211 (2002);

  V.~Shklyar, G.~Penner and U.~Mosel,
  Eur.\ Phys.\ J.\ A {\bf 21}, 445 (2004);

  V.~Shklyar, H.~Lenske and U.~Mosel,
  Phys.\ Rev.\ C {\bf 72}, 015210 (2005).

\bibitem{Tanabe85}
H. Tanabe and K. Ohta, Phys.\ Rev.\ C {\bf 31}, 1876 (1985).

\bibitem{Yang85}
S.N. Yang, J. Phys. G {\bf 11}, L205 (1985); Phys.  Rev.\ C {\bf 40},
1810 (1989); Chin. J. Phys. {\bf 29}, 485 (1991).

\bibitem{NBL90}
S. Nozawa, B. Blankleider, and T.-S.H. Lee, Nucl.
Phys. A {\bf 513}, 459 (1990).

\bibitem{Lee91}
C.~C. Lee, S.~N. Yang, and T.-S.H. Lee, J. Phys.
G {\bf 17}, L131 (1991).

\bibitem{SG96}
Y. Surya and F. Gross, Phys.\ Rev.\ C
{\bf 53}, 2422 (1996).

\bibitem{Chen03}
G.~Y. Chen {\it et al.}, Nucl. Phys. A {\bf 723}, 447 (2003).


\bibitem{Pascal04}
V. Pascalutsa and J.~A. Tjon, 
Phys.\ Rev.\ C {\bf 70}, 035209 (2004); 

G.~L. Caia, V. Pascalutsa, J.~A. Tjon, and L.~E. Wright, 
{\it ibid.} {\bf 70}, 032201(R) (2004); 

G.~L. Caia, L.~E. Wright, and V. Pascalutsa, {\it ibid.} {\bf 72}, 035203 (2005);

G.~L. Caia, PhD Thesis (Ohio University, 2004) [http://www.ohiolink.edu/etd].



\bibitem{GWU} 
SAID website, http://gwdac.phys.gwu.edu.


\bibitem{MAID}
MAID website, http://www.kph.uni-mainz.de.



\bibitem{Barker75}
L.~S. Barker, A. Donnachie, and J.~K. Storrow,
Nucl. Phys. B {\bf 95}, 347 (1975).

\bibitem{Chiang97}
W.~T. Chiang and F. Tabakin, Phys. Rev. C {\bf 55}, 2054 (1997).

\bibitem{Knochlein95}
G. Knoechlein, D. Drechsel, and L. Tiator,
Z. Phys. A {\bf 352}, 327 (1995).

\bibitem{Raskin89}
A.~S. Raskin and T.~W. Donnelly, Ann. Phys. (N.Y.) {\bf 191}, 78 (1989).

\bibitem{Akerlof67}
C.~W. Akerlof {\it et al.}, Phys. Rev. {\bf 163}, 1482 (1967).

\bibitem{Hung01}
C.~T. Hung, S.~N. Yang, and T.-S.~H. Lee,
Phys.\ Rev.\ C {\bf 64}, 034309 (2001).

\bibitem{Klein74}
A. Klein and T.-S.~H. Lee, Phys.\ Rev.\ D {\bf 10}, 4308 (1974).

\bibitem{bs}
  R.~Blankenbecler and R.~Sugar,
  Phys.\ Rev.\  {\bf 142}, 1051 (1966).

\bibitem{kady}
  V.~G.~Kadyshevsky,
  Nucl.\ Phys.\ B {\bf 6}, 125 (1968).

\bibitem{thom}
  R.~H.~Thompson,
  Phys.\ Rev.\ D {\bf 1}, 110 (1970).

\bibitem{CJ89}
M. Cooper and B. Jennings, Nucl. Phys. A {\bf 500}, 553 (1989).

\bibitem{Hung1}
C.~T. Hung, S.~N. Yang and T.-S.~H. Lee, J. Phys. {\bf G20,} 1531 (1994).

\bibitem{Watson54}
K. Watson, Phys.\ Rev.\ {\bf 95}, 228 (1954);

E. Fermi, Suppl. Nuovo Cimento {\bf 2}, 58 (1955).

\bibitem{Yang88}
S.~N. Yang, in {\it Progress in Medium-Energy Physics},
eds. W.-Y.~P. Hwang and J. Speth (World Scientific, Singapore,
1988) p. 201-215.

\bibitem{Kamalov01}
S.~S. Kamalov, G.~Y. Chen, S.~N. Yang, D. Drechsel, and L. Tiator,
Phys. Lett. B {\bf 522}, 27 (2001).

\bibitem{SKO92}
T. Sato, M. Doi, N. Odagawa, and H. Ohtsubo, Few-Body
Syst. Suppl. {\bf 5}, 524 (1992); T. Sato, M. Kobayashi, and H.
Ohtsubo (unpublished).

\bibitem{Hsiao98}
S. S. Hsiao {\it et al.}, Few-Body Systems {\bf 25}, 55 (1998).


\bibitem{deForest66}
T. de Forest and J.D. Walecka, Adv. Phys, Vol. {\bf 15}, 1 (1966).

\bibitem{Dennery61}
P. Dennery, Phys.\ Rev.\ {\bf 124}, 2000 (1961).


\bibitem{VPI97} 
R. A. Arndt, I. I. Strakovsky and R. L. Workman, 
Phys. Rev. C {\bf 53}, 430 (1996).


\bibitem{tjon00}
V.~Pascalutsa and J.~A.~Tjon,
  Phys.\ Rev.\ C {\bf 61}, 054003 (2000).

\bibitem{Pa98thesis}
V.~Pascalutsa, PhD Thesis (University of Utrecht, 1998)
  [Hadronic J.\ Suppl.\  {\bf 16}, 1 (2001)].

\bibitem{Donnachie72}
A. Donnachie, in {\it High energy physics}, Vol. V, ed. E.U.S. Burhop
(Academic Press, New York, 1972).

\bibitem{Gehlen68}
G. v. Gehlen, Nucl. Phys. B {\bf 9}, 17 (1968).

\bibitem{Salin}
Ph.~Salin, Nuovo Cimento, {\bf 32}, 521 (1964).

\bibitem{Loub}
J.~P.~Loubaton, Nuovo Cimento, {\bf 39}, 591 (1965).

\bibitem{Walecka}
J.~D.~Walecka, Phys.\ Rev.\ {\bf 162}, 1462 (1967).

\bibitem{Crawford}
R.~L.~Crawford and W.~T.~Morton, Nucl.\
                 Phys. B {\bf B211}, 1 (1983); 

Particle Data Group, Phys.\ Lett. B {\bf 239}, 1 (1990);
                 
R.~L.~Crawford, in Proc.\ of
                 {\it Nstar 2001, Mainz, Germany, March 7--10,
                 2001}, eds. D.~Drechsel and L.~Tiator (World Scientific, Singapore, 2001), p.~163.





\bibitem{Wei95b} 
S. Weinberg,
{\it The quantum theory of fields}, Vol.\ 2,
(Cambridge U.\ P., Cambridge NY, 1995).

\bibitem{Ecker:1994gg}
  G.~Ecker,
  Prog.\ Part.\ Nucl.\ Phys.\  {\bf 35}, 1 (1995).

\bibitem{Kaplan:2005es}
  D.~B.~Kaplan,
   ``Five lectures on effective field theory,''
  %
  arXiv:nucl-th/0510023.

\bibitem{Rarita:1941mf}
  W.~Rarita and J.~S.~Schwinger,
  %
  Phys.\ Rev.\  {\bf 60}, 61 (1941).


\bibitem{Dir64} 
P.A.M. Dirac, {\it Lectures on quantum mechanics},
(Yeshiba U. P., New York, 1964).

\bibitem{Senjanovic:1977vr}
  G.~Senjanovic,
  %
  Phys.\ Rev.\ D {\bf 16}, 307 (1977).

\bibitem{Baaklini:1978qa}
  N.~S.~Baaklini and M.~Tuite,
  %
  J.\ Phys.\ A {\bf 11}, L139 (1978).

\bibitem{Yamada:1985vb}
  M.~Yamada,
  %
  Nuovo Cim.\ A {\bf 91}, 205 (1986).

\bibitem{Pas98}
V.~Pascalutsa,
Phys.\ Rev.\ D {\bf 58}, 096002 (1998).

\bibitem{Johnson:1960vt}
  K.~Johnson and E.~C.~G.~Sudarshan,
  %
  Annals Phys.\  {\bf 13}, 126 (1961).

\bibitem{Hagen:1972ea}
  C.~R.~Hagen,
  Phys.\ Rev.\ D {\bf 4}, 2204 (1971).

\bibitem{Velo:1969bt}
  G.~Velo and D.~Zwanziger,
  Phys.\ Rev.\  {\bf 186}, 1337 (1969).

\bibitem{Singh:1973gq}
  L.~P.~S.~Singh,
  Phys.\ Rev.\ D {\bf 7}, 1256 (1973).

\bibitem{Hagen:1982ez}
  C.~R.~Hagen and L.~P.~S.~Singh,
  Phys.\ Rev.\ D {\bf 26}, 393 (1982).

\bibitem{Haberzettl:1998rw}
  H.~Haberzettl,
  arXiv:nucl-th/9812043.

\bibitem{Deser:2000dz}
  S.~Deser, V.~Pascalutsa and A.~Waldron,
  %
  Phys.\ Rev.\ D {\bf 62}, 105031 (2000).

\bibitem{Pilling:2004cu}
  T.~Pilling,
    Mod.\ Phys.\ Lett.\ A {\bf 19}, 1781 (2004);
  Int.\ J.\ Mod.\ Phys.\ A {\bf 20}, 2715 (2005).

\bibitem{Wies:2006rv}
  N.~Wies, J.~Gegelia and S.~Scherer,
  Phys.\ Rev.\ D {\bf 73}, 094012 (2006).

\bibitem{Napsuciale:2006wr}
  M.~Napsuciale, M.~Kirchbach and S.~Rodriguez,
  %
  arXiv:hep-ph/0606308.

\bibitem{Pascalutsa:1999zz}
V.~Pascalutsa and R.~G.~E.~Timmermans,
Phys.\ Rev.\ C {\bf 60}, 042201(R) (1999).

\bibitem{Wil85}
H.T. Williams,
                Phys.\ Rev.\ C {\bf 29}, 2222 (1985);
                {\it ibid.}, {\bf 31}, 2297 (1985).

\bibitem{Ade86}
R.A. Adelseck, C. Bennhold, and L.E. Wright,
                Phys.\ Rev.\ C {\bf 32}, 1681 (1986).

\bibitem{GrS93}
F. Gross and Y. Surya, Phys. Rev. C {\bf 47}, 703 (1993).

\bibitem{Pascalutsa:2003vz}
  V.~Pascalutsa,
  in Proc.\ of {\it Nstar 2002: Physics of Excited
Baryons}, eds S.~A.~Dytman and E.~S.~Swanson
(World Scientific, Singapore, 2003) [arXiv:nucl-th/0303005].

\bibitem{Nath:1971wp}
  L.~M.~Nath, B.~Etemadi and J.~D.~Kimel,
  %
  Phys.\ Rev.\ D {\bf 3}, 2153 (1971).


\bibitem{Freedman:1976aw}
  D.~Z.~Freedman and A.~Das,
  Nucl.\ Phys.\ B {\bf 120}, 221 (1977).


\bibitem{Pas01}
V.~Pascalutsa,
Phys.\ Lett.\ B {\bf 503}, 85 (2001).

\bibitem{TaE96}
H.-B. Tang and P. Ellis, Phys.\ Lett.\ B {\bf 387}, 9 (1996).


\bibitem{FeM01}
N.~Fettes and U.~G.~Meissner,
  Nucl.\ Phys.\ A {\bf 679}, 629 (2001).

\bibitem{Weinberg:1970bu}
S.~Weinberg in {\it Lectures on Elementary Particles and Quantum Field
Theory}, Volume 1, Brandeis University Summer Institute 1970, eds.\
S.~Deser, M.~Grisaru and H.~Pendleton (M.I.T. Press, Cambridge, 1970).

\bibitem{Ferrara:1992yc}
S.~Ferrara, M.~Porrati and V.~L.~Telegdi,
Phys.\ Rev.\  {\bf D46}, 3529 (1992).

\bibitem{Holstein:2006wi}
  B.~R.~Holstein,
  arXiv:hep-ph/0607187.

\bibitem{GSS89}
J.~Gasser, M.~E.~Sainio and A.~Svarc,
Nucl.\ Phys.\ B {\bf 307}, 779 (1988).

\bibitem{Gegelia:1999gf}
  J.~Gegelia and G.~Japaridze,
  Phys.\ Rev.\ D {\bf 60}, 114038 (1999);

  J.~Gegelia, G.~Japaridze and X.~Q.~Wang,
J.\ Phys.\ G {\bf 29}, 2303 (2003).

\bibitem{Kubis:2000zd}
  B.~Kubis and U.~G.~Meissner,
  Nucl.\ Phys.\ A {\bf 679}, 698 (2001).


\bibitem{Holstein:2005db}
  B.~R.~Holstein, V.~Pascalutsa and M.~Vanderhaeghen,
  Phys.\ Rev.\ D {\bf 72}, 094014 (2005).

\bibitem{Hanhart:2002bu}
  C.~Hanhart and N.~Kaiser,
  Phys.\ Rev.\ C {\bf 66}, 054005 (2002).

\bibitem{Korpa:2004sh}
  C.~L.~Korpa and A.~E.~L.~Dieperink,
  Phys.\ Rev.\ C {\bf 70}, 015207 (2004).

\bibitem{Kaloshin:2004jh}
  A.~E.~Kaloshin and V.~P.~Lomov,
  Mod.\ Phys.\ Lett.\ A {\bf 19}, 135 (2004);
    Phys.\ Atom.\ Nucl.\  {\bf 69}, 541 (2006)
  [Yad.\ Fiz.\  {\bf 69}, 563 (2006)].




\bibitem{Beane:2002wn}
  S.~R.~Beane, M.~Malheiro, J.~A.~McGovern, D.~R.~Phillips and U.~van Kolck,
  Phys.\ Lett.\ B {\bf 567}, 200 (2003)
  [Erratum-ibid.\ B {\bf 607}, 320 (2005)].

\bibitem{Camen:2001st}
  M.~Camen {\it et al.},
  Phys.\ Rev.\ C {\bf 65}, 032202 (2002).

\bibitem{Bernard:1992qa}
  V.~Bernard, N.~Kaiser, J.~Kambor and U.~G.~Meissner,
  Nucl.\ Phys.\ B {\bf 388}, 315 (1992).

\bibitem{Bernard:1991ru}
  V.~Bernard, N.~Kaiser and U.~G.~Meissner,
  Nucl.\ Phys.\ B {\bf 373}, 346 (1992).

\bibitem{Lvov:1993ex}
  A.~I.~Lvov,
  Phys.\ Lett.\ B {\bf 304}, 29 (1993).

\bibitem{Pascalutsa:2004wm}
  V.~Pascalutsa,
  Prog.\ Part.\ Nucl.\ Phys.\  {\bf 55}, 23 (2005).

\bibitem{McGovern:2001dd}
  J.~A.~McGovern,
  Phys.\ Rev.\ C {\bf 63}, 064608 (2001)
  [Erratum-ibid.\ C {\bf 66}, 039902 (2002)].

\bibitem{Beane:2004ra}
  S.~R.~Beane, M.~Malheiro, J.~A.~McGovern, D.~R.~Phillips and U.~van Kolck,
  Nucl.\ Phys.\ A {\bf 747}, 311 (2005).


\bibitem{Kondratyuk:2001qu}
  S.~Kondratyuk and O.~Scholten,
  Phys.\ Rev.\ C {\bf 64}, 024005 (2001).

\bibitem{L'vov:1996xd}
  A.~I.~L'vov, V.~A.~Petrun'kin and M.~Schumacher,
  Phys.\ Rev.\ C {\bf 55}, 359 (1997).

\bibitem{Drechsel:2002ar}
  D.~Drechsel, B.~Pasquini and M.~Vanderhaeghen,
  Phys.\ Rept.\  {\bf 378}, 99 (2003).

\bibitem{Hildebrandt:2003fm}
  R.~P.~Hildebrandt, H.~W.~Griesshammer, T.~R.~Hemmert and B.~Pasquini,
  Eur.\ Phys.\ J.\ A {\bf 20}, 293 (2004).

\bibitem{Hildebrandt:2005ix}
  R.~P.~Hildebrandt, PhD Thesis (University of Munich, 1995),
  [arXiv:nucl-th/0512064].

\bibitem{Hacker:2006gu}
  C.~Hacker, N.~Wies, J.~Gegelia and S.~Scherer,
  arXiv:hep-ph/0603267.

\bibitem{Hal93}
E.~L.~Hallin {\it et al.},
Phys.\ Rev.\ C {\bf 48}, 1497 (1993).

\bibitem{LEGS97CS}
LEGS Collaboration,
Phys.\ Rev.\ Lett.\  {\bf 76}, 1023 (1996).


\bibitem{Kotulla:2002cg}
  M.~Kotulla {\it et al.},
  Phys.\ Rev.\ Lett.\  {\bf 89}, 272001 (2002).

\bibitem{Machavariani:1999fr}
  A.~I.~Machavariani, A.~Faessler and A.~J.~Buchmann,
  Nucl.\ Phys.\ A {\bf 646}, 231 (1999)
  [Erratum-ibid.\ A {\bf 686}, 601 (2001)].

\bibitem{Drechsel:2000um}
  D.~Drechsel, M.~Vanderhaeghen, M.~M.~Giannini and E.~Santopinto,
  Phys.\ Lett.\ B {\bf 484}, 236 (2000).

\bibitem{Drechsel:2001qu}
  D.~Drechsel and M.~Vanderhaeghen,
  Phys.\ Rev.\ C {\bf 64}, 065202 (2001).


\bibitem{Chiang:2004pw}
  W.~T.~Chiang, M.~Vanderhaeghen, S.~N.~Yang and D.~Drechsel,
  Phys.\ Rev.\ C {\bf 71}, 015204 (2005).

\bibitem{Lein91}
D.~B.~Leinweber, T.~Draper, and R.~M.~Woloshyn, Phys.\ Rev.\ D {\bf 46}, 3067 (1992);
I.~C.~Cloet, D.~B.~Leinweber and A.~W.~Thomas,
Phys.\ Lett.\ B {\bf 563}, 157 (2003).

\bibitem{Lee:2005ds}
  F.~X.~Lee, R.~Kelly, L.~Zhou and W.~Wilcox,
  Phys.\ Lett.\ B {\bf 627}, 71 (2005); 
private communication (revised values). 

\bibitem{Pascalutsa:2004ga}
  V.~Pascalutsa, B.~R.~Holstein and M.~Vanderhaeghen,
  Phys.\ Lett.\ B {\bf 600}, 239 (2004).

\bibitem{CB}
M. Kotulla, to appear in Proc.~ of the Workshop {\it Shape of Hadrons, 
Athens, 2006}, eds. C.~N. Papanicolas and A.~M. Bernstein, AIP (2006).  




\bibitem{chiral96}
See, e.g., {\it Chiral Dynamics: Theory and Experiment}, eds.
A.~M. Bernstein, D. Drechsel, and Th. Walcher (Springer-Verlag, New York, 1997).

\bibitem{Bernard91} 
V. Bernard, J. Gasser, N. Kaiser, and Ulf-G.
Mei{\ss}ner, Phys. Lett. B {\bf 268} (1991) 291.

\bibitem{Bernard} 
V. Bernard, N. Kaiser, and Ulf-G. Mei{\ss}ner, 
Z. Phys. C70 (1996) 483; Nucl. Phys. A {\bf 607} (1996) 379, 
and references therein.

\bibitem{Beck90}
R. Beck {\it et. al.}, Phys. Lett. Lett. {\bf 65}, 1841 (1990).

\bibitem{Fuchs96} 
M. Fuchs {\it et al.}, Phys. Lett.  B {\bf 368}, 20 (1996).

\bibitem{Bergstrom} 
J.~C. Bergstrom {\it et al.}, Phys. Rev. C {\bf 53}, R1052 (1996); 
{\it ibid.} {\bf 55}, 2016 (1997).

\bibitem{NIKHEF} 
H.~B. van den Brink {\it et al.}, Phys. Rev. Lett. {\bf 74}, 3561 (1995); 
Nucl. Phys.  A {\bf 612}, 391 (1997).

\bibitem{Distler} 
M.~O. Distler {\it et al.}, Phys. Rev. Lett. {\bf 80}, 2294 (1998).

\bibitem{Bernstein97}
A.~M. Bernstein {\it et al.}, Phys. Rev. C {\bf 55}, 1509 (1997).

\bibitem{Schmidt}
A.~Schmidt {\it et al.}, 
Phys.\ Rev.\ Lett.\  {\bf 87}, 232501 (2001).

\bibitem{Sauer86}
P.~U. Sauer, Prog. Part. Nucl. Phys. {\bf 16}, 35 (1986).

\bibitem{pearce}
B.~C. Pearce and B. K. Jennings, Nucl. Phys. A {\bf 528}, 655 (1991).

\bibitem{afnan99}
A.~D. Lahiff and I. R. Afnan, Phys. Rev. C {\bf 60}, 024608 (1999).


\bibitem{Cloudy80}
S. Th\'eberge {\it et al.}, Phys. Rev. D {\bf 22}, 2838 (1980).


\bibitem{McPherson64}
D.~A.~McPherson {\it et al.},
Phys.\ Rev. {\bf 136}, B1465 (1964).

\bibitem{MacCormick:1996jz}
  M.~MacCormick {\it et al.},
  Phys.\ Rev.\ C {\bf 53}, 41 (1996).

\bibitem{Ahrens:2000bc}
  J.~Ahrens {\it et al.}  [GDH and A2 Collaborations],
  Phys.\ Rev.\ Lett.\  {\bf 84}, 5950 (2000).


\bibitem{Ahrens:2004pf}
  J.~Ahrens {\it et al.}  [GDH and A2 Collaboration],
  Eur.\ Phys.\ J.\ A {\bf 21}, 323 (2004).


\bibitem{Leukel01}
R. Leukel, PhD Thesis (University of Mainz, 2001).


\bibitem{Belyaev:1983xf}
  A.~A.~Belyaev {\it et al.},
  Nucl.\ Phys.\ B {\bf 213}, 201 (1983).


\bibitem{Bock:1998rk}
  A.~Bock {\it et al.},
  Phys.\ Rev.\ Lett.\  {\bf 81}, 534 (1998).


\bibitem{Getman:1981qt}
  V.~A.~Getman {\it et al.},
  Nucl.\ Phys.\ B {\bf 188}, 397 (1981).


\bibitem{Dutz:1996uc}
  H.~Dutz {\it et al.},
  Nucl.\ Phys.\ A {\bf 601}, 319 (1996).


\bibitem{Belyaev:1985sp}
  A.~A.~Belyaev {\it et al.},
  Yad.\ Fiz.\  {\bf 40}, 133 (1984).

\bibitem{Belyaev85}
A.~A. Belyaev {\it et al.}, Proc. Int. Sym. High
Energy Lepton-Photon Interaction, Kyoto, Japan, August 19-24, 1985.

\bibitem{Ahrens:2005zq}
  J.~Ahrens {\it et al.},
  Eur.\ Phys.\ J.\ A {\bf 26}, 135 (2005).

\bibitem{Ahrens}
J. Ahrens {\it et al.}, Phys.\ Rev.\ Lett. {\bf 88}, 232002 (2002).

\bibitem{Sandorfi04}
A. Sandorfi, in Proc.\ of {\it Nstar 2004}, 
eds. J-P. Bocquet, V. Kuznetsov, D. Rebreyend (World Scientific,
Singapore, 2004).




\bibitem{Foster83}
F. Foster and G. Hughes, Rept. Prog. Phys. {\bf 46}, 1445 (1983).

\bibitem{Merkel02}
H. Merkel {\it et al.}, Phys.\ Rev.\ Lett. {\bf 88}, 012301 (2002).

\bibitem{DeVita02}
R. De Vita {\it et al.}, Phys.\ Rev.\ Lett. {\bf 88}, 082001 (2002).

\bibitem{Klimenkao06}
A.V. Klimenkao Phys.\ Rev.\ C {\bf 73}, 035212 (2002).

\bibitem{Warren98}
G.A. Warren {\it et al.}, Phys. Rev. C {\bf 58}, 3722 (1998).

\bibitem{Bartsch02}
P. Bartsch {\it et al.}, Phys.\ Rev.\ Lett. {\bf 88}, 142001 (2002).

\bibitem{Laveissiere04}
G. Laveissiere {\it et al.}, Phys.\ Rev.\ C {\bf 69}, 045203 (2004).


\bibitem{Joo02b}
K. Joo {\it et al.}, Phys.\ Rev.\ C {\bf 68}, 035202 (2003).

\bibitem{Buuren02}
L.D. van Buuren Phys.\ Rev.\ Lett. {\bf 89}, 012001 (2002).



\bibitem{Kunz:2003we}
  C.~Kunz {\it et al.},
  Phys.\ Lett.\ B {\bf 564}, 21 (2003).




\bibitem{Arrington:2003df}
  J.~Arrington,
  Phys.\ Rev.\ C {\bf 68}, 034325 (2003).

\bibitem{Guichon:2003qm}
  P.~A.~M.~Guichon and M.~Vanderhaeghen,
  Phys.\ Rev.\ Lett.\  {\bf 91}, 142303 (2003).

\bibitem{BMT03}
P.~G.~Blunden, W.~Melnitchouk and J.~A.~Tjon,
Phys.\ Rev.\ Lett.\  {\bf 91}, 142304 (2003).


\bibitem{YCC04}
Y.C. Chen, A. Afanasev, S.J. Brodsky, C.E. Carlson and M. Vanderhaeghen,
Phys.\ Rev.\ Lett.\  {\bf 93}, 122301 (2004);

A.~V.~Afanasev, S.~J.~Brodsky, C.~E.~Carlson, Y.~C.~Chen and M.~Vanderhaeghen,
Phys.\ Rev.\ D {\bf 72}, 013008 (2005).

\bibitem{Pascalutsa:2005es}
  V.~Pascalutsa, C.~E.~Carlson and M.~Vanderhaeghen,
  Phys.\ Rev.\ Lett.\  {\bf 96}, 012301 (2006).

\bibitem{Kondratyuk:2006ig}
  S.~Kondratyuk and P.~G.~Blunden,
  Nucl.\ Phys.\ A {\bf 778}, 44 (2006).

\bibitem{Adl75}
S. Adler, Ann. Phys. (N.Y.) \textbf{50}, 189 (1968).

\bibitem{Tvaskis:2005ex}
V.~Tvaskis, J.~Arrington, M.~E.~Christy, R.~Ent, C.~E.~Keppel, Y.~Liang 
and G.~Vittorini,
  Phys.\ Rev.\ C {\bf 73}, 025206 (2006).

\bibitem{Tvaskis}
V. Tvaskis, talk at the JLab/INT Workshop
{\it Precision Electroweak Physics}, College of William and Mary,
August 15-17, 2005.




\bibitem{JW59}
M. Jacob and G.C. Wick, Ann. Phys. {\bf 7}, 404 (1959).


\end{thebibliography}
